\documentclass[rmp,twocolumn,aps,showpacs,floatfix, nofootinbib]{revtex4-1}
\usepackage{amsmath}
\usepackage{amssymb,ams fonts,times,natbib}
\usepackage{graphicx}
\usepackage{epstopdf}
\usepackage{color}

\begin{document}

\title{Nonlinear optical signals and spectroscopy with quantum light}
\author{Konstantin E. Dorfman$^{1}$}
\email{dorfmank@gmail.com}
\author{Frank Schlawin$^{1}$}
\thanks{also at Physikalisches Institut, Albert-Ludwigs-Universit\"at Freiburg Hermann-Herder-Stra\ss e 3, 79104 Freiburg, Germany}
\email{email: frank.schlawin@physik.uni-freiburg.de}
\author{Shaul Mukamel}
\email{smukamel@uci.edu}
\address{Department of Chemistry, University of California, Irvine, California 92697, USA}
\date{\today}
\pacs{123}
\footnote{These authors contributed equally to this manuscript}

\begin{abstract}
Conventional nonlinear spectroscopy uses classical light to detect  matter properties through the variation of its response with frequencies or time delays. Quantum light opens up  new avenues for  spectroscopy by utilizing   parameters of the quantum state of light  as novel control knobs and through the variation of photon statistics by coupling to matter.   We present an  intuitive diagrammatic approach for calculating  ultrafast spectroscopy signals induced by quantum light, focusing  on applications involving entangled photons with nonclassical bandwidth properties - known as ``time-energy entanglement''.  Nonlinear optical signals induced by quantized light fields are expressed using time ordered multipoint correlation functions of superoperators. These are different from Glauber's $g$- functions for photon counting which use normally ordered products of ordinary operators. Entangled photon pairs are not subjected to the classical Fourier limitations on the joint temporal and spectral resolution. After a brief survey of properties of entangled photon pairs relevant to their spectroscopic applications, different optical signals, and photon counting setups are discussed and illustrated for simple multi-level model  systems.
\end{abstract}
\maketitle

\tableofcontents

\section{Introduction}

Nonlinear optics is most commonly and successfully formulated using a semiclassical approach whereby the matter degrees of freedom are treated quantum mechanically, but the radiation field is classical \cite{Boyd03,Scully97}. Spectroscopic signals are then obtained by computing the polarization induced in the medium and expanding it perturbatively in the impinging field(s) \cite{Mukamel_book, HammZanni, Ginsberg09}. This level of theory is well justified in many applications, owing to the typically large intensities required to generate a nonlinear response from the optical medium, which can only be reached with lasers. Incidentally, it was shortly after Maiman's development of the ruby laser that the first nonlinear optical effect was observed  \cite{Franken61}.

Recent advances in quantum optics extend nonlinear signals down to the few-photon level where the quantum nature of the field is manifested, and must be taken into account: The enhanced light-matter coupling in cavities \cite{Raimond01, Walther06, Schwartz11}, the enhancement of the medium's nonlinearity by additional driving fields \cite{Peyronel12, Chen13}, large dipoles in highly excited Rdberg states \cite{He14, Gorniaczyk14}, molecular design \cite{Loo12, Castet13}, or strong focussing \cite{Potoschnig11, Rezus12, Faez14} all provide possible means to observe and control nonlinear optical processes on a fundamental quantum level. Besides possible technological applications such as all-optical transistors \cite{Shomroni14} or photonic quantum information processing \cite{Franson89,Kok07a, URen03a, Kni01, OBrien03, Jen00, Bra00}, these also show great promise as  novel applications as spectroscopic tools. Parameters of the photon field wavefunction can serve as control knobs that supplement classical parameters such as frequencies and time delays. This review surveys these emerging applications, and introduces a systematic diagrammatic perturbative approach to their theoretical description.  

One of the striking features of quantum light is photon entanglement. This occurs between two beams of light (field amplitudes \cite{enk05}) when the quantum state of each field cannot be described in the individual parameter space of that field. Different degrees of freedom can become entangled. Most common types  of entanglement are: their spin \cite{dol13}, polarization \cite{Shih88}, position and momentum \cite{how04}, time and energy \cite{tit99}, etc. Entangled photon pairs constitute an invaluable tool in fundamental tests of quantum mechanics - most famously in the violation of Bell's inequalities \cite{Aspect1, Aspect3,Aspect2} or in Hong, Ou and Mandel's photon correlation experiments \cite{Hong87,Ou88, Shih88}.  
Besides, their nonclassical bandwidth properties have long been recognized as a potential resource in various ``quantum-enhanced" applications, where the quantum correlations shared between the photon pairs are thought to offer an advantage. For example
when one photon from an entangled pair is sent through a dispersive medium, the leading-order dispersion is compensated in photon coincidence measurements - an effect called \textit{dispersion cancellation} \cite{Franson92a, Steinberg92, Steinberg92b, Larchuk95, Abouraddy02, Minaeva09a}.
In the field of quantum-enhanced measurements, entanglement may be employed to enhance the resolution beyond the Heisenberg limit \cite{Michell04, Giovannetti04, Giovannetti06}.
Similarly, the spatial resolution may be enhanced in quantum imaging applications \cite{Pittman95a,Bennink04a}, quantum-optical coherence tomography \cite{Abouraddy02, Nasr03a,Esposito16}, as well as in quantum lithographic applications \cite{Boto00a, Dangelo01a}.

However, it is now recognized that many of these applications may also be created in purely classical settings: Some two-photon interference effects originally believed to be a hallmark of quantum entanglement can be simulated by post-selecting the signal \cite{Kaltenbaek08, Kaltenbaek09}. This had enabled quantum-optical coherence tomography studies with classical light \cite{Lavoie09}. Similarly, quantum imaging can be carried out with thermal light \cite{Valencia05}, albeit with reduced signal-to-noise ratio. When proposing applications of quantum light, it is thus imperative to carefully distinguish genuine entanglement from classical correlation effects.



\begin{figure}[t]
\centering
\includegraphics[width=0.45\textwidth]{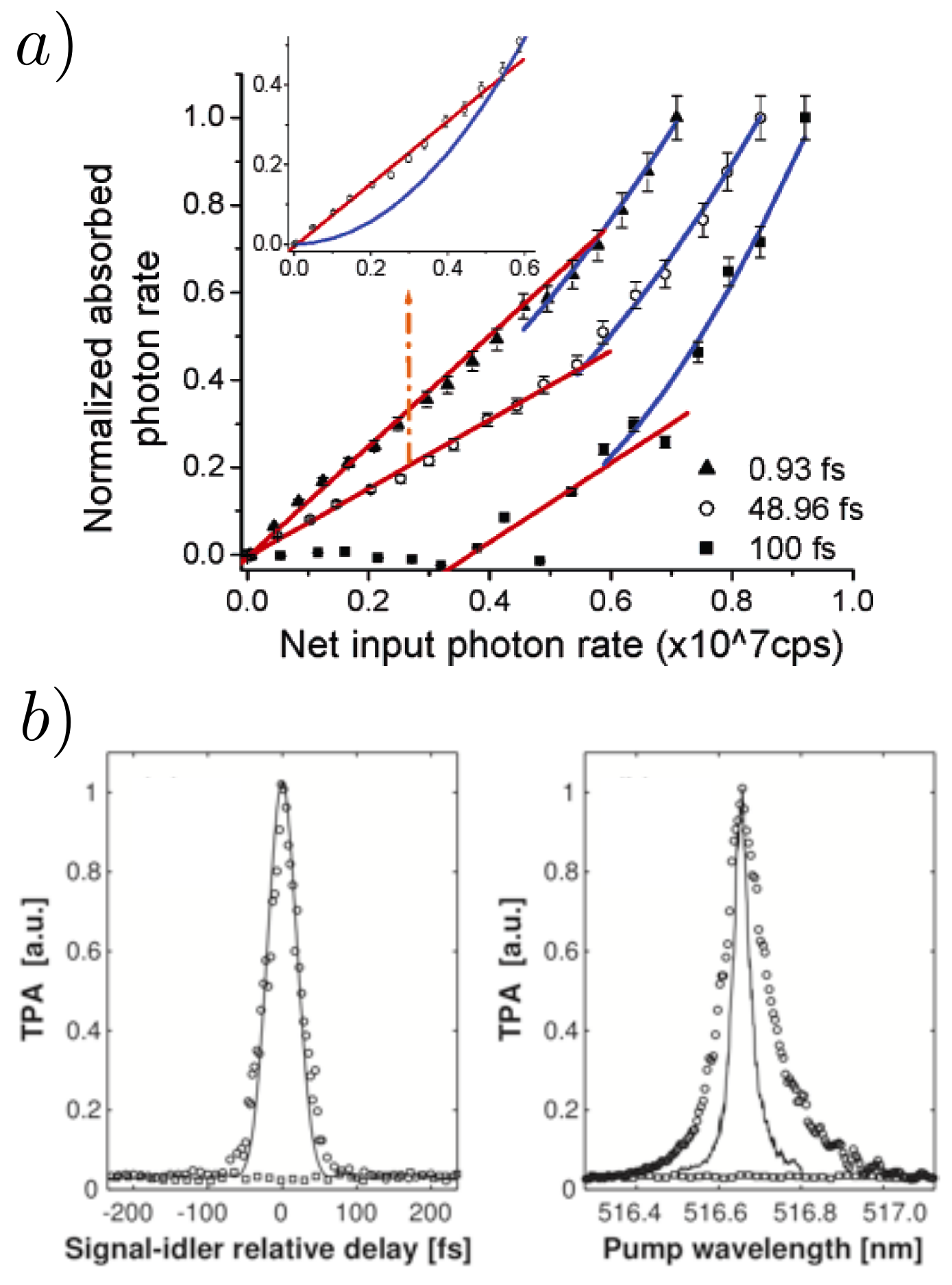}
\caption{a) Linear entangled TPA rate and quadratic nonlinear random TPA rate in a  porphyrin dendrimer at different entanglement times. The inset shows the dominant effect of the  entangled TPA at low input flux of correlated photons. 
b) Two-photon absorption $5S_{1/2} \rightarrow 4D_{5/2,  3/2}$ in atomic Rubidium \cite{Dayan04a} vs. the time delay between the two photons (left panel), and vs. the pump frequency (right panel).}
\label{fig.intro}
\end{figure}

A first clear signature of the quantumness of light is the scaling of optical signals with light intensities: Classical heterodyne $\chi^{(3)}$ signals such as two photon absorption scale quadratically with the intensity, and therefore require a high intensity to be visible against lower-order linear-scaling processes. With entangled photons, such signals scale linearly \cite{geo95,Friberg:OptCommun:85, Javainen90a, Dayan04a, Dayan05a}. This allows to carry out microscopy \cite{Teich:USPatent:98} and lithography \cite{Boto00a} applications at much lower photon fluxes. The different intensity scaling with entangled photons has been first demonstrated in atomic systems by \cite{geo95} and later by \cite{Dayan04a, Dayan05a}, as well as in organic molecules \cite{Lee06a}. An entangled two-photon absorption (TPA) experiment performed in a porphyrin dendrimer is shown in Fig. \ref{fig.intro}a). The linear scaling can be rationalized as follows: entangled photons come in pairs, as they are generated simultaneously. At low light intensity, the different photon pairs are temporally well separated, and the two photon absorption process involves two entangled photons of the same pair. It thus behaves as a linear spectroscopy with respect to the pair. At higher intensities, it becomes statistically more plausible for the two photons to come from different pairs, which are not entangled, and the classical quadratic scaling is recovered [Fig. \ref{fig.intro}a)]. This is related to the Poisson distribution of photons.

The presence of strong time and frequency correlations of entangled photons is a second important feature, which we shall discuss extensively in the course of this review. Fig.~\ref{fig.intro}b) shows the two-photon absorption signal of entangled photons in Rubidium vapor \cite{Dayan04a}. In the left panel, a delay stage is placed into one of the two photon beams, creating a narrow resonance as if the TPA resonance was created by a $23$~fs pulse. However, as the frequency of the pump pulse which creates the photons is varied in the right panel, the resonance is also narrow in the frequency domain, as if it was created by a ns pulse. This \textit{simultaneous} time and frequency resolution along non Fourier conjugate axes is a hallmark of the time-energy entanglement, and its exploitation as a spectroscopic tool offers novel control knobs to manipulate the excited state distribution, and thereby enhance or suppress features in nonlinear spectroscopic signals.

In a different line of research, the seminal photon coincidence counting experiments were turned into a spectroscopic tool by placing a sample into the beam line of one of the two entangled photons, and recording the change of the coincidence count rate \cite{Yabushita04, Kalachev07a, Kalachev08a, Kalashnikov14a}. We will examine related schemes for utilizing entanglement in nonlinear spectroscopy.

This review is structured as follows: In section~\ref{sec.Liouville-notation}, we briefly give a the background and introduce the superoperator notation used in the following. In section~\ref{sec.quantum light}, we discuss properties of entangled photons, and present their impact on excited state distributions upon their absorption in a complex quantum system. In section~\ref{sec.nonlinear-signals} we provide general superoperator expressions for nonlinear optical signals, and review the proposed available setups. Finally, section~\ref{sec.generation-&-response-fct} presents a general classification of nonlinear optical processes induced by quantum light, clarifying under which conditions its quantum nature may play a role. 

\subsection{Liouville space superoperator notation}
\label{sec.Liouville-notation}
The calculation of nonlinear optical signals becomes most transparent by working in Liouville space \cite{Mukamel_book, Nagata11}, $i.e.$ the space of bounded operators on the combined matter-field Hilbert space. It offers a convenient bookkeeping device for matter-light interactions where signals are described as time-ordered products of superoperators.

Here, we introduce the basic notation. With each operator $A$ acting on the Hilbert space, we associate two superoperators \cite{Harbola08a, Roslyak09a,ros10}
\begin{align}
A_L X &\equiv A X, \label{eq.A_L}
\end{align}
which represents the action from the left, and
\begin{align}
A_R X &\equiv X A, \label{eq.A_R}
\end{align}
the action from the right. We further introduce the linear combinations, the commutator superoperator
\begin{align}
A_- &\equiv A_L - A_R, \label{eq.commutator}
\end{align}
and the anti-commutator
\begin{align}
A_+  &\equiv \frac{1}{2} \left( A_L + A_R \right). \label{eq.anti-commutator}
\end{align}
This notation allows to derive compact expressions for spectroscopic signals. At the end of the calculation, after the time-ordering is taken care of, we can switch back to ordinary Hilbert space operators. 

The total Hamiltonian of the field-matter system is given by
\begin{align}
H_{\text{tot}} &= H_{0} + H_{\text{field}} + H_{\text{int}}. \label{eq.H_tot}
\end{align}
$H_0$ describes the matter.
The radiation field Hamiltonian is given by 
\begin{align}
H_{\text{field}} &= \hbar \sum_s \; \omega_s \; a_s^{\dagger} (\omega_s)  a_s (\omega_s). \label{eq.H_field}
\end{align}
Here we introduced the creation (annihilation) operators for mode $s$ which satisfy bosonic commutation relations $[{a}_s,{a}^{\dagger}_{s'}]=\delta_{s,s'}$,  $[{a}_s,{a}_{s'}] = [{a}^{\dagger}_s,{a}^{\dagger}_{s'}]= 0$. In many situations, we will replace the discrete sum over modes by a continuous integral $\sum_s\to\frac{V}{(2\pi)^3}\int d\omega_s\tilde{D}(\omega_s)$ with $\tilde{D}(\omega_s)$ being the density of states that we assume to be flat within the relevant bandwidths $\tilde{D}(\omega_s) \simeq \tilde{D}$.

We assume a dipole light-matter interaction Hamiltonian,
\begin{align}
H_{\text{int}} &= \epsilon \sum_{\nu} \mathcal{V}_{\nu}, \label{eq.H_int}
\end{align}
where $\mathcal{V}_{\nu} = V_{\nu} + V^{\dagger}_{\nu}$ the dipole operator of molecule $\nu$ with the summation running over all the molecules contained in the sample. $V$ ($V^{\dagger}$) is the excitation lowering (raising) part of the dipole. $\epsilon = E + E^{\dagger}$ denotes the electric field operator and E ($E^{\dagger}$) its positive (negative) frequency components which can be written in the interaction picture with respect to the field Hamiltonian \cite{Loudon_book} as
\begin{align}
E &= \int \frac{d\omega}{2 \pi} \; e^{- i \omega t}  a (\omega) \label{eq.E-definition}
\end{align}
which is written in the slowly-varying envelope approximation. Unless specified otherwise, $E$ will denote the sum of all relevant field modes.

In the following applications, we will neglect rapidly oscillating terms, by employing the dipole Hamiltonian in the rotating wave approximation,
\begin{align}
H_{\text{int, RWA}} &= E V^{\dagger} + E^{\dagger} V, \label{eq.H_int_RWA}
\end{align} 
where



\subsection{Diagram construction}

We adopt the diagram representation of nonlinear spectroscopic signals which can be found, for instance, in \cite{Mukamel10a}. It bears close similarity to analogous methods in quantum electrodynamics \cite{Cohen-Tannoudji92}. We will employ two types of diagrams which are associated with calculating the expectation value of an operator $A(t)$ using the density  matrix or the wavefunction. For details see appendix \ref{app:Diag}. First, we evaluate it by propagating the density matrix $\rho(t)$,
\begin{align}
\big\langle A (t) \big\rangle_{\text{DM}} &\equiv \text{tr} \left\{ A (t) \varrho (t) \right\} \label{eq.signal-DM}
\end{align}

Eq.~(\ref{eq.signal-DM}) can be best analyzed in Liouville space~(\ref{eq.commutator}): We write the time evolution of the joint matter plus field density matrix using a time-ordered exponential which can be expanded as a Dyson series,
\begin{align}
\varrho (t) &= \mathcal{T} \exp \left[ - \frac{i}{\hbar} \int^t_{t_0} \!\!\!d\tau H_{\text{int}, -} (\tau) \right] \varrho (t_0), \label{eq.Dyson-series}
\end{align}
where  $H_{\text{int}-}$ is a superoperator (\ref{eq.commutator}) that corresponds to the interaction Hamiltonian (\ref{eq.H_int}). The time-ordering operator $\mathcal{T}$ is defined by its action on a product of arbitrary superoperators $A (t)$ and $B (t)$,
\begin{align}
\mathcal{T} A (t_1) B (t_2) &\equiv \theta (t_1 - t_2) A (t_1) B (t_2) \notag \\
&+ \theta (t_2 - t_1) B (t_2) A (t_1).
\end{align}
$\mathcal{T}$ orders the following products of superoperators so that time increases from right to left.
The perturbative expansion of Eq.~(\ref{eq.Dyson-series}) to $n$-th order in $H_{\text{int}}$ generates a number of pathways - successions of excitations and deexcitations on both the bra or the ket part of the density matrix. These pathways are represented by double-sided ladder diagrams, which represent convolutions of fully time-ordered \textit{superoperator nonequilibrium Green's functions (SNGF)} of the form
\begin{align}
&\int_0^{\infty} dt_1 \cdots \int_0^{\infty} dt_n
\mathbb{V}^{(n)}_{\nu_n\ldots\nu_1}\left({t - t_n\ldots,t - t_n \cdots - t_1}\right) \notag \\
&\qquad \times \mathbb{E}^{(n)}_{\nu_n\ldots\nu_1}\left({t - t_n\ldots,t - t_n \cdots - t_1}\right). \label{eq.Liouville-G}
\end{align}  
The field SNGF  in Eq.~(\ref{eq.Liouville-G})
\begin{align}\label{eq:Esngf}
&\mathbb{E}^{(n)}_{\nu_n\ldots\nu_1}\left({t - t_n\ldots,t - t_n \cdots - t_1}\right)\notag\\
&=\big\langle \mathcal{T}E_{\nu_n} (t - t_n) \cdots E_{\nu_1} (t - t_n \cdots - t_1) \big\rangle
\end{align}
 is evaluated with respect to the initial quantum state of the light. The indices are $\nu_j=L,R$ or $\nu_j=+,-$. When replacing the field operators by classical amplitudes, we recover the standard semiclassical formalism \cite{Mukamel_book}. Similarly, we may also convert the field operators into classical random variables to describe spectroscopic signals with stochastic light \cite{Beach84, Asaka84, Morita84, Turner13}.

The material SNGF's  in Eq.~(\ref{eq.Liouville-G}) are similarly defined:
\begin{align}\label{eq:Vsngf}
&\mathbb{V}^{(n)}_{\nu_n\ldots\nu_1}\left({t - t_n\ldots,t - t_n \cdots - t_1}\right)\notag\\
&=\big\langle V^{(\dagger)}_{\nu_n} \mathcal{G} (t_n) \cdots \mathcal{G} (t_1) V^{(\dagger)}_{\nu_1} \big\rangle
\end{align}
where 
\begin{align}
\mathcal{G} (t) &= - \frac{i}{\hbar} \theta (t) \; \exp \left[ - \frac{i}{\hbar} H_0 t \right] \label{eq.Greens-fct}
\end{align}
denotes the propagator of the free evolution of the matter system, $\nu_j=L,R$. Similarly SNGF may be obtained by replacing some $V^{\dagger}$  by $V$  and $E$  by $E^{\dagger}$. This representation further allows for reduced descriptions of open systems where bath degrees of freedom are eliminated. 

\begin{figure*}[ht]
\centering
\includegraphics[width=0.95\textwidth]{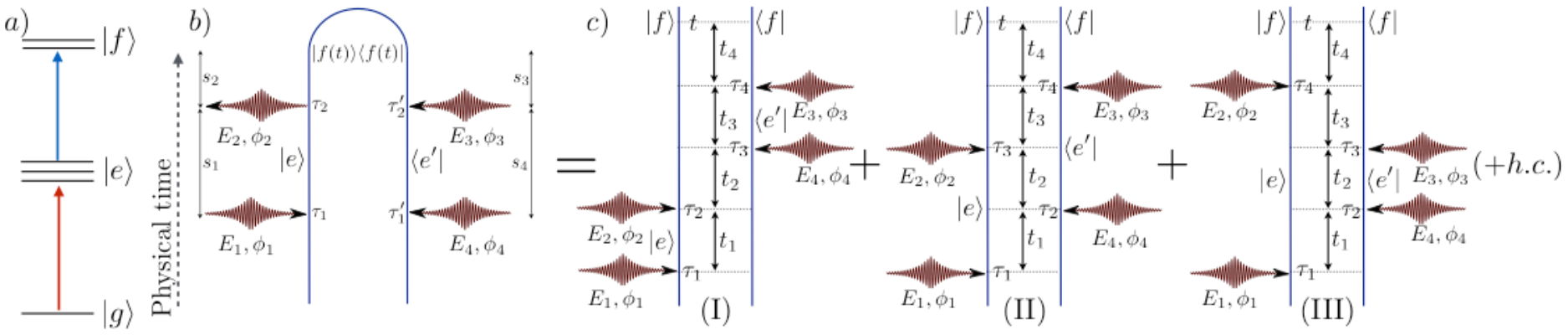}
\caption{Diagrammatic construction of the double excitation probability $\varrho_{ff}$in a three level  model a). b) The single loop diagram representing the evolution of the wavefunction. On the left hand side, the vertical line represents ket and the right hand it corresponds to the bra. The wave function in this closed time path loop diagram propagates forward in a loop from the bottom branch of the diagram along the ket branch to the top of the diagram. It then evolves backward in time from the top to the bottom of the right branch (bra). This forward and backward propagation is similar to the Keldysh contour diagram rules \cite{Keldysh65}. Horizontal arrows represent field-matter interactions.  c) The corresponding three ladder diagrams for the evolution of the density matrix. Here the density matrix evolves forward in time upward from the bottom to the top of the diagram. In order to achieve doubly excited state population starting from the ground state the density matrix has to undergo four interactions represented by the absorption of two photons represented by four inwardly directed arrows. The three diagrams represent different time orderings of ket  vs bra interactions. These are lumped together in the single loop diagram b)}
\label{fig.lop-vs-lap}
\end{figure*}

As an example, the set of fourth-order pathways for the  population of state $f$ in a three-level scheme shown in Fig.~\ref{fig.levelscheme}  are given in Fig.~\ref{fig.lop-vs-lap}b. We will refer to these pathways repeatedly in the course of this review. This ladder diagram representation is most suitable for impulsive experiments involving sequences of short, temporally well-separated pulses ranging from NMR to the X-ray regimes \cite{Mukamel_book, HammZanni}. In such multidimensional experiments, the time variables used to represent the delays between successive pulses  \cite{Abr09} $t_1$, $t_2$, $t_3$, ... serve as the primary control parameters. Spectra are displayed vs. the Fourier conjugates $\tilde{\Omega}_1$, $\tilde{\Omega}_2$, $\tilde{\Omega}_3$, ... of these time variables. 



As indicated in Fig.~\ref{fig.lop-vs-lap}b), each interaction with a field also imprints the phase $\phi$ onto the signal. Filtering the possible phase combinations $\pm \phi_1 \pm \phi_2 \pm \phi_3 \pm \phi_4$ of the signal, known as phase  cycling, allows for the selective investigation of specific material properties \cite{Keu99,Scheurer01,Tia03,Domcke13,Tan08,Abr09,Zha12,Zha121}. Acousto-optical modulation discussed in Fig.~\ref{fig.Raymer2D} offers the possibility to achieve this selectivity even at the single-photon level. 
Phase cycling techniques have been successfully demonstrated as control tools for the selection of fixed-phase components of  optical signals generated by multiwave mixing \cite{Keu99,Tia03,Tan08,Zha12,Zha121}. Phase cycling can be easily implemented by varying the relative inter-pulse phases using a pulse shaper, which is cycled over 2$\pi$ radians in a number of equally spaced steps \cite{Keu99,Tia03}. 

Alternatively, rather than propagating the density matrix Eq. (\ref{eq.signal-DM}), we can follow the evolution of the wave function by expanding 
\begin{align}
\big\langle A (t) \big\rangle_{\text{WF}} &\equiv \langle \psi (t) \vert A (t) \vert \psi (t) \rangle, \label{eq.signal-WF}
\end{align}
with
\begin{align}
|\psi(t)\rangle=\exp\left(-\frac{i}{\hbar}\int_{t_0}^td\tau H'_{int}(\tau)\right)|\psi(t_0)\rangle.
\end{align}
Keeping track of the wave function results in different pathways which can be represented by loop diagrams. 
Rather than propagating of both the bra and the ket, we then place the entire burden of the time evolution on the ket, and write
\begin{align}
\langle A(t)\rangle=\langle\psi(t_0)|\tilde{\psi}(t)\rangle,
\end{align} 
where
\begin{align}
|\tilde{\psi}(t)\rangle&=\mathcal{T}^{-1}\exp\left(\frac{i}{\hbar}\int_{t_0}^td\tau H'_{int}(\tau)\right)A(t)\notag\\
&\times \mathcal{T}\exp\left(-\frac{i}{\hbar}\int_{t_0}^td\tau H'_{int}(\tau)\right)|\psi(t_0)\rangle,
\end{align}
and $\mathcal{T}^{-1}$ denotes the ati-time-ordering operator.
Here the ket first evolves forward and then backward in time, eventually returning to the initial time \cite{Schwinger61,Keldysh65}. Back propagation of the ket is equivalent to forward propagation of the bra. The resulting terms are represented by loop diagrams, as is commonly done in many-body theory \cite{Ram07, Mukamel08a, Rahav10a, Hansen12}. 

As can be seen from Fig.~\ref{fig.lop-vs-lap}, this representation yields a more compact description (fewer pathways) of the signals, since the relative time ordering of ket and bra interactions is not maintained. In this example, the f-state population is given by the single loop diagram, which is the sum of the three ladder diagrams (as well as their complex conjugates). It is harder to visualize short pulse experiments in this representation, and due to the backward time propagation the elimination of bath degrees of freedom in an open system is not possible. Nevertheless, this representation proves most useful and compact for frequency domain techniques involving long pulses where time ordering is not maintained anyhow  \cite{Rahav10a} and for many-body simulations that are usually carried out in Hilbert space \cite{Dlaibard92}. 

The double sided (ladder) and the loop diagrams are two book keeping devices for field-matter interactions. The loop diagrams suggest various wavefunction based simulation strategies for signals \cite{Dor13}. The first is based on the numerical propagation of the wavefunction, which includes all relevant electronic and nuclear (including bath) degrees of freedom explicitly. A second protocol uses a Sum Over States (SOS) expansion of the signals. In the third, semiclassical approach a small subsystem is treated quantum mechanically and is coupled to a classical bath, which causes a time dependent modulation of the system Hamiltonian. The third approach for a wave function is equivalent to the stochastic Liouville equation for the density matrix \cite{Tanimura1}, which is  based on ladder diagram book keeping.

\section{Quantum light}
\label{sec.quantum light}

In this section, we first discuss classical light which is the basis for the semiclassical approximation of nonlinear spectroscopy. We then briefly survey the main concepts from Glauber's photon counting theory for a single mode of the electromagnetic field. Finally, we discuss in detail the multimode entangled states of light, which will be repeatedly used in this review.

\subsection{Classical vs. Quantum Light}

Clearly, our world is governed by quantum mechanics, so on a fundamental level light is \textit{always} quantum. Yet, in many situations a classical description of the light field may be sufficient \cite{Boyd03,Scully97}. For pedagogical reasons, before we describe properties of quantum light, we first discuss in some detail how the classical description of the field emerges from quantum mechanics, and - more importantly - under which circumstances this approximation may break down.

The coherent state of the field, $i.e.$ the eigenstate of the photon annihilation operator at frequency $\omega$, is generally considered as ``classical". In this state the annihilation operator, and hence the electric field operator -  has a non-vanishing expectation value, $\langle a (\omega) \rangle \neq 0$. 
In general, we can write a multimode coherent state as
\begin{align}
\vert \phi_{\text{coh}}(t) \rangle &= \int \!\! d\omega \; e^{ \alpha_{\omega} (t) a^{\dagger} (\omega) - \alpha^{\ast}_{\omega} (t) a (\omega) } \vert 0 \rangle,
\end{align}
where the mode amplitudes $\alpha$ are given by
\begin{align}
\alpha_{\omega} (t) &= \alpha_{\omega} e^{- i \omega t}.
\end{align}
In a normally ordered correlation function (all $a^{\dagger}$ are to the right of all $a$), we may then simply replace the field operators in the correlation functions by by classical amplitudes \cite{mandel1995oca}, 
\begin{align}
E_{\nu} (t) \rightarrow \int \!\! d\omega \; \alpha_{\omega}^{\nu} e^{\pm i \omega t}.
\end{align}

But a coherent state is not yet sufficient for the field to be classical: 
As we have seen in Eq.~(\ref{eq.Liouville-G}), the light field enters into our formalism through its multipoint correlation function $\mathbb{E}^{(n)}_{\nu_n\ldots\nu_1}\left({t - t_n\ldots,t - t_n \cdots - t_1}\right)$. These correlation functions can always be rewritten as the sum of a normally ordered term and lower order normal correlation functions multiplied with commutator terms
\begin{align}
&\mathbb{E}^{(n)}_{\nu_n\ldots\nu_1}\left({t - t_n\ldots,t - t_n \cdots - t_1}\right)\notag\\
&=\big\langle : E_{\nu_n} (t - t_n) \cdots E_{\nu_1} (t - t_n \cdots - t_1) : \big\rangle \notag \\
&+ [E_{\nu_n} (t - t_n), E_{\nu_{n-1}} (t - t_n - t_{n-1})] \notag \\
&\times \mathbb{E}^{(n)}_{\nu_{n-2}\ldots\nu_1}\left({t - t_n - t_{n-1} - t_{n-2}\ldots,t - t_n \cdots - t_1}\right) + \cdots \label{eq.field-ordering}
\end{align}
For the semiclassical limit to hold, we must be able to neglect all the terms containing field commutators. This is typically the case when the intensity of the coherent state, $i.e.$ its mean photon number is very large. 

More generally, probability distributions of coherent states are also classical \cite{mandel1995oca}: States, whose density matrix has a diagonal Glauber-Sudarshan P-representation, are considered classical. This includes classical stochastic fields, which show non-factorizing field correlation functions (for instance, for Gaussian statistics with $\langle E \rangle = 0$ and $\langle E^{\dagger} E \rangle \neq 0$). Such classical correlations can also be employed in spectroscopy \cite{Turner13}, but are not the focus of the present review.

According to this strict criterion, any other quantum state must therefore be considered nonclassical. However, that does not mean that such state will also show nonclassical features, and the elucidation of genuine quantum effects in a given experiment is often very involved; see e.g. the discussion around dispersion cancellation with entangled photons \cite{Steinberg92, Franson92a}, or quantum imaging \cite{Bennink04a, Valencia05}. 

The interest in quantum light for spectroscopy is twofold: As stated in \cite{Walmsley15a}, ``the critical features of quantum light [$\ldots$] are exceptionally low noise and strong correlations".  In the following section, we first discuss the noise properties of quantum light, $i.e.$ quantum correlations \textit{within} an electromagnetic mode. The rest of the chapter is devoted to the strong time-frequency correlations, $i.e.$ quantum correlations between \textit{different} modes, which is the main focus of this review.



\subsection{Single mode quantum states}
\label{sec.cavities}
In confined geometries, the density of states of the electromagnetic field can be altered dramatically. In a cavity, it is often sufficient to describe the field by a single field mode with a sharp frequency $\omega_0$, which may be strongly coupled to dipoles inside the cavity.  In this section, we shall be concerned with this situation, where we can write the cavity field Hamiltonian simply as
\begin{align}
H_{\text{cav}} &= \hbar \omega_0 a^{\dagger} a,
\end{align}
where $a$ ($a^{\dagger}$) describes the photon annihilation (creation) operator of the cavity mode. We introduce the dimensionless cavity field quadratures
\begin{align}
X &= \frac{1}{\sqrt{2}} \left( a + a^{\dagger} \right), \\
P &= \frac{1}{\sqrt{2} i} \left( a - a^{\dagger} \right),
\end{align}
which represent the real and imaginary part of the electric field, respectively. A whole family of quadratures $X (\theta)$ and $P (\theta)$ may be obtained by rotating the field operators as $a \rightarrow a \; e^{- i \theta}$.

Quantum mechanics dictates the Heisenberg uncertainty
\begin{align}
\left\langle \Delta X^2 \right\rangle \left\langle \Delta P^2 \right\rangle \geq \frac{1}{4} \label{eq.Heisenberg-uncertainty}
\end{align}
for \textit{any} quantum state. 

Coherent states form minimal uncertainty states, in which the lower bound in Eq.~(\ref{eq.Heisenberg-uncertainty}) applies
\begin{align}
\left\langle \Delta X^2 \right\rangle = \left\langle \Delta P^2 \right\rangle = \frac{1}{2}.
\end{align}
This coincides with the quantum fluctuations of the vacuum state. Any quantum state that features fluctuations in one of its quadratures below this fundamental limit $1/2$ is therefore said to be \textit{squeezed}. Note that due to the Heisenberg uncertainty~(\ref{eq.Heisenberg-uncertainty}) the conjugate quadrature must show larger fluctuations. 

The simplest example thereof is the squeezed vacuum state
\begin{align}
\vert \xi \rangle &= e^{ \frac{1}{2} \xi a^{2} - \frac{1}{2} \xi^{\ast} a^{\dagger 2} } \vert 0 \rangle,
\end{align}
where we have 
\begin{align}
\left\langle \Delta X^2 \right\rangle &= \frac{1}{2} e^{-2 | \xi |}, \\
\left\langle \Delta P^2 \right\rangle &= \frac{1}{2} e^{2 | \xi |}.
\end{align}
Thus, a squeezed vacuum state is also a minimum uncertainty state.


These fluctuations also show up in nonlinear signals, as we will discuss in section~\ref{sec.nl-signals_cavities}. In the remainder of this section, we are interested in multimode analogues of such squeezed states, and show how new correlations arise due to the multimode structure of the field. This is known as time-energy entanglement.


\subsection{Photon Entanglement in multimode states}
\label{sec.photon-entanglement}

Before presenting some specific models for quantum light, we first discuss time-energy entanglement  in a more general setting. An exhaustive review of continuous variable entanglement theory can be found in a number of specialized reviews \cite{Kok07a, Lvovsky09}. Here, we briefly introduce some important quantities of entangled photon pairs, in order to later distinguish genuine entanglement from other bandwidth properties, which could also be encountered with classical light sources.

We shall be concerned with pairs of \textit{distinguishable} photons that live in separate Hilbert spaces $\mathcal{H}_{\text{field}}^{(1)}$ and $\mathcal{H}_{\text{field}}^{(2)}$, respectively. They may be distinguished, $e.g.$, by their polarization, their frequencies, or their wavevectors and  propagation direction. A two-photon state may  be generally expanded as
\begin{align}
\vert \psi \rangle &=  \int \!\! d\omega_a \!\!\int \!\! d\omega_b \; \tilde{\Phi} (\omega_a, \omega_b) a^{\dagger}_1 (\omega_a) a^{\dagger}_2 (\omega_b) \vert 0 \rangle_{1} \vert 0 \rangle_2, \label{eq.ent-state1}
\end{align}
where $\tilde{\Phi} (\omega_a, \omega_b)$ is the \textit{two-photon amplitude}. 

To analyze the properties of this state, it is convenient to map it into onto a discrete basis. This can be achieved by the singular value decomposition of the two-photon amplitude \cite{Law00,Law04, McKinstrie13a,McKinstrie13b}
\begin{align}
\tilde{\Phi} (\omega_a, \omega_b) &= \sum_{k = 1}^{\infty} \tilde{r}_k \psi^{\ast}_k (\omega_a) \phi^{\ast}_k (\omega_b), \label{eq.decomposition}
\end{align}
where $\{ \psi_k \}$ and $\{ \phi_k \}$ form orthonormal bases, known as the Schmidt modes, and $ \tilde{r}_k$ are the mode weights. These are obtained by solving the eigenvalue equations
\begin{align}
\int d\omega' \kappa_1 (\omega, \omega') \psi_k (\omega') &\equiv \tilde{r}_k^2 \psi_k (\omega),
\end{align} 
with 
\begin{align}
\kappa_1 (\omega, \omega') = \int d\omega'' \tilde{\Phi}^{\ast} (\omega, \omega'') \tilde{\Phi} (\omega', \omega''),
\end{align} 
and 
\begin{align}
\int d\omega' \kappa_2 (\omega, \omega') \phi_k (\omega') &\equiv \tilde{r}_k^2 \phi_k (\omega),
\end{align} 
with 
\begin{align}
\kappa_2 (\omega, \omega') = \int d\omega'' \tilde{\Phi}^{\ast} (\omega'', \omega) \tilde{\Phi} (\omega'', \omega').
\end{align}
We then recast Eq. (\ref{eq.ent-state1} using Eq. (\ref{eq.decomposition})
\begin{align}
\vert \psi \rangle &=  \sum_{k = 1}^{\infty} \tilde{r}_k\int \!\! d\omega_a \!\!\int \!\! d\omega_b \;\psi^{\ast}_k (\omega_a) \phi^{\ast}_k (\omega_b) a^{\dagger}_2 (\omega_b) \vert 0 \rangle_{1} \vert 0 \rangle_2. \label{eq.ent-state100}
\end{align}

The Schmidt decomposition ~(\ref{eq.decomposition}) may be used to quantify the degree of entanglement between photon pairs. Rewriting the singular values as $\tilde{r}_k = \sqrt{B} \lambda_k$, where $B$ denotes the amplification factor of the signal and $\lambda_k$ the normalized set of singular values of the normalized two-photon state, with $\sum_k \lambda_k^2 = 1$. A useful measure is provided by the entanglement entropy \cite{Law00}
\begin{align} \label{eq.ent-entropy}
E (\psi ) &= - \sum_k \lambda_k^2 \ln (\lambda_k^2).
\end{align}
In quantum information applications, $E (\psi)$ represents the effective dimensionality available to store information in the state.

We call the state~(\ref{eq.ent-state1}) \textit{separable}, if the two-photon amplitude factorizes into the product of single-photon amplitudes $\tilde{\Phi} (\omega_a, \omega_b) = \tilde{\Phi}^{(1)} (\omega_a) \tilde{\Phi}^{(2)} (\omega_b)$. This implies that $\lambda_1 = 1$, and the entanglement entropy is $E = 0$. In this situation, no correlations exist between the two photons: Measuring the frequency of photon $1$ will not alter the wavefunction of the other photon. Otherwise, the state is entangled. We then have $\lambda_i < 1$, and correspondingly $E > 0$; measuring the frequency of photon $1$ now reduces photon $2$ to a mixed state described by the density matrix,
\begin{align}
\varrho_2 &\sim \sum_k \vert \psi_k (\omega_a^{(0)}) \vert^2 \notag \\
&\times \int \!\! d\omega_b \int \!\! d\omega'_b \; \phi_k^{\ast} (\omega_b) \phi_k (\omega_b') a^{\dagger}_2 (\omega_b) \vert 0 \rangle \langle 0 \vert a_2 (\omega'_b),
\end{align}
the measurement outcome $\omega_a^{(0)}$ thus influences the quantum state of photon $2$.

\subsection{Entangled photons generation by parametric down conversion}

Entangled photon pairs are routinely created, manipulated, and detected in a variety of experimental scenarios. These include decay of the doubly excited states in semiconductors \cite{Edamatsu04a, Stevenson07a}, four-wave mixing in optical fibers \cite{Palmett07a, Palmett08a} or cold atomic gases \cite{Balic05a, Cho14a}. Here, we focus on the oldest and most established method for their production - parametric downconversion (PDC) in birefringent crystals \cite{Wu86a, Kwiat95a}.

We will introduce the basic Hamiltonian which governs the PDC generation process, and discuss the output fields it creates. These will then be applied to calculate spectroscopic signals. We restrict our attention to squeezed vacua, where the field modes are initially in the vacuum state, and only become populated through the PDC process (see also Eqs. (\ref{V5llrrppcoh}) - (\ref{CQQ})).

In the PDC setup, a photon from a strong pump pulse is converted into an entangled photon pair by the interaction with the optical nonlinearity in the crystal (as will be discussed in detail in section~\ref{sec.generation-&-response-fct}). A birefringent crystal features two ordinary optical axes ($o$), and an extraordinary optical axis ($e$) in which the group velocity of optical light is different. The PDC process can be triggered in different geometries: One distinguishes \textit{type-I} ($e \rightarrow oo$) and \textit{type-II} ($o \rightarrow eo$) downconversion \cite{Shih03a}, and more recently \textit{type-0} \cite{Abolghasem10a, Lerch13a}.
The two photons show strong time and frequency correlations stemming from the conservation of energy and momentum: Since they are created simultaneously, the two photons are strongly correlated in their arrival time at the sample or detector, such that each individual photon wavepacket has a very broad bandwidth. At the same time, the sum of the two photon frequencies has to match the energy of the annihilated pump photon, which may be more sharply defined than the individual photons (pump photon bandwidth is typically below $100$ MHz in visible range).

The created photon pairs are entangled in their frequency, position and momentum degree of freedom \cite{Walborn10a}. The setup may be exploited to control the central frequencies of the involved fields \cite{Grice97a}.
For simplicity, we consider a collinear geometry of all fields. This greatly simplifies our notation, while retaining the time-frequency correlations, which are most relevant in spectroscopic applications. After passing through the nonlinear crystal, the state of the light field is given by \cite{Christ11a, Christ13a}
\begin{align}
\vert \psi_{\text{out}} \rangle &= \exp \left[ - \frac{i}{\hbar} H_{\text{PDC}} \right]  \vert 0 \rangle_{1} \vert 0 \rangle_2 \equiv U_{\text{PDC}} \vert 0 \rangle_{1} \vert 0 \rangle_2. \label{eq.psi_infty}
\end{align}
The propagator $U_{\text{PDC}}$ is given by the effective Hamiltonian 
\begin{align}
H_{\text{PDC}} &= \int \!\! d\omega_a \!\!\int \!\! d\omega_b \; \Phi (\omega_a, \omega_b) a^{\dagger}_1 (\omega_a) a^{\dagger}_2 (\omega_b) + h.c.
\end{align}
which creates or annihilates pairs of photons, whose joint bandwidth properties are determined by the two-photon amplitude
\begin{align}
\Phi (\omega_a, \omega_b) &= \alpha A_p (\omega_a + \omega_b) \; \text{sinc} \left( \frac{\Delta k (\omega_a, \omega_b) L}{2} \right) e^{i \Delta k L / 2}. \label{eq.two-photon-amplitude}
\end{align}
Here, $A_p$ is the normalized pump pulse envelope, $\text{sinc} (\Delta k L / 2)$ denotes the phase-matching function, which originates from wavevector mismatch inside the nonlinear crystal, $\Delta k (\omega_a, \omega_b) = k_p (\omega_a + \omega_b) - k_1 (\omega_a) - k_2 (\omega_b)$. $k_i (\omega)$ denotes the wavevector of either the pump or beams $1$ or $2$ at frequency $\omega$. The prefactor $\alpha$ which is proportional to the pump amplitude determines the strength of the PDC process (and hence the mean photon number). Hence, in contrast to $\tilde{]Phi}$ in the enangled state~(\ref{eq.ent-state1}), the two-photon amplitude $\Phi$ is not normalized, but increases with the pump intensity.

Typically, the wavevector mismatch $\Delta k (\omega_a, \omega_b)$ depends very weakly on the frequencies $\omega_a$ and $\omega_b$. It is therefore possible to expand it around the central frequencies $\omega_1$ and $\omega_2$ of the two downconverted beams. 
For type-I downconversion, the group velocities $dk_1 / d\omega_1$ and $dk_2 / d\omega_2$ are identical [unless the PDC process is triggered in a strongly non-degenerate regime \cite{Kalachev08a}], and the expansion around the central frequencies yields \cite{Joobeur94a, Wasilewski06a}
\begin{align}
\Delta k (\omega_a, \omega_b) L /2 &= \left( \frac{dk_p}{d\omega_p} - \frac{dk}{d\omega} \right) L /2 \left( \omega_a + \omega_b - \omega_p  \right) \notag \\
&+ \frac{1}{2} \frac{d^2 k_p}{d\omega_p^2} L /2 \left( \omega_a + \omega_b - \omega_p \right)^2 \notag \\
&- \frac{1}{2} \frac{d^2k}{d\omega^2} L /2 \left[ \left( \omega_a - \omega_1 \right)^2 + \left( \omega_b - \omega_2 \right)^2 \right] \label{eq.typeI-phase-matching}
\end{align}
The first two terms of Eq.~(\ref{eq.typeI-phase-matching}) create correlations between the two photon frequencies, while the third determines the bandwidth of the individual photons.

In type-II downconversion, the group velocities of the two beams differ, and the wavevector mismatch may be approximated to  linear order \cite{Rubin94, Keller97}
\begin{align}
\Delta k (\omega_a, \omega_b) L / 2 &= (\omega_a - \omega_1) T_1  / 2 + (\omega_b - \omega_2) T_2 / 2, \label{eq.typeII-phase-matching}
\end{align}
where the two time scales $T_1 = L( dk_p / d\omega_p - dk_1 / d\omega_1)$ and $T_2 = L (dk_p / d\omega_p - dk_2 / d\omega_2)$ denote the maximal time delays the two photon wavepackets can acquire with respect to the pump pulse during their propagation through the crystal. Without loss of generality, we assume $T_2 > T_1$.
In the following applications we adopt type-II phase matching~(\ref{eq.typeII-phase-matching}). The corresponding wavefunctions may be controlled to maintain frequency correlations, as we will elaborate in section \ref{sec.ent-light-control}.

The following applications will focus on the weak downconversion regime, in which the output light fields are given by entangled photon pairs, 
\begin{align}
\vert \psi_{\text{twin}} \rangle &\approx - \frac{i}{\hbar} H_{\text{PDC}} \vert 0 \rangle_{1} \vert 0 \rangle_2. \label{eq.ent-state}
\end{align}
Eq.~(\ref{eq.ent-state}) now takes on the form of the entangled two-photon state, Eq.~(\ref{eq.ent-state1}), discussed above, where $\tilde{\Phi}$ denotes the normalized two-photon amplitude. We will review the quantum correlations of the created fields upon excitation by cw pump, which simplifies the discussion, or finite-bandwith pump pulses. The two cases require different tools.

\subsection{Narrowband pump}

One important class of time-frequency entangled photon pairs is created by pumping the nonlinear crystal with a narrow bandwidth laser, where the pump spectral envelope in Eq.~(\ref{eq.two-photon-amplitude}) may be approximated as
\begin{align}
A_p (\omega_a + \omega_b) \simeq \delta (\omega_a + \omega_b - \omega_p). \label{eq.cw-pump}
\end{align}
Using Eqs.~(\ref{eq.typeII-phase-matching}) and (\ref{eq.cw-pump}), the phase mismatch in Eq.~(\ref{eq.two-photon-amplitude}) can be expressed as \cite{Perina98a}
\begin{align}
 &\text{sinc} \left( \frac{\Delta k (\omega_a, \omega_b) L}{2} \right)  e^{i \Delta k L / 2} \notag \\ = &\text{sinc} \left( (\omega_1 - \omega_a) T / 2 \right) e^{i (\omega_1 - \omega_a) T / 2}, \label{eq.phase-matching-cw}
\end{align}
where $T \equiv T_2 - T_1$ is the \textit{entanglement time}, which represents the maximal time delay between the arrival of the two entangled photons. For a cw pump laser, the first photon arrives at a completely random time, but the second photon necessarily arrives within the entanglement time. This property has been exploited in several proposals to probe ultrafast material processes using a cw photon pair source \cite{Roslyak09b, Raymer13a}. 

As mentioned before, entangled photons affect optical signals via their multi-point correlation functions \cite{Roslyak09c, Nagata11}. The relevant field quantity in most applications discussed below - describing the interaction of pairs of photons with a sample - is the four-point correlation function.
Using Eqs.~(\ref{eq.cw-pump}) and (\ref{eq.phase-matching-cw}), for the entangled state~(\ref{eq.ent-state}) this correlation function can be factorized as
\begin{align}
&\big\langle E^{\dagger} (\omega'_a) E^{\dagger} (\omega'_b) E (\omega_b) E (\omega_a) \big\rangle \notag \\
= & \langle \psi_{\text{twin}} \vert E^{\dagger} (\omega'_a) E^{\dagger} (\omega'_b) \vert 0 \rangle \langle 0 \vert E (\omega_b) E (\omega_a) \vert \psi_{\text{twin}} \rangle, \label{eq.four-pt-cw}
\end{align}
Here,
\begin{align}
&\langle 0 \vert E (\omega_b) E (\omega_a) \vert \psi_{\text{twin}} \rangle = \mathcal{N} \delta (\omega_a + \omega_b - \omega_p) \notag \\
\times &\big[ \text{sinc} \left( (\omega_1 - \omega_a) T / 2 \right) e^{i (\omega_1 - \omega_a) T / 2} \notag \\
+& \text{sinc} \left( (\omega_1 - \omega_b) T / 2 \right) e^{i (\omega_1 - \omega_b) T / 2} \big]. \label{eq.two-photon-wavefunction-cw}
\end{align}
is known as the \textit{two-photon wavefunction} \cite{Rubin94}.  

Upon switching to the time domain, we obtain
\begin{align}
&\int d\omega_a \int d\omega_b \; e^{- i (\omega_a t_1 + \omega_b t_2)} \langle 0 \vert E (\omega_b) E (\omega_a) \vert \psi_{\text{twin}} \rangle \notag \\
= &\mathcal{N'} e^{- i ( \omega_1 t_1 + \omega_2 t_2 )} \text{rect} \left( \frac{t_2 - t_1}{T} \right) \notag \\
+ &\mathcal{N'} e^{- i ( \omega_2 t_1 + \omega_1 t_2 )} \text{rect} \left( \frac{t_1 - t_2}{T} \right), \label{eq.two-photon-wavefunction-cw-time}
\end{align}
where rect($x$)$ = 1$ for $0 \leq x \leq 1$ and zero otherwise, and $\mathcal{N}, \mathcal{N'}$ denote the normalization of the two-photon wavefunction. The physical significance of the entanglement time is now clear: it sets an upper bound for the arrival of the second photon, given the absorption of the first one. Note that Eq.~(\ref{eq.two-photon-wavefunction-cw-time}) is symmetric with respect to $t_1$ and $t_2$ because each interaction occurs with the entire field $E = E_1 + E_2$. In a situation where, e.g., $\langle 0 \vert E_2 (t_2) E_1 (t_1) \vert \psi_{\text{twin}} \rangle$ is measured the two-photon wavefunction is not symmetric.

\subsection{Broadband pump}
\label{sec.pulsed-entangled-light}
\begin{figure*}[ht]
\centering
\includegraphics[width=\textwidth]{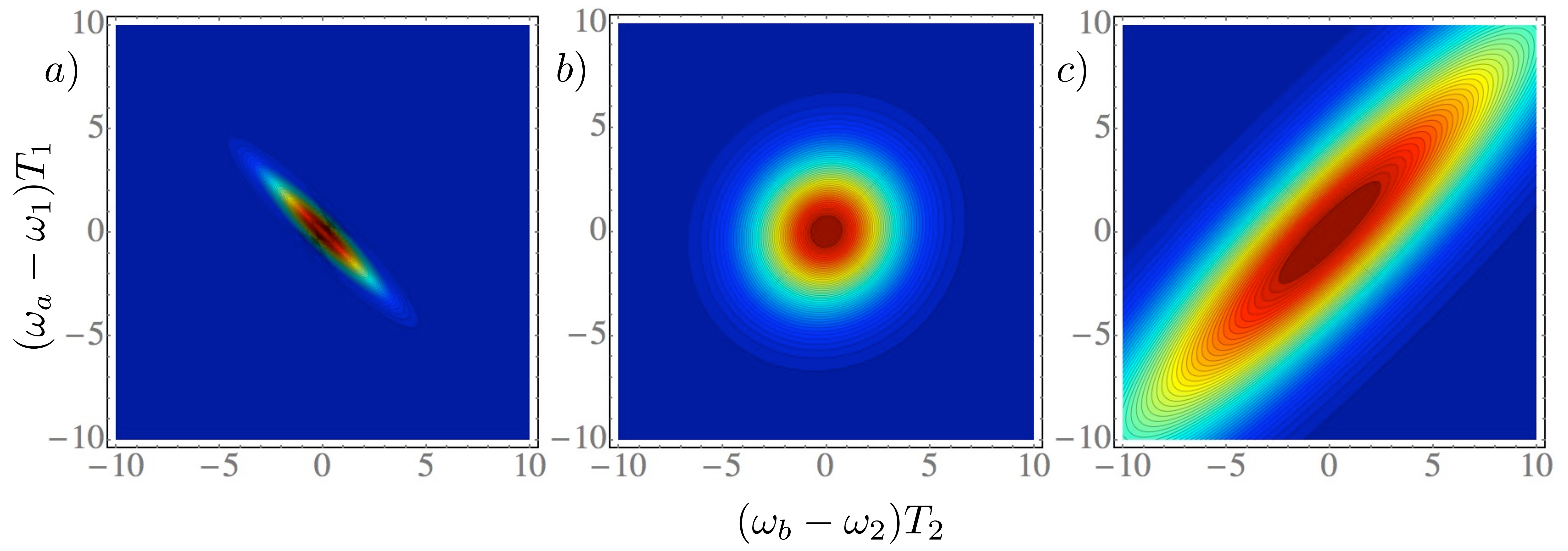}
\caption{Absolute value of the two-photon correlation function~(\ref{eq.f12-definition}) of entangled photon pairs with a) strong frequency anti-correlations, $\sigma_p = 0.6$ / $T_2$ [entanglement entropy~(\ref{eq.ent-entropy}) $E = 1.9$], b) very weak correlations ($E  =0.018$), $\sigma_p = 3.5$ / $T_2$, and c) strong positive frequency correlations, $\sigma_p = 50$ / $T_2$ ($E = 1.7$). }
\label{fig.freq-correlations}
\end{figure*}

We next consider entangled light created by a pump pulse with a normalized Gaussian envelope,
\begin{align}
A_p (\omega_a + \omega_b) &= \frac{1}{\sqrt{2 \pi \sigma_p^2}} \exp \left( - \frac{- (\omega_a + \omega_b - \omega_p)^2}{2 \sigma_p^2} \right). \label{eq.pump-envelope}
\end{align}
This is a realistic model in many experimental scenarios.

The analysis of Eq.~(\ref{eq.psi_infty}) becomes most transparent by switching to a basis obtained by the Schmidt decomposition~(\ref{eq.decomposition}) of the two-photon wavefunction. The mode weights $r_k$ are positive and form a monotonically decreasing series, such that in practical applications the sum in Eq.~(\ref{eq.decomposition}) may be terminated after a finite number of modes (For a cw pump defined above, an infinite number of Schmidt modes is required to represent the delta-function).
In appendix~\ref{sec.analytic-decomposition}, we present approximate analytic expressions for the eigenfunctions $\{ \psi_k \}$ and $\{ \phi_k \}$. The following analysis is restricted to a PDC regime in which six- (and higher) photon processes may be neglected. Such corrections can affect the bandwidth properties at very high photon numbers \cite{Christ13a}.
The linear to quadratic intensity crossover of signals will be further discussed in section \ref{sec.pump-intensity}.

We next introduce the Schmidt mode operators 
\begin{align}
A_k &= \int d\omega_a \; \psi_k (\omega_a) a_1 (\omega_a),
\end{align}
and 
\begin{align}
B_k &= \int d\omega_b \; \phi_k (\omega_b) a_2 (\omega_b),
\end{align}
which inherit the bosonic commutation relations from the orthonormality of the eigenfunctions $\{ \psi_k \}$ and $\{ \phi_k \}$. The transformation operator $U_{\text{PDC}}$ now reads \cite{Christ11a, Christ13a}
\begin{align}
U_{\text{PDC}} &= \exp \left( \sum_k r_k A^{\dagger} B^{\dagger}_k - h.c. \right). \label{eq.U_PDC}
\end{align}
The output state $\vert \psi_{\text{out}} \rangle$ is thus a multimode squeezed state with squeezing parameters $r_k$,
\begin{align}
\vert \psi_{\text{out}} \rangle &= \prod_{k = 1}^{\infty} \frac{1}{\sqrt{\cosh (r_k)}} \sum_{n_k = 0}^{\infty} \left( \tanh (r_k) \right)^{n_k} \vert n_k \rangle_1 \vert n_k \rangle_2, \label{eq.psi_out}
\end{align}
in which fluctuations of the collective quadrature of the two states  $Y_k = (A_k^{\dagger} + B_k^{\dagger} - A_k - B_k) / (2 i)$ may be squeezed below the coherent state value ($\langle Y_k^2 \rangle = 1/2$),
\begin{align}
\big\langle Y_k^2 \big\rangle &= \frac{1}{2} e^{- 2 r_k}.
\end{align}
The mean photon number in each of the two fields is given by
\begin{align}
\overline{n} &= \big\langle \sum_k A^{\dagger}_k A_k \big\rangle = \big\langle \sum_k B^{\dagger}_k B_k \big\rangle = \sum_k \sinh^2 (r_k). \label{eq.mean-photon-number}
\end{align}
We will be mostly concerned with the time-frequency correlations in the weak downconversion regime, $i.e.$ $\overline{n} \leq 1$, when the output state is dominated by temporally well separated pairs of time-frequency entangled photons. 


The multi-point correlation functions of state~(\ref{eq.psi_out}), which are the relevant quantities in nonlinear spectroscopy, may be most conveniently evaluated by switching to the Heisenberg picture, in which the Schmidt mode operators become \cite{Christ11a, Christ13a}
\begin{align}
A_k^{\text{out}} &= \cosh (r_k) A_k^{\text{in}} + \sinh (r_k) B^{\dagger \text{in}}_k, \label{eq.in-out1}\\
B_k^{\text{out}} &= \cosh (r_k) B_k^{\text{in}} + \sinh (r_k) A^{\dagger \text{in}}_k. \label{eq.in-out2}
\end{align}
The four-point correlation function then reads \cite{Schlawin13a}
\begin{align}
&\big\langle E^{\dagger} (\omega'_a) E^{\dagger} (\omega'_b) E (\omega_b) E (\omega_a) \big\rangle \notag \\
= &\left( h^{\ast}_{12} (\omega'_a, \omega'_b) + h^{\ast}_{21} (\omega'_a, \omega'_b) \right) \left( h_{12} (\omega_a, \omega_b) + h_{21} (\omega_a, \omega_b)\right) \notag \\
+ &\left( g_1 (\omega_a, \omega'_a) + g_2 (\omega_a, \omega'_a) \right) \left( g_1 (\omega_b, \omega'_b) + g_2 (\omega_b, \omega'_b) \right)  \notag \\
+ &\left( g_1 (\omega_a, \omega'_b) + g_2 (\omega_a, \omega'_b) \right) \left( g_1 (\omega_b, \omega'_a) + g_2 (\omega_b, \omega'_a) \right), \label{eq.four-pt-full-text}
\end{align}
with
\begin{align}
h_{12} (\omega_a, \omega_b) &= \sum_k \cosh (r_k) \sinh (r_k) \psi_k (\omega_a) \phi_k (\omega_b), \label{eq.f12-definition} \\
g_1 (\omega, \omega') &= \sum_k \sinh^2 (r_k) \psi_k (\omega) \psi_{k'}^{\ast} (\omega'), \label{eq.g1-definition}
\end{align}
and 
\begin{align}
g_2 = \sum_k \sinh^2 (r_k) \phi_k (\omega) \phi_{k'}^{\ast} (\omega'). \label{eq.g2-definition}
\end{align}
The first line in Eq.~(\ref{eq.four-pt-full-text}) shows the same structure as Eq.~(\ref{eq.four-pt-cw}), in that the two absorption events at frequencies $\omega_a$ and $\omega_b$ (and at $\omega'_a$ and $\omega'_b$) are correlated. Indeed, in the weak pump regime, when $r_k \ll 1$, Eq.~(\ref{eq.f12-definition}) reduces to the two-photon wavefunction of the pulsed entangled pairs, and $\sinh (r_k) \cosh (r_k) \simeq r_k$ \cite{Schlawin13a},
\begin{align}
h_{12} (\omega_a, \omega_b)  &\Rightarrow_{r_k \ll 1} \sum_k r_k \psi_k (\omega_a) \phi_k (\omega_b) \notag \\
&= \Phi^{\ast} (\omega_a, \omega_b) = \langle 0 \vert E_2 (\omega_b) E_1 (\omega_a) \vert \psi_{\text{twin}} \rangle.
\end{align}
$h_{12}$ denotes the \textit{two-photon contribution} to the correlation function , which should be distinguished from the \textit{autocorrelation contributions} $g_1$ and $g_2$.

The ratio of the inverse entanglement time $T^{-1}$ defined above and the pump bandwidth $\sigma_p$ determines the frequency correlations in Eq.~(\ref{eq.f12-definition}): As shown in Fig.~\ref{fig.freq-correlations}a), for $\sigma_p \ll T_2^{-1}$ we recover the cw regime with strong frequency \textit{anti-correlations}. In the opposite regime, when $\sigma_p \gg T_2^{-1}$ [panel c)], the two photons show \textit{positive} frequency correlations, and in between, there is a regime shown in panel b), in which the photonic wavefunction factorizes, and no frequency correlations exist. The field correlation function as depicted in Fig.~\ref{fig.freq-correlations} can be measured experimentally \cite{Kim05}.

\begin{figure}[t]
\centering
\includegraphics[width=0.45\textwidth]{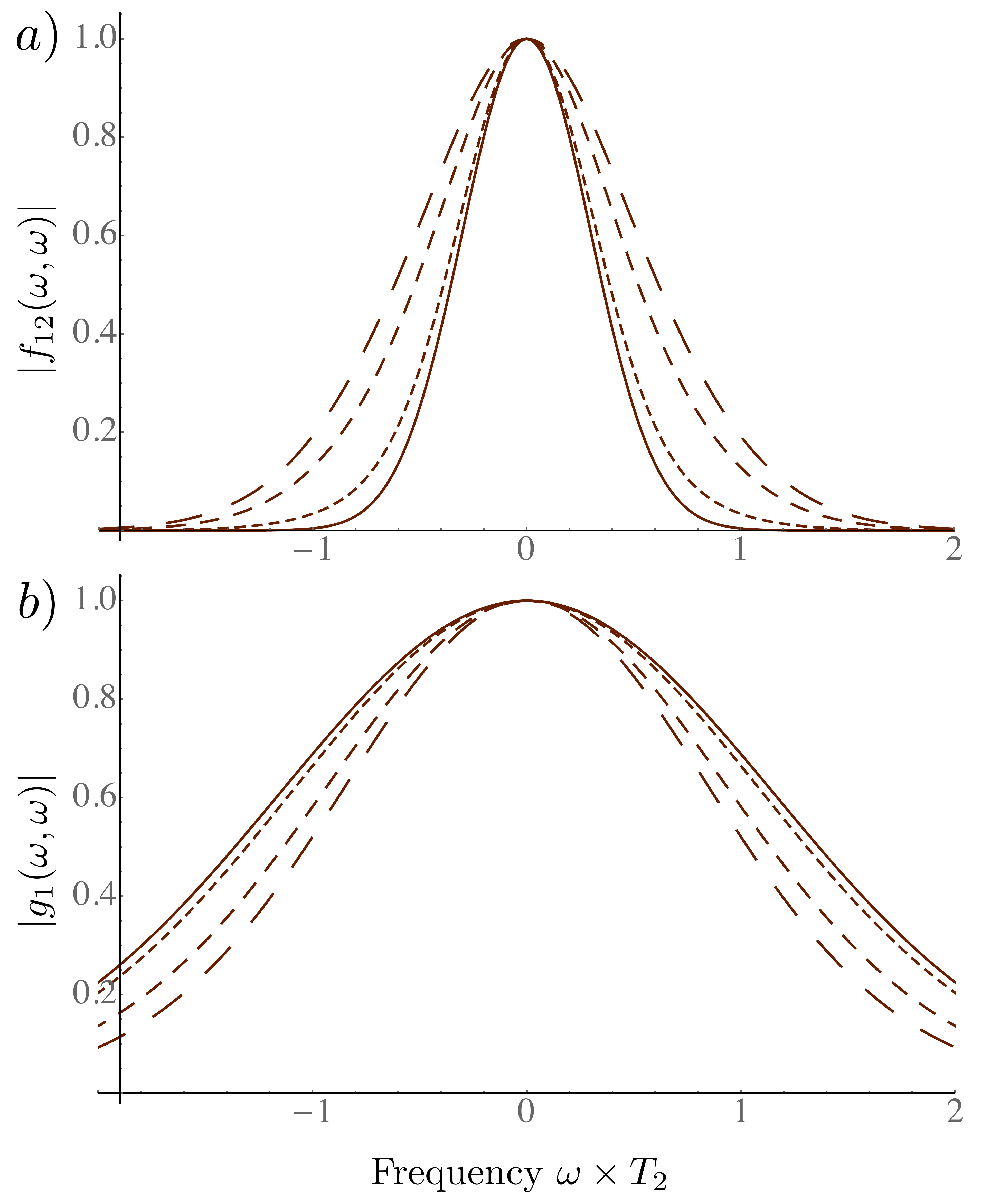}
\caption{a) The two-photon correlation function $h_{12} (\omega, \omega)$, Eq.~(\ref{eq.f12-definition}), plotted vs the frequency $\omega$ in units of $T_2$, and for mean photon numbers (with increasing dashing) $\bar{n} \sim 0.1, 1, 10$, and $100$. b) the same for the autocorrelation function $g_1 (\omega, \omega)$, Eq.~(\ref{eq.g1-definition}). }
\label{fig.intensity_corr-fcts}
\end{figure}

In spectroscopic applications, two-photon events involving uncorrelated photons from different pairs become more likely at higher pump intensities. These events are described by the functions $g_1$ and $g_2$, which at low intensities scale as $\sim r_k^2$, so that Eq.~(\ref{eq.f12-definition}) with $h_{12} \sim r_k$ dominates the signal. As the pump intensity is increased, events involving photons from different pairs must be taken into account as well. 

The various contributions to the correlation function behave differently with increasing photon number: They depend nonlinearly on the mode weights $r_k$ which in turn depend linearly on the pump amplitude. Thus, with increasing pump amplitude (and photon number) the few largest eigenvalues get enhanced nonlinearly compared to the smaller values, and fewer Schmidt modes contribute to Eqs.~(\ref{eq.f12-definition}) and (\ref{eq.g1-definition}). This is shown in Fig.~\ref{fig.intensity_corr-fcts} where the two correlation functions are plotted for different mean photon numbers, $\bar{n} = 0.1, \cdots 100$. As the number of participating Schmidt modes is decreased, the frequency correlations encoded in $h_{12}$ are weakened, and $h_{12}$ broadens. Conversely, the bandwidth of the beams $g_1$ is reduced with increasing $\bar{n}$.

\subsection{Shaping of entangled photons}
\label{sec.ent-light-control}



\begin{figure}[t]
\begin{center}
\includegraphics[trim=0cm 0cm 0cm 0cm,angle=0, width=0.5\textwidth]{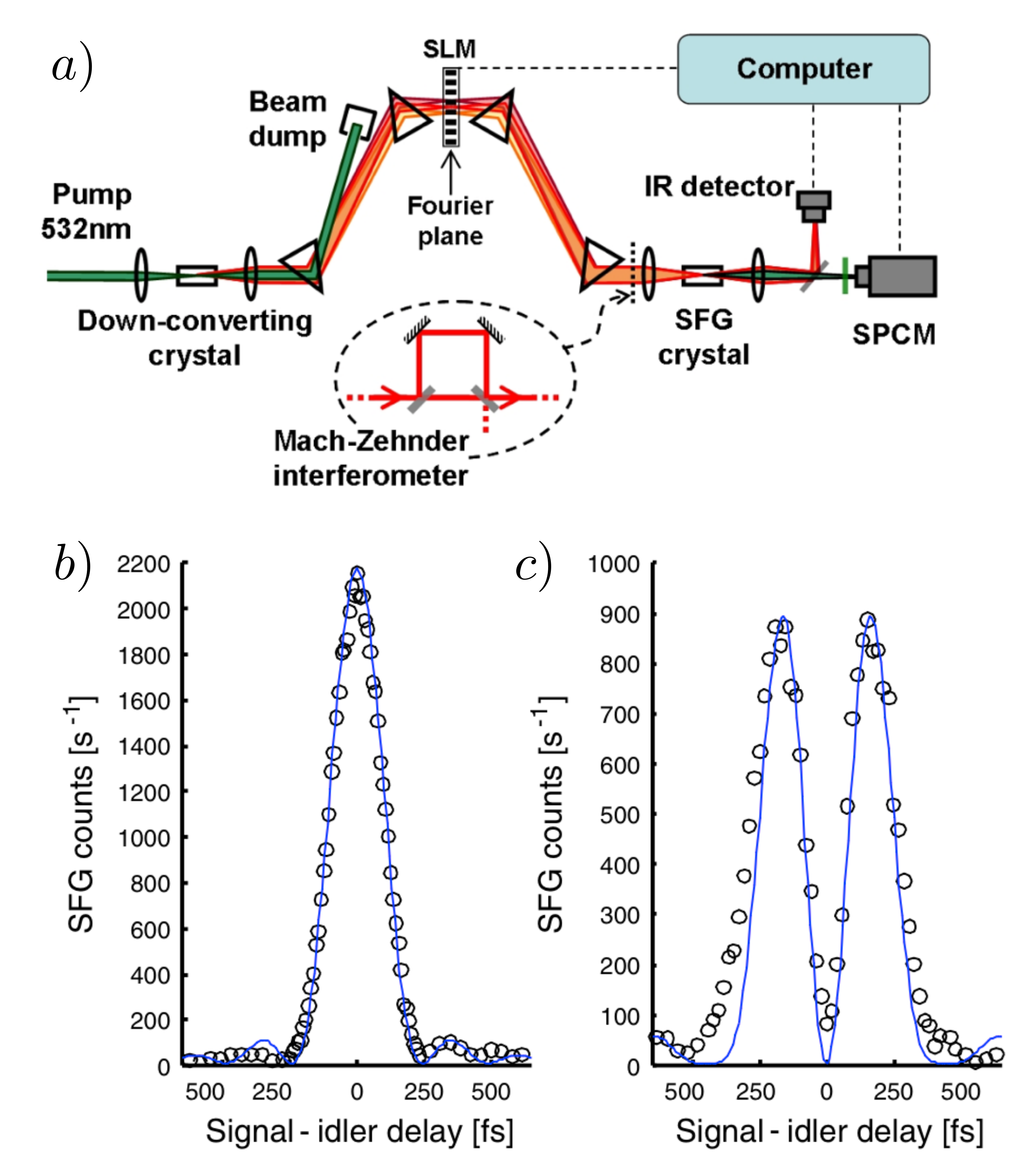}
\end{center}
\caption{(Color online) a) Experimental layout of the entangled photon shaper: A computer-controlled spatial light modulator (SLM) is used to manipulate the spectral phase of the entangled photons. The photon pairs are detected in the inverse process of PDC - sum frequency generation (SFG). The SFG photons are subsequently counted in a single-photon counting module. In order to demonstrate two-photon interference oscillations, a Mach-Zehnder interferometer is placed between the last prism and the SFG crystal. b) The SFG counts (circles) and the calculated second-order correlation function (line) of the unperturbed wavefunction as a function of the signal-idler delay. c) the same for the shaped wavefunction. Figure is taken from \cite{Peer05a}.}
\label{fig:shaping}
\end{figure}

The ability to manipulate amplitude and phase of the ultrashort pulses allows to coherently control matter information in chemical reactions and other dynamical processes \cite{Wol07}. Pulse shaping allows to drive a quantum system from an initial state to a desired final state by exploiting constructive quantum-mechanical interferences that build up the state amplitude, while avoiding undesirable final states through destructive interferences \cite{sil09}. The most common experimental pulse shaping configuration is based on spatial dispersion and is often based on back-to-back optical grating spectrometer which contains two gratings (see the top part of Fig. \ref{fig:shaping}a). The first grating disperses the spectral components of the pulse in space, and the second grating packs them back together, while a pixelated spatial light modulator (SLM) applies a specific transfer function (amplitude, phase, or polarization mask), thereby modifying the amplitudes, phases, or polarization states of the various spectral components. Originally developed for strong laser beams, these pulse shaping techniques have been later extended to single photon regime \cite{Bellini03a, Peer05a, car06, Zah08, Defienne15a} allowing to control the amplitude and phase modulation of entangled photon pairs, thereby providing additional spectroscopic knobs. 

An example of a pulse shaping setup is shown in Fig. \ref{fig:shaping}a). A symmetric phase profile in SLM yields a single unshaped Gaussian pulse [Fig. \ref{fig:shaping}b) and c)]. The phase step results in a shaped pulse that mimics two pulses with $\sim 400$~fs delay (Fig. \ref{fig:shaping}d,e). By employing more complex functions in the SLM, one may produce a replica of multiple well separated pulses or more complex shapes. This may be also achieved using e.g. the Franson interferometer with variable phases and delays in both arms of the interferometer as proposed in \cite{Raymer13a} (see section~\ref{sec.MD-signals}). Beam splitters in the two arms can create four pulses, when starting with a single entangled photon pair. 

\subsection{Polarization entanglement}

So far we have ignored the photon polarization degrees of freedom. In type-II PDC the two photons are created with orthogonal polarizations. Using a suitable setup, this allows for the preparation of Bell states of the form \cite{Pan12a}
\begin{align}
\vert \text{Bell} \rangle &= \frac{1}{\sqrt{2}} \left( \vert H \rangle_1 \vert V \rangle_2 \pm \vert V \rangle_1 \vert H \rangle_2 \right),
\end{align}
where $(\vert H\rangle, |V \rangle)$ denote the horizontal (vertical) polarization, respectively. The fidelity for this state preparation is maximal, when the two photon wavepackets factorize. Similarly, type-I PDC allows for the creation of states of the form $( \vert HH \rangle \pm \vert VV \rangle ) / \sqrt{2}$ \cite{Pan12a}. 

The polarization degrees of freedom offer additional control knobs which may be used to suppress or enhance the signal from (anti-)parallel or orthogonal dipoles in a sample system - in a quantum mechanical extension of polarized photon echo techniques \cite{Voronine06a, Voronine07a}.

\subsection{Matter correlations in noninteracting two-level atoms induced by quantum light}
\label{sec.matter-correlations}

The entangled photon correlation functions, may be used to prepare desired distributions of excited states in matter. In an insightful article \cite{Muth04}, which triggered other work \cite{aki06,das081}, it was argued that using time-ordered entangled photon pairs, two-body two-photon resonances, where two noninteracting  particles are excited simultaneously, can be observed in two-photon absorption. The surprising consequence is that the nonlinear response is non additive and does not scale as number of atoms $N$. Such cooperative (nonadditive) response is not possible with classical or coherent light fields. Arguments were made that the cooperatively is induced in two-photon absorption by the manipulation of the interference among pathways. One consequence of the interference in nonlinear response is that the fluorescence from one atom can be enhanced by the presence of a second atom, even if they do not interact. If true this effect could be an interesting demonstration of quantum nonlocality and the Einstein-Podolsky-Rosen (EPR) paradox.  
In this section, we shall calculate this two photon process with quantum light  and show that  quantum locality  is not violated.



\subsubsection{Collective resonances induced by entangled light}

 \begin{figure}[t]
\begin{center}
\includegraphics[trim=0cm 0cm 0cm 0cm,angle=0, width=0.5\textwidth]{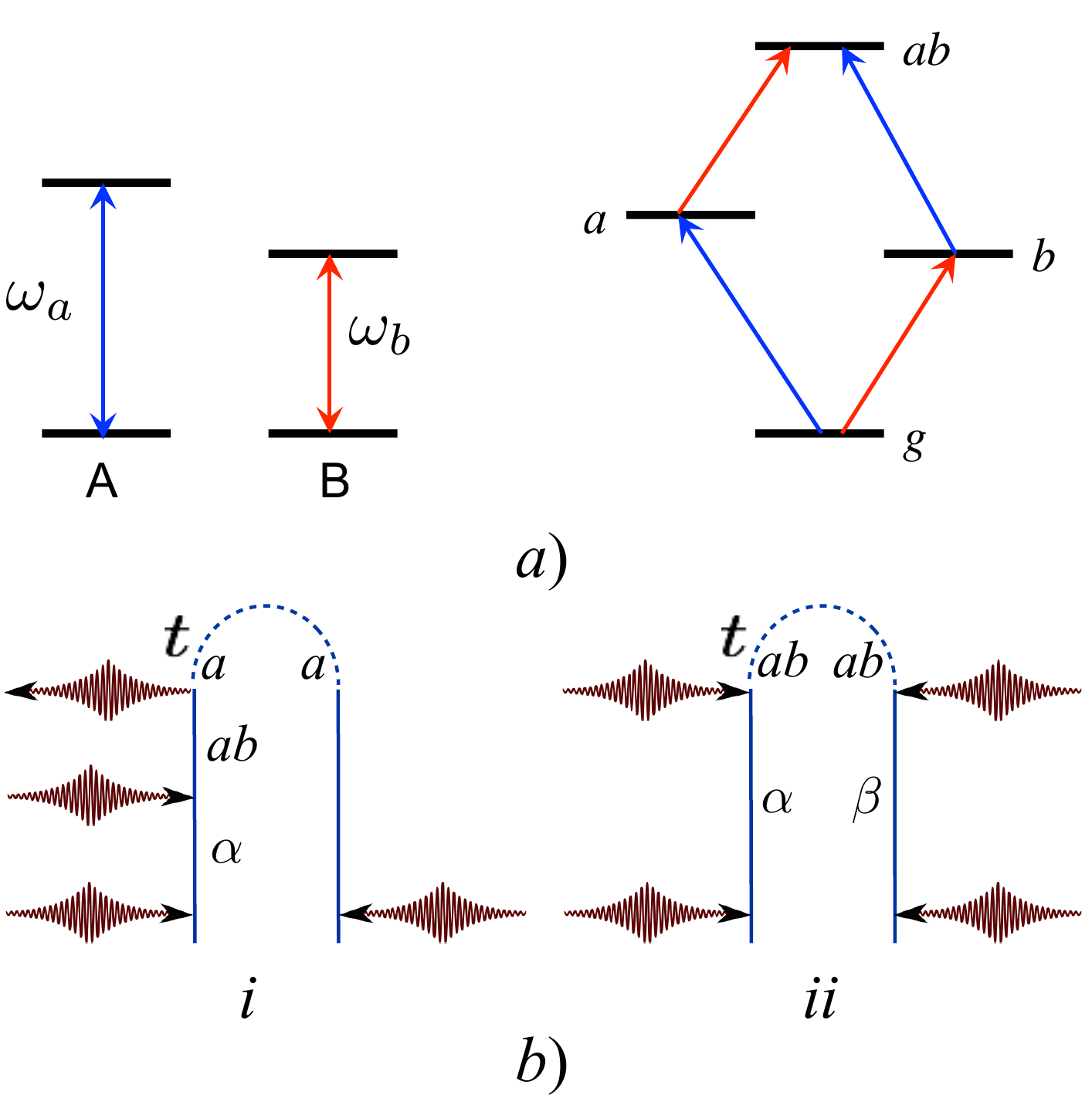}
\end{center}
\caption{(Color online) a) Scheme of two noninteracting two-level atoms $A$ and $B$ and corresponding many-body state diagram where ground state $g$ corresponds to both atom in the ground state, $a$ - atom $A$ is excited and $B$ is in the ground state, $b$ - atom $B$ is in the excited state and atom $A$ is in the ground state, and $ab$ - both atoms are in the excited state. Arrow directed upward represent two-photon excitation. b) Relevant set of diagrams corresponding to atom $A$ being in excited state to $\sim |\mu_A|^2|\mu_B|^2$ in field-matter interactions. $\alpha,\beta$ run over $a$ and $b$ to account for all possible permutations in the excitation pathways.}
\label{fig:twoatom}
\end{figure}

We use superoperator formalism \cite{Marx:PhysRevA:08} to investigate how the two-atom excitation cross section depends on the properties of the photon wave function.



Consider two noninteracting two-level atoms $A$ and $B$ coupled to
the radiation field (see Fig. \ref{fig:twoatom})a. 
We assume that the total field-matter density matrix is
initially in a factorizable form:
\begin{align}
\rho(t_0)=\rho_{A,0}\otimes\rho_{B,0}\otimes\rho_{ph,0},
\end{align}
where the $\rho_{A,0}$ ($\rho_{B,0}$) corresponds to the density matrix of the atom $A$ ($B$) and $\rho_{ph,0}$ is the density matrix of the field.
The time-dependent density matrix is given by Eq.~(\ref{eq.Dyson-series}), which in the present case reads \cite{Richter11a}
\begin{align}\label{eq:rhogen}
&\rho(t)=\notag\\
&\mathcal{T}\exp\left(-\frac{i}{\hbar}\int_{t_0}^tH_{int-}^A(\tau)d\tau-\frac{i}{\hbar}\int_{t_0}^tH_{int-}^{B}(\tau)d\tau\right) \rho(t_0).
\end{align}
If the radiation field is classical then the matter density matrix factorizes and atoms $A$ and $B$ remain uncorrelated at all times:
\begin{align}\label{eq:rhoab}
\rho(t)=\rho_A(t) \otimes \rho_B(t),
\end{align}
with
\begin{align}
\rho_A(t)=\mathcal{T}\exp\left(-\frac{i}{\hbar}\int_{t_0}^tH_{int-}^A(\tau)d\tau\right)\rho_{A,0},
\end{align}
\begin{align}
\rho_B(t)=\mathcal{T}\exp\left(-\frac{i}{\hbar}\int_{t_0}^tH_{int-}^A(\tau)d\tau\right)\rho_{B,0}.
\end{align}
This result remains valid for quantum fields as long as all relevant  normally ordered field modes are in a coherent state, and cooperative spontaneous emission is neglected so that all field modes behave classically \cite{gla63,Marx:PhysRevA:08}. Eq. (\ref{eq:rhoab}) does not hold for a general quantum state. We define the reduced matter density matrix in the joint space $w=\text{tr}_{ph}(\rho)$. Upon expanding Eq. (\ref{eq:rhogen}) order by order in the
field operators and tracing over the field modes, we obtain for the matter density matrix
\begin{align}\label{eq:wt}
w(t)&=\sum_\nu\int_{t_0}^td\tau_1...\int_{t_0}^td\tau_{n_\nu}\int_{t_0}^td\tau_1'...\int_{t_0}^{t}d\tau_{m_\nu}'\notag\\
&\times\rho_A^\nu(\tau_1,...\tau_{n_\nu})\rho_B^\nu(\tau_1',...\tau_{m_\nu}')F_\nu(\tau_1,...\tau_{n_\nu},\tau_1',...,\tau_{m_\nu}'),
\end{align}
where $\nu$ is summed over all possible pathways. Pathway $\nu$ has $n_\nu$ $\tilde{V}^A$ interactions and $m_\nu$ $\tilde{V}^B$ interactions. $\rho_A^\nu$ ($\rho_B^\nu$) are time-ordered products of system $A$ (system $B$) operators and $F_\nu(\tau_1,...\tau_{n_\nu},\tau_1',...,\tau_{m_\nu}')$ are time-ordered field correction functions. In each order of this perturbative order-by-order calculation, all the correlation functions are factorized between the three spaces. The factorization Eq. (\ref{eq:rhoab}) no longer holds in general, and atoms $A$ and $B$ may become correlated or even entangled. Eq. (\ref{eq:wt}) will be used in the following. Note that pathways with $n_\nu = 0$ or $m_\nu = 0$ are single-body pathways, where all interactions occur either with system $A$ or with $B$. Our interest is in the two-body pathways, where both $n_\nu$ and $m_\nu$ contribute to collective response.

\subsubsection{Excited state populations generated by nonclassical light}

So far, we have not discussed how the overall excitation probability by entangled light sources compares to classical light with similar photon flux. This is relevant to recent demonstration of molecular internal conversion \cite{Oka12a} and two-photon absorption with the assistance of plasmonics \cite{Oka15a}.  This is a very important practical point, if quantum spectroscopy is to be carried out at these very low photon fluxes, in the presence of additional noise sources. We now discuss the correlations on the excitation probability induced by the interaction with nonclassical light for the present of two noninteracting two-level systems.

The doubly excited state population is given - to leading-order perturbation theory in the interaction Hamiltonian~(\ref{eq.H_int_RWA}) - by the loop diagram in Fig.~\ref{fig.lop-vs-lap}a) which may be written as a modulus square of the corresponding transition amplitude
\begin{align} \label{eq:Pab1}
p_{ab} (t) &= \sum_{\psi'} \vert T_{ab; \psi'} (t) \vert^2,
\end{align}
with the transition amplitude between initial state $\vert \psi \rangle$ and final state $\vert \psi' \rangle$,
\begin{align}\label{eq:Tabdef}
T_{ab}(t)=\int_{t_0}^tdt_1\int_{t_0}^tdt_2\mu_A\mu_Be^{-i\epsilon_at_1-i\epsilon_bt_2}\langle\psi| E(t_2)E(t_1)|\psi'\rangle.
\end{align}
In the following, we only consider a unique final state and will drop the subscript $\psi'$.

\subsubsection{Classical vs entangled light}

We first evaluate Eq. (\ref{eq:Pab1}) for a classical field composed of two modes $\alpha$ and $\beta$ , which is switched on at $t = t_0$:
\begin{align}
\tilde{E}(t)=\theta(t-t_0)\left(E_\alpha e^{i\omega_\alpha t}+E_\beta e^{i\omega_\beta t}\right)+c.c.
\end{align}
$T_{ab}^{(c)}(t)$ is then given by \cite{Muth04}
\begin{align}\label{eq:Tabc}
T_{ab}^{(c)}(t)=T_a^{(c)}(t)T_b^{(c)}(t),
\end{align}
where   
\begin{align}
T_m^{(c)}(t)=\sum_{\mu=\alpha,\beta}\frac{iA_{\mu m}(t)}{\epsilon_m-\omega_\mu-i\gamma},~~m=a,b
\end{align}
$A_{\nu m}=\mu_mE_\nu\left(e^{i(\omega_\nu-\epsilon_m)t_0-\gamma t_0}-e^{i(\omega_\nu-\epsilon_m)t-\gamma t}\right)$ and $\gamma\to 0$.
A straightforward calculation of Eq. (\ref{eq:Tabdef}) shows several terms, where each of the pathway contains single-particle resonances $\epsilon_{a,b}=\omega_{\alpha,\beta}$ as well as a collective resonance $\epsilon_a+\epsilon_b=\omega_\alpha+\omega_\beta$. While combining these terms the collective resonance disappear due to destructive interference and the final result factorizes as a product of two amplitudes that contain only single-particle resonances $\epsilon_j=\omega_\mu$, $j=a,b$ and $\mu=\alpha,\beta$ and no two-photon resonances $\epsilon_a+\epsilon_b=\omega_\alpha+\omega_\beta$ (see \cite{Richter11a}).

 \begin{figure}[t]
\begin{center}
\includegraphics[trim=0cm 0cm 0cm 0cm,angle=0, width=0.35\textwidth]{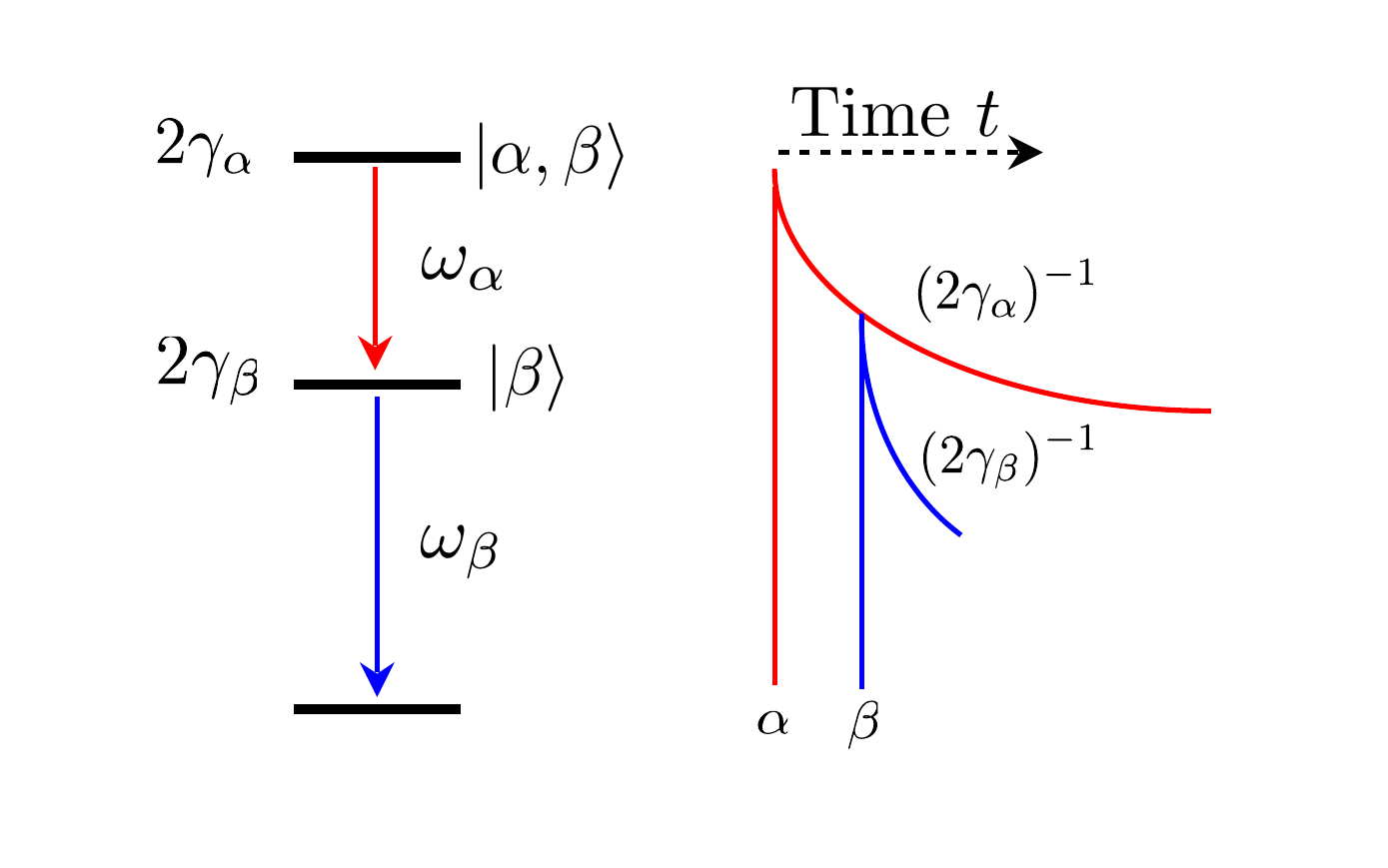}
\end{center}
\caption{(Color online) The temporal profiles of two photons emitted by a cascade source illustrate time-frequency entanglement: the red curve represents the marginal probability $P(\tau_\alpha)$ Eq. (\ref{eq:condprob}), and the blue curve corresponds to the conditional probability $P(\tau_\beta|\tau_\alpha)$ Eq. (\ref{eq:margprob}). The intrinsic time ordering of the photons, $\alpha$ first, followed by $\beta$, suppress the excitation pathway where $\beta$ is absorbed first, followed by $\alpha$, inducing joint two-atom excitation.}
\label{fig:enttempprof}
\end{figure}

We first  describe the properties of the entangled source before discussing the spectroscopy applications.
Consider a field made of entangled photon pairs of a cascade state $|\psi_{ent}\rangle$ depicted in Fig. \ref{fig:enttempprof}  which was considered by  \cite{Muth04}. This state can be prepared when an atom is promoted to the doubly excited state which consequently decays spontaneously back to the ground state by emitting a cascade of two photons and described by the wave function:
\begin{align}\label{eq:psic}
&|\psi_{ent}\rangle=\sum_{p,q}\phi_{p,q}|1_p,1_q\rangle\notag\\
&\phi_{p,q}=\frac{g_{p\alpha}g_{q\beta}e^{i(\mathbf{p}+\mathbf{q})\cdot \mathbf{r}_R}}{(\omega_p+\omega_q-\omega_\alpha-\omega_\beta+i\gamma_\alpha)(\omega_q-\omega_\beta+i\gamma_\beta)}.
\end{align}
Here $\gamma_\alpha$ is the lifetime of the upper level of the three-level cascade and $\gamma_\beta$ is the lifetime of the intermediate state. $\mathbf{p}$ and $\mathbf{q}$ are the wave vectors of different modes in vacuum  and $g_{p\alpha}$ and $g_{q\beta}$ are coupling constants.  $\omega_\alpha$ is the transition frequency from the highest to the intermediate state and $\omega_\beta$ is the transition
frequency from the intermediate state to the ground state.  It is important that photon with momentum $\mathbf{p}$ comes first and interacts with upper transition whereas photon with $\mathbf{q}$ comes later and interacts with lower $\beta$ - transition. The two-photon frequency $\omega_p+\omega_q$ is narrowly distributed around $\omega_\alpha+\omega_\beta$ with a width $\gamma_\alpha$, the lifetime of the upper level, whereas the single-photon frequencies $\omega_p$, $\omega_q$ are distributed around $\omega_\beta(\omega_\alpha)$ with a width of $\gamma_\beta(\gamma_\alpha)$, the lifetime of the intermediate (highest) level. Maximum entanglement occurs for $\gamma_\beta\gg\gamma_\alpha$. Using Eq. (\ref{eq:psic}), and assuming that the atoms $A$ and $B$ have the same distance from the cascade source, so that $t_R$ is the time retardation (with $t_R = |\mathbf{r}_R|/c$) we obtain \cite{Zheng13}
\begin{align}\label{eq:Tabent}
&T_{ab}^{ent}(t)=\sum_{m\neq n=a,b}\frac{A_{m\alpha,n\beta}(t)}{(\epsilon_m-\omega_\alpha-i\gamma_\alpha)(\epsilon_n-\omega_\beta-i\gamma_\beta)}\notag\\
&+\sum_{m=a,b}\frac{A_{TPA}(t)}{(\epsilon_a+\epsilon_b-\omega_\alpha-\omega_\beta-i\gamma_\alpha-i\gamma_\beta)(\epsilon_m-\omega_\beta-i\gamma_\beta)},
\end{align}
where 
\begin{align}
A_{n\nu,m\mu}(t)=\mu_A\mu_BA(e^{i\omega_{n\nu}(t-t_R)-\gamma_\nu(t-t_R)}-e^{-i\epsilon_mt_R-i\epsilon_nt_R}),
\end{align}
 \begin{align}
 A_{TPA}(t)&=\mu_A\mu_BA\left(e^{-i\epsilon_at_R-i\epsilon_bt_R}\right.\notag\\
 &\left.-e^{-i(\epsilon_a+\epsilon_b)t+i(\omega_\alpha-\omega_\beta-\gamma_\alpha-\gamma_\beta)(t-t_R)}\right).
 \end{align}

The first term in Eq. (\ref{eq:Tabent}) contains single-particle resonances, where the two systems are separately excited. The second term represents collective two-photon resonances $\epsilon_a+\epsilon_b-\omega_\alpha-\omega_\beta$ which can only be distinguished for non identical atoms $\epsilon_a\neq\epsilon_b$. 

Under two photon - two atom resonance condition $\omega_a+\omega_b=\omega_\alpha+\omega_\beta$ for $\gamma_{\alpha}t\gg 1$, $\gamma_{\beta}t\gg 1$ the probability $P_{ab}^{ent}(t)$ reads
\begin{align}\label{eq:Paberes}
p_{ab}^{ent}&= p_0\frac{\gamma_\beta}{\gamma_\alpha\Delta^2},
\end{align}
where $\Delta=\epsilon_b-\omega_\beta$ is a single atom detuning and $p_0=|\mu_A|^2|\mu_B|^2\epsilon_a\epsilon_b/\hbar^2\epsilon_0^2c^2S^2$ under normalization of $g_\alpha g_\beta=2c\sqrt{\gamma_\alpha\gamma_\beta}/L$.

Note, that the classical amplitude (\ref{eq:Tabc}) scales quadratically with the field amplitude $A_{\nu m}$, whereas the entangled amplitude (\ref{eq:Tabent}) scales linearly. This reflects the fact that at low intensity (entangled case) two photon absorption is effectively a linear absorption of the entangled pair as discussed earlier. More detailed analysis of the scaling is presented in Section \ref{sec.pump-intensity}.

Comparing Eqs. (\ref{eq:Tabc}) and (\ref{eq:Tabent}), we see that two-photon resonances are induced by the lack of time ordering in the photonic field. To explain this, we calculate the marginal probability
\begin{align}\label{eq:margprob}
P(\tau_\alpha)=2\gamma_\alpha\theta(\tau_\alpha)e^{-2\gamma_\alpha\tau_\alpha},
\end{align}
 and the conditional probability 
 \begin{align}\label{eq:condprob}
 P(\tau_\beta|\tau_\alpha)=2\gamma_\beta\theta(\tau_\beta-\tau_\alpha)e^{-2\gamma_\beta(\tau_\beta-\tau_\alpha)}
\end{align}
 for the two photon absorption process displayed in Fig. \ref{fig:enttempprof} \cite{Muth04}. These probabilities have the following meaning: the absorption of $\alpha$ is turned on at $\tau_\alpha=0$ and decays slowly at the rate $\gamma_\alpha$, while the absorption of $\beta$ turns on at $\tau_\beta=\tau_\alpha$ and decays rapidly at the rate $\gamma_\beta$. Thus, the two photons arrive in strict succession, $\alpha$ followed by $\beta$, with the time interval between the absorption going to zero when $\gamma_\beta\gg\gamma_\alpha$, i.e. in the limit of large frequency entanglement. Another related effect, which is discussed later in this section and which eliminates one-particle observables, is based on the lack of time ordering of the absorption of the two systems, while this interference effect originally described in \cite{Muth04} is based on a lack of time ordering of the two photons. Only the single-body single-photon resonances remain.

We now turn to the spectroscopy carried out using the model of entangled light given by Eq. (\ref{eq:psic}). To that end we turn to the excited state population of atom $A$, which is given by 
\begin{align}
p_a(t ) \equiv \text{tr}[|a\rangle\langle a|\rho(t )]=p_{a0}(t)+p_{ab}(t),
\end{align}
where the two terms correspond to the final state of atom $B$: $p_{a0}$ and $p_{ab}$ represent $B$ being in the ground and excited state, respectively. We shall only calculate the $\sim\mu_A^2\mu_B^2$ contributions to $p_a$, which are relevant to our discussion. The relevant set of diagrams is depicted in Fig. \ref{fig:twoatom}b. Diagram $i$ contains three interactions with ket- and one with bra- and represents $p_{a0}$ whereas diagram $ii$ corresponds to $p_{ab}$. The exact formulas read off these diagrams are presented in \cite{Richter11a}. Unlike the $p_{a1}(t)$ which contains only normally ordered field operators [see Eq.~(\ref{eq:Pab1})], $p_{a0}(t)$ contains also non normally ordered contributions. These can be recast as
a normally ordered correlation plus a term that includes a
commutator:
\begin{align}\label{eq:comm}
\langle E^{\dagger}(t_2)&E(t_3)E^{\dagger}(t_4)E(t_1)\rangle=\langle E^{\dagger}(t_2)E^{\dagger}(t_4)E(t_3)E(t_1)\rangle\notag\\
&+\langle E^{\dagger}(t_2)[E(t_3),E^{\dagger}(t_4)]E(t_1)\rangle.
\end{align}
The second term involves a commutator of the field which is a c-number. For certain type of states of the field, the commutator remains dependent upon the state of the field and this term has to be evaluated exactly (for instance, in the case of coherent state this effect is responsible for the revival of damped Rabi oscillations discussed below \cite{rem87}. For other types of states of the field, the commutator becomes independent of the external field (e.g. Fock state) and therefore represents spontaneous emission. The spontaneous emission pathways introduce a coupling between the two systems, since a photon emitted by system $B$ can be absorbed by system $A$. This coupling has both real (dipole-dipole) and imaginary (superradiance) parts. These couplings will obviously result in collective signals which involve several atoms.

In the case of a classical field, one can neglect the spontaneous contributions and only include the stimulated ones which results in $p_{a0}(t)=-p_{ab}(t)$. The population $p_a(t)$ therefore vanishes, and is not affected by the collective resonances. We thus do not expect any enhanced fluorescence from $A$. The two-photon absorption signal is a product  of individual excitation probabilities for atoms $A$ and $B$, and shows no collective resonances.

\subsubsection{Collective two-body resonances generated by illumination with entangled light}

We shall focus on the change of the photon number for the two-body part $\Delta n_{ph}$. This will depend on the following probabilities: the excitation probability of $p_{ab}$,which means that two photons are absorbed (counts twice), and the excitation probabilities of only system $A$ $p_a$ and of only system $B$ $p_b$ (we assume that both photons have equal frequency and are resonant to the two-photon absorption):
\begin{align}\label{eq:dn}
\Delta n_{ph}=-2p_{ab}(t)- p_{a0}(t)- p_{b0}(t).
\end{align}
Now, in the stimulated emission and TPA pathway absorption, we know that $p_{ab}(t )$ and $p_{a0}(t )$ cancel, and that $p_{ab}(t )$ and $p_{b0}(t )$ cancel. So for the stimulated pathways, Eq. (\ref{eq:dn}) vanishes.
 Since the field-matter interaction Hamiltonian connects the photon number with the excitation probability
  number, the photon number itself is
a single-particle observable like the population of state $|a\rangle$ or state $|b\rangle$, and therefore vanishes. However, collective resonances in $p_{ab}(t )$ can be revealed in two-photon counting (Hanbury-Brown-Twiss measurements) \cite{bro56}. For our entangled photon state, Eq.~(\ref{eq:psic}), the change in the photon-photon correlation
$\Delta'n_{ph}$ is attributed to any buildup of probability, that the
either atom $A$, or atom $B$ is excited or both atoms  are excited,
which will cause a reduction of the photon-photon correlation:
\begin{align}
\Delta'n_{ph}=- p_{ab}(t)- p_{a0}(t)- p_{b0}(t).
\end{align}
Since $p_{ab}(t)$ enters twice [unlike
Eq. (\ref{eq:dn})] the stimulated contributions can only cancel with
one of the two other contributions $p_{a0}(t)$ or $p_{b0}(t )$, and we have
\begin{align}
\Delta'n_{ph}= p_{ab}(t).
\end{align}
The interference mechanism, which caused the cancellation for the stimulated signal and the photon number, does not lead to a full cancellation, and two-photon absorption involving both systems might be observed.

\paragraph{Two-photon two-atom problem with non entangled quantum states}

We  now discuss whether  entangled light is essential to  the collective resonances.
Whether or not photon entanglement is essential for these properties is the subject of current debate. In \cite{Zheng13}, the authors presented a detailed analysis of the cross section created by entangled pulses, and compared this with ``correlated-separable" states in which the entanglement is replaced by classical frequency correlations. They concluded that it is the frequency anti-correlations [see Fig.~\ref{fig.intensity_corr-fcts}a)], and not the entanglement per se, which is responsible for the enhancement of the cross section. The effect of enhanced two-photon absorption probability has been later shown to come from the entangled matter/field evolution that occurs with any quantum light (which is not necessarily entangled) \cite{Zheng13} which is consistent with earlier demonstrations of \cite{geo97}. 
Starting from entangled pure quantum state having
a density matrix  $\rho_0=|\psi_{ent}\rangle\langle\psi_{ent}|$  of matrix elements $\rho_{kk'qq'}=\langle 1_k,1_q|\rho_0|1_{k'}1_{q'}\rangle$ one can construct others that have the
same mean energy and the same single-photon spectrum, and
hence that would give the same transition probabilities for a
single-photon resonance. We choose a special case of the states that originates from entangled state (\ref{eq:psic}) that
will allow a quantitative evaluation of the role of correlations. It is defined as
\begin{align}\label{eq:rho1}
\rho_1=\sum_{k,q}\rho_{kkqq}|1_k,1_q\rangle\langle 1_k,1_q|,
\end{align}
It is the diagonal part of $\rho_0$. It has lost any temporal field coherence and is time-independent. It is actually a correlated separable state \cite{dua00}, in which the quantum correlations are replaced by a purely classical frequency distribution. It gives rise, however, to correlations between its two parties. Using the entangled state~(\ref{eq:psic}), Eq.~(\ref{eq:rho1}) reads
\begin{align}\label{eq:rho11}
\rho_1&=\left(\frac{2c}{L}\right)^2\sum_{k,q}\frac{\gamma_\beta}{(\omega_{q\beta}^2+\gamma_\beta^2)}\frac{\gamma_\alpha}{[(\omega_{q\beta}+\omega_{k\alpha})^2+\gamma_\alpha^2]}\notag\\
&\times|1_k,1_q\rangle\langle 1_k,1_q|.
\end{align}
The state (\ref{eq:rho11}) corresponds to an atomic cascade for which
the starting time is random, thereby averaging to zero all the
off-diagonal time-dependent terms in the density matrix.
It gives rise to the following transition probability:
\begin{align}
p_1\simeq p_0\frac{\gamma_\alpha\gamma_\beta}{\delta^2+\gamma_\alpha^2}\left(\frac{1}{(\omega_1-\omega_\beta)^2}+\frac{1}{(\omega_2-\omega_\beta)^2}\right)\frac{t^2}{(L/c)^2}.
\end{align}
At exact two-photon two-atom resonance, we have $p_1\simeq p_0\gamma_\beta\gamma_\alpha^{-1}c^2t^2/(\Delta L)^2$.
Note that $P_1$ depends on time, as can be expected in a
situation where the detecting atoms, which have an infinite
lifetime, are submitted to a stationary quantum state, and
therefore to cw light. In order to compare $p_1$ to corresponding probability (\ref{eq:Paberes}) for entangled light, which is induced by a pulse of light, we need to fix an interaction time $t$.
One then obtains at $t=L/c$ and at exact resonance
\begin{align}
p_1\simeq p_0\frac{\gamma_\beta}{\gamma_\alpha\Delta^2}\simeq p_{ab}^{ent}.
\end{align}
We thus find that a correlated-separable state like $\rho_1$
can induce the two-photon two-atom transition similar to the entangled
cascade state. 
Note that even though $\rho_1$ is not entangled, it has
 quantum properties, being a mixture of single-photon
states which are highly nonclassical.

In summary, the two-photon absorption probability of noninteracting atoms can be enhanced compared to classical light  by using special types of quantum light. This quantum light does not require entanglement, rather it is necessary to have spectral anticorrelations which can be achieved in e.g. correlated separable states. This is a consequence of the stationary system of two noninteracting atoms which has no dynamical properties.

 In the following section we consider more complex material system that is subject to various relaxation channels, e.g. transport between excited states therefore adding temporal scale to the problem. We show that in this case, entanglement is required in order to achieve both high spectral and high temporal correlations which is not possible by  classical correlated light.


\subsection{Quantum light induced correlations between two-level systems with dipole-dipole coupling}\label{sec:aggregate}

\subsubsection{Model system}

\begin{figure}[t]
\centering
\includegraphics[width=0.1\textwidth]{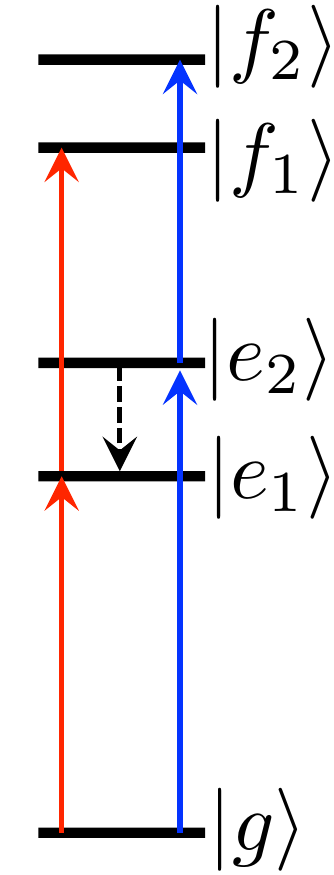}
\caption{Level scheme of the multilevel model employed in this review: Two singly excited states $e_1$ and $e_2$ with energies $11,000$~cm$^{-1}$ and $11,500$~cm$^{-1}$ are coupled to doubly excited states $f_1$ and $f_2$, as indicated. Furthermore, $e_2$ decays to $e_1$ within $1 / k \simeq 30$~fs.}
\label{fig.levelscheme}
\end{figure}

\label{sec.model-system}

The standard calculation of the nonlinear response to classical light assumes that the matter is made up of $N$ noninteracting particles (atoms or molecules) in the active zone, such that the matter Hamiltonian may be written as the sum over the individual particles [see Eq.~(\ref{eq.H_int})]
\begin{align}
H_0 &= \sum_{\nu} H_{\nu}. \label{eq.H_0}
\end{align} 
The individual nonlinear susceptibilities or response functions of these atoms then add up to give the total response. The nonlinear response becomes then a single-body problem and no cooperative resonances are expected. It is not obvious how to rationalize the $\sim N$ scaling for noninteracting atoms had we chosen to perform the calculation in the many-body space. Massive cancellations of most $\sim N(N-1)$ scaling light-matter pathways recover in the end the final $\sim N$ signal scaling \cite{spa89}.

When the atoms are coupled, the calculation must be carried out in their direct-product many-body space whose size grows exponentially with $N$ ($\sim n^N$ dimensions for $n$-level atoms). The interatomic coupling can be induced by the exchange of virtual photons leading to dipole-dipole and cooperative spontaneous emission, or superradiance \cite{das08}.  In molecular aggregates, the dipole interaction between its constituents creates inherent entanglement on the level of the quasiparticles \cite{Mukamel10b}, which shifts the doubly excited state energies, and redistributes the dipole moments, in which the individual Hamiltonians $H_{\nu}$ are \cite{Abr09}
\begin{align}
 H_{\nu} &=\hbar \sum_i \varepsilon_m B^\dagger_{m \nu} B_{m\nu}+\hbar \sum_{m\neq n} J_{mn} B_{m \nu}^\dagger B_{n \nu} \nonumber\\
&\quad+\hbar \sum_m  \frac{\Delta_m}{2} B^{\dagger}_{m \nu} B^{\dagger}_{m \nu} B_{m \nu} B_{m \nu}, \label{eq.H_nu}
\end{align}
where $B_m$ ($B_m^{\dagger}$) describes an excitation annihilation (creation) operators for chromophore $m$. These excitations are hard-core bosons with Pauli commutation rules \cite{lee57}:
\begin{align}
[B_m,B_n^{\dagger}]=\delta_{mn}(1-2B_n^{\dagger}B_n).
\end{align}
To describe two level sites, which cannot be doubly excited according to the Pauli exclusion we set  $\Delta_m\rightarrow\infty$. In condensed matter physics and molecular aggregates the many particle delocalized states are called excitons. The model Hamiltonian (\ref{eq.H_nu}) can represent e.g. Rydberg atoms in optical lattice or Frenkel excitons in molecular aggregates. 
For considering four-wave mixing processes, we have truncated the Hamiltonian $H_{\nu}$ in Eq.~(\ref{eq.H_nu}) at the doubly excited level, such that it diagonalization reads
\begin{align}\label{eq.dipole-RWA}
&H_{\nu}=\hbar \omega_{g\nu}|g_{\nu}\rangle \langle g_{\nu} |+ \hbar \sum_e \omega_{e \nu} |e_{\nu}\rangle \langle e_{\nu}|+ \hbar \sum_f \omega_{f \nu} |f_{\nu}\rangle \langle f_{\nu}|,
\end{align}
where $g$ indexes the ground state, $e$ the singly, and $f$ the doubly excited states, which will be the central concern. We will consider processes in which the signals factorize into ones stemming from individual constituents, as well as collective signals from two or more constituents. 
We illustrate basic properties of the interaction of entangled photons with complex quantum systems at the hand of the simple multilevel model depicted in Fig.~\ref{fig.levelscheme}. It consists of two excited states $e_1$ and $e_2$, a two doubly excited states $f_1$ and $f_2$. The higher-energy excited state $e_2$ decays to $e_1$ within few tens of femtoseconds. We shall focus on doubly-excited states created by the absorption of pairs of photons.

\begin{figure}[t]
\centering
\includegraphics[width=0.3\textwidth]{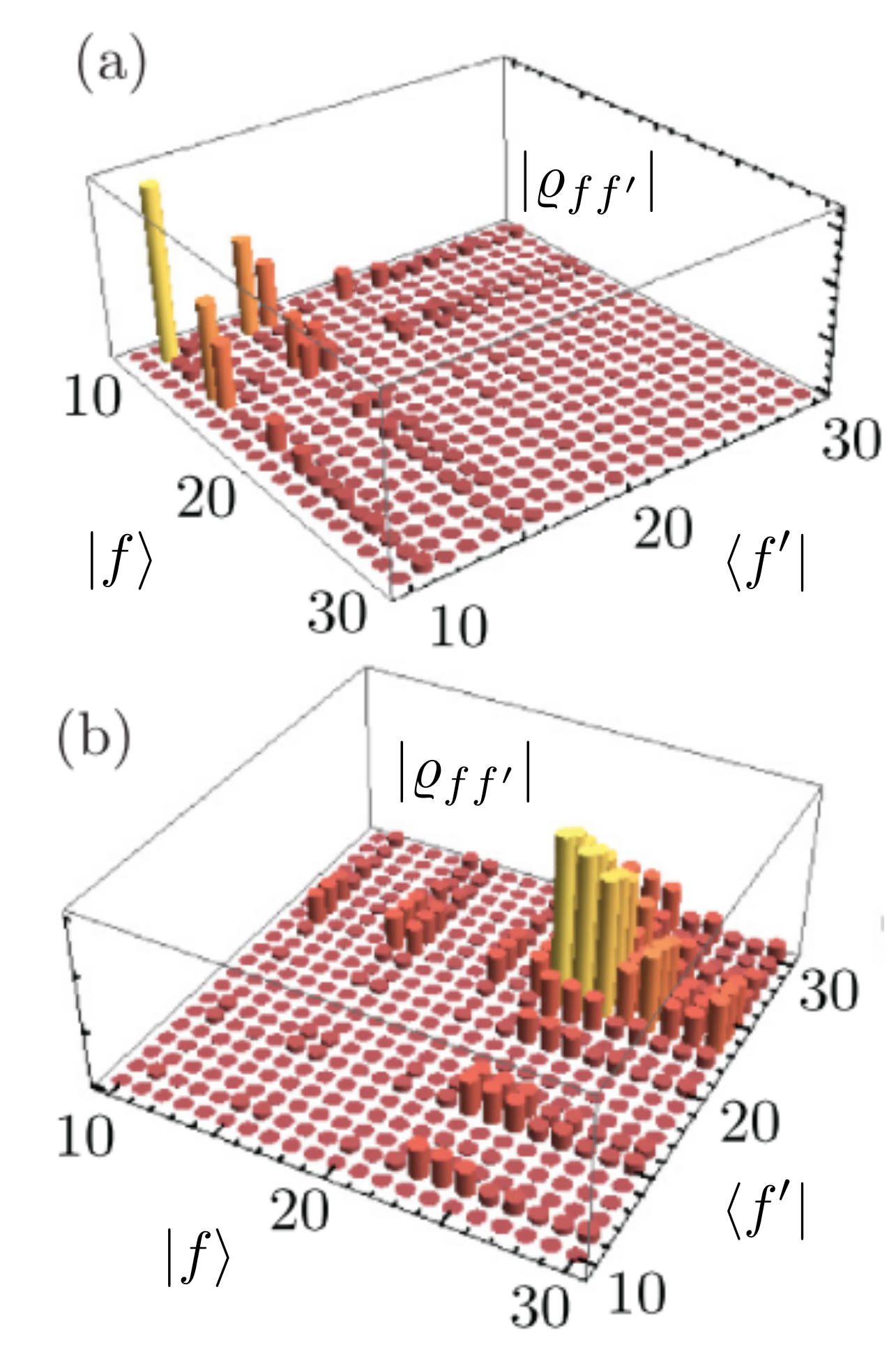}
\caption{a) The density matrix for doubly excited states $\varrho_{f f'} (t) = T^{\ast}_{f' g} (t) T_{fg} (t)$, Eq.~(\ref{eq.T_fg-mol-agg}), prepared by the absorption of entangled photons with pump frequency $\omega_p = 22,160$ cm$^{-1}$. b) the same, with $\omega_p = 24,200$ cm$^{-1}$. [from \cite{Schlawin12a}]}
\label{fig.density-matrix_mol-agg}
\end{figure}


Similarly to the noninteracting case~(\ref{eq:Tabdef}), the population in the final state $f$ at time t is given by the loop diagram in Fig.~\ref{fig.lop-vs-lap}, which can be written as the modulus square of a transition amplitude. Tracing out the matter degrees of freedom, the amplitude can be written as a nonlinear field operator,
\begin{align}
T_{fg} (t; \Gamma) &= - \frac{1}{\hbar^2} \int^t_{t_0} \!\! d\tau_2 \int^{\tau_2}_{t_0} \!\!\! d\tau_1  \notag E (\tau_2) E (\tau_1) \\
&\times \langle f(t) \vert V^{\dagger} (\tau_2) V^{\dagger} (\tau_1) \rangle \vert g (t_0) \rangle, \label{eq.T_fg}
\end{align}
which may be evaluated for arbitrary field input states. Specifically, using Eq.~(\ref{eq.two-photon-wavefunction-cw-time}), we obtain
\begin{align}
T_{fg}^{(ent)} (t; \Gamma) &= \frac{1}{\hbar^2 T} \frac{\mu_{ge} \mu_{ef}\mathcal{N}''A_p}{\omega_1 + \omega_2 - \omega_f + i \gamma_f} e^{- i (\omega_1 + \omega_2) t} \notag \\
&\times \big[ \frac{e^{i (\omega_1 - \omega_e + i \gamma_e) T} - 1}{\omega_1 - \omega_e + i \gamma_e} + \frac{e^{i (\omega_2 - \omega_e + i \gamma_e) T} - 1}{\omega_2 - \omega_e + i \gamma_e}  \big], \label{eq.T_fg-mol-agg}
\end{align}
where we used $\mathcal{N}=\mathcal{N}''A_p$ which is linear with respect to classical pump amplitude.
For comparison, the classical probability governed by excitation by two classical fields with amplitudes $A_1$ and $A_2$ is given by
\begin{align}
T_{fg}^{(c)} (t; \Gamma) &= \frac{1}{\hbar^2}\frac{\mu_{ge} \mu_{ef}A_1A_2e^{- i (\omega_1 + \omega_2) t}}{[\omega_1- \omega_{eg} + i \gamma_{eg}][\omega_2-\omega_{fe}+i\gamma_{fe}]} \notag \\
&+ \big( \omega_1 \leftrightarrow \omega_2 \big), \label{eq.T_fg-class}
\end{align}
which is quadratic in the field amplitudes, unlike Eq. (\ref{eq.T_fg-mol-agg}). The second line denotes the same quantity, but with the beam frequencies $\omega_1$ and $\omega_2$ interchanged. We will further discuss more on the intensity scaling.

Eq.~(\ref{eq.T_fg-mol-agg}) reflects the entangled photon structure described earlier: The pump frequency $\omega_p =  \omega_1 + \omega_2$ is sharply defined such that the $\omega_f$-resonance is broadened only by the state's lifetime broadening and pure dephasing rate $\gamma_f$. The strong time correlations in the arrival time creates a resonance of the form $(\exp (i \omega T) - 1) / \omega$, which, for very short entanglement times is independent of the frequency: $(\exp (i \omega T) - 1) / \omega \simeq T + \mathcal{O} (T^2)$. This implies that the intermediate $e$-states effectively interact with broadband light, whereas the $f$-states interact with cw light. Put differently, thanks to their large bandwidth, the entangled photons can induce all possible excitation pathways through the $e$-manifold leading to a specific selected $f$-state \cite{Schlawin12a}. Tuning the pump frequency $\omega_p$ allows to select the excitation of specific wave packets: Fig.~\ref{fig.density-matrix_mol-agg} depicts the density matrices for doubly excited states induced by the absorption of entangled photon pairs with different pump frequencies. This behavior remains the same for pulsed excitation in the strong frequency anti-correlations regime \cite{Schlawin15b}.

\subsubsection{Control of energy transfer}
\label{sec.control-transport}

When the material system is coupled to an environment which causes relaxation among levels, the  distributions of excited states populations may no longer be described by  the wavefunction and transition amplitudes, and the density matrix must be used instead. 
As shown in Fig.~\ref{fig.lop-vs-lap}, the loop diagram, which represent the transition amplitudes needs to be broken up into three fully time-ordered ladder diagrams (and their complex conjugates). These are given in the time domain \cite{Schlawin13b},
\begin{widetext}
\begin{align}
p_{f, (I)} (t; \Gamma) &=\left( - \frac{i}{\hbar} \right)^4 \int_{- \infty}^{t} \!\! d\tau_4 \int_{- \infty}^{\tau_4} \!\! d\tau_3 \int_{- \infty}^{\tau_3} \!\! d\tau_2 \int_{- \infty}^{\tau_2} \!\! d\tau_1 \notag \\
&\times \big\langle V_R (\tau_4) V_R (\tau_3) V_L^{\dagger} (\tau_2) V_L^{\dagger} (\tau_1) \big\rangle \big\langle E^{\dagger} (\tau_3) E^{\dagger} (\tau_4) E (\tau_2) E (\tau_1) \big\rangle, \label{eq.p_fI} \\
p_{f, \text{(II)}} (t; \Gamma) &= \left( - \frac{i}{\hbar} \right)^4 \int_{- \infty}^{t} \!\! d\tau_4 \int_{- \infty}^{\tau_4} \!\! d\tau_3 \int_{- \infty}^{\tau_3} \!\! d\tau_2 \int_{- \infty}^{\tau_2} \!\! d\tau_1 \notag \\
&\times\big\langle V_R (\tau_4) V_L^{\dagger} (\tau_3)  V_R (\tau_2)  V_L^{\dagger} (\tau_1) \big\rangle \big\langle E^{\dagger} (\tau_2) E^{\dagger} (\tau_4) E (\tau_2) E (\tau_1) \big\rangle, \label{eq.p_fII} \\
p_{f, \text{(III)}} (t; \Gamma) &= \left( - \frac{i}{\hbar} \right)^4  \int_{- \infty}^{t} \!\! d\tau_4 \int_{- \infty}^{\tau_4} \!\! d\tau_3 \int_{- \infty}^{\tau_3} \!\! d\tau_2 \int_{- \infty}^{\tau_2} \!\! d\tau_1\notag \\
&\times \big\langle V_L^{\dagger} (\tau_4)  V_R (\tau_3)  V_R (\tau_2)  V_L^{\dagger} (\tau_1) \big\rangle \big\langle E^{\dagger} (\tau_2) E^{\dagger} (\tau_3) E (\tau_4) E (\tau_1) \big\rangle, \label{eq.p_fIII}
\end{align}
\end{widetext}
Here, $\Gamma$ collectively denotes the set of control parameters of the light field. The matter correlation functions in Eqs.~(\ref{eq.p_fI}) - (\ref{eq.p_fIII}) are given by Liouville space superoperator correlation functions, which translate for the integration variables in Eqs.~(\ref{eq.p_fI})-(\ref{eq.p_fIII}) in Hilbert space into

\begin{align}
&\big\langle V_R (\tau_4) V_R (\tau_3) V_L^{\dagger} (\tau_2) V_L^{\dagger} (\tau_1) \big\rangle \notag \\ 
= &\big\langle V (\tau_3) V (\tau_4) V^{\dagger} (\tau_2) V^{\dagger} (\tau_1) \big\rangle, \\
&\big\langle V_R (\tau_4) V_L^{\dagger} (\tau_3)  V_R (\tau_2)  V_L^{\dagger} (\tau_1) \big\rangle \notag \\
= &\big\langle  V (\tau_2)  V (\tau_4) V^{\dagger} (\tau_3) V^{\dagger} (\tau_1) \big\rangle, \\
&\big\langle V_L^{\dagger} (\tau_4)  V_R (\tau_3)  V_R (\tau_2)  V_L^{\dagger} (\tau_1) \big\rangle \notag \\
= &\big\langle V (\tau_2)  V (\tau_3)  V^{\dagger} (\tau_4)  V^{\dagger} (\tau_1) \big\rangle.
\end{align}
The entangled photon field correlation functions are given by the Fourier transform of Eq.~(\ref{eq.four-pt-full-text}). We first restrict our attention to the weak downconversion regime, $\bar{n} \ll 1$, in which the autocorrelation functions $g_{1, 2}$, Eqs.~(\ref{eq.g1-definition}) and (\ref{eq.g2-definition}), may be neglected.

\begin{figure}[t]
\centering
\includegraphics[width=0.45\textwidth]{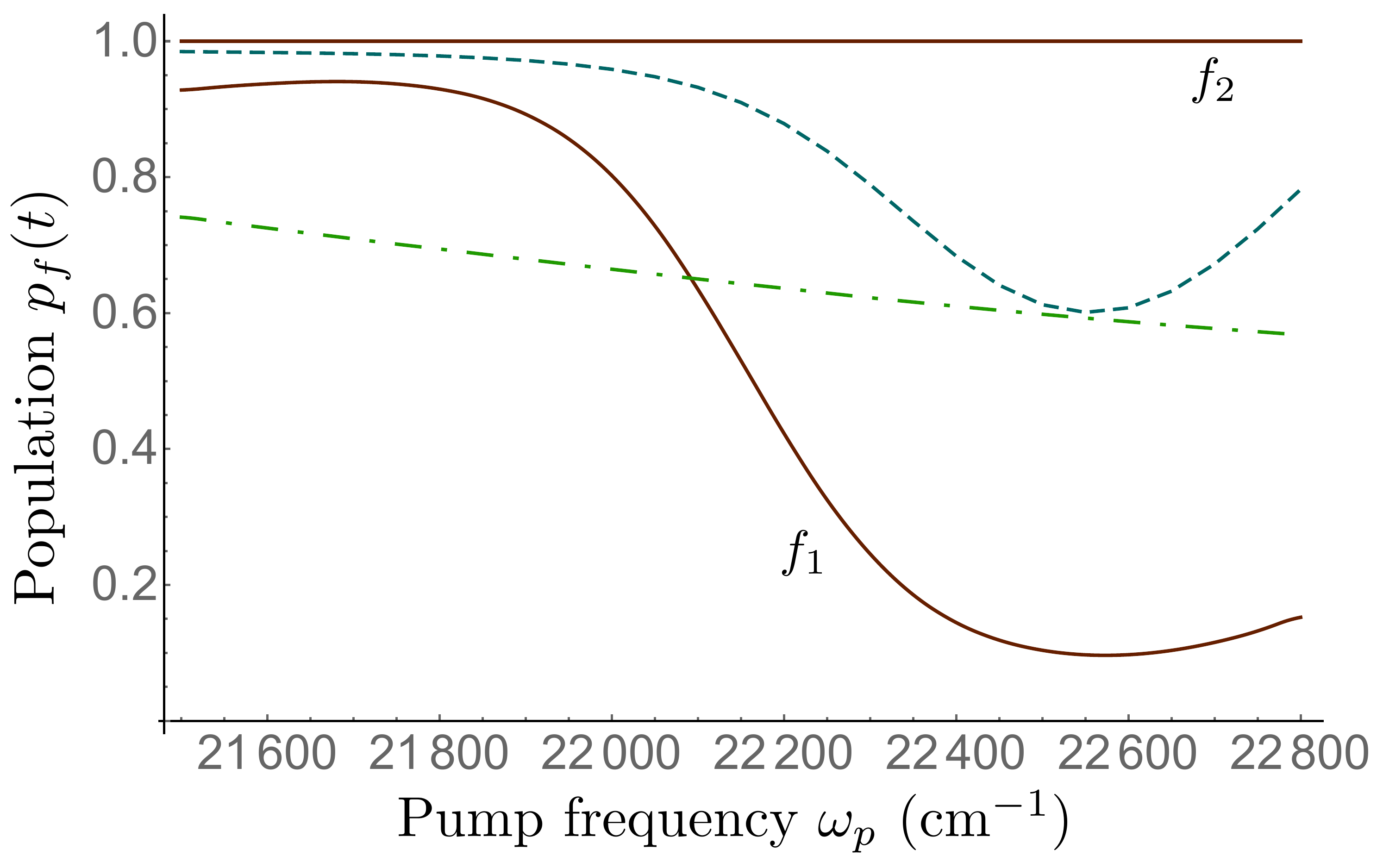}
\caption{Variation of the double-excited state populations $p_f (t)$, Eqs.~(\ref{eq.p_fI})-(\ref{eq.p_fIII}) with the pump frequency $\omega_p$ after excitation by entangled photons with strong frequency anti-correlations (solid, red), by laser pulses matching the pump bandwidth $\sigma_p$ (blue, dashed), and laser pulses matching the photon bandwidths $\sim 1 / T$ (green, dot-dashed). The total $f$-population is normalized to unity at each frequency, $i.e.$ $\sum_f p_f (t) = 1$.}
\label{fig.f-pop}
\end{figure}

The essential properties of entangled photon absorption may be illusttrated by using the simple model system introduced in section~\ref{sec.model-system} where the correlation functions~(\ref{eq.p_fI}) - (\ref{eq.p_fIII}) can be written as the sum-over-states expressions given in appendix~\ref{sec.model-system-appendix}. To excite state $f_2$ faithfully, the intermediate decay process $e_2 \rightarrow e_1$ needs to be blocked (see Fig.~\ref{fig.levelscheme}), and one needs to select the final state $f_2$ spectrally. This can be achieved with entangled photons as shown in Fig.~\ref{fig.f-pop}, where we depict the relative population in each final state vs. the pump frequency $\omega_p$. At each frequency  the total population is normalized to unity, $i.e.$ $p_{f_1} + p_{f_2} = 1$. By choosing a spectrally narrow pump bandwidth $\sigma_p = 100$~cm$^{-1}$ in combination with a short entanglement time $T = 10$~fs, almost 90 \% of the total $f$-population can be deposited in the state $f_2$ at $\omega_p = 22 500$~cm$^{-1}$ (and $\sim 95$~\% in $f_1$ at $\omega_p = 21 800$~cm$^{-1}$).

The same degree of state selectivity may not be achieved with classical laser pulses: For comparison, the $f$-manifold populations created by classical laser pulses with bandwidths $\sigma = 100$~cm$^{-1}$ (same spectral selectivity) and with $\sigma = 1000$~cm$^{-1}$ (same time resolution in the manifold of singly excited states) are shown as well in Fig. \ref{fig.f-pop}. In the former case, the intermediate relaxation process limits the maximal yield in $f_2$ to $\sim 35$~\% (with ca. $65$~\% population in $f_1$ at $\omega_p = 22 500$~cm$^{-1}$), in the latter case, the lower spectral resolution limits the achievable degree of control over the $f$-populations. 

For very short entanglement times, transport between various excited states in $e$-manifold may be neglected, and the $f$-populations may as well be calculated using transition amplitudes - thus greatly reducing the computational cost \cite{Schlawin15b}. At the same time, by varying the entanglement time becomes possible to probe subsets of transport pathways via the selection of specific $f$-states in the detected optical signal.

The properties described above may also be observed in more complex systems such as molecular aggregates \cite{Schlawin13b}, where the number of excited states is much higher, but the described physics is essentially the same: Ultrafast relaxation processes limit the classically achievable degree of selectivity due to the trade-off between spectral and time resolution for every single absorption process. With entangled photons, this trade-off only limits the overall two-photon process, but not each individual transition. This may be understood in the following way: In the absorption of entangled photons, the light-matter system becomes entangled in between the two absorption events, but it remains separable at all times for classical pulses so that the energy-time uncertainty only applies to the entire system.

Finally it is worth noting an additional advantage of using
entangled photons for  two-photon excitation in molecules pointed out by  \cite{Raymer13a}.  In many aggregates the doubly-excited state undergoes rapid nonradiative decay to the singly excited states. This may occur on the same time scale of the excited-state transport \cite{van00}. For weakly anharmonic aggregates where the transition frequencies $f\to e$ and $e\to g$ are close, it is hard to discriminate between the two-photon-excited  and single photon excited fluorescence. Entangled light can help isolate the doubly-excited state population by
monitoring the transmitted single-photon pathways with a high-quantum-efficiency detector, that
 can partially rule out one-photon absorption.

\begin{figure}[t]
\centering
\includegraphics[width=0.45\textwidth]{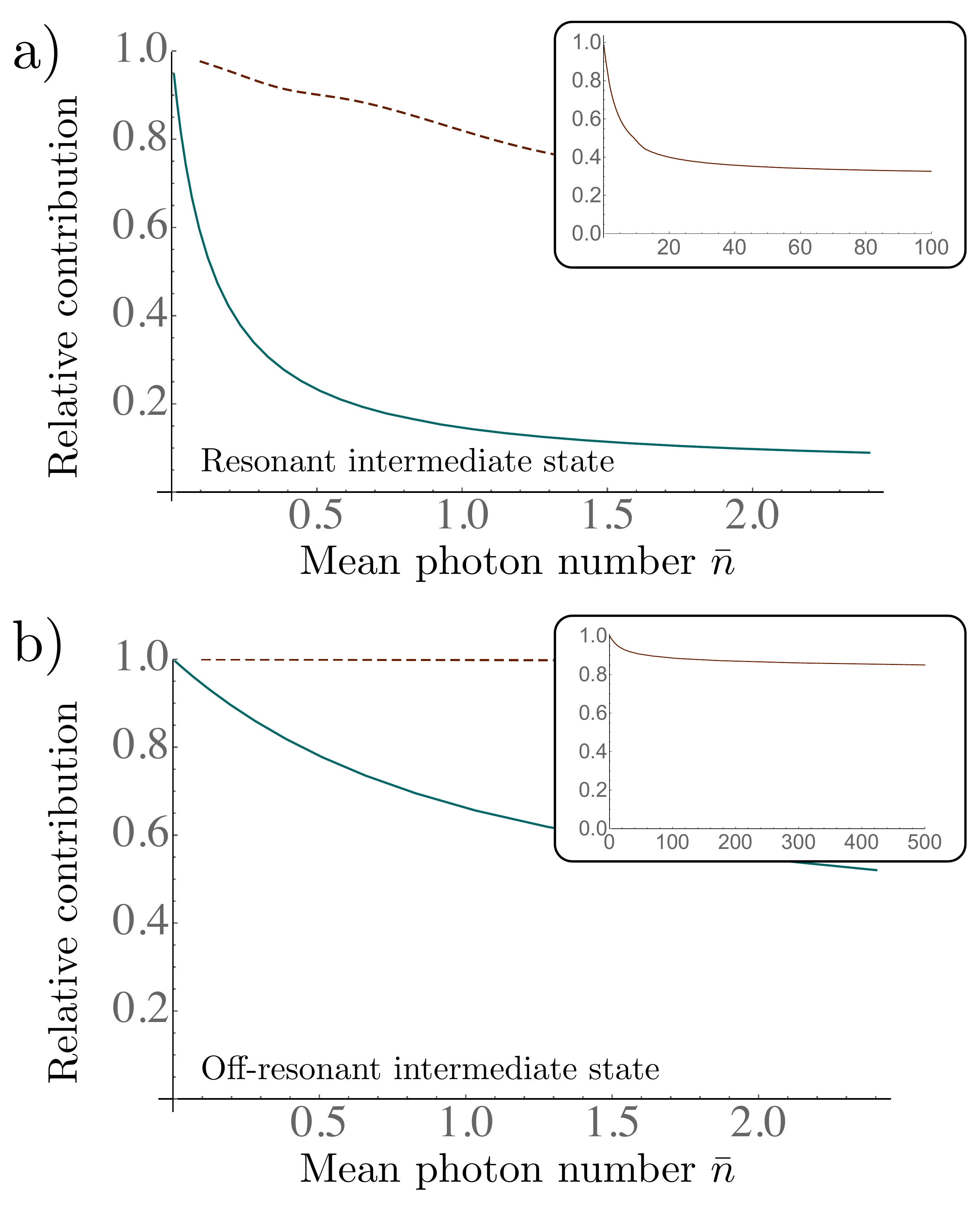}
\caption{a) Relative contribution of the coherent two-photon contribution $h_{12}$, Eq.~(\ref{eq.f12-definition}), to the full time-integrated population of $f_2$ plotted vs. the mean photon number $\bar{n}$, Eq.~(\ref{eq.mean-photon-number}). The dashed line shows strong entanglement with entanglement time $T = 10$~fs, and the solid line weaker entanglement with $T = 100$~fs. b) same for an intermediate state moved to $\omega_{e_2} = 18,500$ cm$^{-1}$.}
\label{fig.intensity}
\end{figure}

\subsubsection{Scaling of two photon absorption with pump intensity}
\label{sec.pump-intensity}

Eq. (\ref{eq.T_fg-class}) shows that TPA probability $|T_{fg}|^2$ scales quadratically with pump intensity $A_1^2A_2^2$ for classical light but Eq. (\ref{eq.T_fg-mol-agg}) shows linear scaling $A_p^2$ with weak entangled light. As the pump pulse intensity is increased, the two-photon state~(\ref{eq.ent-state1}) no longer represents the output state~(\ref{eq.psi_out}), and the autocorrelation contributions $g_{1, 2}$ to Eq.~(\ref{eq.four-pt-full-text}) must be taken into account. To understand how this affects the excited state distributions, we first simulate in Fig.~\ref{fig.intensity} the relative contribution of the two-photon term $h_{12}$ in Eq.~(\ref{eq.four-pt-full-text}) to the population in $f_2$, when the pump frequency is set at resonance, $\omega_p = 22 500$~cm$^{-1}$, and its variation with the mean photon number~(\ref{eq.mean-photon-number}). Clearly, in the weakly entangled case this contribution drops rapidly, and already at $\bar{n} = 0.5$ the autocorrelation contribution ($i.e.$ excitations by uncorrelated photons from different downconversion events) dominates the signal. In contrast, in the strongly entangled case the contribution drops much more slowly, and - as can be seen in the inset - still accounts for roughly a third of the total absorption events even at very high photon number $\bar{n} \simeq 100$. As described in \cite{Dayan05a}, this may be attributed to the strong time correlations in the former case: For $T = 10$ fs, one can fit ten times as many photon pairs into a time period compared to $T = 100$ fs, before different pairs start to overlap temporally. In addition, the strong frequency correlations discussed in the previous section imply that, on resonance with the final state $f_2$, the coherent contribution is enhanced even further.



The situation becomes even more striking  in panel b when the intermediate state $e$ is shifted away from the entangled photon frequencies), where we set it to $\omega_{e_2} = 18,500$ cm$^{-1}$, and only the two-photon transition $g \rightarrow f_2$ is resonant. In the weakly entangled case (solid line), little has changed, and parity between the two-photon and the autocorrelation contribution is reached again for $\overline{n} \sim 2$. In the strongly entangled regime, however, the coherent contribution remains close to unity, and dominates the signal even at $\bar{n} \simeq 500$, as can be seen in the inset.

The strong frequency correlations of entangled photon pairs lead to the enhancement of the two-photon contribution $h_{12}$ compared to the autocorrelation contributions $g_{1,2}$ - even for very high photon fluxes. Consequently, The nonclassical excited states distributions created by entangled photons can still dominate the optical signal  even when the mean photon number in each beam greatly exceeds unity \cite{Schlawin15b}. The linear to a quadratic crossover of the two-photon absorption rate with the pump intensity (Fig. \ref{fig.intro}a) which is traditionally regarded as a crossover from a quantum to a classical regime \cite{Lee06a}, is not necessarily a good indicator for this transition after all, as pointed out by \cite{geo97}. The time-frequency correlations of entangled pairs may be harnessed at much higher photon fluxes - as long as the coherent contribution $h_{12}$ dominates the incoherent contributions $g_{1,2}.$


\section{Nonlinear optical signals obtained with entangled light}
\label{sec.nonlinear-signals}

We now revisit  the excited state populations created in matter following the absorption of entangled photon pairs in section~\ref{sec.quantum light}, and present optical signals associated with these distributions. We first derive compact superoperator expressions for arbitrary field observables, which naturally encompass standard expressions based on a semiclassical treatment of the field \cite{Mukamel_book}.

Using Eq.~(\ref{eq.Dyson-series}), the expectation value of an arbitrary field operator $A (t)$ is given by 
\begin{align}
S (t; \Gamma) &= \big\langle A_+ (t) \big\rangle_{\text{final}} \\
&=  \big\langle \mathcal{T} A_+ (t) \exp \left[ - \frac{i}{\hbar} \int^t_{t_0} \!\! d\tau H_{\text{int},-} (\tau) \right] \big\rangle \label{eq.general-signal} \\
&= S_0 - \frac{i}{\hbar} \int^t_{t_0} \!\! d\tau \; \big\langle  \mathcal{T} A_+ (t) H_{\text{int}, -} (\tau) \notag \\
&\times \exp \left[ - \frac{i}{\hbar} \int^{\tau}_{t_0}\!\! d\tau' H_{\text{int}, -} (\tau') \right] \big\rangle.
\end{align}
Here, the first term describes the expectation value in the absence of any interaction with the sample, which we set to zero. The second term describes the influence of the sample on the expectation value. It may be represented more compactly, when the interaction Hamiltonian is not written in the RWA approximation \cite{Schlawin15b},
\begin{align}
S (t; \Gamma) &= - \frac{i}{\hbar} \int^t_{t_0} \!\! d\tau \; \big\langle \left[ A (t), \epsilon (\tau) \right]_+ \mathcal{V}_+ (\tau) \big\rangle_{\text{final}}, \label{eq.first-order-signal}
\end{align}
where we recall $\epsilon (t) = E (t) + E^{\dagger} (t)$ and $\mathcal{V} (t) = V (t) + V^{\dagger} (t)$. 

Eq.~(\ref{eq.first-order-signal}) may be used to calculate arbitrary optical signals induced by entangled photons, as will be demonstrated in the following.

\subsection{Stationary Nonlinear signals}
\label{sec.nl-signals_cavities}
Here, we consider the situation described in section~\ref{sec.cavities}, where the matter system is either coupled to a single cavity mode, and where only a single quantum mode of the light is relevant. This is most prominently the case in the strong coupling regime of cavity quantum electrodynamics \cite{Walther06}, or when one is interested in steady state solutions, where only a single (or few) frequency(ies) of the field is(are) relevant.

The intricate connection between nonlinear optical signals and photon counting was first made explicit by \cite{Mollow68}, who connected the two-photon absorption rate of stationary fields with their $G^{(2)}$-function\footnote{We change the notation to match the remainder of this review.}
\begin{align}
w_2 &= 2 \vert g (\omega_0) \vert^2 \int_{-\infty}^{\infty} \!\!\! dt \; e^{2 i \omega_f t - \gamma_f | t |} G^{(2)} (- t, - t; t, t),
\end{align}
where $g (\omega_0) \sim \mu_{eg} \mu_{eg} / (\omega_0 - \omega_e + i \gamma_e)$ is the coupling element evaluated at the central frequency $\omega_0$ of the stationary light field. The transition rate loosely corresponds to the time-integrated, squared two-photon transition amplitude~(\ref{eq.T_fg}) we derived earlier. 
If the lifetime of the final state $\gamma_f$ is much smaller than the bandwidth of the field $\Delta \omega$, $\gamma_f \ll \Delta \omega$, one may even neglect the time integration, and replace $G^{(2)} (-t, -t; t, t) \sim G^{(2)} (0)$.

This implies, as pointed out in \cite{Gea89}, that strongly bunched light can excite two-photon transitions more efficiently than classical light with identical mean photon number. It further implies that the two-photon absorption rate scales linearly in the low gain regime, even though the single mode squeezed state does not show time-energy entanglement in the sense of section~\ref{sec.photon-entanglement}. An experimental verification thereof is reported in \cite{geo95, geo97}. 

A more recent proposal \cite{Lopez15} aims to employ nonclassical fluctuations contained in the fluorescence of a strongly driven two-level atom: As is well known, the driven two-level atom's fluorescence develops side peaks - known as the Mollow triplet \cite{Scully97}. By driving polaritons - strongly coupled light-matter states in a cavity - with this light, it is predicted to allow for precise measurements of weak interactions between polaritons ``even in strongly dissipative environments".

\subsection{Fluorescence detection of nonlinear signals}

Fluorescence is given by the (possibly time-integrated) intensity $A (t) = E^{\dagger} (t) E (t)$, when the detected field mode is initially in the vacuum. 
It is obtained by expanding Eq.~(\ref{eq.first-order-signal}) to second order in the interaction Hamiltonian with the vacuum field \cite{Mukamel_book}, 
\begin{align}
S_{\text{FLUOR}} &= \frac{1}{\hbar^2}\big\langle V_L (t) V^{\dagger}_R (t) \big\rangle_{\text{final}}. \label{eq.fluorescence-signal}
\end{align}
We first investigate the time-integrated $f \rightarrow e$ fluorescence signal~(\ref{eq.fluorescence-signal}), following excitation by either entangled photons or classical pulses. Using Eqs.~(\ref{eq.fluorescence-signal}), and (\ref{eq.p_fI})-(\ref{eq.p_fIII}), we may readily evaluate the signal as
\begin{align}
S_{\text{TPIF}} (\Gamma) &= \int dt \; \sum_{e, f} \vert \mu_{ef} \vert^2 p_f (t; \Gamma). \label{eq.TPIF-definition}
\end{align}
In a different scenario, the two-excitation state may rapidly decay nonradiatively - $e.g.$ via internal conversion - and the $e - g$ fluorescence is detected,
\begin{align}
\tilde{S}_{\text{TPIF}} (\Gamma) &\propto \int \!\! dt \; p_f (t; \Gamma).
\end{align}
Fluorescence signals are proportional to excited state populations. Therefore, they are closely related to the excited state distributions discussed in sections~\ref{sec.matter-correlations} to \ref{sec.pump-intensity}.
This is not necessarily the case for absorption measurements, as will be shown in section~\ref{sec.N-photon-counting}.


\subsubsection{Two-photon absorption vs. two-photon-induced fluorescence}
\label{sec.TPA-vs-TPIF}
Before reviewing fluorescence signals induced by entangled photons, we comment on some ambiguity in the nomenclature of nonlinear signals such as fluorescence that depend on the doubly excited state population $p_f$. Since population is created by the absorption of two photons,  the signal is often termed \textit{two-photon absorption}. However, it does not pertain to a $\chi^{(3)}$-absorption measurement, which will be discussed in section~\ref{sec.N-photon-counting}. We will refer to signals measuring $p_f$ as \textit{two-photon-induced fluorescence} and to two-photon resonances in $\chi^{(3)}$ as two-photon absorption.

In section \ref{sec.TPA-vs-Raman}, we will further show that for off-resonant intermediate state(s) $e$ - as is the case in most experimental studies to date \cite{Dayan04a, Dayan05a, Lee06a, Harpham09a, Guzman10a, Upton13a} - the two signals carry the same information.



\subsubsection{Two-photon induced transparency and entangled-photon virtual-state spectroscopy}
\label{sec.virtual-state-spec}
\begin{figure}[t]
\centering
\includegraphics[width=0.45\textwidth]{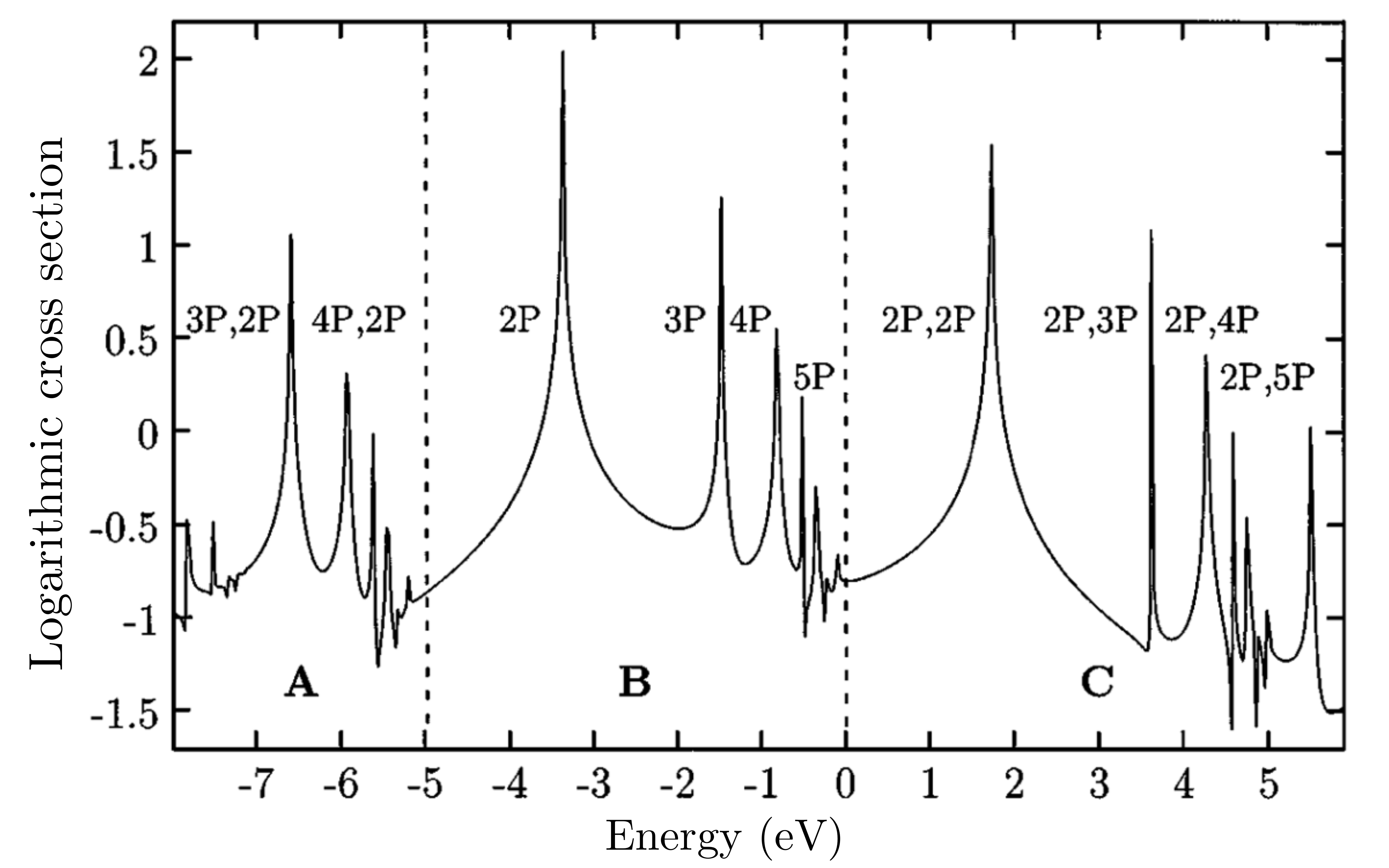}
\caption{Entangled virtual-state spectroscopy: TPIF signal vs. the Fourier transform of an additional time delay between the two photons (see text) in hydrogen [taken from \cite{Saleh98a}].}
\label{fig.virtual-state-spec}
\end{figure}

\textit{Entangled virtual-state spectroscopy}, proposed in \cite{Saleh98a}, suggests a means to detect far off-resonant intermediate states in the excitation of $f$-states by employing \textit{entanglement-induced two-photon transparency} \cite{Fei97a}: 
The transition amplitude~(\ref{eq.T_fg-mol-agg}) of entangled photons created by a cw-source oscillates as a function of the entanglement time $T$. For infinite excited-state lifetimes ($\gamma_e = 0$), the excitation of $f$ is even completely suppressed whenever $(\omega_1 - \omega_e) T = n \times 2 \pi$ $\forall n \in \mathbb{N}$ - the medium becomes transparent. 

Saleh et al. had proposed to turn this counterintuitive effect into a spectroscopic tool by sending one of the two photons through a variable delay stage $\tau$, which simply amounts to evaluating the two-photon wavefunction~(\ref{eq.two-photon-wavefunction-cw-time}) $\langle 0 \vert E (\tau_2 + \tau) E (\tau_1) \vert \psi_{\text{twin}} \rangle$. For degenerate downconversion the transition amplitude~(\ref{eq.T_fg-mol-agg}) then reads
\begin{align}
&T_{fg} (\tau, t; \Gamma) = \frac{\mathcal{N}' A_p}{\hbar^2 T} \sum_e \frac{\mu_{ge} \mu_{ef}}{\omega_p - \omega_f + i \gamma_f} \frac{ e^{- i \omega_p t} e^{- i \omega_p \tau /2}}{\Delta_e + i \gamma_e} \notag \\
&\times \big[ e^{i (\Delta_e + i \gamma_e) (T - \tau)} + e^{i (\Delta_e + i \gamma_e) (T + \tau)} - 2 \big], 
\end{align}
where we introduced the detuning $\Delta_e = \omega_p / 2 - \omega_e$.
Fourier transform of the TPIF signal $\propto \vert T_{fg} (\tau) \vert^2$ with respect to $\tau$ reveals different groups of resonances shown in Fig.~\ref{fig.virtual-state-spec} for the TPIF signal in hydrogen: Group A denotes resonances of differences between intermediate states, $\Delta_e - \Delta_{e'}$, group B of detunings $\Delta_e$, and group C of the sum of detunings $\Delta_e + \Delta_{e'}$. Such signals were simulated in molecular system \cite{Kojima04a}, and similar resonances can be identified in absorption TPA measurements, as we will discuss in section~\ref{sec.TPA-vs-Raman}.

However, it turns out that the effect depends crucially on the tails of the spectral distribution in Eq.~(\ref{eq.phase-matching-cw})  \cite{Leon-Montiel13a}. The sinc-function in the two-photon wavefunction~(\ref{eq.two-photon-wavefunction-cw}) has a long Lorentzian tail $\sim 1 / \omega$, which covers extremely far off-resonant intermediate states. When the sinc is replaced by Gaussian tails, the resonances vanish. The effect is thus caused by the long spectral tails, and is not intrinsically connected to entanglement.

\subsubsection{Fluorescence from multi-level systems}

\begin{figure}[t]
\centering
\includegraphics[width=0.45\textwidth]{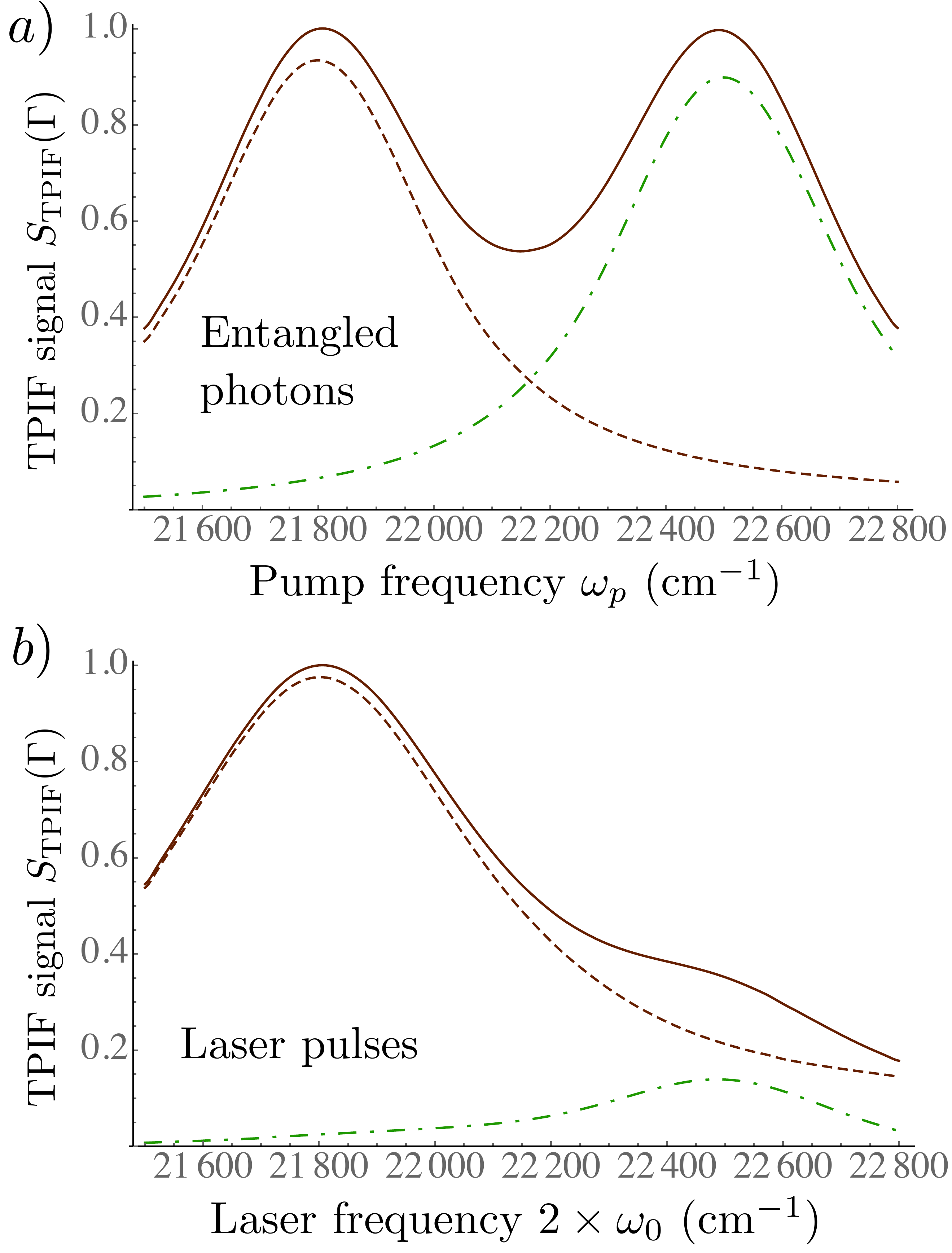}
\caption{a) TPIF action spectrum $S_{\text{TPIF}} (\Gamma)$, Eq.~(\ref{eq.TPIF-definition}), induced by entangled photon pairs with the same parameters as in Fig.~\ref{fig.f-pop}. The fluorescence created by the state $f_1$ is shown separately as a dashed plot, and the signal from $f_2$ as a dot-dashed line.
 b) TPIF action spectrum induced by laser pulses with bandwidth $\sigma_0 = 100$~cm$^{-1}$.}
\label{fig.actionspectrum}
\end{figure}

The TPIF signal, Eq.~(\ref{eq.TPIF-definition}), directly reflects the doubly excited state population distributions created by the absorption of entangled photon pairs. This may be seen in simulations of the TPIF signal from Frenkel excitons in a molecular aggregates \cite{Schlawin12a}. In Fig.~\ref{fig.actionspectrum}a), we present the TPIF signal resulting from the decay of the $f$-state distributions in Fig.~\ref{fig.f-pop}. Clearly, the signal peaks whenever a two-excitation state is on resonance with the pump frequency $\omega_p$. The signal from the two $f$-states has approximately the same strength. In contrast, the TPIF signal created by classical pulses shown in panel b) has a strong resonance only when the light is on resonance with the state $f_1$, whereas $f_2$ can only be observed as a weak shoulder of the main resonance. As discussed in chapter~\ref{sec.matter-correlations}, the fast decay of the intermediate state $e_2$ limits the excitation of $f_2$ with classical light, and therefore the state can hardly be observed, even though it has the same dipole strength as $f_1$.

\begin{figure}[t]
\centering
\includegraphics[width=0.4\textwidth]{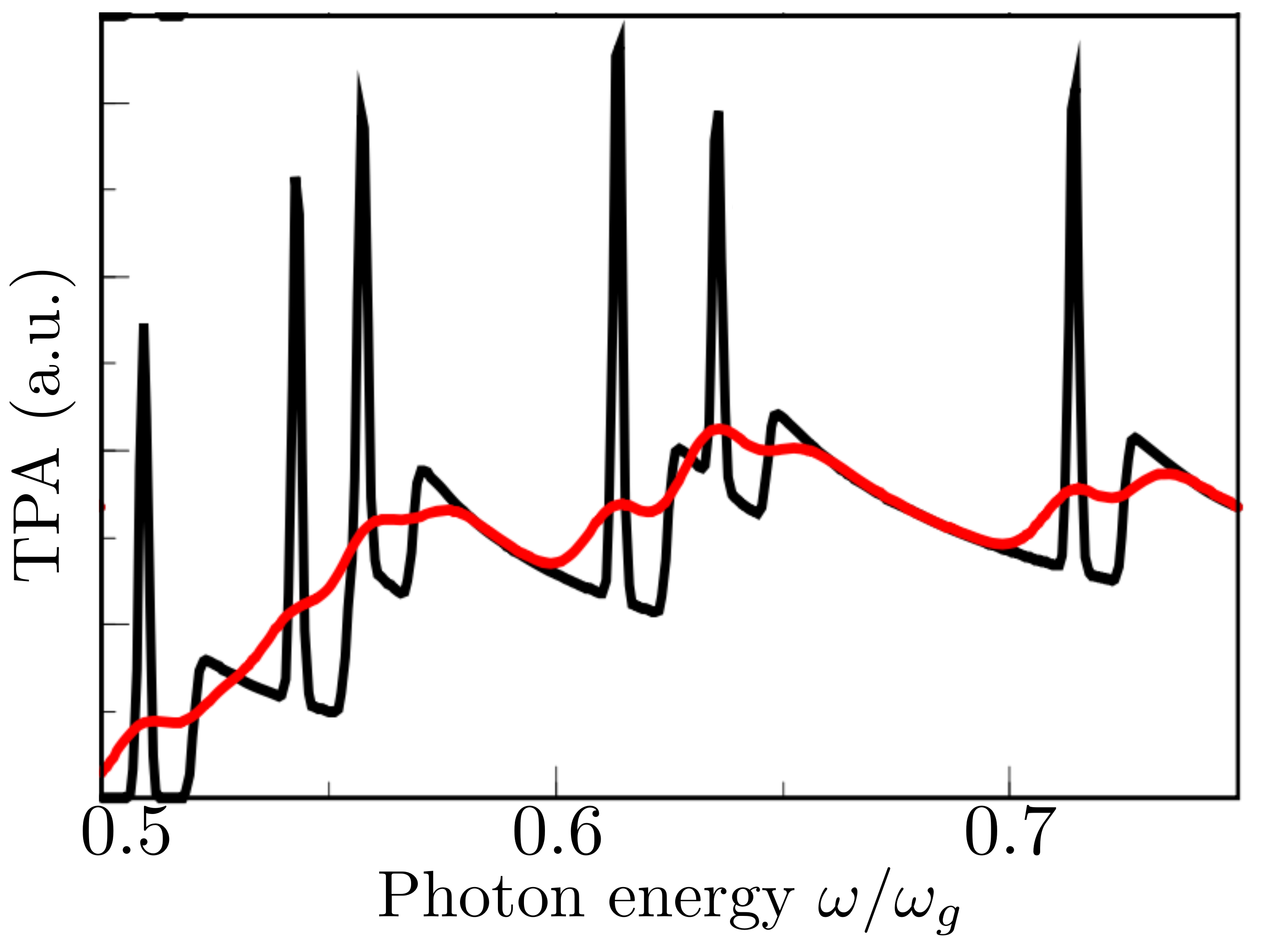}
\caption{Two-photon absorption (TPIF in our notation, see section.~\ref{sec.TPA-vs-TPIF}) in a semiconductor quantum well [taken from \cite{Salazar12a}]: excited state resonances within a continuum of delocalized states may be enhanced with frequency anti-correlations (black) or suppressed with positive frequency correlations (red).}
\label{fig.quantum-well}
\end{figure}

These nonclassical bandwidth features of entangled photon pairs may be further exploited to control vibronic states, as reported in \cite{Oka11a, Oka11b}. They were also investigated in semiconductor quantum wells. In this system, the absorption of excited state states competes with the absorption into a continuum of excited electronic states of free electrons and holes. By interpolating between negative and positive frequency correlations (see Fig.~\ref{fig.freq-correlations}), it was shown in \cite{Salazar12a} how the absorption of the excited states may be either enhanced or suppressed, as shown in Fig.~\ref{fig.quantum-well}.

\subsubsection{Multidimensional signals}
\label{sec.MD-signals}

\begin{figure*}
\centering
\includegraphics[width=0.8\textwidth]{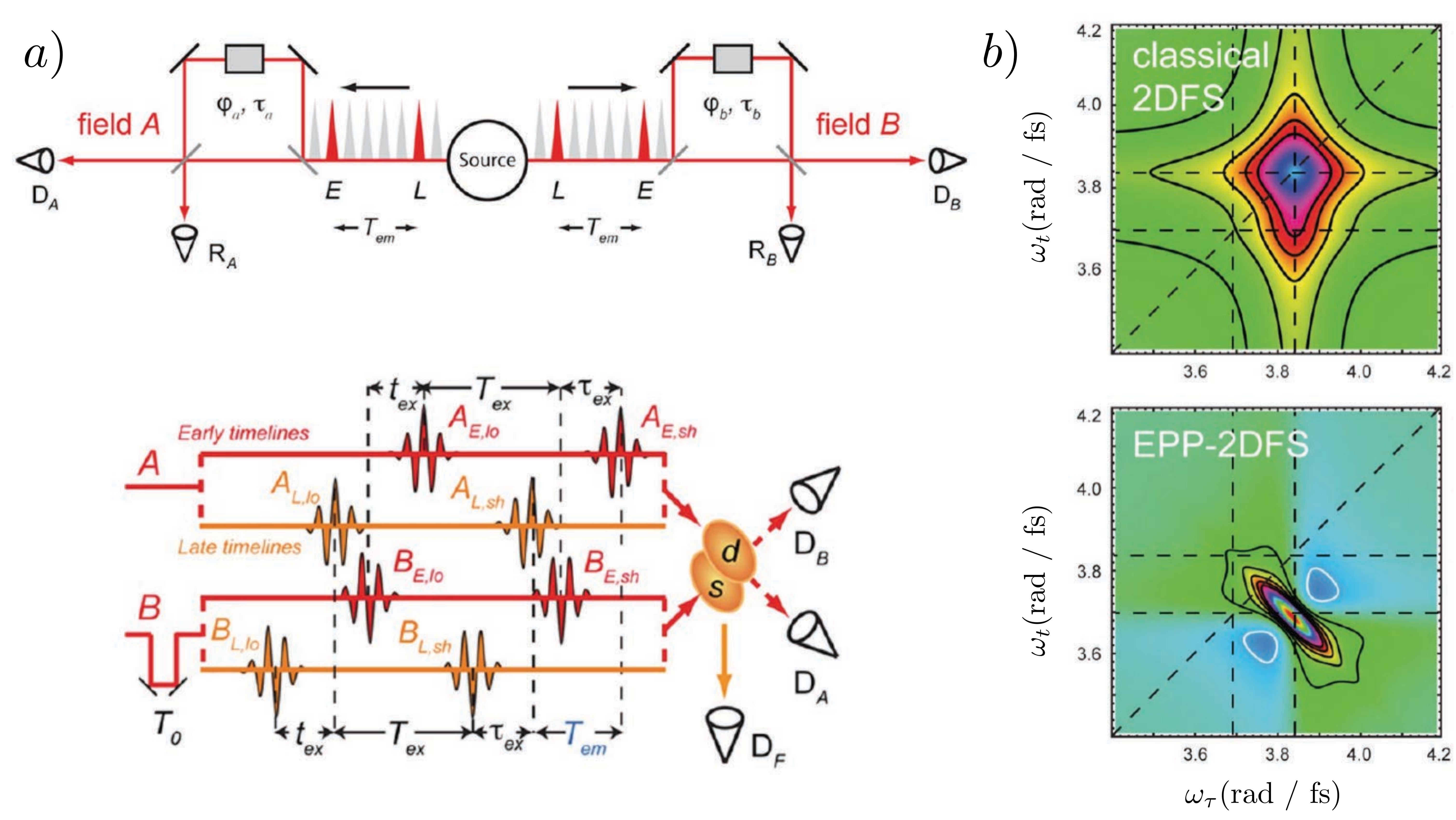}
\caption{a) Experimental setup for the LOP protocol: By sending each beam through a Franson interferometer, the photon excitation may occur either via a delayed or an ``early" photon. b) Top panel: two-dimensional fluorescence spectrum induced by classical light. Bottom panel: the same spectrum induced by entangled photons. [taken from \cite{Raymer13a}]}
\label{fig.Raymer2D}
\end{figure*}

In Fig.~\ref{fig.lop-vs-lap}, we had explained how each light-matter interaction event also imprints the light phase $\phi$ onto the matter response. We now exploit this fact to create multidimensional spectroscopic signals.
Phase-cycling is essential for partially non-collinear or collinear geometry 2D spectroscopic experiments
\cite{Yan2009,Keu99,Scheurer01,Tia03,Tan08,Zha12,Domcke13}. In phase-cycling protocol, the desired 2D signals are retrieved by the weighted summation of data collected using different interpulse phases, $\phi_{21}$, which are cycled over $2\pi$ radians in a number of equally spaced steps. In the pump-probe configuration, Myers et al. \cite{Myers2008} have shown that other than the pure absorptive signal, the rephasing and nonrephasing contributions may also be retrieved. 
This phase difference detection can be defined as a two-step phase-cycling scheme. For the phase difference detection, the Ogilvie group needs to collect signals with interpulse phases of $\phi_{21}=0$, $\pi$, $\pi/2$ and $3\pi/2$. The same signal can be obtained with chopping, but this is only half as intense compared to the phase difference method. In the context of this article, what they have performed is similar to a four-step phase-cycling scheme, where four sets of data need to be collected and linearly combined.

Phase cycling provides a means to post-select signals with a certain phase signature, as will be shown in the next section. Using additional delay stages - like in the entangled-photon virtual state spectroscopy discussed above - these signals can be spread to create multidimensional frequency correlation maps, which carry information about couplings between different resonances or relaxation mechanisms \cite{Ginsberg09}.

\cite{Raymer13a} had exploited the formal similarity between the TPIF and a photon coincidence signal to propose the setup shown in Fig.~\ref{fig.Raymer2D}a). Here, the successive absorption events promoting the molecule into the $f$-state are modulated by phase cycling and delay stages in both photon beams, creating interference effects between absorption events in which the photon takes the short path through the interferometer, and those in which it takes the long path. The Fourier transform creates two-dimensional signals which are shown in panel b). In contrast to the classical signal shown in the upper panel, the proposed scheme only shows resonances at the cross-peak between the two electronic states. This could provide a useful tool to study conformations of aggregates. We next discuss this strategy in more detail.


\subsubsection{Loop (LOP) vs ladder (LAP) delay scanning protocols for multidimensional fluorescence detected signals}
\label{sec.lop-vs-lap}
\begin{figure*}[t]
\begin{center}
\includegraphics[trim=0cm 0cm 0cm 0cm,angle=0, width=0.89\textwidth]{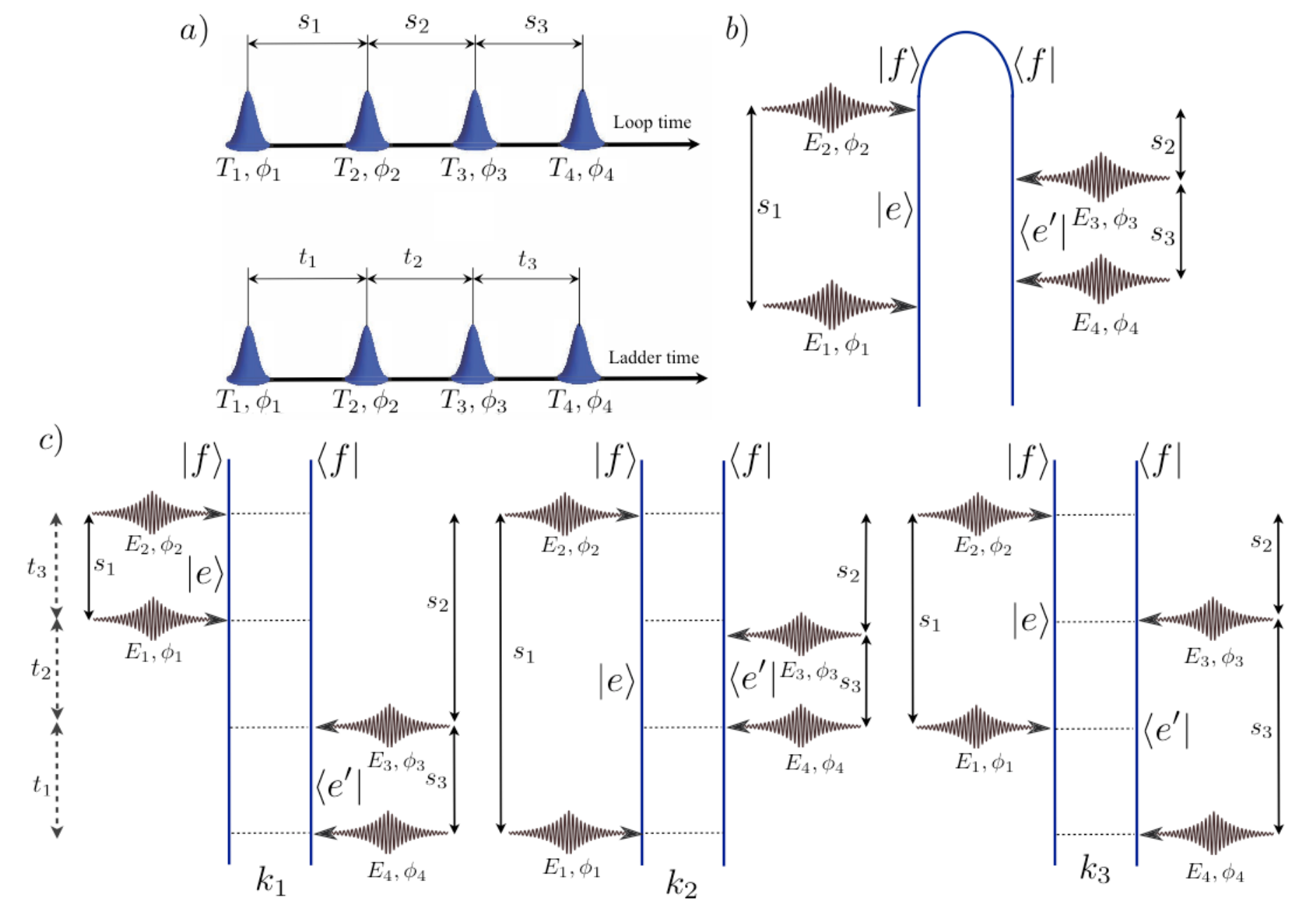}
\end{center}
\caption{(Color online) a) Pulse sequence for LOP (top panel) and LAP (bottom panel) scanning protocols. b) Loop diagram for the TPA process with indicated loop delays for the non specified phase cycling that depends on chronological time ordering between pulses $1$ to $4$ c). Ladder diagrams for the TPA signal with selected phase cycling component corresponding to double quantum coherence term $\mathbf{k}_I$: $-\phi_d-\phi_c+\phi_b+\phi_a$, rephasing $\mathbf{k}_{II}$: $-\phi_4+\phi_2+\phi_1-\phi_3$, and nonrephasing $\mathbf{k}_{III}$: $-\phi_4+\phi_2-\phi_3+\phi_1$. Both loop $s_j$ and ladder $t_j$ delays, $j=1,2,3$ are indicated. The transformation between two is different for each diagram. Time translation invariance implies $\omega_1+\omega_2-\omega_3-\omega_4=0$.}
\label{fig:diag}
\end{figure*}
Since the loop  and the ladder diagrams involve different time  variables they suggest different multidimensional signals obtained by scanning the  corresponding time delays.
We consider the TPIF signal created by a train of four pulses centered at times $T_1$, $T_2$, $T_3$, and $T_4$ with phases $\phi_1$, $\phi_2$, $\phi_3$, and $\phi_4$ \cite{Tek07,Pes09}.
We first analyze signals obtained by the LOP/LAP delay scanning protocols with classical light. The two protocols highlight different resonances and processes. Below we demonstrate  what type of information can be extracted from each protocol for excited electronic states in a model molecular aggregate. We then show some benefits of LOP protocols when applied to entangled light.

For the system model shown in Fig. \ref{fig.lop-vs-lap}a) the signal (\ref{eq.TPIF-definition}) is given by the single loop diagram in Fig. \ref{fig.lop-vs-lap}b). $a,b,c,d$ denote the pulse sequence {\it ordered along the loop} (but not in real time); $a$ represents ``first''.  on the loop etc. Chronologically-ordered pulses in real time will be denoted $1,2,3,4$ which are permutations of $a,b,c,d$, as will be shown below. One can scan various delays $T_\alpha-T_\beta$, $\alpha,\beta=1, \cdots 4$ and control the phases $\pm\phi_1\pm\phi_2\pm\phi_3\pm\phi_4$. In standard multidimensional techniques the time variables represent the pulses as they interact with sample in chronological order \cite{Mukamel_book}. These are conveniently given by the ladder delays $t_i$ shown in Fig.~\ref{fig:diag}c). The LAP maintains complete time-ordering of all four pulses.  The arrival time of the various pulses in chronological order is $T_1<T_2<T_3<T_4$. 
The ladder delays are defined as $t_1=T_2-T_1$, $t_2=T_3-T_2$, $t_3=T_4-T_3$. 
One can then use phase cycling to select the rephasing and nonrephasing diagrams shown in Fig. \ref{fig:diag}c, and read off the signal in a sum-over-states expansion
\begin{align}\label{eq:Sk22}
S_{\mathbf{k}_{II}}^{(LAP)}(t_1,t_2,t_3)&=\mathcal{I}\mathcal{E}_1^{*}\mathcal{E}_2^{*}\mathcal{E}_3\mathcal{E}_4\sum_{e,e',f}\mu_{ge'}\mu_{e'f}\mu_{fe}^{*}\mu_{eg}^{*}\notag\\
&\times\mathcal{G}_{ef}(t_3)\mathcal{G}_{ee'}(t_2)\mathcal{G}_{eg}(t_1),
\end{align}
\begin{align}\label{eq:Sk32}
S_{\mathbf{k}_{III}}^{(LAP)}(t_1,t_2,t_3)&=\mathcal{I}\mathcal{E}_1^{*}\mathcal{E}_2^{*}\mathcal{E}_3\mathcal{E}_4\sum_{e,e',f}\mu_{ge'}\mu_{e'f}\mu_{fe}^{*}\mu_{eg}^{*}\notag\\
&\times\mathcal{G}_{ef}(t_3)\mathcal{G}_{ee'}(t_2)\mathcal{G}_{ge'}(t_1).
\end{align}
The LAP signals (\ref{eq:Sk22}) - (\ref{eq:Sk32}) factorize into a product of several Green's functions with uncoupled time arguments. This implies that the corresponding frequency domain signal will also factorize into a product of individual Green's functions, each depending on a single frequency argument $\tilde{\Omega}_j$, $j=1,2,3$ which yields 
 \begin{align}\label{eq:Sk23}
&S_{\mathbf{k}_{II}}^{(LAP)}(\tilde{\Omega}_1,t_2 = 0,\tilde{\Omega}_3) \notag \\
=&\mathcal{R}\sum_{e,e',f}\frac{\mu_{ge'}\mu_{e'f}\mu_{fe}^{*}\mu_{eg}^{*}\mathcal{E}_1^{*}\mathcal{E}_2^{*}\mathcal{E}_3\mathcal{E}_4}{[\Omega_3-\omega_{ef}+i\gamma_{ef}][\Omega_1-\omega_{eg}+i\gamma_{eg}]},
\end{align}
\begin{align}\label{eq:Sk33}
&S_{\mathbf{k}_{III}}^{(LAP)}(\tilde{\Omega}_1,t_2 = 0,\tilde{\Omega}_3)\notag \\
= &\mathcal{R}\sum_{e,e',f}\frac{\mu_{ge'}\mu_{e'f}\mu_{fe}^{*}\mu_{eg}^{*}\mathcal{E}_1^{*}\mathcal{E}_2^{*}\mathcal{E}_3\mathcal{E}_4}{[\Omega_3-\omega_{fe}+i\gamma_{fe}][\Omega_1-\omega_{e'g}+i\gamma_{e'g}]}.
\end{align}
This factorization holds only in the absence of additional correlating mechanisms the frequency variables caused by $e.g.$ dephasing (bath) or the state of light.

In the LOP the time ordering of pulses is maintained only on each branch of the loop but not between branches. To realize the LOP experimentally the indices $1$ to $4$ are assigned as follows: first by phase cycling we select a signal with phase $\phi_1+\phi_2-\phi_3-\phi_4$. The two pulses with positive phase detection are thus denoted $1$, $2$ and with negative phase - $3$, $4$. In the $1$, $2$ pair pulse $1$ comes first. In the $3$, $4$ pair pulse $4$ comes first. The time variables in Fig.~\ref{fig:diag}b) are $s_1=T_2-T_1$, $s_2=T_3-T_2$, $s_3=T_3-T_4$. With this choice $s_1$ and $s_3$ are positive whereas $s_2$ can be either positive or negative. This completely defines the LOP experimentally. 

When the electronic system is coupled to a bath, it cannot be described by a wave function in the reduced space where the bath is eliminated. As described in section~\ref{sec.control-transport}, the loop diagram must then be broken into several ladder diagrams shown in Fig. \ref{fig.lop-vs-lap}b) which represent the density matrix. The full set of diagrams and corresponding signal expressions are given in \cite{Dorfman14b}. In Fig.~\ref{fig:diag}c) we present simplified expressions for the rephasing $\mathbf{k}_{II}$ and non-rephasing $\mathbf{k}_{III}$ signals in the limit of well separated pulses:
\begin{align}\label{eq:Sk20}
S_{\mathbf{k}_{II}}^{(LOP)}&(s_1,s_2,s_3)=\mathcal{I}\mathcal{E}_1^{*}\mathcal{E}_2^{*}\mathcal{E}_3\mathcal{E}_4\sum_{e,e',f}\mu_{ge'}\mu_{e'f}\mu_{fe}^{*}\mu_{eg}^{*}\notag\\
&\times\mathcal{G}_{ef}(s_2)\mathcal{G}_{ee'}(s_3)\mathcal{G}_{eg}(s_1-s_2-s_3)
\end{align}
\begin{align}\label{eq:Sk30}
S_{\mathbf{k}_{III}}^{(LOP)}&(s_1,s_2,s_3)=\mathcal{I}\mathcal{E}_1^{*}\mathcal{E}_2^{*}\mathcal{E}_3\mathcal{E}_4\sum_{e,e',f}\mu_{ge'}\mu_{e'f}\mu_{fe}^{*}\mu_{eg}^{*}\notag\\
&\times\mathcal{G}_{ef}(s_2)\mathcal{G}_{ee'}(s_1-s_2)\mathcal{G}_{ge'}(s_2+s_3-s_1),
\end{align}
where $\mathcal{I}$ denotes the imaginary part and $\mathcal{G}_{\alpha\beta}(t)=(-i/\hbar)\theta(t)e^{-[i\omega_{\alpha\beta}+\gamma_{\alpha\beta}]t}$ is the Liouville space Green's function [see Eq.~(\ref{eq.Greens-fct})]. Note that the loop delays $s_j$, $j=1,2,3$ are coupled and enter e.g. in the Green's function $\mathcal{G}_{eg}$ in Eq. (\ref{eq:Sk20}). Due to the Heaviside-theta function in this Green's function, the delays $s_j$ are not independent but rather couple the dynamics of the system during these delay times, which eventually result in cross-resonances in multidimensional spectra. To see the effect on the mixing of the frequency variables that yield these cross-peaks we take a Fourier transform of Eqs. (\ref{eq:Sk20}) - (\ref{eq:Sk30}) with respect to loop delay variable $s_1$ and $s_3$ keeping $s_2=0$ and obtain the resonant component of the signal analogous to the frequency-domain LAP signals (\ref{eq:Sk23}) - (\ref{eq:Sk33})

 \begin{align}\label{eq:Sk21}
&S_{\mathbf{k}_{II}}^{(LOP)}(\Omega_1,s_2 = 0,\Omega_3)=\notag\\
&\mathcal{R}\sum_{e,e',f}\frac{\mu_{ge'}\mu_{e'f}\mu_{fe}^{*}\mu_{eg}^{*}\mathcal{E}_1^{*}\mathcal{E}_2^{*}\mathcal{E}_3\mathcal{E}_4}{[\Omega_1+\Omega_3-\omega_{ee'}+i\gamma_{ee'}][\Omega_1-\omega_{eg}+i\gamma_{eg}]},
\end{align}
\begin{align}\label{eq:Sk31}
&S_{\mathbf{k}_{III}}^{(LOP)}(\Omega_1,s_2 = 0,\Omega_3)=\notag\\
&\mathcal{R}\sum_{e,e',f}\frac{\mu_{ge'}\mu_{e'f}\mu_{fe}^{*}\mu_{eg}^{*}\mathcal{E}_1^{*}\mathcal{E}_2^{*}\mathcal{E}_3\mathcal{E}_4}{[\Omega_1+\Omega_3+\omega_{ee'}+i\gamma_{ee'}][\Omega_3-\omega_{e'g}+i\gamma_{e'g}]},
\end{align}
where $\mathcal{R}$ denotes real part. Eqs. (\ref{eq:Sk21}) - (\ref{eq:Sk31})  yield  cross-peaks $\Omega_1+\Omega_3=\omega_{ee'}$. The time correlations therefore result in the frequency mixing of arguments.

\begin{figure}[t]
\begin{center}
\includegraphics[trim=0cm 0cm 0cm 0cm,angle=0, width=0.45\textwidth]{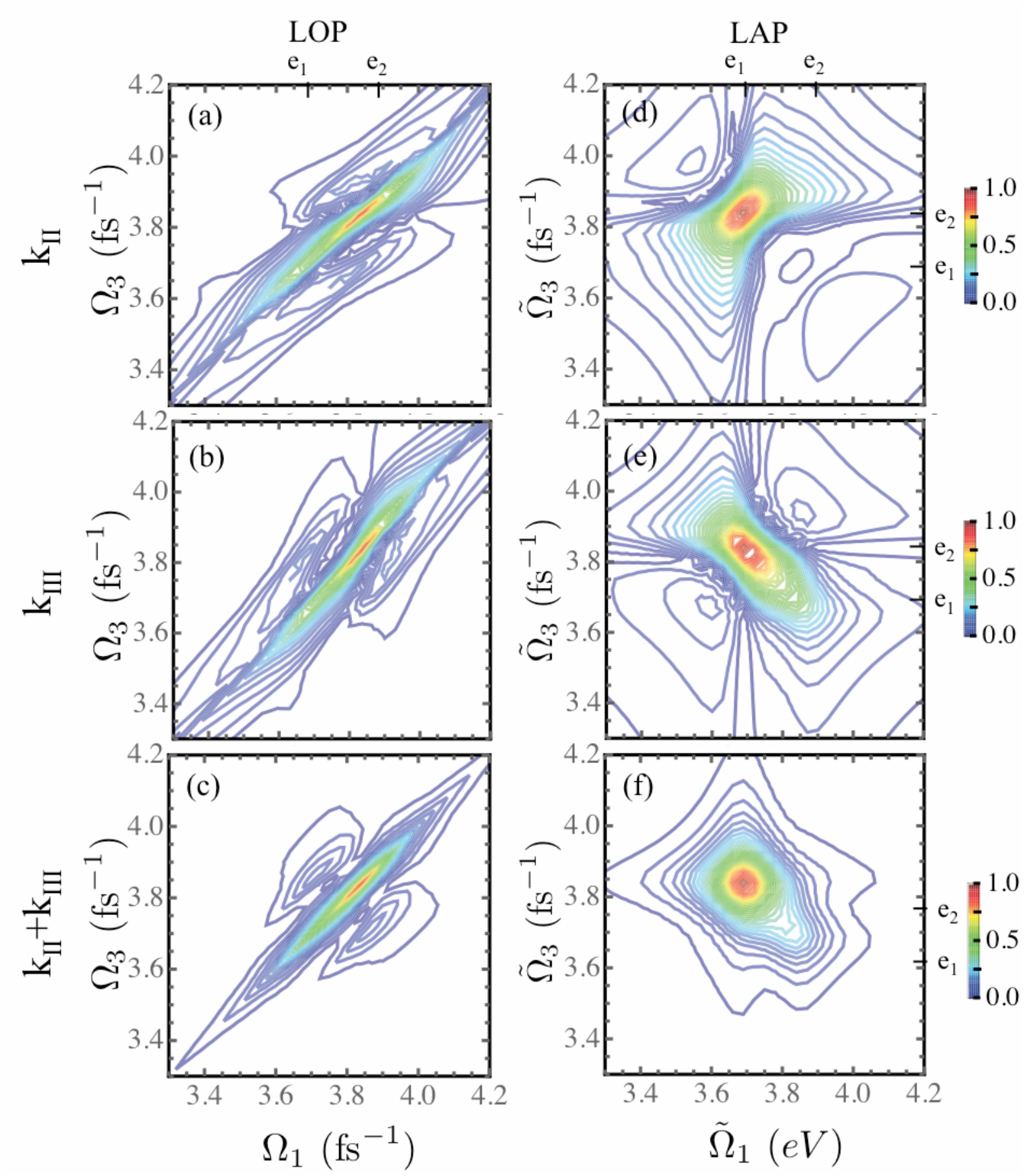}
\end{center}
\caption{(Color online) Left column: $S_{LOP}(\Omega_1,\tau_2=0,\Omega_3)$   for the molecular dimer model of Ref. \cite{Raymer13a} calculated using classical light for rephasing $\mathbf{k}_{II}$ Eq. (\ref{eq:Sk21}) -  (a), nonrephasing $\mathbf{k}_{III}$ Eq. (\ref{eq:Sk31}) - (b) and the sum of two $\mathbf{k}_{II}+\mathbf{k}_{III}$ - (c). Right column: same for  $S_{LAP}(\tilde{\Omega}_1,t_2=0,\tilde{\Omega}_3)$ Eq. (\ref{eq:Sk23}) - (\ref{eq:Sk33}). Note that the the sum of rephasing and nonrephasing components for LAP signal we flipped the rephasing component to obtain absorptive peaks as is typically done in standard treatment of photon echo signals. The difference between the two columns stems from the display protocol and is unrelated to entanglement.}
\label{fig:loplap}
\end{figure}

Fig. \ref{fig:loplap} compares  the LOP and LAP signals for the model dimer parameters of  \cite{Raymer13a} under two photon excitation by classical light. The LOP spectra for rephasing $\mathbf{k}_{II} = - \mathcal{k}_1 + \mathcal{k}_2 + \mathcal{k}_3$, nonrephasing $\mathbf{k}_{III}= + \mathcal{k}_1 + \mathcal{k}_2 - \mathcal{k}_3$ and their sum are shown in Fig.~\ref{fig:loplap}a)-c), respectively. The corresponding LAP spectra are shown in  Fig.  \ref{fig:loplap}d-f. We see that the scanning protocol makes a significant difference as seen by the two columns. The LOP resonances are narrow and clearly resolve the $e_1$ and $e_2$ states whereas the corresponding LAP signals are broad and featureless. This is a consequence of the display variables chosen in each protocols. LOP variables are coupled in a very specific fashion that allows to extract the intraband dephasing rate $\gamma_{ee'}$ in $(\Omega_1,\Omega_3)$ plot with higher precision compared to LAP case. Of course, one can extract similar information using LAP if displayed using $(\tilde{\Omega}_1,\tilde{\Omega}_2)$ or $(\tilde{\Omega}_2,\tilde{\Omega}_3)$ which we will discuss below in the context of entangled light. This difference in two protocols has been originally attributed to entanglement by \cite{Raymer13a}. However, it was explicitly demonstrated in \cite{Dorfman14b} supported by Fig. \ref{fig:loplap} that they calculated entangled signals using LOP whereas they used LAP for classical signals.

\paragraph{Entangled vs classical light}

\begin{figure}[t]
\begin{center}
\includegraphics[trim=0cm 0cm 0cm 0cm,angle=0, width=0.45\textwidth]{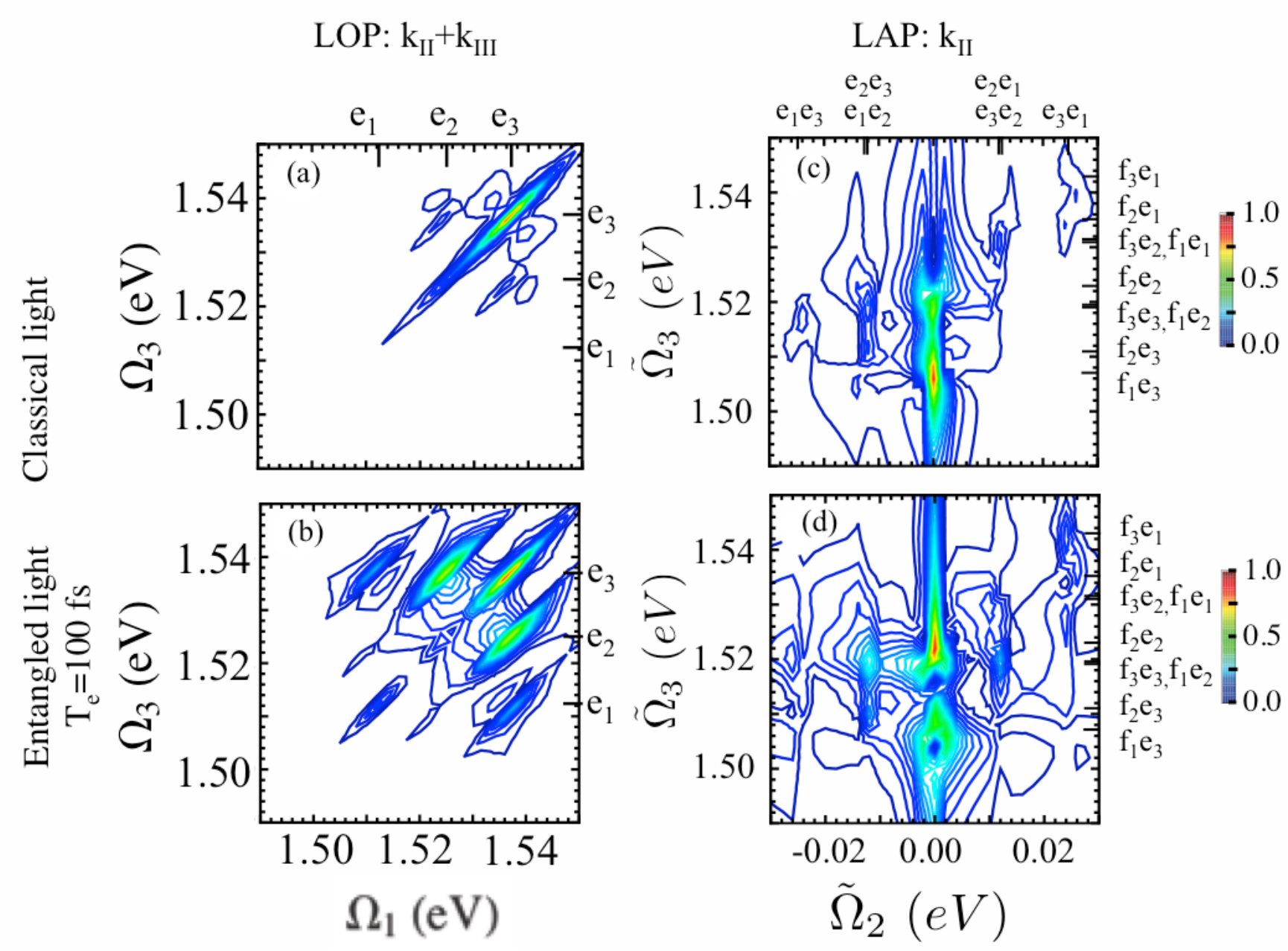}
\end{center}
\caption{(Color online) Left column: LOP signal $S_{\mathbf{k}_{II}+\mathbf{k}_{III}}(\Omega_1,\tau_2=0,\Omega_3)$  for a molecular trimer with classical light Eq. (\ref{eq:Sk21}) - (\ref{eq:Sk31}) -(a), entangled light (\ref{eq:S60}) with $T_e=100$ fs -(b). Right: column: LAP signal $S_{\mathbf{k}_{II}}(\tilde{\Omega}_1,t_2=0,\tilde{\Omega}_3)$   using classical light Eq. (\ref{eq:Sk23}) -(c), entangled light with $T_e=100$ fs, Eq.~(\ref{eq:S80}) -(d). Intraband dephasing $\gamma_{ee'}=1$ meV. All other parameters are given in Section 5 of \cite{Dorfman14b}.}
\label{fig:w1w3}
\end{figure}

So far, we presented the LOP and the LAP delay scanning protocols for classical light. We now turn to the LOP protocol with entangled light. Similarly to Eqs. (\ref{eq:Sk20}) - (\ref{eq:Sk30}) we calculate the signals for the CW-pump model~(\ref{eq.two-photon-wavefunction-cw}) discussed earlier, and obtain

\begin{align}\label{eq:S60}
S_{LOP}^{(j)}&(s_1,s_2,s_3)=\mathcal{I}\int_{-\infty}^{\infty}\frac{d\omega_a}{2\pi}\frac{d\omega_b}{2\pi}\frac{d\omega_d}{2\pi}\notag\\
&\times R_j(\omega_a,\omega_b,\omega_d) D_{LOP}^{(j)}(s_1,s_2,s_3;\omega_a,\omega_b,\omega_d) \notag\\
&\times\langle E^{\dagger}(\omega_d)E^{\dagger}(\omega_a+\omega_b-\omega_d)E(\omega_b)E(\omega_a)\rangle,
\end{align}
where $j=\mathbf{k}_{II}, \mathbf{k}_{III}$ and the display function for both rephasing and nonrephasing contributions reads 
\begin{align}\label{eq:Dlop1}
D_{LOP}^{(j)}(s_1,s_2,s_3;\omega_a,\omega_b,\omega_d)&=\theta(s_1)\theta_j(\pm s_2)\theta(s_3)\notag\\
&\times e^{-i\omega_a s_1+i\omega_d s_3-i(\omega_a+\omega_b) s_2},
\end{align}
and matter responses are
\begin{align}\label{eq:S61}
R_{\mathbf{k}_{II}}(\omega_a,\omega_b,\omega_d)&= \sum_{e,e',f}\mu_{ge'}^a\mu_{e'f}^b\mu_{fe}^{c*}\mu_{eg}^{d*}\notag\\
&\times\mathcal{G}_{ef}(-\omega_b)\mathcal{G}_{ee'}(\omega_a-\omega_d)\mathcal{G}_{eg}(\omega_a),\notag\\
R_{\mathbf{k}_{III}}(\omega_a,\omega_b,\omega_d)&=\sum_{e,e',f}\mu_{ge'}^a\mu_{e'f}^b\mu_{fe}^{c*}\mu_{eg}^{d*}\notag\\
&\times\mathcal{G}_{ef}(-\omega_b)\mathcal{G}_{ee'}(\omega_a-\omega_d)\mathcal{G}_{ge'}(-\omega_d).
\end{align}
Similarly we obtain for LAP signals Eqs. (\ref{eq:Sk22}) - (\ref{eq:Sk32})
\begin{align}\label{eq:S80}
S_{LAP}^{(j)}&(t_1,t_2,t_3)=\mathcal{I}\int_{-\infty}^{\infty}\frac{d\omega_a}{2\pi}\frac{d\omega_b}{2\pi}\frac{d\omega_d}{2\pi}\notag\\
&\times R_j(\omega_a,\omega_b,\omega_d)D_{LAP}^{(j)}(t_1,t_2,t_3;\omega_a,\omega_b,\omega_d)\notag\\
&\times \langle E^{\dagger}(\omega_d)E^{\dagger}(\omega_a+\omega_b-\omega_d)E(\omega_b)E(\omega_a)\rangle,
\end{align}
where the display function for the rephasing signal is
\begin{align}\label{eq:S82}
D_{LAP}^{(\mathbf{k}_{II})}(t_1,t_2,t_3;\omega_a,\omega_b,\omega_d)&=\theta(t_1)\theta(t_2)\theta(t_3)\notag\\
&\times e^{i\omega_bt_3-i\omega_at_1+i(\omega_d-\omega_a)t_2},
\end{align}
and for nonrephasing signal
\begin{align}\label{eq:S83}
D_{LAP}^{(\mathbf{k}_{III})}(t_1,t_2,t_3;\omega_a,\omega_b,\omega_d)&=\theta(t_1)\theta(t_2)\theta(t_3)\notag\\
&\times e^{i\omega_bt_3+i\omega_dt_1+i(\omega_d-\omega_a)t_2}.
\end{align}
The complete set of signals along with frequency domain signals for entangled light can be found in Section 2 of \cite{Dorfman14b}.

The Fourier transform of the signal (\ref{eq:S60}) was simulated using the LOP protocol and compared it with the standard fully time ordered LAP protocol given by Eq. (\ref{eq:S80}) for a model trimer with parameters discussed in Section 5 of \cite{Dorfman14b}. Fig. \ref{fig:w1w3} shows the simulated $S_{LOP}(\Omega_1,\tau_2=0,\Omega_3)$ for a trimer using classical light (top row) and entangled light (bottom row). Fig. \ref{fig:w1w3}a shows classical light which gives a diagonal cross peak $e=e'$ and one pair of weak side peaks parallel to the main diagonal at $(e,e')=(e_2,e_3)$. The remaining two pairs of side peaks at $(e,e')= (e_1,e_2)$ and $(e,e')=(e_1,e_3)$ are too weak to be seen. Fig. \ref{fig:w1w3}d) depicts the  signal obtained  using entangled photons where we observe two additional strong side cross peak pairs with $(e,e')=(e_1,e_3)$ and $(e,e')=(e_1,e_2)$. The weak peak at $(e,e')=(e_2,e_3)$ is significantly enhanced as well. 

As has been shown in Fig. \ref{fig:loplap} the LAP signal displayed vs $(\tilde{\Omega}_1,\tilde{\Omega}_3)$ does not effectively reveal intraband dephasing. This can, however, be done by the LAP signal displayed vs $(\tilde{\Omega}_2,\tilde{\Omega}_3)$. Fig. \ref{fig:w1w3}c) reveals several $\tilde{\Omega}_3=\omega_{fe}$ peaks  that overlap in $\tilde{\Omega}_3$ axes due to large dephasing $\gamma_{fe}$. With entangled light, the LAP protocol yields some enhancement in several peaks around $\pm 0.01$ eV but overall the LOP yields much cleaner result. The pathways for the density matrix (LAP) and the wavefunction (LOP) are different and suggest different types of resonances.

\subsection{Heterodyne detection of nonlinear signals}

So far we have considered fluorescence (homodyne) detection. Homodyne and heterodyne are two complementary detection schemes for nonlinear optical signals. In terms of classical fields, if the sample radiation is detected for different modes than  the incident radiation, the signal is proportional to $|E|^2$. This is known as homodyne detection whereby the intensity is the square modulus of the emitted field itself. If the emitted field coincides with a frequency of the incident radiation $E_{in}$, then the signal intensity is proportional to $|E + E_{in}|^2$. Consequently, the detected intensity contains a mixed interference term, i.e., $E^*E_{in}+c.c.$. This is defined as the heterodyne signal, since the emitted field is mixed with another field. In terms of quantized fields, the signal denoted homodyne if detected at a field mode that is initially vacant and heterodyne when detected at a field mode that is already occupied. Note that the current definition, which is commonly used in nonlinear multidimensional spectroscopy, is different from the definition used in quantum optics and optical engineering where homodyne and heterodyne refer to mixing with a field with the same or different frequency of the signal. In this review, we will use the spectroscopy terminology \cite{pot13}. 
 Fluorescence detection is often more sensitive than heterodyne as the latter is limited by the pulse duration so there are fewer constraints on the laser system. In addition the low intensity requirements for biological samples limit the range of heterodyne detection setups. This have been demonstrated by \cite{Tek07,War07,War09} even in single molecule spectroscopy \cite{Bri10}. Historically, Ramsey fringes constitute the first example of incoherent detection \cite{Ram50,Coh11,Schlawin13b}. Information similar to coherent spectroscopy can be extracted from the parametric dependence on various pulse sequences applied prior to the incoherent detection \cite{Ric11,Rahav10a}. Possible incoherent detection modes include fluorescence \cite{Elf07, Bar08}, photoaccoustic \cite{Pat81}, AFM \cite{Mam05,Pog09,Raj10,Raj11} or photocurrent detection \cite{Che13,Nar13}.

\subsubsection{Heterodyne Intensity measurements - Raman vs. TPA pathways}

\label{sec.TPA-vs-Raman}
\begin{figure}[t]
\centering
\includegraphics[width=0.49\textwidth]{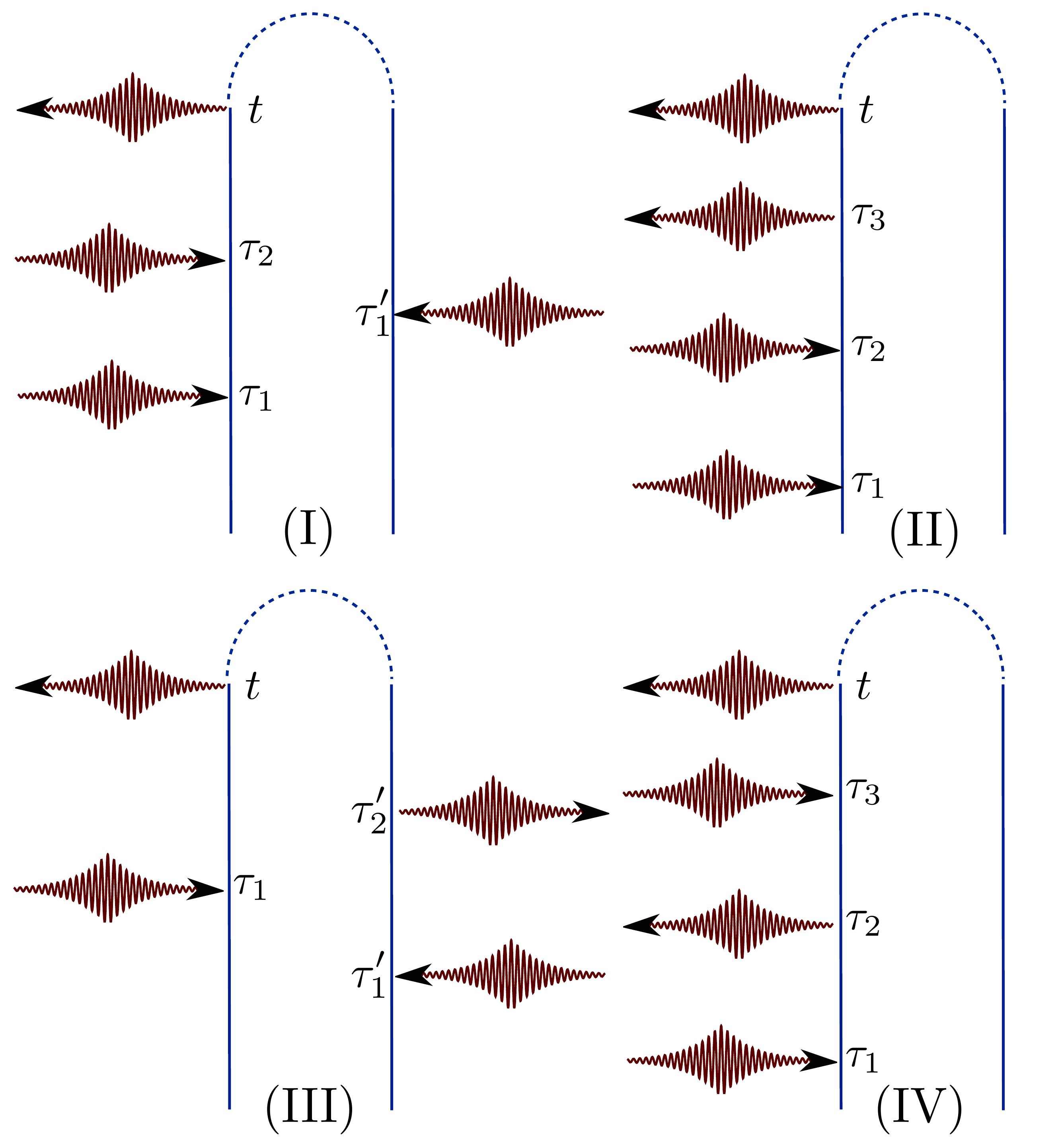}
\caption{Diagrams representing the third-order contributions to Eq.~(\ref{eq.intensity-signal}).}
\label{fig.intensity-diagrams}
\end{figure}

With $A (t) = E^{\dagger}(t)E(t)$, Eq.~(\ref{eq.first-order-signal}) yields the rate of change of the transmitted photon number,
\begin{align}
S_1 (t; \Gamma)  &= \frac{2}{\hbar} \Im \big\langle E^{\dagger}_+ (t) V_+ (t) \big\rangle_{\text{final}}.  \label{eq.intensity-signal}
\end{align}
The semiclassical signal may be obtained from Eq.~(\ref{eq.intensity-signal}) by simply replacing the field operator $E^{\dagger} (t)$ by a classical field amplitude $A^{\ast} (t)$. Similarly, by spectrally dispersing the time-integrated intensity, which amounts to measuring $A (\omega) = E^{\dagger} (\omega) E (\omega)$, we obtain
\begin{align}
S (\omega; \Gamma)  &= \frac{2}{\hbar} \Im \big\langle E^{\dagger} (\omega) V (\omega) \big\rangle_{\text{final}}. \label{eq.intensity-signal-dispersed}
\end{align}

The third-order contribution of the time-integrated absorption signal~(\ref{eq.intensity-signal}) is given by the four loop diagrams in Fig.~\ref{fig.intensity-diagrams},
\begin{align}
S_{1, \text{(I)}} (\Gamma) = - \frac{1}{\hbar} \Im \bigg[ &\left( - \frac{i}{\hbar} \right)^3 \int_{- \infty}^{\infty} \!\!\! dt \int^t_{-\infty} \!\!\! d\tau_2 \int^{\tau_2}_{-\infty} \!\!\! d\tau_1 \int^t_{-\infty} \!\! d\tau'_1 \notag \\
&\times \big\langle V (\tau'_1) V (t) V^{\dagger} (\tau_2) V^{\dagger} (\tau_1) \big\rangle \notag \\
\times & \big\langle E^{\dagger} (\tau'_1) E^{\dagger} (t) E (\tau_2) E (\tau_1) \big\rangle \bigg], \label{eq.S_1I} \\
S_{1, \text{(II)}} (\Gamma) =  \frac{1}{\hbar} \Im \bigg[ &\left( - \frac{i}{\hbar} \right)^3 \int_{- \infty}^{\infty} \!\!\! dt \int^t_{-\infty} \!\! d\tau_3 \int^{\tau_3}_{-\infty} \!\! d\tau_2 \int^{\tau_2}_{-\infty} \!\!\! d\tau_1 \notag \\
&\times \big\langle V (t) V (\tau_3) V^{\dagger} (\tau_2) V^{\dagger} (\tau_1) \big\rangle \notag \\
\times & \big\langle E^{\dagger} (t) E^{\dagger} (\tau_3) E (\tau_2) E (\tau_1) \big\rangle \bigg], \label{eq.S_1II} \\
S_{1, \text{(III)}} (\Gamma) = \frac{1}{\hbar} \Im \bigg[ &\left( - \frac{i}{\hbar} \right)^3 \int_{- \infty}^{\infty} \!\!\! dt \int^t_{-\infty} \!\!\! d\tau_1 \int^t_{- \infty} \!\!\! d\tau'_2 \int^{\tau'_2}_{-\infty} \!\! d\tau'_1 \notag \\
&\times \big\langle V (\tau'_1) V^{\dagger} (\tau'_2)  V (t) V^{\dagger} (\tau_1) \big\rangle \notag \\
\times & \big\langle E^{\dagger} (\tau'_1) E (\tau'_2) E^{\dagger} (t) E (\tau_1) \big\rangle \bigg], \label{eq.S_1III} \\
S_{1, \text{(IV)}} (\Gamma) = \frac{1}{\hbar} \Im \bigg[ &\left( - \frac{i}{\hbar} \right)^3 \int_{- \infty}^{\infty} \!\!\! dt \int^t_{-\infty} \!\! d\tau_3 \int^{\tau_3}_{-\infty} \!\! d\tau_2 \int^{\tau_2}_{-\infty} \!\!\! d\tau_1\notag \\
&\times \big\langle V (t) V^{\dagger}  (\tau_3) V(\tau_2) V^{\dagger} (\tau_1) \big\rangle \notag \\
\times & \big\langle E^{\dagger} (t) E (\tau_3) E^{\dagger}  (\tau_2) E (\tau_1) \big\rangle \bigg]. \label{eq.S_1IV} 
\end{align}

An identical expansion of Eq.~(\ref{eq.intensity-signal-dispersed}) yields the frequency-resolved third-order signal (its sum-over-state expansion is shown in Appendix \ref{app:countingsos})
\begin{align}
S_{1, \text{(I)}} (\omega; \Gamma) = - \frac{1}{\hbar} \Im \bigg[ &\left( - \frac{i}{\hbar} \right)^3 \int_{- \infty}^{\infty} \!\!\!\! dt \int^t_{-\infty} \!\!\!\! d\tau_2 \int^{\tau_2}_{-\infty} \!\!\!\! d\tau_1 \int^t_{-\infty} \!\!\!\! d\tau'_1 \notag \\
&\times e^{i \omega t} \big\langle V (\tau'_1) V (t) V^{\dagger} (\tau_2) V^{\dagger} (\tau_1) \big\rangle \notag \\
\times & \big\langle E^{\dagger} (\tau'_1) E^{\dagger} (\omega) E (\tau_2) E (\tau_1) \big\rangle \bigg], \label{eq.S_1I_dispersed}
\end{align}
\begin{align}
S_{1, \text{(II)}} (\omega; \Gamma) =  \frac{1}{\hbar} \Im \bigg[ &\left( - \frac{i}{\hbar} \right)^3 \int_{- \infty}^{\infty} \!\!\!\! dt \int^t_{-\infty} \!\!\!\! d\tau_3 \int^{\tau_3}_{-\infty} \!\!\!\! d\tau_2 \int^{\tau_2}_{-\infty} \!\!\!\! d\tau_1 \notag \\
&\times e^{i \omega t} \big\langle V (t) V (\tau_3) V^{\dagger} (\tau_2) V^{\dagger} (\tau_1) \big\rangle \notag \\
\times & \big\langle E^{\dagger} (\omega) E^{\dagger} (\tau_3) E (\tau_2) E (\tau_1) \big\rangle \bigg], \label{eq.S_1II_dispersed}
\end{align}
\begin{align}
S_{1, \text{(III)}} (\omega; \Gamma) = \frac{1}{\hbar} \Im \bigg[ &\left( - \frac{i}{\hbar} \right)^3 \int_{- \infty}^{\infty} \!\!\!\! dt \int^t_{-\infty} \!\!\!\! d\tau_1 \int^t_{- \infty} \!\!\!\! d\tau'_2 \int^{\tau'_2}_{-\infty} \!\!\!\! d\tau'_1 \notag \\
&\times e^{i \omega t} \big\langle V (\tau'_1) V^{\dagger} (\tau'_2)  V (t) V^{\dagger} (\tau_1) \big\rangle \notag \\
\times & \big\langle E^{\dagger} (\tau'_1) E (\tau'_2) E^{\dagger} (\omega) E (\tau_1) \big\rangle \bigg], \label{eq.S_1III_dispersed}
\end{align}
\begin{align}
S_{1, \text{(IV)}} (\omega; \Gamma) = \frac{1}{\hbar} \Im \bigg[ &\left( - \frac{i}{\hbar} \right)^3 \int_{- \infty}^{\infty} \!\!\!\! dt \int^t_{-\infty} \!\!\!\! d\tau_3 \int^{\tau_3}_{-\infty} \!\!\!\! d\tau_2 \int^{\tau_2}_{-\infty} \!\!\!\! d\tau_1\notag \\
&\times e^{i \omega t} \big\langle V (t) V^{\dagger}  (\tau_3) V(\tau_2) V^{\dagger} (\tau_1) \big\rangle \notag \\
\times & \big\langle E^{\dagger} (\omega) E (\tau_3) E^{\dagger}  (\tau_2) E (\tau_1) \big\rangle \bigg]. \label{eq.S_1IV_dispersed} 
\end{align}
It is instructive to relate the four diagrams corresponding to Eqs.~(\ref{eq.S_ppI})-(\ref{eq.S_ppIV}) to the transition amplitudes between the initial and final matter states to gain some intuition for this signal. This is only possible by the total photon counting signal, $\int d\omega \; S (\omega; \Gamma) / (2\pi)$, which represents the full energy exchanged between the light field and the matter system. We now define the \textit{transition amplitude operators}
\begin{align}
T^{(1)}_{e'g} (\omega) &= \frac{\mu_{e'g}}{\hbar} E (\omega), \label{eq.T^1-definition}
\end{align}
\begin{align}
T^{(2)}_{fg} (\omega_{\text{sum}}) &= \frac{1}{\hbar^2} \sum_e \int \!\! \frac{d\omega_a}{2 \pi} \frac{\mu_{ge} \mu_{ef}E (\omega_{\text{sum}} - \omega_a) E (\omega_a)}{\omega_a - \omega_e + i \gamma_e}, \label{eq.T^2-definition}
\end{align}
\begin{align}
T^{(2)}_{g' g} (\omega) &= \frac{1}{\hbar^2} \sum_{e} \int \!\! \frac{d\omega_a}{2 \pi} \; \frac{\mu_{ge} \mu_{e g'}E^{\dagger} (\omega_a - \omega) E (\omega_a)}{\omega_a - \omega_e + i \gamma_e}, \label{eq.T_gg}
\end{align}
\begin{align}
T^{(3)}_{e'g}& (\omega) = \notag\\
&\frac{1}{\hbar^3} \sum_{e, f} \int \!\! \frac{d\omega_a}{2 \pi} \int \!\! \frac{d\omega_x}{2 \pi} \frac{\mu_{ge}}{\omega_a - \omega_e + i \gamma_e} \frac{\mu_{ef} \mu_{f e'}}{\omega_x - \omega_f + i \gamma_f} \notag \\
&\times E^{\dagger} (\omega_x - \omega) E (\omega_x - \omega_a) E (\omega_a), \label{eq.T^3-definition}
\end{align}
\begin{align}
T'^{(3)}_{e' g}& (\omega) =\notag\\
& \frac{1}{\hbar^3} \sum_{e, g'} \int \!\! \frac{d\omega_a}{2 \pi} \int \!\! \frac{d\omega_-}{2 \pi} \frac{\mu_{ge}}{\omega_a - \omega_e + i \gamma_e} \frac{\mu_{eg' \mu{g' e'}}}{\omega_- - \omega_{g'} + i \gamma_g} \notag \\
&\times E (\omega - \omega_-) E^{\dagger} (\omega_a - \omega_-) E (\omega_a), \label{eq.T'^3-definition}
\end{align}
When we assume unitary time evolution, we can replace the dephasing rates in Eqs.~(\ref{eq.S_ppI})-(\ref{eq.S_ppIV}) by infinitesimal imaginary factors $\gamma \rightarrow \epsilon$. This allows us to use the identity $1 / (\omega + i \epsilon) = \text{PP} 1/ \omega + i \pi \delta (\omega)$, and carry out the remaining frequency integrals. We arrive at
\begin{align}
\int \!\! \frac{d\omega}{2 \pi} S_{\text{1, (I)}} (\omega; \Gamma) &= \sum_f \big\langle T^{(2) \dagger}_{fg} (\omega_{f}) T^{(2) }_{fg} (\omega_{f}) \big\rangle, \\
\int \!\! \frac{d\omega}{2 \pi} S_{\text{1, (II)}} (\omega; \Gamma) &= \sum_e \big\langle T^{(1) \dagger}_{eg} (\omega_e) T^{(3)}_{eg} (\omega_e) \big\rangle, \\
\int \!\! \frac{d\omega}{2 \pi} S_{\text{1, (III)}} (\omega; \Gamma) &= \sum_{g'}  \big\langle T^{(2) \dagger}_{g' g} (\omega_{g'}) T^{(2)}_{g' g} (\omega_{g'}) \big\rangle,\\
\int \!\! \frac{d\omega}{2 \pi} S_{\text{1, (IV)}} (\omega; \Gamma) &= \sum_e  \big\langle T^{(1) \dagger}_{eg} (\omega_e) T'^{(3)}_{eg} (\omega_e) \big\rangle.
\end{align}
The details of sum-over-state expansion are presented in Appendix \ref{app:countingsos}.
These results clarify the statement we made at the end of section~\ref{sec.TPA-vs-TPIF}: The $\chi^{(3)}$-absorption signal comprises matter transitions from the ground to the $f$-state - just like the TPIF signal -, but it also contains transitions to $e$- and $g$-states. The absorption measurement and the TPIF signal contain the same information, only when transitions to the $e$- and $g$-states can be neglected, as is the case when $e$ is off-resonant.  

\subsubsection{Heterodyne detected four-wave mixing; the double-quantum-coherence technique}


Time domain two dimensional (2D) spectroscopic techniques \cite{Mukamel:AnnuRevPhysChem:00}, provide a versatile tool for exploring the properties of molecular  systems, such as photosynthetic aggregates \cite{Engel:Nature:07,Abramavicius:PNAS:08} or coupled (Hybrid-)nanostructures  to semiconductor quantum wells \cite{Zhang:ProcNatlAcadSci:07,Yang:JChemPhys:08,Pasenow:SolStateComm:08,Vogel:PhysRevB:09}.
These techniques use sequences of short coherent pulses in comparison to the dephasing times of the system. 

Earlier we had presented different delay scanning protocols (LOP and LAP) for multidimensional spectroscopy with entangled photons. These protocols can be experimentally realized using entangled photon pulse shaping, using collinear geometry and precise control of the phase, phase cycling. These protocols allow to extract inter- and intraband dephasing with high resolution by exciting doubly excited state distributions. 
We now present a different multidimensional time-domain spectroscopy that involve higher electronic states manifold, but does not require excited populations. This is so called double-quantum-coherence (DQC) signals \cite{Mukamel:JChemPhys:07,Yang:PhysRevB:08,Kim:AccChemRes:09,Palmieri:JChemPhys:09} where the system evolves in the coherence between ground and doubly excited state $\vert f \rangle \langle g \vert$ rather than in a population $\vert f \rangle \langle f \vert$. This technique renders the energies of single and doubly excited state energies accessible and reveals the correlations between single and doubly excited states .
We show how pulsed entangled photons affect the two photon resonances. Some bandwidth limitations of classical beams are removed and selectivity of quantum pathways is possible.

\paragraph{The DQC signal}

\begin{figure}[t]
\center
\includegraphics[width=0.45\textwidth]{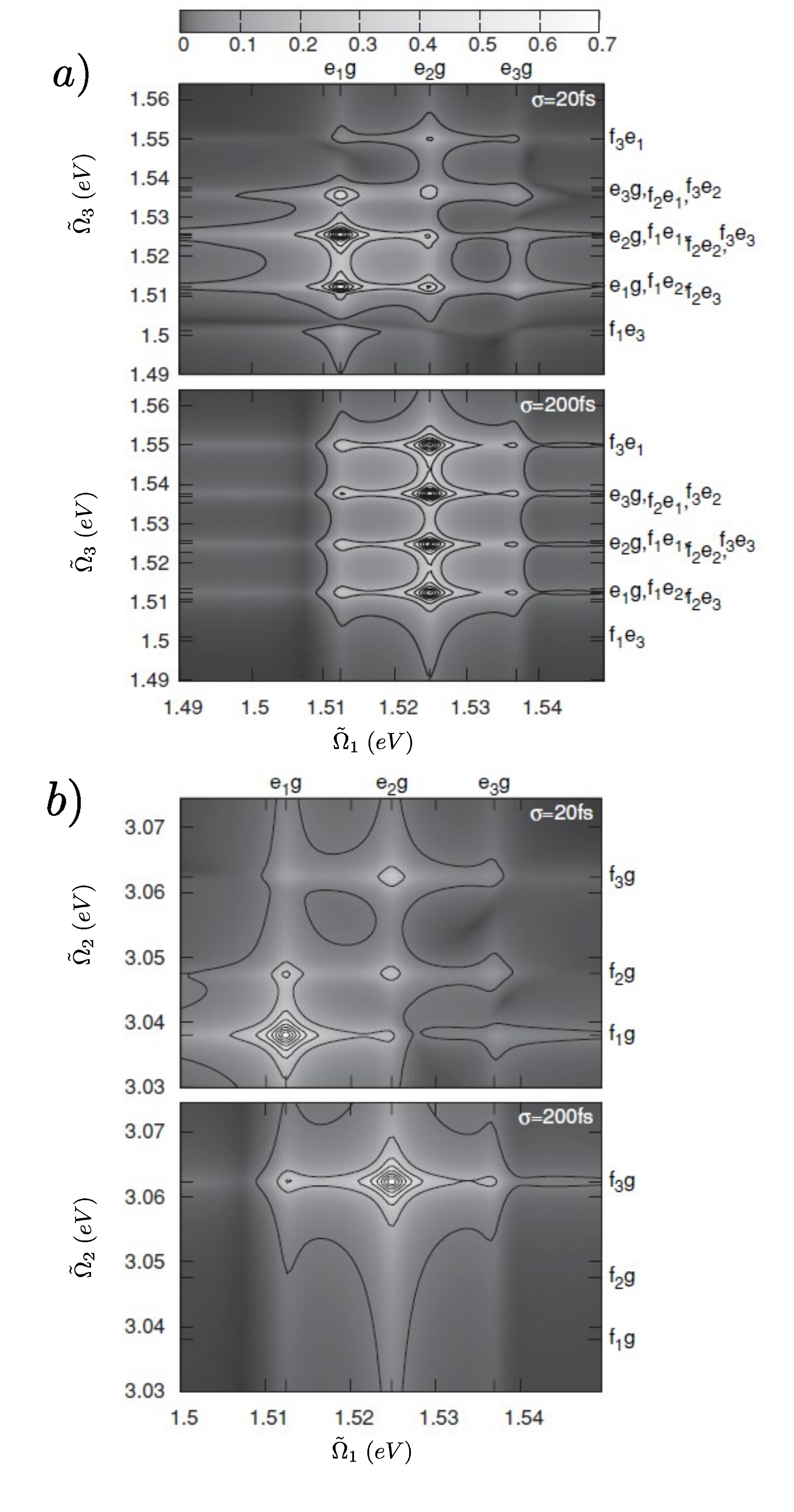}
\caption{(Color online) 
a) 2D signal $ S_{DQC}^{(LAP)}(\tilde{\Omega}_1,\tilde{\Omega}_3)$ Eq. (\ref{SignalDefT1T3}) (absolute value), showing correlation plots with different pump bandwidths $\sigma$ as indicated. The bottom panel is multiplied by x6. b) Same as a) but for 2D signal $ S_{DQC}^{(LAP)}(\tilde{\Omega}_1,\tilde{\Omega}_2)$ Eq. (\ref{SignalDefT1T2}). Parameters for simulations are given in Ref. \cite{Richter10a}. }\label{fig:dqc}
\end{figure}

In the following we use the LAP delay scanning protocol which monitors the density matrix. A pulse shaper creates a sequence of four well separated chronologically ordered pulses described by field operator $E_j(t)=\int\frac{d\omega}{2\pi}E_j(\omega)e^{-i\omega(t-T_j)}$, $j=1,2,3,4$. The control parameters are their central times $T_1<T_2<T_3<T_4$ and phases $\phi_1$, $\phi_2$, $\phi_3$, and $\phi_4$ (see Fig. \ref{fig:dqc}a). The DQC signal selects those contributions with the phase signature $\phi_1+\phi_2-\phi_3-\phi_4$.
The signal is defined as the change in the time-integrated transmitted intensity in component $\phi_4$, which is given by
\begin{eqnarray}\label{eq:Sdqc0}
 S=\frac{2}{\hbar}\int\!\! \mathrm{d} t \; \langle E^\dagger_4(t) V (t)\rangle
\end{eqnarray}
We thus have a similar  configuration to an impulsive experiment with four short well separated classical fields.
Introducing the LAP delays $t_3=T_4-T_3$, $t_2=T_3-T_2$ and $t_1=T_2-T_1$ we can calculate Eq. (\ref{eq:Sdqc0}) by expanding perturbatively in the dipole field-matter interaction Hamiltonian~(\ref{eq.H_int}). The two contributions to the signal are  represented by the ladder diagrams shown in Fig. \ref{fig:dqc}b.  The corresponding signal (\ref{eq:Sdqc0}) can be read off these diagrams and is given by  $S_{DQC}^{(LAP)}(\Gamma)=S_{DQCi}^{(LAP)}(\Gamma)+S_{DQCii}^{(LAP)}(\Gamma)$ where
\begin{align}\label{eq:Sdqc10}
&S_{DQCi}^{(LAP)}(\Gamma)=\frac{1}{\hbar^3}\mathrm{Re}\int_{-\infty}^{\infty} \mathrm{d}t \int_0^{\infty}ds_1\int_0^{\infty}ds_2\int_0^{\infty}ds_3\times \nonumber\\
&\langle \Psi| E_3^{\dagger}(t-s_3)E_4^{\dagger} (t)E_2(t-s_3-s_2)E_1 (t-s_3-s_2-s_1)|\Psi\rangle\notag\\
&\times\sum_{ee'f}V_{e'f}V_{ge'}V_{ef}^*V_{ge}^* e^{-i \xi_{fe'}s_3-i \xi_{fg}s_2-i \xi_{eg} s_1}
\end{align}
\begin{align}\label{eq:Sdqc20}
&S_{DQCii}^{(LAP)}(\Gamma)=-\frac{1}{\hbar^3}\mathrm{Re}\int_{-\infty}^{\infty} \mathrm{d}t \int_0^{\infty}ds_1\int_0^{\infty}ds_2\int_0^{\infty}ds_3\times \nonumber\\
&\langle \Psi| E_4^{\dagger}(t)E_3^{\dagger} (t-s_3)E_2(t-s_3-s_2)E_1 (t-s_3-s_2-s_1)|\Psi\rangle\notag\\
&\times\sum_{ee'f}V_{e'f}V_{ge'}V_{ef}^*V_{ge}^* e^{-i \xi_{eg}s_3-i \xi_{fg}s_2-i \xi_{e'g} s_1}
\end{align}
We have introduced the complex frequency variables $\xi_{ij}=\omega_{ij}+i \gamma_{ij}$, where $\omega_{ij}=\varepsilon_i-\varepsilon_j$ are the transition frequencies and
$\gamma_{ij}$ are the dephasing rates.
The signal may be depicted by its variation with various parameters of the field wavefunction. These are denoted collectively as $\Gamma$. Various choices of $\Gamma$ lead to different types of 2D signals. These will be specifiedbelow (see Eqs. (\ref{SignalDefT1T3}) and (\ref{SignalDefT1T2})).

\paragraph{The field correlation function for entangled photon pairs} 

We consider the pulsed entangled photon pairs described in section~\ref{sec.pulsed-entangled-light}. 
Specifically, we here consider the wavefunction introduced in \cite{Keller97}, where the two-photon wavefunction $\Phi (\bm{k}_i,\bm{k}_j)$ takes the form 
\begin{eqnarray}
 \Phi (\bm{k}_i,\bm{k}_j)&=&g\, \hat{u}((\omega(\bm{k}_j)-\omega(\bm{k}_i))/2)e^{-i (\omega(\bm{k}_i)+\omega(\bm{k}_j)) \hat{\tau}_{ij}}\notag\\ 
&&\quad \cdot e^{- i (\omega(\bm{k}_j)) \tau_{ij}}A_P(\omega(\bm{k}_i)+\omega(\bm{k}_j)-\Omega_P) \\
\hat{u}(\omega)&=&e^{i T_{ij}/2\cdot \omega}\mathrm{sinc}(T_{ij}/2\cdot \omega)
\end{eqnarray}
where  $A_P(\cdots)$ is the pulse envelope and $\Omega_p$ the central frequency of the pump pulse used to generate the pairs.

The correlation functions of the entangled fields reads \cite{Keller97}
\begin{align}
&\langle 0| E_1(s_1-T_1)E_2 (s_2-T_2)|\Psi\rangle= V_0 e^{-\frac{\Omega_p}{2} (s_1+s_2-t_1)} \nonumber\\&\qquad\; \text{rect} (s_2-s_1-t_1) A_p(\frac{s_1+s_2-t_1}{2} ) \label{photoncorrfct}
\end{align}
\begin{align}
&&\text{rect}(t)=\begin{cases}
        \frac1{T} &0<t<T \nonumber \\
0 &\text{otherwise}
       \end{cases} \nonumber \\
&&\quad A_p(t)=\mathrm{exp}(-t^2/(2\sigma^2)), \label{photonexpect}
\end{align}
where $V_0$ is given by ...
It describes the pulsed counterpart of the cw-correlation function~(\ref{eq.two-photon-wavefunction-cw-time}) we had described earlier, where the step-function rect $(t)$ is amended by the finite pulse amplitude $A_p$.

\paragraph{Simulated 2D Signals}
Below we present simulated DQC signal for a model trimer system with parameters given in Ref. \cite{Richter10a}.
By inserting Eq. (\ref{photoncorrfct}) in Eq. (\ref{eq:Sdqc10}) - (\ref{eq:Sdqc20}) and carrying out all integrations, assuming that $\gamma\ll\sigma$, we arrive at the final expression for the two contributions to the signal:
\begin{eqnarray}
 S_{DQCi}^{(LAPe)}(\Gamma)&=&\frac{1}{\hbar^3}\mathrm{Re}\sum_{ee'f}V_{e'f}V_{ge'}V_{ef}^*V_{ge}^* |V_0|^2 |A_p(\omega_{fg}-\Omega_p)|^2 \nonumber\\
&&\qquad e^{-i \xi_{eg}t_1} e^{-i \xi_{fe'}t_3}  e^{-i \xi_{fg}t_2}\nonumber\\
&&\qquad \frac{(e^{i (\omega_{fg}/2-\xi_{eg})T}-1) e^{-i \xi_{fg}T/2}}{i (\omega_{fg}/2-\xi_{eg})T} \nonumber\\
&&\qquad \frac{(e^{i (\omega_{fg}/2-\xi_{fe'})T}-1) e^{-i \xi_{fg}T/2}}{i (\omega_{fg}/2-\xi_{fe'})T}\label{final1} 
\end{eqnarray}
\begin{eqnarray}
 S_{DQCii}^{(LAPe)}(\Gamma)&=&\frac{1}{\hbar^3}\mathrm{Re}\sum_{ee'f}V_{ge'}V_{e'f}V_{ef}^*V_{ge}^* |V_0|^2 |A_p (\omega_{fg}-\Omega_p)|^2 \nonumber\\
&&\qquad e^{-i \xi_{eg}t_1} e^{-i \xi_{e'g}t_3}  e^{-i \xi_{fg}t_2}\nonumber\\
&&\qquad \frac{(e^{i (\omega_{fg}/2-\xi_{eg})T}-1) e^{-i \xi_{fg}T/2}}{i (\omega_{fg}/2-\xi_{eg})T} \nonumber\\
&&\qquad \frac{(e^{i (\omega_{fg}/2-\xi_{e'g})T}-1) e^{-i \xi_{fg}T/2}}{i (\omega_{fg}/2-\xi_{e'g})T} 
\label{final2}\\
&&|A_p(\omega)|^2=\mathrm{exp}(-\sigma^2\omega^2)
\end{eqnarray}
The control parameters $\Gamma$ now include the delay times $(t_1,t_2,t_3)$ and the entanglement time $T$.

For comparison, we present the same signal obtained with four impulsive classical pulses with envelopes $E_j(\cdot)$ $j=1,\dots,4$ $E_2(\cdot)$, $E_3(\cdot)$ and $E_4(\cdot)$ and carrier frequency $\Omega_p^0$ \cite{Abr09}:
\begin{eqnarray}
 S^{(LAPc)}_{DQCi}&=&\frac{1}{\hbar^3}\mathrm{Re}\sum_{ee'f}V_{e'f}V_{ge'}V_{ef}^*V_{
ge}^* \nonumber\\
&&\qquad E^{\ast}_{4}(\omega_{fe'}-\Omega_p^0) E^{\ast}_{3}(\omega_{e'g}-\Omega_p^0)
\nonumber\\
&&\qquad
 E_{2}(\omega_{fe}-\Omega_p^0) E_{1}(\omega_{eg}-\Omega_p^0)
\nonumber\\
&&\qquad e^{-i \xi_{eg}\tau_{12}} e^{-i \xi_{fe'}\tau_{34}}  e^{-i \xi_{fg}\tau_d}
\label{classical1} 
\end{eqnarray}
\begin{eqnarray}
 S^{(LAPc)}_{DQCii}&=&\frac{1}{\hbar^3}\mathrm{Re}\sum_{ee'f}V_{ge'}V_{e'f}V_{ef}^*V_{
ge}^*  \nonumber\\
&&\qquad E^{\ast}_4(\omega_{e'g}-\Omega_p^0) E^{\ast}_3(\omega_{fe'}-\Omega_p^0)
\nonumber\\
&&\qquad
 E_{2}(\omega_{fe}-\Omega_p^0) E_{1}(\omega_{eg}-\Omega_p^0)
\nonumber\\
&&\qquad e^{-i \xi_{eg}\tau_{12}} e^{-i \xi_{e'g}\tau_{34}}  e^{-i \xi_{fg}\tau_d}
\label{classical2}
\end{eqnarray}
With  Eqs.~(\ref{final1}) and (\ref{final2}), we can compare the entangled photon and classical DQC signals. 
We first note that
Eq. (\ref{final1}) scales with the intensity of the generating pump pulse, in contrast with the intensity square of the classical case Eq. (\ref{classical1}). 
In the classical case the signal is limited by the bandwidths of the four pulses (cf. Eqs. (\ref{classical1}-\ref{classical2})), which control the four transitions ($\omega_{eg}$, $\omega_{fe}$, $\omega_{e'g}$, $\omega_{fe'}$) in the two photon transitions inside the pulse bandwidth \cite{Abr09}. 
In Eq.(\ref{final1})  bandwidth limitations of the envelopes are only imposed through the bandwidth of entangled photon pair $A_p(\cdot)$ and the limitation is only imposed on the two photon transition $\omega_{fg}$, leading to a much broader bandwidth for the $\omega_{eg}$ and $\omega_{fe}$ transitions, if the  $\omega_{fg}$ transition is within the generating pump pulse bandwidth.

We  illustrate this effect in Fig. \ref{fig:dqc}a for the following 2D signal:
\begin{eqnarray}
 S_{DQC}^{(LAP)}(\tilde{\Omega}_1,\tilde{\Omega}_3)=\int_0^\infty\mathrm{d}t_1\int_0^\infty\mathrm
{d}t_3S(t_1,t_3)e^{i t_1\tilde{\Omega}_1+i
t_3\tilde{\Omega}_3}, \label{SignalDefT1T3}
\end{eqnarray}
 $\omega_{eg}$ resonances are seen along $\tilde{\Omega}_1$ and  $\omega_{eg}$ and $\omega_{fe}$ on  axis $\tilde{\Omega}_3$. As the bandwidth is reduced, we only get contributions from the doubly excited state resonant to the generating pump pulse. This results in four identical patterns along the $\Omega_{{1}}$ axis. All peaks are connected to the same doubly excited states.
More precisely we get four contributions along the  $\tilde{\Omega}_1$ axis connected to the transitions $\omega_{f_3e_1}$, $\omega_{f_3e_2}$ overlapping with $\omega_{e_3g}$,  $\omega_{e_2g}$ overlapping with $\omega_{f_3e_3}$ and $\omega_{e_1g}$.
  The remaining transitions are not affected by the reduced pump bandwidth. Here  the narrow bandwidth of the pump can be used to select contributions in the spectra connected to a specific doubly excited state. 
In Fig. \ref{fig:dqc}b we display a different signal:
\begin{eqnarray}
 S_{DQC}^{(LAP)}(\tilde{\Omega}_1,\tilde{\Omega}_2)=\int_0^\infty\mathrm{d}t_1\int_0^\infty\mathrm{d}t_2S(t_1,t_2)e^{i t_1\tilde{\Omega}_1+i t_2\tilde{\Omega}_2}, \label{SignalDefT1T2}
\end{eqnarray}
$\omega_{eg}$ resonances are now seen along $\tilde{\Omega}_1$ and  $\omega_{fg}$  in $\tilde{\Omega}_2$.  This is similar to Fig. \ref{fig:dqc}f except that here we see the singly excited state contributions $\omega_{e_1g}$,$\omega_{e_2g}$, $\omega_{e_3g}$ to the selected doubly excited state $f_3$ along a single row.

Bandwidth limitations on the  of the singly excited state transitions $\omega_{eg}$ and $\omega_{fe}$ are only imposed indirectly by the factors in Eqs. (\ref{final1}-\ref{final2}) which depend on the entanglement time $T$. These become largest for $\omega_{fg}=2\omega_{fe}$. 

The factors  in Eq.(\ref{final1}-\ref{final2}) which depend on the entanglement time $T$  contain an interference term of the form $(e^{i (\omega-\gamma)T}-1)$, where  $\omega$ is a material frequency (see sections~\ref{sec:aggregate} and \ref{sec.virtual-state-spec}). If we now vary the entanglement time, some resonances will interfere destructively for values of the entanglement time which match the period of $\omega$. One can therefore  use the entanglement times to control selected resonances. 
This holds only as long the  entanglement times are not much bigger than the dephasing time, since in this case  the signal will be weak.
The frequencies $\omega$ can be in the contributing diagrams $\omega_{fg}/2-\omega_{eg}$ or $\omega_{fg}/2-\omega_{fe}$ (which is differs from the first frequency only by a sign) for different combination of the  states $e$, $e'$ and $f$.
 By varying  $T$, we expect an oscillation of the magnitudes of  resonances with different frequencies.The details of manipulation of entanglement time as a control parameters have been studied in Refs. \cite{Richter10a, Schlawin13b}

Previously, to utilize entangled photons for DQC signal we used two pulsed entangled photon pairs \cite{Richter10a} ($\bm{k}_1$, $\bm{k}_2$) and ($\bm{k}_3$, $\bm{k}_4$). With the pulse shaping described in section~\ref{sec.ent-light-control}, a single shaped entangled photon pair is sufficient to realize any time-domain four-wave mixing signals. 

\subsection{Multiple photon counting detection}
\label{sec.N-photon-counting}





Another class of multidimensional signals is possible by detecting sequences of individual photons emitted by an optically driven system. With proper gating, each photon $j$ can be characterized by a frequency $\omega_j$ and a time $t_j$. By detecting $N$ photons we thus obtain a $2N$ dimensional signal parametrized by $\omega_1$, $t_1$, ..., $\omega_N$, $t_N$. Unlike coherent multidimensional heterodyne signals which are parametrized by delays of the incoming fields, these incoherent signals are parametrized by the emitted photons. The photon time $t_j$ and frequency $\omega_j$ are not independent and can only be detected to  within a Fourier uncertainty $\Delta\omega_j\Delta t_j>1$. This poses a fundamental limit on the temporal and spectral resolution. We derive these signals and connect them to multipoint $2N$ dipole correlation functions of matter. The Fourier uncertainty is naturally built in by a proper description of photon gating and need not be imposed in an ad hoc matter as is commonly done.

A semiclassical formalism for photon counting was first derived by \cite{Man58,Man59}. The full quantum mechanical description of the field and photon detection was developed by Glauber \cite{Glau07}. The theory of the electromagnetic field measurement through photoionization and the resulting photoelectron counting has been developed by Kelley and Kleiner \cite{kel64}. The experimental application to normal and time ordered intensity correlation measurements was given in the seminal work of Kimble et al. \cite{kim77}. According to these treatments free-field operators, in general, do not commute with source-quantity operators. This is the origin of the fact that the normal and time ordering of the measured field correlations, according to the Kelley-Kleiner theory \cite{kel64}, are transformed into normal and time ordered source quantities occurring inside the integral representations of the filtered source-field operators. This constitutes the back reaction of the detector on the field state \cite{Cohen-Tannoudji92}.  An ideal photon detector is a device that measures the radiation field intensity at a single point in space. The detector size should be much smaller than spatial variations of the field. The response of an ideal time-domain photon detector is independent of the frequency of the radiation. 

The resolution of simultaneous frequency and time domain measurements is limited by the Fourier uncertainty $\Delta\omega\Delta t>1$. A naive calculation of signals without proper time and frequency gating can work for slowly varying spectrally broad optical fields but otherwise may yield unphysical negative signals \cite{Ebe77}. In Ref. \cite{Muk96}, the mixed time-frequency representation for the coherent optical measurements with interferometric or autocorrelation detection were calculated in terms of a mixed material response functions and a Wigner distribution for the incoming pulses, the detected field and the gating device. Multidimensional gated fluorescence signals for single-molecule spectroscopy have been calculated in Ref. \cite{Muk11}.

The standard Glauber's theory of photon counting and correlation measurements \cite{Glauber07,Scully97,Mol70} is formulated solely in the radiation field space (matter is not considered explicitly). Signals are related to the multi-point normally-ordered field correlation function, convoluted with time and frequency gating spectrograms of the corresponding detectors. This approach assumes that the detected field is given. Thus, it does not address the matter information and the way this field has been generated. Temporally and spectrally resolved measurements can reveal important matter information. Recent single photon spectroscopy of single molecules \cite{Fle00,Let10,Rezus12} call for an adequate microscopic foundation where joint matter and field information could be retrieved by a proper description of the detection process. 

A microscopic diagrammatic approach may be used for calculating time-and-frequency gated  photon counting measurements \cite{Dorfman12a}. The observed signal can be represented by a convolution of the bare signal and a detector spectrogram that contains the time and frequency gate functions. The bare signal is given by the product of two transition amplitude superoperators \cite{Mukamel10a} (one for bra and one for ket of the matter plus field joint density matrix), each creating a coherence in the field between states with zero and one photon. By combining the transition amplitude superoperators from both branches of the loop diagram we obtain the photon occupation number that can be detected. The detection process is described in the joint field and matter space by a sum over pathways each involving a pair of quantum modes with different time orderings. The signal is recast using time ordered superoperator products of matter and field. In contrast to the Glauber theory that uses normally ordered field operator, the microscopic approach of \cite{Dorfman12a} is based on time-ordered superoperators.  Ideal frequency domain detection only requires  a single mode \cite{Muk11}. However, maintaining any time resolution requires a superposition of several field modes that contain the pathway information. This information is not directly accessible in the standard detection theory that operates in the field space alone \cite{Glau07}.

\subsubsection{Photon correlation measurements using gated  photon number operators}
\label{sec.photon-measurement}
Time-and-frequency gated $N$- th order photon correlation measurement performed at $N$ detectors characterized by central time $t_j$ and central frequency $\omega_j$, $j=1,...,N$ is defined as
\begin{align}\label{eq:gN}
&g^{(N)}(t_1,\omega_1,\Gamma_1;...,t_N;\omega_N,\Gamma_N) \notag \\
= &\frac{\langle \mathcal{T} \hat{n}_{t_1,\omega_1}...\hat{n}_{t_N,\omega_N}\rangle_T}{\langle \mathcal{T}\hat{n}_{t_1,\omega_1}\rangle_T...\langle \mathcal{T}\hat{n}_{t_N,\omega_N}\rangle_T},
\end{align}
where  $\langle ...\rangle_T=\text{Tr}[...\rho_T(t)]$ and $\rho_T(t)$ represents the total density matrix of the entire system in joint field plus matter space and contains information about system evolution prior to the detection (e.g. photon generation process, etc.). $\Gamma_j$, $j=1,...N$ represents other parameters of the detectors such as e.g. bandwidth ($\sigma_T^j$ and $\sigma_\omega^j$ are the time gate, and frequency gate bandwidths, respectively). The time-and-frequency gated photon number superoperator is given by
\begin{align}\label{eq:nDn}
\hat{n}_{t,\omega}=\int dt' \int d\tau D(t,\omega;t',\tau)\hat{n}(t',\tau).
\end{align}
Here $D(t,\omega,t',\tau)$ is a detector time-domain spectrogram (the ordinary function, not an operator) which takes into account the detector's parameters, and is given by
\begin{align}\label{eq:Ddef}
&D(t,\omega,t',\tau)=\notag\\
&\int\frac{d\omega''}{2\pi}e^{-i\omega''\tau}|F_f(\omega'',\omega)|^2F_t^{*}(t'+\tau,t)F_t(t',t),
\end{align}
where $F_t$ and $F_f$ are time and frequency gating functions that are characterized by central time $t$ and frequency $\omega$ and detection bandwidths $\sigma_T$ and $\sigma_\omega$, respectively. Note that in Eq.~(\ref{eq:Ddef}) the time gate is applied first, followed by the frequency gate. Similar expression can be written if the order in gating is reversed.
$\hat{n}(t,t')$ is a bare photon number superoperator defined in terms of the bare field operators as
\begin{align}\label{eq:ntt}
\hat{n}(t',\tau)= \sum_{s,s'}\hat{E}_{sR}^{\dagger}(t'+\tau)\hat{E}_{s'L}(t')\rho(t').
\end{align}
Details may be found in Appendix~\ref{sec.gating}. 

\subsubsection{Photon  counting and matter dipole correlation functions}

\begin{figure*}[t]
\begin{center}
\includegraphics[trim=0cm 0cm 0cm 0cm,angle=0, width=0.9\textwidth]{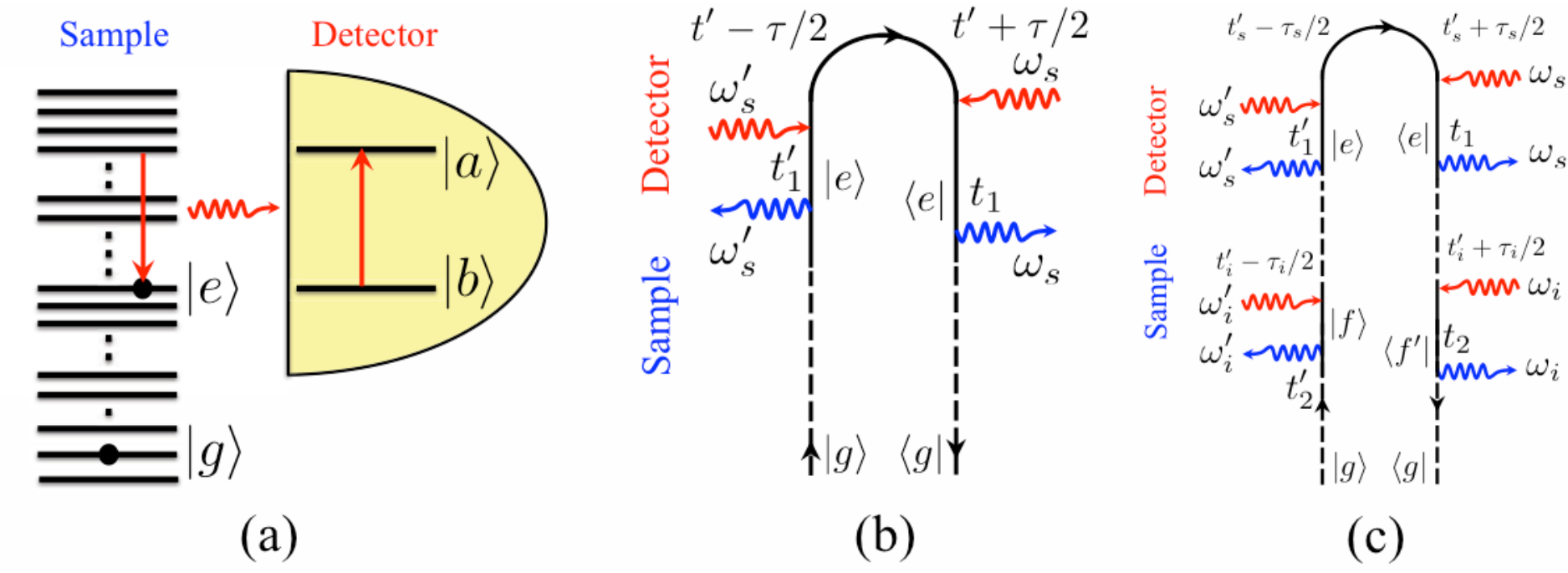}
\end{center}
\caption{(Color online) (a) Schematic of time-and-frequency resolved photon coincidence measurement. (b) Loop diagram for the bare signal~(\ref{eq:WE11}) in a gated measurement. (c) - Loop diagram for correlated two photons measurement~(\ref{eq:Bm1}). Dashed lines represent the the dynamics of the system driven by the field modes. $\tau_i$ and $\tau_s$ can be either positive or negative.}
\label{fig:gate}
\end{figure*}

To connect the photon coincidence counting (PCC) signals to matter properties one needs to expand the density operator in Eq. (\ref{eq:ntt}) in perturbative series over field-matter interactions. We first calculate the time-and-frequency resolved emission spectra
\begin{align}\label{eq:emis}
n_{t,\omega}=\int dt' \int d\tau D(t,\omega;t',\tau)n(t',\tau),
\end{align}
where the bare photon number $n(t',\tau)\equiv\langle\mathcal{T}\hat{n}(t',\tau)\rangle_T$ is an expectation value of the bare photon number operator with respect to the total density matrix. The leading contribution is coming from the second-order in field matter interactions with vacuum modes [see diagram in Fig. \ref{fig:gate}b]
\begin{align}
n(t',\tau)&=\frac{1}{\hbar^2}\int_{-\infty}^{t'}dt_1\int_{-\infty}^{t'+\tau}dt_2\langle V^{\dagger}(t_2)\langle V(t_1)\rangle\notag\\
&\times\sum_{s,s'}\langle \hat{E}_{s'}(t_2)\hat{E}_{s'}^{\dagger}(t'+\tau)\hat{E}_s(t')\hat{E}_s^{\dagger}(t_1)\rangle_v,
\end{align}
where we had utilized superoperator time ordering and $\langle ...\rangle=\text{Tr}[...\rho(t)]$ where $\rho(t)$ is the density operator that excludes vacuum modes and $\langle ...\rangle_v=\text{Tr}[...\rho_v(t)]$ with $\rho_v(t)=|0\rangle\langle 0|$ is the density matrix of the vacuum modes. 
Using the bosonic commutation relations introduced in section~\ref{sec.Liouville-notation} [see Eq.~(\ref{eq.H_field})], and moving to the continuous density of states, one can obtain
\begin{align}\label{eq:nVV}
&n(t',\tau)=\mathcal{D}^2(\omega)\langle V^{\dagger}(t'+\tau)V(t')\rangle,
\end{align}
where $\mathcal{D}(\omega)=\frac{1}{2\pi}\tilde{\mathcal{D}}(\omega)$ is a combined density of states evaluated at the central frequency of the detector $\omega$ for smooth enough distribution of modes. The corresponding detected signal (\ref{eq:nDn}) is given by
\begin{align}\label{eq:S1g}
S^{(1)}(t,\omega)&\equiv n_{t,\omega}=\notag\\
&\int dt' \int d\tau D(t,\omega;t',\tau)\mathcal{D}^2(\omega)\langle V^{\dagger}(t'+\tau)V(t')\rangle.
\end{align}

One can similarly calculate the second-order bare correlation function
\begin{align}\label{eq:Tnn}
\langle \mathcal{T} \hat{n}_{t_1,\omega_1}\hat{n}_{t_2,\omega_2}\rangle_T&=\int dt_1' \int d\tau_1 D^{(1)}(t_1\omega_1;t_1',\tau_1)\notag\\
&\times\int dt_2' \int d\tau_2 D^{(2)}(t_2,\omega_2;t_2',\tau_2)\notag\\
&\times\langle \mathcal{T}\hat{n}(t_1',\tau_1)\hat{n}(t_2',\tau_2')\rangle_T.
\end{align}
The bare PCC rate $\langle \mathcal{T}\hat{n}(t_1',\tau_1)\hat{n}(t_2',\tau_2')\rangle_T$ can be read off the diagram shown in Fig. \ref{fig:gate}c. The leading contribution requires a fourth-order expansion in field-matter interactions
\begin{align}
&\langle \mathcal{T}\hat{n}(t_1',\tau_1)\hat{n}(t_2',\tau_2')\rangle_T=\frac{1}{\hbar^4}\int_{-\infty}^{t_1'}dt_1\int_{-\infty}^{t_1'+\tau_1}dt_3\notag\\
&\times\int_{-\infty}^{t_2'}dt_2\int_{-\infty}^{t_2'+\tau_2}dt_4\langle V^{\dagger}(t_4)V^{\dagger}(t_3)V(t_1)V(t_2)\rangle\notag\\
&\times\sum_{s,s'}\sum_{r,r'}\langle E_{r'}(t_4)E_{s'}(t_3)E_{r'}^{\dagger}(t_2'+\tau_2)E_{s'}^{\dagger}(t_1'+\tau_1)\notag\\
&\times E_s(t_1')E_r(t_2')E_s^{\dagger}(t_1)E_r^{\dagger}(t_2)\rangle_v.
\end{align}
After tracing over the vacuum modes we obtain
\begin{align}\label{eq:nnVVVV}
&\langle \mathcal{T}\hat{n}(t_1',\tau_1)\hat{n}(t_2',\tau_2')\rangle_T=\mathcal{D}^2(\omega_1)\mathcal{D}^2(\omega_2)\notag\\
&\times\langle V^{\dagger}(t_2'+\tau_2)V^{\dagger}(t_1'+\tau_1)V(t_1')V(t_2')\rangle.
\end{align}
The gated coincidence signal (\ref{eq:Tnn}) is finally given by
\begin{align}\label{eq:S2g}
&S^{(2)}(t_1,\omega_1;t_2,\omega_2)\equiv\langle \mathcal{T} \hat{n}_{t_1,\omega_1}\hat{n}_{t_2,\omega_2}\rangle_T\notag\\
&=\mathcal{D}^2(\omega_1)\mathcal{D}^2(\omega_2)\int dt_1' \int d\tau_1 D^{(1)}(t_1\omega_1;t_1',\tau_1)\notag\\
&\times\int dt_2' \int d\tau_2 D^{(2)}(t_2,\omega_2;t_2',\tau_2)\notag\\
&\times\langle V^{\dagger}(t_2'+\tau_2)V^{\dagger}(t_1'+\tau_1)V(t_1')V(t_2')\rangle.
\end{align}
Therefore, the fundamental material quantity that yields the emission spectra (\ref{eq:S1g}) is a two-point dipole correlation function given in Eq. (\ref{eq:nVV}), and for the coincidence $g^{(2)}$-measurement (\ref{eq:S2g}) it is the four-point dipole correlation function in Eq. (\ref{eq:nnVVVV}). Sum-over-state  expansion of these expressions is given in appendix~\ref{sec.gating}. 

\subsubsection{Connection to the physical spectrum}

An early work \cite{Ebe77}, had argued that detector gating with finite bandwidth must be added to describe the real detector. In recent work \cite{del12,gon15} used a two-level model detector with a single parameter $\Gamma$ that characterize both time-and-frequency detection is comsidered. This was denoted the  physical spectrum (\ref{eq:gdV}) which we shall derive in the following. It can be recovered from our model by removing the time gate $F_t=1$ and using a Lorentzian frequency gate
\begin{align}
F_f(\omega,\omega')=\frac{i}{\omega'+\omega+i\Gamma/2}.
\end{align}
Using the physical spectrum, the time-and-frequency resolved photon coincidence signal is given by
\begin{align}\label{eq:g2dV}
&g_{\Gamma_1\Gamma_2}^{(2)}(\omega_1,\omega_2;\tau)=\notag\\
&\text{lim}_{t\to\infty}\frac{\langle \hat{A}_{\omega_1,\Gamma_1}^{\dagger}(t)\hat{A}_{\omega_2,\Gamma_2}^{\dagger}(t+\tau)A_{\omega_2,\Gamma_2}(t+\tau)\hat{A}_{\omega_1,\Gamma_1}(t)\rangle}{\langle \hat{A}_{\omega_1,\Gamma_1}^{\dagger}(t)\hat{A}_{\omega_1,\Gamma_1}(t)\rangle\langle \hat{A}_{\omega_2,\Gamma_2}^{\dagger}(t+\tau)\hat{A}_{\omega_2,\Gamma_2}(t+\tau)\rangle},
\end{align}
where
\begin{align}\label{eq:gdV}
\hat{A}_{\omega,\Gamma}(t)=\int_{-\infty}^tdt_1e^{(i\omega-\Gamma/2)(t-t_1)}\hat{E}(t_1),
\end{align}
is the gated field. This model provides a simple benchmark for finite-band detection, which has several limitations. First, the time and frequency gating parameters are not independent, unlike the actual experimental setup, where frequency filters and avalanche photodiodes are two independent devices. Second, this method does not address the generation and photon bandwidth coming from the emitter, as the analysis is performed solely in the field space. Finally, the multi-photon correlation function presented in \cite{gon15} is stationary. For instance, the four-point bare correlation function
\begin{align}\label{eq:bare0}
A_B(\omega_1,\omega_2,t_1,t_2)=\langle E_{\omega_1}^{\dagger}(t_1)E_{\omega_2}^{\dagger}(t_2)E_{\omega_2}(t_2)E_{\omega_1}(t_1)\rangle
\end{align}
depends on four times and four frequencies. After gating suggested in \cite{del12,gon15} the correlation function (\ref{eq:bare0}) is recast using $C_B(\omega_1,\omega_2,t_2-t_1)$, which only depends on the time difference $t_2-t_1$, which is an approximation for stationary fields. This model also works if $t\gg\Gamma^{-1}$, which means that $\Gamma$ cannot approach zero (perfect reflection in Fabri Perot cavity). It also works when $\Gamma\tau_0\ll1$ where $\tau_0$ is the scale of change in the field envelope. For comparison, the photon coincidence counting  (PCC) (\ref{eq:gN}) for $N=2$ reads
\begin{align}\label{eq:g2}
&g^{(2)}(t_1,\omega_1,\Gamma_1;t_2;\omega_2,\Gamma_2)\frac{\langle \mathcal{T} \hat{n}_{t_1,\omega_1}\hat{n}_{t_2,\omega_2}\rangle}{\langle \mathcal{T}\hat{n}_{t_1,\omega_1}\rangle\langle \mathcal{T}\hat{n}_{t_2,\omega_2}\rangle},
\end{align}
which depends on two time $t_1$, $t_2$ and two frequency $\omega_1$, $\omega_2$ arguments.
The theory summarized above which gives rise to Eq.~(\ref{eq:g2}) has several advantages compared to the physical spectrum used by \cite{del12,gon15}. First, independent control of time and frequency gates (with guaranteed Fourier uncertainty for the time and frequency resolution) along with the fact that the bare photon number operator depends on two time variables $\hat{n}(t,\tau)$ allows to capture any dynamical process down to the very short scale dynamics in ultrafast spectroscopy applications.  Second, the gating (\ref{eq:nDn}) provides a unique tool that can capture nonequlibrium and non-stationary states of matter which can be controlled by gating bandwidths. In this case a series of frequency $\omega_1$, $\omega_2$ correlation plots (keeping the central frequencies of the spectral gates as variables) for different time delays $t_1-t_2$ yields a 2D spectroscopy tool capable of measuring ultrafast dynamics. Third, the superoperator algebra allows to connect  the gated field correlation function
\begin{align}\label{eq:gated0}
A_G(\bar{\omega}_1,\bar{\omega}_2,\bar{t}_1,\bar{t}_2)=\langle E_{\bar{\omega}_1}^{(tf)\dagger}(\bar{t}_1)E_{\bar{\omega}_2}^{(tf)\dagger}(\bar{t}_2)E_{\bar{\omega}_2}^{(tf)}(\bar{t}_2)E_{\bar{\omega}_1}^{(tf)}(\bar{t}_1)\rangle
\end{align}
with the bare correlation function (\ref{eq:bare0}), with time-and-frequency gates (arbitrary, not necessarily Lorentzian) as well as material response that precedes the emission and detection of photons. The superoperator expressions require time-ordering, and can be generalized to  correlation functions of field operators that are not normally ordered. Superoperators provide an effective bookkeeping tool for field-matter interactions prior to the spontaneous emission of photons. We next apply it to the detection of photon correlations. Finally, as we show in the next section, PCC can be recast in terms of matter correlation functions by expanding the total density matrix operator in a perturbation series, and tracing the vacuum modes. This way, photon counting measurements can be related to the matter response which is the standard building block of nonlinear spectroscopy.

\subsection{Interferometric detection of photon coincidence signals}

Photon coincidence signals also known as biphoton signals \cite{Sca03,Yabushita04,Kalachev07a,Slattery13} became recently available as a tool for nonlinear spectroscopy. In a typical setup, a pair of entangled photons denoted as $E_s$, and $E_r$ generated by PDC are split on a beam splitter [see Fig. \ref{fig:pcclin}a)]. One photon $E_s$ is transmitted through the molecular sample and then detected in coincidence with $E_r$. In order to use it as a spectroscopic tool, a frequency filter can be placed in front of one of the detectors which measures the spectrum. This type of signal shows a number of interesting features: First, coincidence detection improves the signal-to-noise ratio \cite{Kalashnikov14a}. Second, it is possible
to make the two detectors operate in very different spectral regions and at different spatial locations \cite{Kalachev07a}.
For example, to measure the spectroscopic properties of a sample in vacuum ultraviolet (VUV) range, it is not necessary to set a spectrometer in a vacuum chamber and control it under the vacuum condition.
Instead, provided that a entangled photon pair consists of a VUV  and a visible photon, only the latter is to be resolved by a spectrometer under usual conditions. 
Another advantage is when spectroscopic measurements are to be performed in infrared range. The power of the light
source must often be very low to prevent possible damage of a sample, but an infrared photodetector is usually
noisy. Photon coincidence measurements involve the lowest intensities of light - single photons, and can overcome the
noise. 

In the following, we present photon coincidence version of three signals: linear absorption, pump-probe, and Femtosecond Stimulated Raman Signals (FSRS).

\subsubsection{Coincidence detection of linear absorption}

\begin{figure}[t]
\centering
\includegraphics[width=0.35\textwidth]{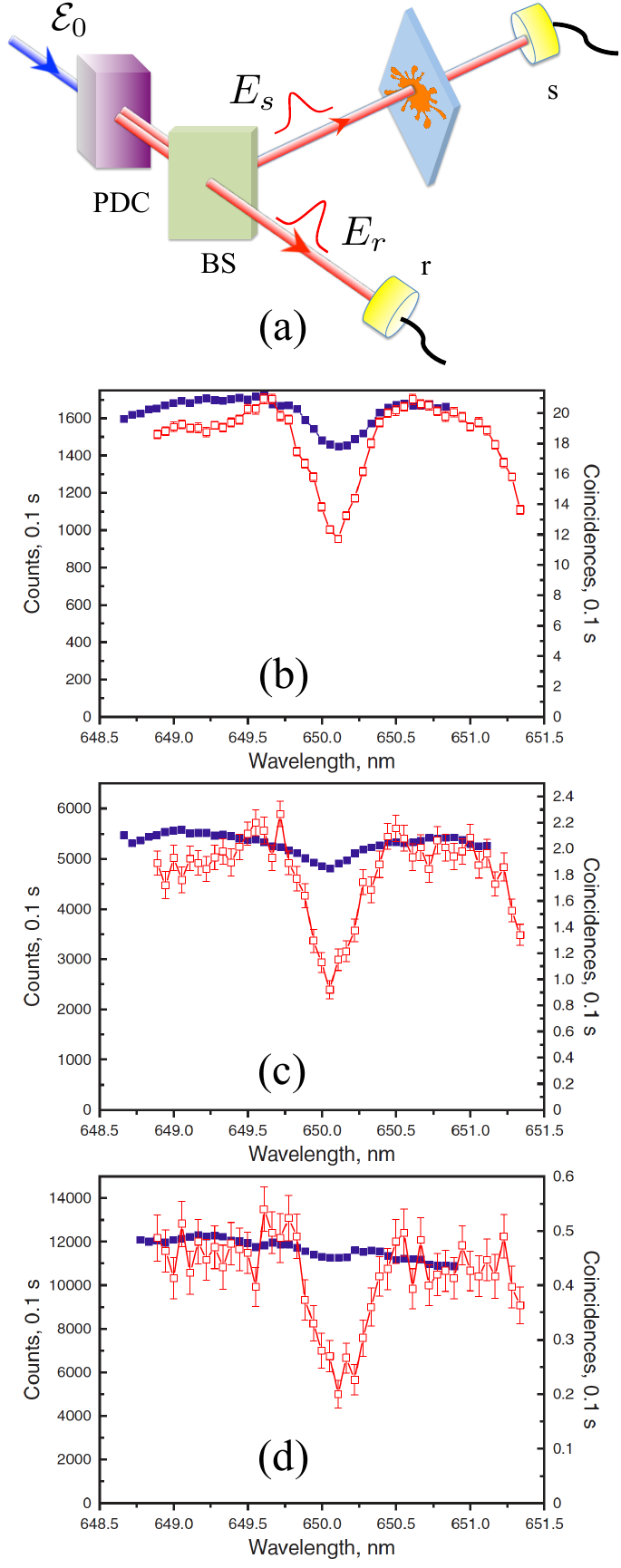}
\caption{(Color online) Linear absorption experiment of \cite{Kalachev07a} with coincidence detection. The entangled photon pair in beams $E_s$ and $E_r$ are split on a beam splitter.  $E_s$ is employed as a probe transmitted through the sample, while $E_r$ is detected in coincidence - (a).The absorption spectrum
of $Er^{3+}$ ion in YAG crystal around 650 nm obtained by
single photon counting (blue filled squares, left Y-axis) and coincidence
counting (red open squares, right Y-axis) at various values
of signal-to-noise ratio in the channel with the sample: 1/2
(b), 1/5 (c), and 1/30 (d)}
\label{fig:pcclin}
\end{figure}

Linear absorption is the most elementary spectroscopic measurement. Combined with photon coincidence detection [see Fig. \ref{fig:pcclin}a)],  it yields interesting results: Below we present the simplest intuitive phenomenological approach. In following sections we present more rigorous microscopic derivation for pump-probe and Raman signals.

In the case of the linear signal, the joint detection of two entangled photons provides linear absorption information if one of the photons is transmitted through a molecular sample. 
 If the coincidence gate window accepts counts for a time $T$, then the joint detection counting rate, $R_c$, between
detectors $r$ and $s$ is proportional to 
\begin{align}
 R_c\propto\int_0^Tdt_1\int_0^Tdt_2|\langle 0| a_s^{\dagger}(t_1) a_r^{\dagger}(t_2)|\psi\rangle|^2,
 \end{align}
where $\psi$ is a two-photon entangled state (\ref{eq.ent-state1}). 
The gating window $T$ is typically much larger than the
reciprocal bandwidth of the light or the expected dispersive
broadening 
\begin{align}
R_c\propto\int d\omega_1\int d\omega_2|\langle 0|a_s(\omega_1)a_r(\omega_2)|\psi\rangle|^2.
\end{align}
Denoting the spectral transfer functions of
the sample and monochromator $H_S(\omega)$ and $H_M(\omega)$, respectively. In this case,
\begin{align}
a_s(\omega_1)=\frac{1}{\sqrt{2}}\tilde{a}_s(\omega_1)H_S(\omega_1)
\end{align}
and
\begin{align}
a_r(\omega_2)=\frac{i}{\sqrt{2}}\tilde{a}_r(\omega_2)H_M(\omega_2)
\end{align}
provided that the signal and idler photons are separated by a 50/50 beam splitter. For narrowband down conversion $\Phi(\omega_1,\omega_2)=F\left(\frac{\omega_1-\omega_2}{2}\right)\delta(\omega_1+\omega_2-2\omega_0)$. The coincidence counting rate is
then given by
\begin{align}
R_c\propto \int d\Omega|H_S(\omega_0+\Omega)H_M(\omega_0-\Omega)F(\Omega)|^2.
\end{align}
Now, assume that $H_M(\omega)$ is much narrower than $H_S(\omega)$ and $\tilde{\Phi}(\omega_1,\omega_2)$, we can set $H_M(\omega) = \delta(\omega-\omega_M)$, and the frequency $\omega_M$ does not exceed the frequency
range in which function $F(\omega)$ is essentially nonzero. Dividing the coincidence counting rate with a sample, $R_{c,\text{sample}}$, by
one without the sample, $R_c$, we obtain  the absorption spectrum
\begin{align}
S_{ILA}(\omega_M)=\frac{R_{c,\text{sample}}}{R_c}\propto |H_S(2\omega_0-\omega_M)|^2,
\end{align}
where subscript $ILA$ marks the interferometric nature of the photon coincidence detection combined with linear absorption measurement. Thus, the joint detection counting rate reproduces the spectral function of the sample, which is reversed in frequency
with respect of the pump frequency, provided that the line-width of the pumping field as well as bandwidth
of the monochromator are narrow enough for resolving the absorption spectrum features. 

In \cite{Kalachev07a} the coincidence signals were used to measure the spectroscopic properties of YAG:Er$^{3+}$ crystal. In order to demonstrate the advantage of coincidence detection  in the presence of an enhanced background noise. Figs. \ref{fig:pcclin}b)-d) show the central part of the absorption spectrum measured in two ways: using the coincidence counting as described before, and using the single-photon counting, when the sample was placed above the monochromator in the idler channel. The signal-to-noise ratio in the channel was changed from 1/2 to 1/30 in both cases. It is evident from these experimental data that the standard classical method which suffers from a high noise level does not allow one
to obtain any spectroscopic information, but the coincidence counting measurement in
contrast does not undergo the reduction in resolution and is resistant to noise. The example of YAG:Er$^{3+}$ has been later extended to plasmonic nanostructures \cite{Kalashnikov14a}.

\subsubsection{Coincidence detection of pump-probe signals}
\label{sec.pp-coincidence}
\begin{figure*}[t]
\centering
\includegraphics[width=\textwidth]{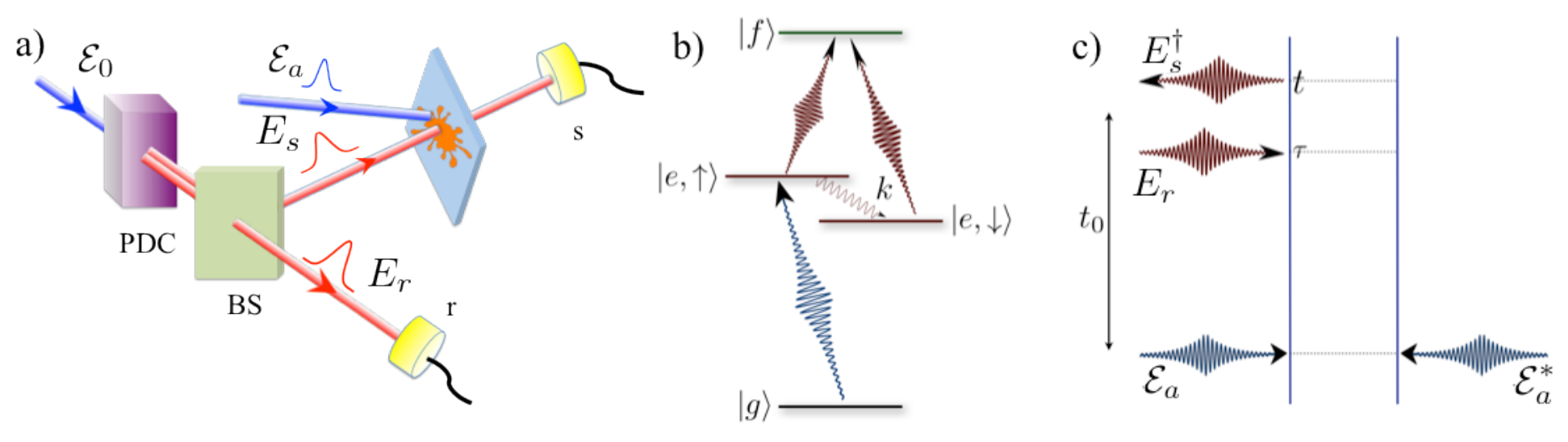}
\caption{(Color online) a) The IPP setup: the entangled photon pair in beams $E_s$ and $E_r$ are split on a beam splitter. A classical, actinic, ultrafast pulse $\mathcal{E}_a$ excites the sample, and $E_s$ is employed as a probe in a pump-probe measurement, while $E_r$ is detected in coincidence. b) The level scheme for the two-state jump (TSJ) model considered in this work. The $g - e$ transition is far off-resonant from the spectral range of the entangled photon wavepacket, which only couples to the $e - f$ transition. c) Diagram representing the pump-probe measurement. [taken from \cite{Schlawin15a}]}
\label{fig.TPC-setup}
\end{figure*}

Consider the setup depicted in Fig.~\ref{fig.TPC-setup}a). Unlike the linear absorption case here the probe photon is sent through the sample, which has previously been excited by a classical ultrafast laser pulse, and then detected \cite{Schlawin15a}.

The interferometric pump probe (IPP) signal with photon coincidence detection is dispersed spectrally by placing spectral filters in front of both detectors, and our signal is given by the change to this two-photon counting rate that is governed by a four-point correlation function of the field \cite{Walmsley2, Cho14a}
\begin{align}
\big\langle E^{\dagger}_r (\omega_r) E^{\dagger}_s (\omega) E_s (\omega) E_r (\omega_r) \big\rangle. \label{eq.TPC-definition}
\end{align}
Here, $\omega / \omega_r$ denotes the detected frequency of the respective spectral filter, and the brackets $\langle \cdots \rangle$ represent the expectation value with respect to the transmitted fields.  To obtain the desired pump-probe signal in a three-level system, we used a third order perturbation theory \cite{Schlawin15a}. The first two interactions are with the classical pump pulse, which is taken to be impulsive, $\mathcal{E}_a (t) = \mathcal{E}_a \delta (t)$, and the third is with the probe centered at $t=t_0$. Assuming $E_s$ to be far off-resonant from the $e - g$ transition, we obtain only the single diagram shown in Fig.~\ref{fig.TPC-setup}c), which reads
\begin{align}
&S_{IPP} (\omega,\omega_r;t_0) = - \frac{2}{\hbar} \Im \left( - \frac{i}{\hbar} \right)^3 \vert \mathcal{E}_a \vert^2 \int^{\infty}_0 \!\! dt\; e^{i \omega (t - t_0)}  \notag \\
&\times\int^t_0 \!\! d\tau F (t - \tau, \tau)\big\langle E^{\dagger}_r (\omega_r) E^{\dagger}_s (\omega) E_s (\tau) E_r (\omega_r) \big\rangle. \label{eq.S^TPC} 
\end{align}
We have defined the matter correlation function,
\begin{align}
F (t - \tau, \tau) &= \big\langle \vert \mu_{ge} \vert^2 \vert \mu_{ef} \vert^2 G_{fe} (t - \tau) G_{ee} (\tau) \big\rangle_{\text{env}}, \label{eq.field-corr-fct}
\end{align}
where $\mu_{ge}$ and $\mu_{ef}$ denote the dipole moments connecting ground state with the singly excited states manifold, as well as single with doubly excited state manifold, respectively. $\langle \cdots \rangle_{\text{env}}$ denotes the average with respect to environmental degrees of freedom, obtained from tracing out the bath. 
Here, we employ a stochastic Liouville equation \cite{Tanimura1} which represents the two-state jump (TSJ) model: A ground state $g$ is dipole-coupled to an electronic excited state $e$, which is connected to two spin states $\uparrow$ and $\downarrow$ undergoing relaxation \cite{Sanda1}. We additionally consider a doubly excited state $f$, which is dipole-coupled to both $\vert e, \uparrow \rangle$ and $\vert e, \downarrow \rangle$ [see Fig.~\ref{fig.TPC-setup}b)]. The electronic states are damped by a dephasing rate $\gamma$. We assume the low-temperature limit, where only the decay from $\uparrow$ to $\downarrow$ is allowed \cite{Konstantin2}. The decay is entirely incoherent, such that the description may be restricted to the two spin populations $\vert \! \uparrow \rangle \langle \uparrow \!\vert \hat{=} (1, 0)^T$ and $\vert \!\downarrow \rangle \langle \downarrow \!\vert \hat{=} (0, 1)^T$. The field correlation function (\ref{eq.field-corr-fct}) is then given by
\begin{align}
&F (t_2, t_1) = \vert \mu_{ge} \vert^2 \vert \mu_{ef} \vert^2 e^{- \gamma (t_1 + 2 t_2)} \notag \\
\times &\left( e^{- i \omega_+ t_2} + \frac{2 i \delta}{k + 2 i \delta} e^{- k t_1} \left[ e^{- (k + i \omega_-) t_2} - e^{- i \omega_+ t_2} \right] \right), \label{eq.F} 
\end{align}
where $\delta$ the energy difference between the two spin states, and $\omega_{\pm} = \omega_{fe} \pm \delta$. Note that, since we monitor the $f - e$ transition, the detected frequency will increase in time, from $\omega_-$ to $\omega_+$.


In the following, we use these results to first simulate the classical pump-probe signal, and then the two-photon counting signal with entangled photons. For the former case, we consider a classical Gaussian probe pulse
\begin{align}
E_{pr} (\omega) &= \frac{1}{\sqrt{2 \pi \sigma^2}} \exp \left[ - (\omega - \omega_0)^2 / 2 \sigma^2 \right].
\end{align}
We chose the following system parameters: $\omega_{fe} = 11,000$ cm$^{-1}$, $\delta = 200$ cm$^{-1}$, $k = 120$ cm$^{-1}$, and $\gamma = 100$ cm$^{-1}$. 

The center frequency $\omega_0$ is fixed at the transition frequency $\omega_{fe}$, and we vary the probe bandwidth. Panel a) shows the signal for $\sigma = 1,000$ cm$^{-1}$. 
Two peaks at $\omega_{fe} \pm \delta$ correspond to the detected frequency, when the system is either in the upper state (at $\omega_{fe} - \delta$), or in the lower state ($\omega_{fe} + \delta$). Due to the spectrally dispersed detection of the signal, the resonance widths are given by the linewidth $\gamma$, and not the much broader probe pulse bandwidth $\sigma$. For very short time delays $t_0$, both resonances increase, until the probe pulse has fully passed through the sample. Then the resonance at $10,800$ cm$^{-1}$, $i.e.$ the state $\vert e, \uparrow \rangle$, starts to decay rapidly, while the  resonance at $11,200$ cm$^{-1}$ peaks at longer delay times due to its initial population by the upper state. For longer delays, both resonance decay due to the additional dephasing.

\begin{figure}[t]
\centering
\includegraphics[width=0.5\textwidth]{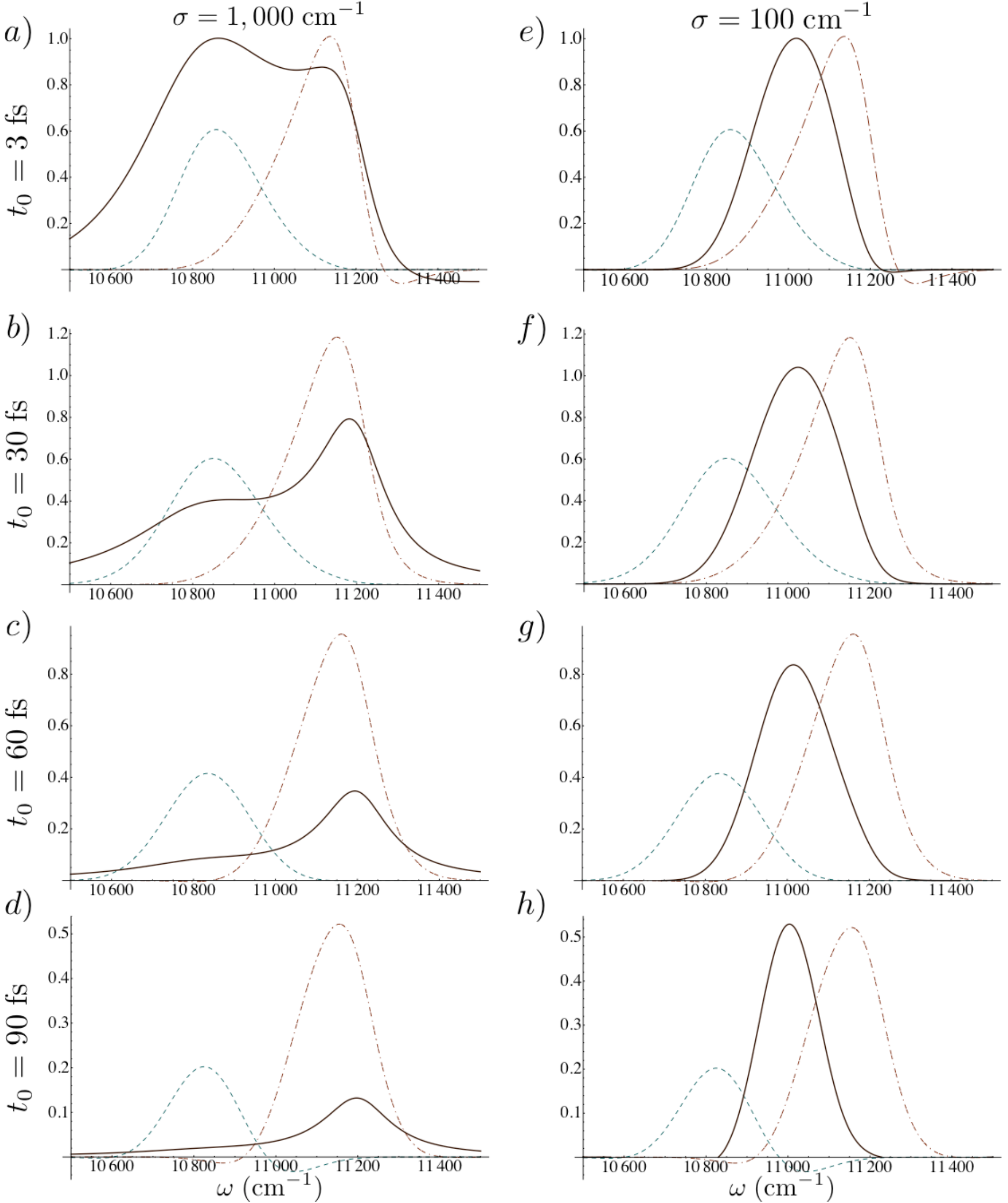}
\caption{(Color online) a) TPC signals with $\sigma_p = 2,000$~cm$^{-1}$ and $T = 90$~fs,  $\omega_r = 10,400$~cm$^{-1}$ (blue, dashed) and $11,400$~cm$^{-1}$ (red, dot-dashed) as well as the classical pump-probe signal (black, solid) with $\sigma = 1,000$~cm$^{-1}$ are plotted vs. the dispersed frequency $\omega$ with a time delay set to $t_0 = 3$~fs. b) Same for $t_0 = 30$~fs, c) $60$~fs, d) $90$~fs. e)-h) Same as a)-d), but with classical bandwidth $\sigma = 100$~cm$^{-1}$. 
The classical signal is normalized, such that its maximum value at $t_0 = 0$ is equal to one. Similarly, the TPC signal are normalized to the maximum value of the signal with $\omega_r = 11,400$~cm$^{-1}$ at $t_0 = 0$.[taken from \cite{Schlawin15a}]}
\label{fig.TPC-slices}
\end{figure}

The two-photon counting signal with entangled photons offers novel control parameters: The dispersed frequency $\omega$ of beam $1$, the pump frequency $\omega_p$ and its bandwidth $\sigma_p$ loosely correspond to the classical control parameters, $i.e.$ the central frequency $\omega_0$ and bandwidth $\sigma$. In addition, we may vary the entanglement time $T$ and the detected frequency of the reference beam $\omega_r$. 

Fig.~\ref{fig.TPC-slices} depicts the signal (\ref{eq.S^TPC}) obtained from entangled photons with $T = 90$ fs for different time delays $t_0$. For comparison, we show the classical pump-probe with bandwidth $\sigma = 1,000$~cm$^{-1}$ in the left column, and with $100$~cm$^{-1}$ in the right column. The TPC signals are normalized with respect to the maximum value of the signal at $t_0 = 3$~fs and $\omega_r = 11,400$~cm$^{-1}$.  The classical signal is normalized to its peak value at zero time delay, and the TPC signals to the signal with $\omega_r = 11,400$~cm$^{-1}$ at zero delay. As becomes apparent from the figure, a broadband classical probe pulse (left column) cannot excite specific states, such that the two resonances merge into one band. A narrowband probe (right column), on the other hand, cannot resolve the fast relaxation at all, and only shows the unperturbed resonance at $\omega_{fe}$. Interferometric signals, however, can target the relaxation dynamics of the individual states.

\subsubsection{Coincidence detection of Femtosecond Stimulated Raman Signals}

So far, we have demonstrated how  coincidence detection
can enhance linear absorption and pump-probe signals.  We now demonstrate the power of this interferometric detection for stimulated Raman signals commonly used to probe molecular vibrations. Applications include probing time-resolved photophysical and photochemical processes \cite{Kukura:AnnurevPhysChem:2007,Schreier:Science:2007,Adamczyk:Science:2009,Kuramochi:JPCLett:2012}, chemically specific biomedical imaging \cite{Xin02}, remote sensing \cite{Pes08,Aro12}. Considerable effort has been devoted to increasing the sensitivity and eliminating off-resonant background, thus improving the signal-to-noise ratio and enabling the detection of small samples and even single molecules. Pulse shaping \cite{Oro02,Pes07} and the combination of broad and narrow band pulses (technique known as Femtosecond Stimulated Raman Spectroscopy (FSRS) \cite{Die16}) have been employed. Here, we present an Interferometric FSRS (IFSRS) technique that combines quantum entangled light with interferometric detection \cite{Sca03,Yabushita04,Kalachev07a,Slattery13} in order to enhance the resolution and selectivity of Raman signals  \cite{Dorfman14a}. The measurement uses a pair of entangled photons, one (signal) photon interacts with the molecule, and acts as the broadband probe, while the other (idler) provides a reference for the coincidence measurement. By counting photons,  IFSRS can measure separately the gain and loss contributions to the Raman signal \cite{Har13} which is not possible with classical FSRS signals that only report their sum previousy (i.e. the net gain or loss). We showed how the entangled twin photon state may be used to manipulate  two-photon absorption $\omega_1+\omega_2$ type resonances in aggregates \cite{Saleh98a,Lee06a, Schlawin13b,Dorfman14b} but these ideas do not apply to Raman $\omega_1-\omega_2$ resonances. 

\begin{figure*}[t]
\begin{center}
\includegraphics[trim=0cm 0cm 0cm 0cm,angle=0, width=0.9\textwidth]{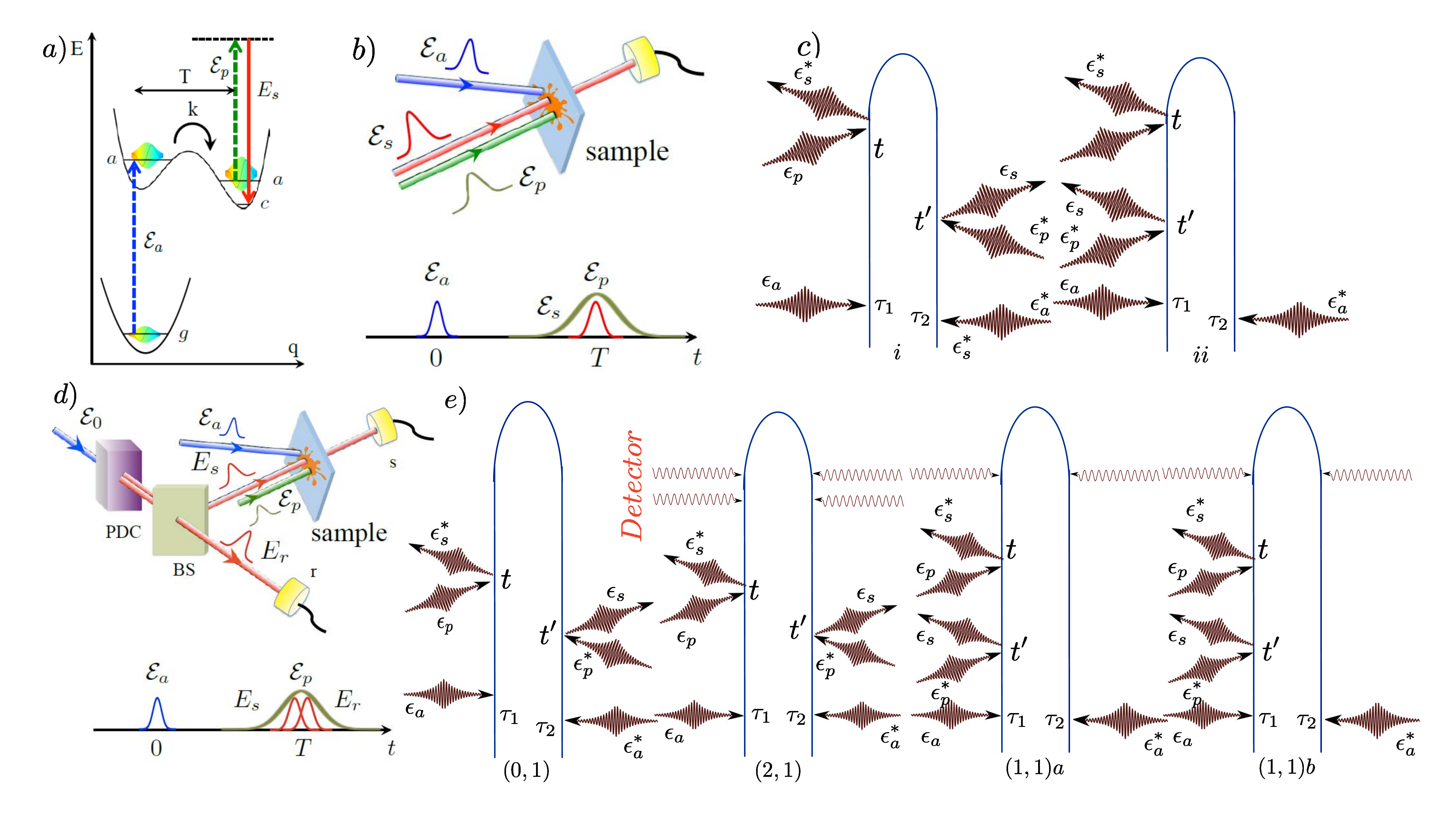}
\end{center}
\caption{(Color online) Top row: FSRS level scheme for the tunneling model - $\textbf{a}$, pulse configuration  - $\textbf{b}$, loop diagrams (for diagram rules see Appendix \ref{app:Diag})  - $\textbf{c}$. $\textbf{d}$ and $\textbf{e}$ - the same as $\textbf{b}$, and $\textbf{c}$ but for IFSRS. The pairs of indices  $(0,1)$ etc.  indicate number of photons registered by detectors $s$ and $r$ in each photon counting signal:  $(N_s,N_r)$}
\label{fig:setup}
\end{figure*}

In FSRS, an actinic resonant pulse $\mathcal{E}_a$ first creates a vibrational superposition state in an electronically excited state (see  Fig. \ref{fig:setup}a,b). After a variable delay $\tau$, the frequency resolved transmission of a broadband (femtosecond) probe $E_s$ in the presence of a narrowband (picosecond) pump $\mathcal{E}_p$ shows excited state vibrational resonances  generated by an off-resonant stimulated Raman process.  The FSRS signal is given by \cite{Dor13}
\begin{align}\label{eq:Sc0}
&S_{FSRS}(\omega,\tau)\notag\\
&=\frac{2}{\hbar}\mathcal{I}\int_{-\infty}^{\infty}dte^{i\omega(t-\tau)}\langle\mathcal{T}\mathcal{E}_s^{*}(\omega)\mathcal{E}_p(t)\alpha(t)e^{-\frac{i}{\hbar}\int H'_-(\tau)d\tau}\rangle,
\end{align}
where $\alpha$ is the electronic polarizability, $\mathcal{I}$ denotes the imaginary part, and $\mathcal{E}_s=\langle E_s\rangle$ is expectation value of the probe field operator with respect to classical state of light (hereafter $\mathcal{E}$  denotes classical fields  and $E$ stands for quantum fields). $H_-'$ is the Hamiltonian superoperator in the interaction picture which, for off resonance Raman processes, can be written as
\begin{align}
H'(t)=\alpha E_s^{\dagger}(t)\mathcal{E}_p(t)+ \mathcal{E}_a^{*}(t)V+H.c., 
\end{align}
where  $V$ is the dipole moment, $\alpha$ is the off resonant polarizability. Formally, this is a six-wave mixing process. Expanding the signal (\ref{eq:Sc0}) to sixth order in the fields $\sim \mathcal{E}_s^2\mathcal{E}_p^2\mathcal{E}_a^2$. we obtain the classical FSRS signal 
\begin{align}\label{eq:Sisr1}
S_{FSRS}^{(i)}&(\omega,\tau)=\frac{2}{\hbar}\mathcal{I}\int_{-\infty}^{\infty}dt\int_{-\infty}^td\tau_1\int_{-\infty}^tdt'\int_{-\infty}^{t'}d\tau_2 \notag\\
&\times e^{i\omega(t-\tau)}\mathcal{E}_p(t)\mathcal{E}_p^{*}(t')\mathcal{E}_a^{*}(\tau_2)\mathcal{E}_a(\tau_1)\mathcal{E}_s^{*}(\omega)\mathcal{E}_s(t') \notag\\
&\times F_i(t'-\tau_2,t-t',t-\tau_1),
\end{align}
\begin{align}\label{eq:Siisr1}
S_{FSRS}^{(ii)}&(\omega,\tau)=\frac{2}{\hbar}\mathcal{I}\int_{-\infty}^{\infty}dt\int_{-\infty}^td\tau_2\int_{-\infty}^tdt'\int_{-\infty}^{t'}d\tau_1\notag\\
&\times e^{i\omega(t-\tau)}\mathcal{E}_p(t)\mathcal{E}_p^{*}(t')\mathcal{E}_a(\tau_1)\mathcal{E}_a^{*}(\tau_2)\mathcal{E}_s^{*}(\omega)\mathcal{E}_s(t')\notag\\
&\times F_{ii}(t-\tau_2,t-t',t'-\tau_1).
\end{align}
The two terms correspond to the two diagrams in Fig. \ref{fig:setup}c).
All relevant information is contained in the two four point correlation functions 
\begin{align}\label{eq:Fi}
F_i(t_1,t_2,t_3)=\langle V G^{\dagger}(t_1)\alpha G^{\dagger}(t_2)\alpha G(t_3)V^{\dagger}\rangle,
\end{align}
\begin{align}\label{eq:Fii}
F_{ii}(t_1,t_2,t_3)=\langle V G^{\dagger}(t_1)\alpha G(t_2)\alpha G(t_3)V^{\dagger}\rangle,
\end{align}
where the retarded Green's function $G(t)=(-i/\hbar)\theta(t)e^{-iHt}$ represents forward time evolution with the free-molecule Hamiltonian $H$ (diagrams $(1,1)a$, $(1,1)b$) and $G^{\dagger}$ represents  backward evolution. $F_i$ involves one forward and two backward evolution periods and $F_{ii}$ contains two forward followed by one backward propagation. $F_i$ and $F_{ii}$ differ by the final state of the matter (at the top of each diagram). In $F_i$ ($F_{ii}$) it is different (the same) as the state after preparation by the actinic pulse.

\paragraph{Photon correlation measurements.} 

In IFSRS, the probe pulse $E_s$ belongs to a pair of entangled beams generated in degenerate type-II PDC. The polarizing beam splitter (BS) in  Fig. \ref{fig:setup}d) then separates the orthogonally polarized photons. The horizontally polarized beam $E_s$ is propagating in the $s$  arm of the interferometer, and interacts with the molecule. The vertically polarized beam $E_r$ propagates freely in the $r$ arm, and serves as a reference. IFSRS has the following control knobs: the time and frequency parameters of the single photon detectors, the frequency of the narrowband classical pump pulse $\omega_p$, and the time delay $T$ between the actinic pulse $\mathcal{E}_a$ and the probe $E_s$.  

As discussed in section~\ref{sec.photon-measurement}, the joint time-and-frequency gated detection rate of $N_s$ photons ($N_s = 0,1,2$) in detector $s$ and a single photon in $r$ when both detectors have narrow spectral gating is given by 
\begin{align}\label{eq:S10}
&S_{IFSRS}^{(N_s,1)}(\bar{\omega}_{s_1},...,\bar{\omega}_{s_{N_s}},\bar{\omega}_r,\Gamma_i)\notag\\
&=\langle\mathcal{T} E_r^{\dagger}(\bar{\omega}_r)E_r(\bar{\omega}_r)\prod_{j=1}^{N_s}E_s^{\dagger}(\bar{\omega}_{s_j})E_s(\bar{\omega}_{s_j})e^{-\frac{i}{\hbar}\int_{-\infty}^{\infty}H'_-(\tau)d\tau}\rangle.
\end{align}
where $\Gamma_i$ stands for the incoming light beam parameters. In the standard Glauber's approach \cite{Glauber07}, the correlation function is calculated in the field space using normally-ordered field operators. The present expressions  in contrast  are given in the joint field and matter degrees of freedom, and the bookkeeping of the fields is instead solely based on the time ordering of superoperators. Normal ordering is never used. 

Expansion of Eq. (\ref{eq:S10}) in the number of the field matter interactions depicted by loop diagrams in Fig. \ref{fig:setup}e) yields for $N_s=0$ - Raman loss (no photon in the molecular arm)
\begin{widetext}
\begin{align}\label{eq:S010}
S_{IFSRS}^{(0,1)}(\bar{\omega}_r;\tau)&=\mathcal{I}\frac{1}{\hbar}\int_{-\infty}^{\infty}dt\int_{-\infty}^{\infty}dt'\int_{-\infty}^{t}d\tau_1\int_{-\infty}^{t'}d\tau_2\mathcal{E}_p(t')\mathcal{E}_p^{*}(t)\mathcal{E}_a(\tau_1)\mathcal{E}_a^{*}(\tau_2)\notag\\
&\times\langle \mathcal{T}E_s^{\dagger}(t')\tilde{E}_r^{\dagger}(\bar{\omega}_r)\tilde{E}_r(\bar{\omega}_r)E_s(t)\rangle F_i(t'-\tau_2,t-t',t-\tau_1).
\end{align}
To make sure that there is no photon at detector $s$ we had integrated over its entire bandwidth, thus eliminating the dependence on detector parameters. 

For the Raman gain $N_s=2$ signal ($i.e.$ two photons in the $s$-arm, one photon in the $r$-arm), when both $s$ and $r$ detectors have narrow frequency gates, we get
\begin{align}\label{eq:S2100}
S_{IFSRS}^{(2,1)}(\bar{\omega}_{s_1},\bar{\omega}_{s_2},\bar{\omega}_r;\tau)&=\mathcal{I}\frac{1}{\hbar}\int_{-\infty}^{\infty}d\bar{t}_{s_1}e^{i\bar{\omega}_{s_1}(\bar{t}_{s_1}-\tau)}\int_{-\infty}^{\bar{t}_{s_1}}dt\int_{-\infty}^{t}dt'\int_{-\infty}^{t}d\tau_1\int_{-\infty}^{t'}d\tau_2\mathcal{E}_p(t)\mathcal{E}_p^{*}(t')\mathcal{E}_a(\tau_1)\mathcal{E}_a^{*}(\tau_2)\notag\\
&\times\langle \mathcal{T}E_s(t')\tilde{E}_{s}^{\dagger}(\bar{\omega}_{s_1})\tilde{E}_{s}^{\dagger}(\bar{\omega}_{s_2})\tilde{E}_r^{\dagger}(\bar{\omega}_r)\tilde{E}_r(\bar{\omega}_r)\tilde{E}_{s}(\bar{\omega}_{s_2})\tilde{E}_{s}(\bar{t}_{s_1})E_s^{\dagger}(t)\rangle F_i(t'-\tau_2,t-t',t-\tau_1).
\end{align}
Finally, the $N_s=1$ signal (single photon in each arm) is given by
\begin{align}\label{eq:S11a1}
S_{IFSRS}^{(1,1)a}(\bar{\omega}_s,\bar{\omega}_r;\tau)&=-\mathcal{I}\frac{1}{\hbar}\int_{-\infty}^{\infty}dt_s'e^{i\bar{\omega}_s(t_s'-\tau)}\int_{-\infty}^{t_s'}dt\int_{-\infty}^tdt'\int_{-\infty}^{t'}d\tau_1\int_{-\infty}^{t_s'}d\tau_2\mathcal{E}_p(t)\mathcal{E}_p^{*}(t')\mathcal{E}_a(\tau_1)\mathcal{E}_a^{*}(\tau_2)\notag\\
&\times\langle \mathcal{T}\tilde{E}_s^{\dagger}(\bar{\omega}_s)\tilde{E}_r^{\dagger}(\bar{\omega}_r)\tilde{E}_r(\bar{\omega}_r)\tilde{E}_s(t_s')E_s^{\dagger}(t)E_s(t')\rangle F_{ii}(t-\tau_2,t-t',t'-\tau_1),
\end{align}
\begin{align}\label{eq:S11b1}
S_{IFSRS}^{(1,1)b}(\bar{\omega}_s,\bar{\omega}_r;\tau)&=-\mathcal{I}\frac{1}{\hbar}\int_{-\infty}^{\infty}dt_s'e^{i\bar{\omega}_s(t_s'-\tau)}\int_{-\infty}^{t_s'}dt\int_{-\infty}^tdt'\int_{-\infty}^{t'}d\tau_1\int_{-\infty}^{t_s'}d\tau_2\mathcal{E}_p(t')\mathcal{E}_p^{*}(t)\mathcal{E}_a(\tau_1)\mathcal{E}_a^{*}(\tau_2)\notag\\
&\times\langle \mathcal{T}\tilde{E}_s^{\dagger}(\bar{\omega}_s)\tilde{E}_r^{\dagger}(\bar{\omega}_r)\tilde{E}_r(\bar{\omega}_r)\tilde{E}_s(t_s')E_s(t)E_s^{\dagger}(t')\rangle F_{ii}(t-\tau_2,t-t',t'-\tau_1).
\end{align}
\end{widetext}

\paragraph{Photon counting detection window for the molecular response} 

\begin{figure*}[t]
\begin{center}
\includegraphics[trim=0cm 0cm 0cm 0cm,angle=0, width=0.9\textwidth]{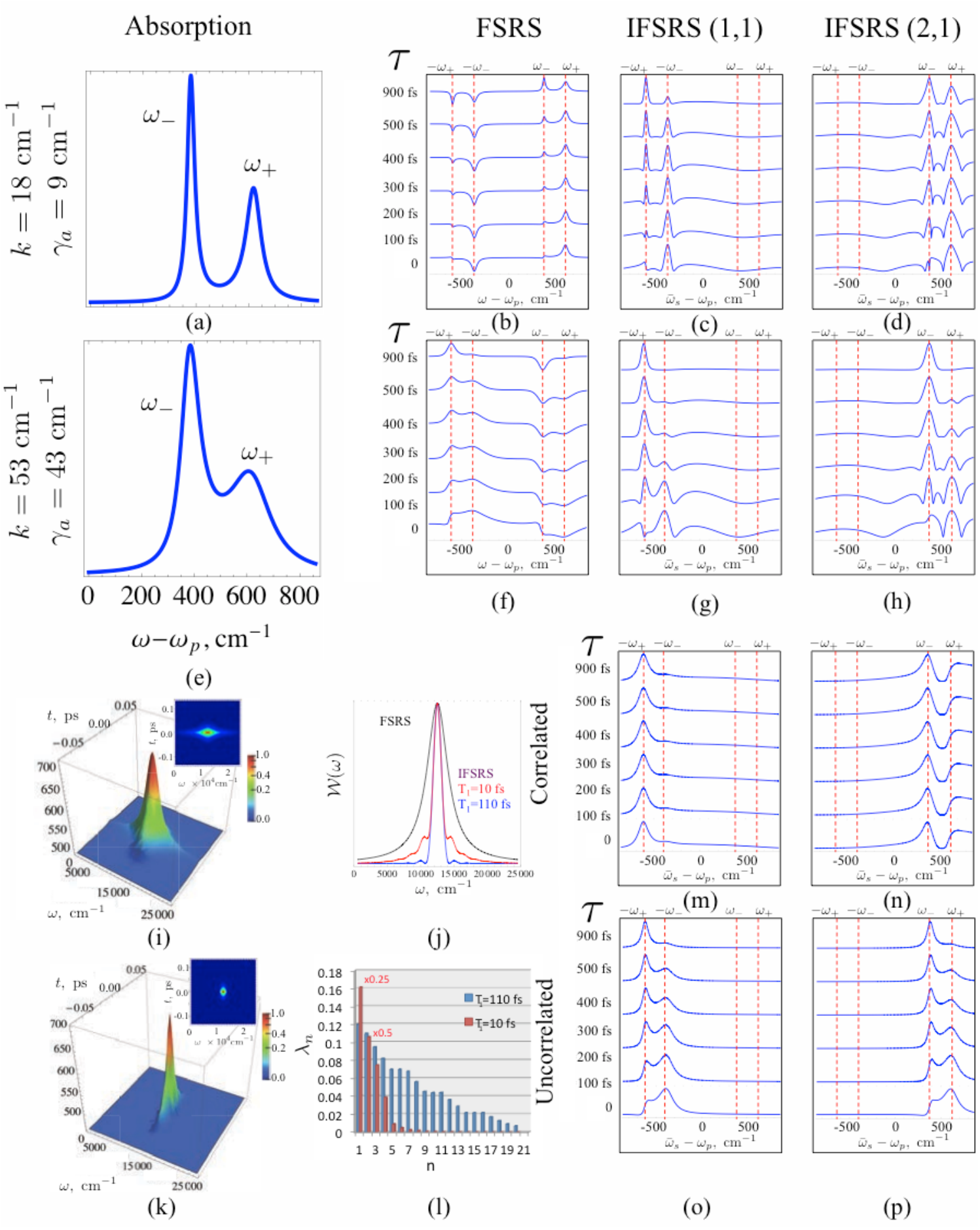}
\end{center}
\caption{(Color online)Top: (a) - Absorption, (b) - classical FSRS, (c) - $S_{IFSRS}^{(1,1)}$, (d) - $S_{IFSRS}^{(2,1)}$ for a time evolving vibrational mode i vs $\omega-\omega_p$ for slow tunneling rate and narrow dephasing, (e) - (h) are the same as (a) - (d) but for fast tunneling rate and broad dephasing. 
Bottom left: (i) - time-frequency Wigner spectrogram for classical light, (k) - same as (i) but for entangled twin state given by Eq. (\ref{eq:twin}). Inserts depict a 2D projection. (j) - window function for FSRS $\mathcal{E}_s^{*}(\omega)\mathcal{E}_s(\omega+i\gamma_a)$ -black, and IFSRS $\Phi^{*}(\omega,\bar{\omega}_r)\Phi(\omega+i\gamma_a,\bar{\omega}_r)$ different values of $T_1$. (l) - spectrum of the eigenvalues $\lambda_n$ in the Schmidt decomposition (\ref{eq.decomposition}) for entangled state with amplitude (\ref{eq:twin1}). Bottom right: $S_{IFSRS}^{(1,1)}$ signal  vs $\bar{\omega}_s-\omega_p$ for correlated - (m) and uncorrelated -(o) separable states. (n), and (p) - same as (m) and (o) but for $S_{IFSRS}^{(2,1)}$ signal. Parameters for simulations are listed in \cite{Dorfman14a}.}
\label{fig:Wig}
\end{figure*}

The input two-photon state has a  single photon in each the $s$- mode and single photon in the $r$ - mode. 
and is described by [compare to Eq.~(\ref{eq.ent-state1})]
\begin{align}\label{eq:twin}
|\psi\rangle=|0\rangle+\int_{-\infty}^{\infty} \!\! d\omega_s \int_{-\infty}^{\infty} \!\!d\omega_r\Phi(\omega_s,\omega_r)a_{\omega_s}^{\dagger}a_{\omega_r}^{\dagger}|0\rangle,
\end{align}
where $a_{\omega_s}^{\dagger}$($a_{\omega_r}^{\dagger}$) is the creation operator of a horizontally (vertically) polarized photon and the two-photon amplitude is given by
\begin{align}\label{eq:twin1}
\Phi(\omega_s,\omega_r)&=\sum_{i\neq j=1}^2\text{sinc}\left(\omega_{s0}T_i/2+\omega_{r0}T_j/2\right)\notag\\
&\times A_p (\omega_s+\omega_r)e^{i\omega_{s0}T_i/2+i\omega_{r0}T_j/2},
\end{align}
where $\omega_{k0}=\omega_k-\omega_0$, $k=s,r$, $A_p (\omega)=A_0/[\omega-\omega_0+i\sigma_0]$ is the classical pump $T_{j}=(1/v_p-1/v_{j})L$, $j=1,2$ is the time delay between the $j-th$ entangled and the classical pump beam after propagation through the PDC crystal. $T=T_2-T_1$ is  the entanglement time. In $A_p (\omega)$, the sum-frequency $\omega_s+\omega_r$ is centered around $2\omega_0$ with bandwidth $\sigma_0$. 
For a broadband classical pump, the frequency difference $\omega_s-\omega_0$ becomes narrow with bandwidth $T_j^{-1}$, $j=1,2$. The output state of light in mode $s$ may contain a varying number of photons, depending on the order of the field-matter interaction. 

As discussed in section~\ref{sec.photon-entanglement}, the twin photon state Eq. (\ref{eq:twin}) is not necessarily entangled. Using the Schmidt decomposition~(\ref{eq.decomposition}), we obtain for the two-photon amplitude in Eq. (\ref{eq:twin1}) the rich spectrum of eigenvalues shown in  Fig. \ref{fig:Wig}l) shows that the state is highly entangled with up to 20 modes contributing with appreciable weight. As will be shown later [Eq.~(\ref{eq.Wigner-two-photon-state})], this entanglement is reflected in the violation of the Fourier uncertainty $\Delta\omega\Delta t\geq 1$ in two-photon transitions. 

We next turn to how entanglement affects  the Raman resonances. Both FSRS and IFSRS signals are governed by a four-point matter correlation function (two polarizabilities $\alpha_{ac}$ and two dipole moments $V_{ag}$ as depicted by the loop diagrams shown in  Fig. \ref{fig:setup}c) and e), respectively. Depending on the number of photons detected, this four-point matter correlation function is convoluted with different field correlation functions. For $N_s=0$, and $N_r=1$  Eq. (\ref{eq:S010}) is given by the four-point correlation function for a quantum field [red arrows in  Fig.~\ref{fig:setup}e)]. For a twin photon state, we recall from Eq.~(\ref{eq.four-pt-cw}) that the   field correlation function can be factorized as 
\begin{align}
\langle\psi|E_s^{\dagger}(\omega_a)E_r^{\dagger}(\omega_b)E_r(\omega_c)E_s(\omega_d)|\psi\rangle=\Phi^{*}(\omega_a,\omega_b)\Phi(\omega_c,\omega_d).
\end{align}
The $N_s=2$ signal is given by an eight-point [see Eq. (\ref{eq:S2100})], and for $N_s=1$ it is governed by a six-point field correlation function as shown in Eqs.~(\ref{eq:S11a1}) - (\ref{eq:S11b1}). The detailed derivation and explicit closed form expressions for multipoint correlation functions of the field are presented in Ref. \cite{Dorfman14a}. 
All three IFSRS signals with $N_s=0,1,2$ eventually scale linearly with the classical pump intensity $S_{IFSRS}\propto |A_0|^2$, similar to classical FSRS even though a different number of fields contribute to the detection.

We next compare the different field spectrograms which represent the temporal and spectral windows created by the fields. Fig. \ref{fig:Wig}i depicts the time-frequency Wigner spectrogram for the classical probe field $\mathcal{E}_s$: 
 \begin{align}
 W_s(\omega,t)=\int_{-\infty}^{\infty}\frac{d\Delta}{2\pi}\mathcal{E}_s^{*}(\omega)\mathcal{E}_s(\omega+\Delta)e^{-i\Delta t}.
 \end{align}
  The Fourier uncertainty $\Delta \omega\Delta t\geq 1$ limits the frequency resolution for a given time resolution. The corresponding Wigner two-photon spectrogram  for the entangled twin photon state
  \begin{align}
  W_q(\omega,t;\bar{\omega}_r)=\int_{-\infty}^{\infty}\frac{d\Delta}{2\pi}\Phi^{*}(\omega,\bar{\omega}_r)\Phi(\omega+\Delta,\bar{\omega}_r)e^{-i\Delta t}, \label{eq.Wigner-two-photon-state}
  \end{align}
    is depicted in  Fig. \ref{fig:Wig}k. For the same temporal resolution of the FSRS, the spectral resolution of IFSRS is significantly higher since the time and frequency variables for entangled light are not Fourier conjugate variables \cite{Schlawin13b}. The high spectral resolution in the entangled case is governed by $T_j^{-1}$, $j=1,2$ which is narrower than the broadband probe pulse such that $\Delta \omega\Delta t\simeq 0.3$.  Fig.~\ref{fig:Wig}j) demonstrates that entangled window function $R_q^{(N_s,1)}$ for $N_s=1,2$ (see  Eqs. (\ref{eq:Rq11}), (\ref{eq:Rq21}))  that enters the IFSRS Signal (\ref{eq:Sq3}) yields a much higher spectral resolution than the classical $R_c$ in Eq. (\ref{eq:Rc}).

The molecular information required for all three measurement outcomes $(N_s=0,1,2)$ is given by two correlation functions $F_i$ and $F_{ii}$ (see  Fig. \ref{fig:setup}c,e and Eqs. (\ref{eq:Fi}) - (\ref{eq:Fii})), which are then convoluted with a different detection window for FSRS and IFSRS. $F_i$ and $F_{ii}$ may not be separately detected by FSRS. However, in IFSRS the loss  $S_{IFSRS}^{(0,1)}$ and the gain $S_{IFSRS}^{(2,1)}$ Raman signals probe $F_i$ (the final state $c$ can be different from initial state $a$) whereas the coincidence counting signal $S_{IFSRS}^{(1,1)}$ depends on $F_{ii}$ (initial and final states are identical). Interferometric signals can thus separately measure the two matter correlation functions.

\paragraph{IFSRS for a vibrational mode in a tunneling system.} 

To demonstrate the effect of entanglement in interferometric measurements, we show the calculated signals for the three-level model system depicted in  Fig. \ref{fig:setup}a. Once excited by the actinic pulse, the initial state with vibrational frequency $\omega_{+}=\omega_{ac}+\delta$ can tunnel through a barrier at a rate $k$ and assume a different vibrational frequency $\omega_-=\omega_{ac}-\delta$ (see Ref. \cite{Konstantin2}). The probability to be in the initial state with $\omega_+$ decreases exponentially $P_+(t)=e^{-kt}$, whereas for $\omega_-$ it grows as $P_-(t)=1-e^{-kt}$. This model represents Kubo's two-state jump model in the low temperature limit \cite{Kub63, Konstantin2} which we had also discussed before in section \ref{sec.pp-coincidence}. The absorption lineshape is given by 
\begin{align}\label{eq:lin}
S_l(\omega)&=-\mathcal{I}\frac{4}{\hbar^2}|\mathcal{E}(\omega)|^2\frac{|\mu_{ac}|^2}{k+2i\delta}\notag\\
&\times\left(\frac{k+i\delta}{\omega-\omega_-+i\gamma_a}+\frac{i\delta}{\omega-\omega_++i(\gamma_a+k)}\right).
\end{align}
This shows two peaks with combined width governed by dephasing $\gamma_a$ and tunneling rate $k$. The corresponding IFSRS signal $S_{IFSRS}^{(N_s,1)}$ with $N_s=0,1,2$  is given by
\begin{align}\label{eq:Sq3}
&S_{IFSRS}^{(N_s,1)}(\bar{\omega}_s,\bar{\omega}_r;\omega_p,\tau)=\mathcal{I}\frac{\mu}{\hbar^4}|\mathcal{E}_p|^2|\mathcal{E}_a|^2\sum_{a,c}\alpha_{ac}^2|\mu_{ag}|^2\notag\\
&\times e^{-2\gamma_a\tau}\left(R_q^{(N_s,1)}(\bar{\omega}_s,\bar{\omega}_r,2\gamma_a,\bar{\nu}\omega_\nu-i\gamma_a)-\frac{2i\delta e^{-k\tau}}{k+2i\delta}\right.\notag\\
&\times\left.[R_q^{(N_s,1)}(\bar{\omega}_s,\bar{\omega}_r,2\gamma_a+k,\bar{\nu}\omega_\nu-i\gamma_a)\right.\notag\\
&\left.-R_q^{(N_s,1)}(\bar{\omega}_s,\bar{\omega}_r,2\gamma_a+k,\bar{\nu}\omega_{\bar{\nu}}-i(\gamma_a+k)]\right),
\end{align}
where $\nu=-$ for $N_s=0,2$ and $\nu=+$ for $N_s=1$, $\mu=-$ for $N_s=1,2$ and $\mu=+$ for $N_s=0$. The Raman response $R_q^{(N_s,1)}$ which depends on the window created by the quantum field for different photon numbers $N_s$ is given by
\begin{align}\label{eq:Rq01}
R_q^{(0,1)}(\bar{\omega}_s,\bar{\omega}_r,\gamma,\Omega)=\int_{-\infty}^{\infty}\frac{d\omega}{2\pi}\frac{\Phi^{*}(\omega,\bar{\omega}_r)\Phi(\omega+i\gamma,\bar{\omega}_r)}{\omega-\omega_p-\Omega},
\end{align}
\begin{align}\label{eq:Rq11}
R_q^{(1,1)}(\bar{\omega}_s,\bar{\omega}_r,\gamma,\Omega)=\frac{\Phi^{*}(\bar{\omega}_s,\bar{\omega}_r)\Phi(\omega_p+\Omega-i\gamma,\bar{\omega}_r)}{\bar{\omega}_s-\omega_p-\Omega},
\end{align}
\begin{align}\label{eq:Rq21}
R_q^{(2,1)}(\bar{\omega}_s,\bar{\omega}_r,\gamma,\Omega)=\frac{\Phi^{*}(\bar{\omega}_s,\bar{\omega}_r)\Phi(\bar{\omega}_s+i\gamma,\bar{\omega}_r)}{\bar{\omega}_s-\omega_p-\Omega}.
\end{align}
For comparison, we give the classical FSRS signal (\ref{eq:Sc0})
\begin{align}\label{eq:Sc3}
&S_{FSRS}^{(c)}(\omega,\tau)=-\mathcal{I}\frac{2}{\hbar^4}|\mathcal{E}_p|^2|\mathcal{E}_a|^2\sum_{a,c}\alpha_{ac}^2|\mu_{ag}|^2e^{-2\gamma_a\tau}\times\notag\\
&\left[R_c(\omega,2\gamma_a,\omega_--i\gamma_a)-\frac{2i\delta e^{-k\tau}}{k+2i\delta}[R_c(\omega,2\gamma_a+k,\omega_--i\gamma_a)\right.\notag\\
&\left.-R_c(\omega,2\gamma_a+k,\omega_+-i(\gamma_a+k)]-(\omega_{\pm}\leftrightarrow-\omega_{\mp})\right],
\end{align}
 where
\begin{align}\label{eq:Rc}
R_c(\omega,\gamma,\Omega)=\frac{\mathcal{E}_s^{*}(\omega)\mathcal{E}_s(\omega+i\gamma)}{\omega-\omega_p-\Omega}
\end{align}
is the Raman response gated by the classical field.

 Figs. \ref{fig:Wig}a)-h) compare the classical FSRS signal [Eq. (\ref{eq:Sc3})] with the IFSRS signals $S_{IFSRS}^{(1,1)}$ and $S_{IFSRS}^{(2,1)}$ [Eq. (\ref{eq:Sq3})]. For slow modulation and long dephasing time $k,\gamma_a\ll\delta$ the absorption spectrum [Fig. \ref{fig:Wig}a)] has two well-resolved peaks at $\omega_{\pm}$. The classical FSRS shown in Fig. \ref{fig:Wig}b) has one dominant resonance at $\omega_+$ which decays with increase of delay $T$, whereas the $\omega_-$ peak slowly builds up and dominates at longer $T$. This signal contains both blue- and red-shifted Raman resonances relative to the narrowband pump frequency: $\omega-\omega_p=\pm\omega_{\pm}$. If the modulation and dephasing rates are comparable to the level splitting $k,\gamma_a\sim\delta$, then the $\omega_{\pm}$ resonances in the absorption [Fig. \ref{fig:Wig}e)] and the classical FSRS [Fig. \ref{fig:Wig}f)] broaden, and become less resolved. 

We next compare this with IFSRS. For small modulation and long dephasing, $S_{IFSRS}^{(1,1)}$ is similar to the classical FSRS [see  Fig. \ref{fig:Wig}c)]. However, both temporal and spectral resolution remains high, even when the modulation is fast and the dephasing width is large, as seen in  Fig. \ref{fig:Wig}g). The same applies to the $S_{IFSRS}^{(2,1)}$ signal depicted for slow -  Fig. \ref{fig:Wig}d) and fast -  Fig. \ref{fig:Wig}h) tunneling. 
 
Apart from the different detection windows, there is another important distinction between IFSRS (Eq. (\ref{eq:Sq3})) and the classical FSRS (\ref{eq:Sc3}) signals. In the latter, the gain and loss contributions both contain red- and blue- shifted features relative to the narrow pump.  The FSRS signal can contain both Stokes and anti Stokes components. FSRS can only distinguish between red and blue contributions. In contrast, the interferometric signal can measure separately the gain $S_{IFSRS}^{(2,1)}$ and the loss contributions $S_{IFSRS}^{(0,1)}$. 
 
 \paragraph{The role of entanglement.} 
 
 We now show that the enhanced resolution of Raman resonances may not be reproduced by classically shaped light and entanglement is essential. To this end, we calculate the IFSRS signals (\ref{eq:Sq3}) for the correlated-separable state \cite{Zheng13} described by the density matrix 
 \begin{align}
 \rho_{cor}=\int_{-\infty}^{\infty}d\omega_sd\omega_r|\Phi(\omega_s,\omega_r)|^2|1_{\omega_s},1_{\omega_r}\rangle\langle 1_{\omega_s},1_{\omega_r}|.
 \end{align}
  This  is a diagonal part of the density matrix corresponding to state Eq. (\ref{eq:twin}) with amplitude Eq. (\ref{eq:twin1}), which results from the disentanglement of the twin state. This state yields the same single-photon spectrum and shows strong frequency correlations similar to entangled case, and is typically used as a benchmark to quantify entanglement in quantum information processing \cite{Law00}. We further compare this with signals from the fully separable uncorrelated Fock state given by Eq. (\ref{eq:twin}) with 
  \begin{align}
  \Phi_{uncor}(\omega_s,\omega_r)=\Phi_s(\omega_s)\Phi_r(\omega_r)
  \end{align}
   with $\Phi_k(\omega_k)=\Phi_0/[\omega_k-\omega_0+i\sigma_0]$, $k=s,r$ with parameters matching the classical probe pulse used in FSRS.

Fig. \ref{fig:Wig}m-p illustrates $S_{IFSRS}^{(1,1)}$ for these two states of light. The separable correlated state shown in Fig. \ref{fig:Wig}m has high spectral and no temporal resolution, as expected from a cw time-averaged state in which the photons arrive at any time \cite{Zheng13}.  The separable uncorrelated state (see Fig. \ref{fig:Wig}o) yields slightly better resolution than in the classical FSRS signal in  Fig. \ref{fig:Wig}f. Similar results can be obtained for the $S_{IFSRS}^{(2,1)}$ (see  Fig. \ref{fig:Wig}n and p respectively). 



\section{Entangled light generation via nonlinear light-matter interactions; nonclassical response functions}
\label{sec.generation-&-response-fct}



The light generation process is typically based on the assumption that the light field is given, or that different optical fields that participate in generation of the desired light state interact with other fields. This way the generation process can be described by an effective Hamiltonian in the field space alone. The material quantities that assist the light conversion are typically set to be constant parameters governed by nonlinear semiclassical susceptibilities.

Superoperator non-equilibrium Green's functions are useful for calculating nonlinear optical
processes involving any combination of classical and quantum optical modes. Closed correlation-function
expressions based on superoperator time-ordering may be derived for the combined effects of causal (response) and non-causal (spontaneous fluctuations) correlation functions \cite{Roslyak09a}. 

Below we survey several wave-mixing schemes for generating quantum light by using a combination of classical and quantum
modes of the radiation field.  Homodyne-detected sum frequency
generation (SFG)\cite{shen1989spp} and difference frequency
generation (DFG)\cite{dick1983sal,Mukamel_book} involve two
classical and one quantum mode.  Parametric down conversion
(PDC)\cite{mandel1995oca, HOM85, klyshko1988pan, Louisell61}
involves one classical and two quantum modes and is one of the primary sources of
entangled photon pairs \cite{Ger05,edamatsu2007epg,uren2006gtp}.
All of these are coherent measurements, and scale as $N(N-1)$ for $N$ active molecules the signals \cite{Marx:PhysRevA:08}.

We further consider incoherent $\sim N$-scaling signals. These include heterodyne detected SFG and DFG, which involve three
classical modes, and two types of two photon fluorescence
\cite{denk1990tpl}: two photon induced fluorescence with one
classical and two quantum modes
(TPIF)\cite{xu1996mtp,callis_theory_1993,rehms1993tpf} and two
photon emitted fluorescence with two classical and one quantum make
(TPEF). The list of the different measurement is summarized in the Table \ref{TABLE1}.

Finally we present a more detailed microscopic theory of entangled light generation in two schemes, which describe light-matter interactions that involve quantum field. The first is based on type-I PDC in a cascade three-level scheme, and in the second scheme an entangled photon pair is generated by two remote molecules assisted by an ideal 50:50 beam splitter.

\begin{figure}
  \includegraphics[width=0.5\textwidth]{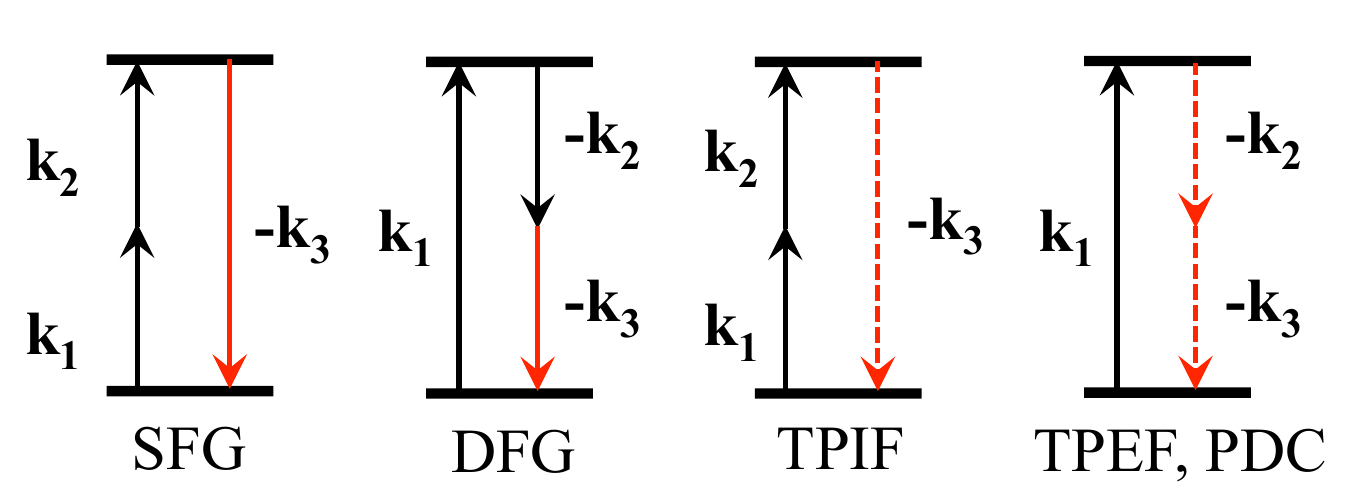}\\
  \caption{Three wave process involuting classical (solid line) and quantum (dashed line) modes. Black lines represent incoming fields, red lines correspond to generated light.}\label{fig:sketch5}
\end{figure}


\begin{table*}
  \begin{tabular}{|c|c|c|c|c|c|c|}
    \hline
    \multicolumn{7}{|c|}{Three wave processes}\\
    \hline
    \multicolumn{3}{|c|}{Heterodyne} & \multicolumn{4}{|c|}{Homodyne}\\
    \hline
    \multicolumn{3}{|c|}{} & \multicolumn{2}{|c|}{Incoherent} & \multicolumn{2}{|c|}{Coherent}\\
    \hline
     Technique & SFG & DFG & TPIF & TPEF & SFG & PDC \\
    \hline
     Modes & c/c/c & c/c/c & c/c/q & c/q/q & c/c/q & c/q/q \\
    \hline
    $\omega_3$ & $\omega_1+\omega_2$ & $\omega_1-\omega_2$ & - & - & $\approx \omega_1+\omega_2$ & $\approx \omega_1-\omega_2$\\
    \hline
    SNGF's & $\chi^{(2)}_{+--}$ & $\chi^{(2)}_{+--}$ & $(\chi^{(5)}_{++----}$ & $(\chi^{(5)}_{++----}+2\chi^{(5)}_{+++---}$ & $\left|{\chi^{(2)}_{+--}}\right|^2$ & $\left|{\chi^{(2)}_{+--}+\chi^{(2)}_{++-}}\right|^2$ \\
    & & & $\chi^{(5)}_{++----})/2$ & $+\chi^{(5)}_{++++--})/4$ & &\\
    \hline
    \small{Expression} & eq.\eqref{ORGSFG} & eq.\eqref{ORFDFG} & eq.\eqref{TPIFChi} & eq.\eqref{TPEF} & eq.\eqref{ORGSFG} & Type I eq.\eqref{PDCTypeI}, Type II eq.\eqref{PDCTII},\eqref{PDCTIII} \\
    \hline
  \end{tabular}
  \caption{SNGF's of three wave mixing techniques: heterodyne-detected SFG and DFG with all classical modes (c); incoherent TPIF with two classical and one quantum (q) mode and corresponding coherent homodyne-detected SFG; incoherent TPEF with one classical and two quantum modes and Type I PDC; Type II PDC SNGF with one classical and four quantum modes.}\label{TABLE1}
\end{table*}

\subsection{Superoperator description of n-wave mixing}

So far we mostly used $L$ and $R$ representation for describing signals. Here, we consider various nonlinear signals which involve two photon resonances \cite{Roslyak09a} (see sketch in Fig. \ref{fig:sketch5}) and describe them using $\pm$ representation. For coherent optical states, all field superoperator nonequilibrium Green's functions (SNGF) in the $L,R$ representation are identical $E'_{L}=E'_{R}$(the superoperator index makes no difference since all operations commute). In the $+,-$ representation "minus"  field indices are not allowed, since $E'_{-}=0$. The general $m$ wave mixing signals are given by $2^m$ products of material and corresponding optical field SNGF's of $m^{th}$ order:
\begin{align}
\label{eq:mwmix}
S^{(m)}_\alpha=\Im \frac{i^m \delta_{m+1,\alpha}}{\pi m! \hbar^{m+1}} \sum \limits_{\nu_m} \ldots \sum \limits_{\nu_1} \int \limits_{-\infty}^{\infty} dt_{m+1} dt_m \ldots dt_1 \\
\nonumber
\Theta(t_{m+1}) \mathbb{V}^{(m)}_{\nu_{m+1}\nu_m\ldots\nu_1}\left({t_{m+1},t_m,\ldots,t_1}\right)\times\\
\nonumber
\mathbb{E}^{(m)}_{\nu_{m+1}\bar{\nu}_m\ldots\bar{\nu}_1}\left({t_{m+1},t_m,\ldots,t_1}\right)
\end{align}
where $t_{m+1},t_m,\ldots,t_1$ are the light/matter interaction times. The factor $\Theta(t_{m+1})=\prod \limits_{i=1}^m \theta(t_{m+1}-t_i)$ insures that the $t_{m+1}$ is the last light-matter interaction. The indexes $\bar{\nu}_j$ are the conjugates to $\nu_{j}$ and defined as follows: the conjugate of $+$ is $-$ and vice versa. Equation \eqref{eq:mwmix} implies that is the excitation in the material are caused by fluctuations in the optical field and vice versa. Equation \eqref{eq:mwmix} also holds in the $L,R$ representation. Here we have $\nu_{m+1}=L$ and $\nu_{j}\in \{{L,R}\}$, $j=m,\ldots,1$. $\bar{\nu}_j$. In the $L,R$ representation the conjugate of "left" is "left" and the conjugate of "right" is "right": $\bar{L}=L, \bar{R}=R$. The material SNGF $\mathbb{V}^{(m)}_{L \underbrace{L \ldots L}_{n} \underbrace{R \ldots R}_{m-n}}$ represent a \emph{Liouville space pathway} with $n+1$ interactions from the left (i.e. with the ket) and $m-n$ interactions from the right (i.e. with the bra).

Field SNGF in Eq. (\ref{eq:mwmix}) is defined in Eq. (\ref{eq:Esngf}). Material SNGF  is defined by Eq. (\ref{eq:Vsngf}). The latter in the form of $\mathbb{V}^{(m)}_{+ \underbrace{-\ldots -}_{m}}$ give causal ordinary response function of $m^{th}$ order. The material SNGF of the form $\mathbb{V}^{(m)}_{+ \underbrace{+\ldots +}_{m}}$ represent $m^{th}$ moment of molecular fluctuations.
The material SNGF of the form $\mathbb{V}^{(m)}_{+ \underbrace{+\ldots +}_{n} \underbrace{-\ldots-}_{m-n}}$ indicates changes in $n^{th}$ moment of molecular fluctuations up by $m-n$ perturbations.
We next recast the material SNGF (\ref{eq:Vsngf}) in the frequency domain by performing a multiple Fourier transform:
\begin{gather}
\label{Rmpmomega}
\chi^{(m)}_{\nu_{m+1}\nu_m\ldots\nu_1}(-\omega_{m+1};\omega_m,\ldots,\omega_1)=\\
\notag
\int_{-\infty}^{\infty} dt_{m+1} \ldots dt_1 \Theta (t_{m+1}) e^{i(\omega_m t_m + \ldots + \omega_1 t_1)}\\
\notag
\delta(-\omega_{m+1}+\omega_m+ \ldots +\omega_1)
\mathbb{V}^{(m)}_{\nu_{m+1}\nu_m\ldots\nu_1}\left({t_{m+1},t_m,\ldots,t_1}\right)
\end{gather}
The SNGF $\chi^{(m)}_{+\underbrace{- \ldots -}_m}(-\omega_{m+1};\omega_m,\ldots,\omega_1)$ (with one $+$ and the rest $-$ indices) are the $m^{th}$ order nonlinear susceptibilities or causal response functions. The rest of SNGF's in the frequency domain can be interpreted similarly to their time domain counterparts (\ref{eq:Vsngf}).

\subsection{Connection to nonlinear fluctuation-dissipation relations}
\label{sec.nl-fluctuation-dissipation}

Spontaneous fluctuations and response are uniquely related in the linear regime \cite{has91} by the fluctuation-dissipation theorem, but not when they are nonlinear. Some nonlinear fluctuation-dissipation relations have been proposed for specific models under limited conditions \cite{lip08,boc81,ber01} but there is no universal relation of this type \cite{kry12}.

A different approach for calculating and analyzing quantum spectroscopy signals  is based on the Glauber Sudarshan $P$-representation which expresses the field density matrix as an integral over coherent state density matrices $|\beta\rangle\langle\beta|$ weighted by a quasi-probability distribution $P(\beta)$
\begin{align}\label{eq:Pb}
\hat{\rho}=\int d^2\beta P(\beta)|\beta\rangle\langle\beta|.
\end{align}
It has been suggested \cite{Kira11a,Almand-Hunter14,Mootz14} that, since the response of a material system to a field initially prepared in a coherent state $|\beta\rangle$ is given by the classical response function CRF, the quantum response  $R_{QM}$ may be recast as an average of the classical response $R_{|\beta\rangle}$ with respect to this quasi-probability  
\begin{align}\label{eq:Rqm}
R_{QM}=\int d^2\beta P(\beta)R_{|\beta\rangle}.
\end{align}
By transforming to the response of a nonclassical state given by $P(\beta)$, in which the fluctuations along one quadrature are squeezed below the classical limit, this form of data processing can uncover otherwise hidden features in the signal. For instance, it was put to use in \cite{Almand-Hunter14} to reveal the ``dropleton", as new kind of quasiparticle excitation in semiconductors. 

While this analysis has proven successful, it misses the multimode nature of the entangled states, which lies at the heart of time-energy entanglement presented here. Based on our analysis of the entangled light state~(\ref{eq.ent-state1}), we have shown that optical signals induced by such entangled states cannot be reduced to a simple sum over the signal of each quantum mode: The quantum correlations depicted in Fig.~\ref{fig.freq-correlations} are necessarily missed in a single-mode approach. For instance, the transition amplitude~(\ref{eq.T_fg}) may be written as a sum over the Schmidt modes,
\begin{align}
T_{fg} (t) &\sim \sum_k \int \!\! d\omega_a \int \!\! d\omega_b \; R_t (\omega_a, \omega_b) \psi_k (\omega_a) \phi_k (\omega_b),
\end{align}
where $R_t$ shall denote the material response. Hence, the TPIF signal $\sim \vert T_{fg} \vert^2$ cannot be reduced to the signal of the indivdual Schmidt modes,  $\vert T_{fg} \vert^2 \sim \sum_{k, k'} \cdots \neq \sum_k \cdots$, and the above transformation fails to capture such a signal.

The idea behind the Glauber-Sudarshan quasi-probability in relation to quantum and classical light outlined in \cite{Kira11a,Almand-Hunter14,Mootz14} makes an equality between classical light and coherent states. This is a potential source of confusion: For instance, the effect of revival of Rabi oscillations demonstrated by Rempe \cite{rem87} which is observed when a coherent state is treated quantum mechanically shows clearly that a coherent state is generally a quantum state of light. Therefore, Eq.~(\ref{eq:Pb}) merely represents a transformation between two different basis sets for the quantum state. Of course, in the nonlinear response case, if the operators are normally ordered as in Eq. (\ref{eq:Pab1}) the result of the coherent state is equivalent to the classical case. In Rempe's example, the nonclassical contributions arise due to commutator terms in Eq. (\ref{eq:comm}) and the higher the order in perturbative expansion - the larger the contribution of the commutator terms. Therefore, in the strong field limit one can observe the revival of Rabi oscillations. 

We now use the superoperator notation introduced in section \ref{sec.Liouville-notation} to show more broadly why the quantum response is different from the classical one so that Eq. (\ref{eq:Rqm}) is violated. 
In Liouville space, the time-dependent density matrix is given by Eq. (\ref{eq:rhogen}). We now make use of the algebraic relation of superoperators \cite{Marx:PhysRevA:08}:
\begin{align}
H_{int-}=E_+V_-+E_-V_+.
\end{align}
Let us first assume that the electric field operators commute
and set $E_- = 0$. We then calculate the expectation value of a
system $A$ operator $O_A$:
\begin{align}
&\text{tr}[O_A\rho(t)]\notag\\
&=\text{tr}[O_A\mathcal{T}\exp\left(-\frac{i}{\hbar}\int_{t_0}^tE_+(t')V_-^A(t')dt'\right)\notag\\
&\times\exp\left(-\frac{i}{\hbar}\int_{t_0}^tE_+(t')V_-^B(t')dt'\right)\rho_{A,0}\rho_{B,0}\rho_{ph,0}].
\end{align}
Since the trace of a commutator vanishes, and since there are only $V_{-}^B$ operators for system $B$, all correlation functions of the form $\langle V_-^BV_-^B...V_-^B\rangle=0$. The nonlinear response function is thus additive, and contains no cooperative terms. The time evolution of two coupled quantum systems and the field is generally given by a sum over Feynman paths in their joint phase space. Order by order in the coupling, dynamical observables can be factorized into products of correlation functions defined in the individual spaces of the subsystems. These correlation functions represent both {\it causal} response and {\it non-causal} spontaneous fluctuations \cite{coh03,ros10}. 

The linear response contains several possible combinations of field superoperators $\langle V_{\pm}V_{\pm}\rangle$.  $\langle V_-V_-\rangle$ represents a commutator, and thus vanishes since its trace is zero. Therefore $\langle V_+V_+\rangle$ (two anticommutators)  and $\langle V_+V_-\rangle$ (commutator followed by anticommutator) are the only two quantities that contribute to the linear response. These two quantities are related by the universal fluctuation-dissipation relation \cite{has91},
\begin{align}\label{eq:fdr}
C_{++}=\frac{1}{2}\coth(\beta\hbar\omega/2)C_{+-}(\omega).
\end{align}
Here
\begin{align}
C_{+-}(\omega)=\int d\tau\langle V_{+}(\tau)V_-(0)\rangle e^{i\omega\tau}
\end{align} 	
is the response function, whereas
\begin{align}
C_{++}(\omega)=\int d\tau\langle V_{+}(\tau)V_+(0)\rangle e^{i\omega\tau}
\end{align} 
denotes spontaneous fluctuations.					

The classical response function $C_{+-}(\omega)$  thus carries all relevant information about linear radiation matter coupling, including the quantum response. In the nonlinear regime the CRF is a specific causal combination of matter correlation functions given by one ``+'' and several ``-'' operators. e.g  $\langle V_+V_-V_-V_-\rangle$  for the third order response. However, the quantum response may also depend on the other combinations. To  $n^{\text{th}}$   order in the external field the CRF $\langle V_{+}(\omega_{n+1})V_{-}(\omega_n)...V_{-}(\omega_2)V_-(\omega_1)\rangle$  is one member of a larger family of  $2^n$ quantities $\langle V_{+}(\omega_{n+1})V_{\pm}(\omega_n)...V_{\pm}(\omega_2)V_\pm(\omega_1)\rangle$  representing various combinations of spontaneous fluctuations (represented by $V_+$ ) and impulsive excitations (represented by  $V_-$). For example, an "all +" quantity such as  $\langle V_+V_+V_+V_+\rangle$ represents purely spontaneous fluctuations. The CRF does not carry enough information to reproduce all  $2^n$  possible quantities which are accessible by quantum spectroscopy. The reason why the CRF and QRF are not simply related in is the lack of a fluctuation-dissipation relation in the nonlinear regime \cite{lip08,boc81,ber01,kry12}.

The field commutator $E_-$ is intrinsically related to vacuum modes of the field which may induce coupling between noninteracting parts of the system. One example where such an effect arising from $E_-$ is combined with the appearance of collective resonances, which occur for $E_+$, has been recently investigated in the context of harmonic systems \cite{gle15}. The response of classical or quantum harmonic oscillators coupled linearly to a classical field is strictly linear; all nonlinear response functions vanish identically. We have recently shown that quantum modes of the radiation field that mediate interactions between harmonic oscillator resulted in nonlinear susceptibilities. A third-order nonlinear transmission of the optical field yields collective resonances that involve pairs of oscillators, and are missed by the conventional quantum master equation treatment \cite{Dor130}.

\subsection{Heterodyne-detected sum and difference frequency generation with classical light}

We first compare two experiments involving three classical modes. The third mode is singled out by
the heterodyne detection which measures its time-averaged photon flux (photons per unit
time). Both techniques represent second order nonlinear signals $S^{(2)}_3$. The initial state of the field is given by a direct product of coherent
states: $|{t=-\infty}\rangle=
|\beta\rangle_1 | \beta \rangle_2
| \beta \rangle_3$, where $| \beta \rangle_\alpha$ are eigenfunctions of the
mode $\alpha$ annihilation operator:
$a_{\alpha}|\beta\rangle_{\alpha} =
\beta_{\alpha}|\beta\rangle_{\alpha}$. Coherent states are
the most classical states of quantum light, hence we shall refer to
them as classical optical fields \cite{Glauber07}.
\par
Only one type of optical SNGF contribute to the classical  third order signalsignal:
\begin{equation*}
\mathbb{E}^{(2)}_{+++}\left({t_3,t_2,t_1}\right)=\mathcal{E'}(t_3)\mathcal{E'}(t_2)\mathcal{E'}(t_1)
\end{equation*}
where $\mathcal{E'}(t)=\langle E'_L(t) \rangle =
\langle E'_R(t) \rangle =\beta_\alpha \sqrt{2 \pi \hbar
\omega_\alpha/\Omega}$ is the classical field amplitude.
\par
The conjugate material SNGF's become:
\begin{equation}
\label{V2}
\mathbb{V}^{(2)}_{+--}\left({t_3,t_2,t_1}\right)=
\mathbb{V}^{(2)}_{LLL}\left({t_3,t_2,t_1}\right)+\mathbb{V}^{(2)}_{LRL}\left({t_3,t_2,t_1}\right)
\end{equation}
where we have assumed that initially the material system is in the ground state, which implies that $\mathbb{V}^{(2)}_{LRR}\left({t_3,t_2,t_1}\right) \equiv 0$.
\par
We assume the same three-level system used earlier.
The $L,R$ representation plus RWA allows the material SNGFs \eqref{V2} to be represented by the loop diagrams 
shown in Fig.\ref{FIG:1}(C), Fig.\ref{FIG:2}(C). The rules for constructing these
partially-time-ordered diagrams are summarized in Appendix \ref{app:Diag} \cite{ros09}. The signal is given by the causal $\chi^{(2)}_{+--}(-\omega_3,\pm \omega_2,\pm \omega_1)$ response function. This result is not new and can be obtained by using various combinations of a classical field from the outset.
To calculate equation \eqref{V2} we have to specify the phase matching conditions (frequencies and wave vectors of the optical modes).
This will be done below. 

\subsection{Difference frequency generation (DFG)}

\begin{figure}
  \includegraphics[width=0.45\textwidth]{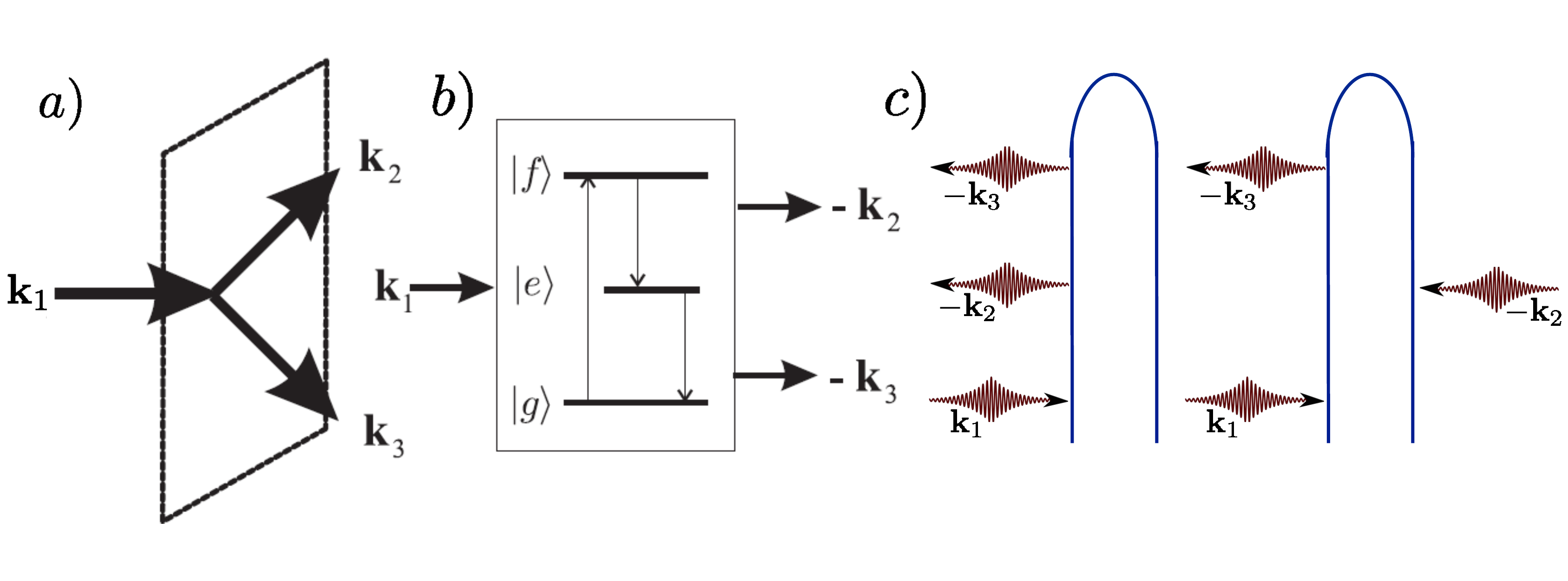}\\
  \caption{Heterodyne detected DFG: (A) phase matching condition $\mathbf{k}_3=\mathbf{k}_1 -\mathbf{k}_2$; (B) molecular level scheme; (C) loop diagrams.}\label{FIG:1}
\end{figure}

In DFG the first mode $\mathbf{k}_1$ promotes the system from its ground state $\left|{g}\right\rangle$ into  state $\left|{f}\right\rangle$. The second mode $\mathbf{k}_2$ induces stimulated emission from $\left|{f}\right\rangle$ to an intermediate $\left|{e}\right\rangle$; and the third mode $\mathbf{k}_3$ stimulates the emission from $\left|{e}\right\rangle$ to $\left|{g}\right\rangle$, as sketched in Fig.\ref{FIG:2} (B). The signal is measured in the phase matching direction: $\mathbf{k}_3=\mathbf{k}_1 -\mathbf{k}_2$ (See Fig.\ref{FIG:1} (A)).
\par
The corresponding loop diagrams corresponding are
shown in Fig.\ref{FIG:1}(C). The optical field SNGF yields:
\begin{equation}
\label{E2pmm}
\mathbb{E}^{(2)}_{+++}\left({t_3,t_2,t_1}\right)=\mathcal{E}_3^{*} (t) \mathcal{E}_2^{*} (t_2)\mathcal{E}_1 (t_1)
\end{equation}
The material SNGF's are:
\begin{gather}
\label{V2pmm}
\mathbb{V}^{(2)}_{+--}\left({t_3,t_2,t_1}\right)=\\
\notag
\langle {\mathcal{T} V^{3}_L(t) V^{2,\dag}_L(t_2) V^{1,\dag}_L(t_1)}
+\langle {\mathcal{T} V^{3}_L(t) V^{2,\dag}_R(t_2) V^{1,\dag}_L(t_1)} \rangle
\end{gather}
\par
The DFG signal in the frequency domain can be written as:
\begin{align}
& S_{DFG}(-\omega_3,\omega_2,\omega_1)=\\
\notag
& \frac{1}{\pi \hbar}\Im \delta(\omega_1-\omega_2-\omega_3) \chi^{(2)}_{+--}\left({-\omega_3;-\omega_2,\omega_1}\right)\mathcal{E}_3^{*} \mathcal{E}_2^{*} \mathcal{E}_1
\end{align}

Utilizing the rules given in Appendix \ref{app:Diag} and the diagrams shown in Fig.\ref{FIG:1}(C) we obtain:
\begin{align}
\label{secondorderresponse}
& \chi^{(2)}_{+--}\left({-(\omega_1-\omega_2);-\omega_2,\omega_1}\right)=\\
\notag
& \frac{1}{2! \hbar^2}(\left\langle{g}\right|{V_3 G(\omega_g+\omega_1-\omega_2) V_2 G(\omega_g+\omega_1) V^\dag_1}\left|{g}\right\rangle-\\
\notag
& \left\langle{g}\right|{V_2 G^\dag(\omega_g+\omega_1-\omega_2) V_3 G(\omega_g+\omega_1) V^\dag_1}\left|{g}\right\rangle).
\end{align}
A sum-over-states  (SOS) expansion then yields
\begin{align}
\label{ORFDFG}
& \chi^{(2)}_{+--}\left({-(\omega_1-\omega_2);-\omega_2,\omega_1}\right)=\\
\notag
& \frac{1}{2! \hbar^2}\frac{\mu^x_{gf} \mu^x_{fe} \mu^x_{eg}}{(\omega_1-\omega_{gf}+i\hbar \gamma_{gf})(\omega_1-\omega_2-\omega_{eg}+i\hbar\gamma_{eg})}-\\
\notag
& \frac{1}{2! \hbar^2}\frac{\mu^x_{gf} \mu^x_{fe} \mu^x_{eg}}{(\omega_1-\omega_{gf}+i\hbar \gamma_{gf})(\omega_2-\omega_{eg}-i\hbar\gamma_{eg})}
\end{align}
\par
Equation \eqref{ORFDFG} indicates that the signal induced by classical optical fields is given by the second order classical response function (CRF).

\subsection{Sum Frequency Generation (SFG)}

\begin{figure}
  \includegraphics[width=8cm]{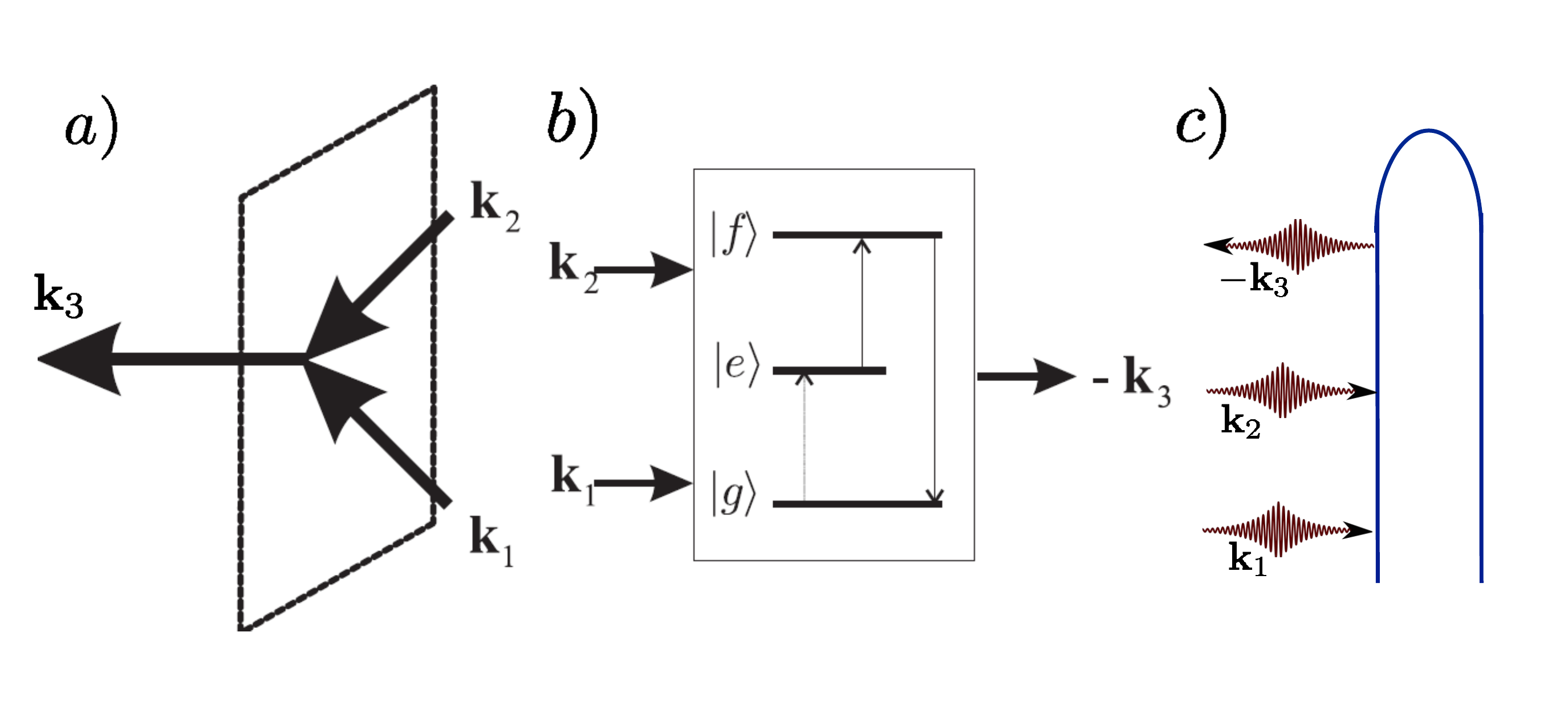}\\
  \caption{Heterodyne detected SFG: (A) phase matching condition $\mathbf{k}_3=\mathbf{k}_1 + \mathbf{k}_2$; (B) level scheme; (C) loop diagrams.}\label{FIG:2}
\end{figure}

In SFG the first two modes promote the molecule from its ground
state $|g\rangle$ through intermediate state $|e\rangle$ into the state $|f\rangle$. The third mode induces stimulated emission
from $\left|{f}\right\rangle$ to the ground state $|g\rangle$ as
sketched in Fig.\ref{FIG:2} (B). The signal is generated in the
 direction: $\mathbf{k}_3=\mathbf{k}_1 +\mathbf{k}_2$
(See Fig.\ref{FIG:2} (A)).
\par
The heterodyne-detected SFG signal can be obtained in an analogous manner to the DFG by utilizing the diagrams shown in Fig.\ref{FIG:2}(C):
\begin{align}
\label{SFG}
& S_{SFG}(\omega_1,\omega_2)=\\
\notag
& \frac{1}{\pi \hbar}\Im \delta(\omega_1+\omega_2-\omega_3) \chi^{(2)}_{+--}\left({-\omega_3;\omega_2,\omega_1}\right)\mathcal{E}_3^{*} \mathcal{E}_2 \mathcal{E}_1
\end{align}
where the CRF is given by
\begin{align}
\label{ORGSFG}
& \chi^{(2)}_{+--}\left({-(\omega_1+\omega_2);\omega_2,\omega_1}\right)=\\
\notag
& \frac{1}{2! \hbar^2}\left\langle{g}\right|{V_3 G(\omega_g+\omega_1+\omega_2) V^\dag_2 G(\omega_g+\omega_1) V^\dag_1}\left|{g}\right\rangle=\\
\notag
& \frac{1}{2! \hbar^2}\frac{\mu^x_{ge} \mu^x_{fe} \mu^x_{eg}}{(\omega_1-\omega_{eg}+i\hbar \gamma_{eg})(\omega_1+\omega_2-\omega_{gf}+i\hbar\gamma_{gf})}
\end{align}
\par
In the coming sections equations \eqref{ORGSFG} and \eqref{ORFDFG} will be compared with other techniques involving various combinations of quantum and classical optical fields.
These include-homodyne detected SFG, DFG and PDC where one or more optical modes are spontaneously generated and must be treated quantum mechanically.

\subsection{Two-photon-induced fluorescence (TPIF) vs. homodyne-detected SFG}

\begin{figure}
  \includegraphics[width=0.5\textwidth]{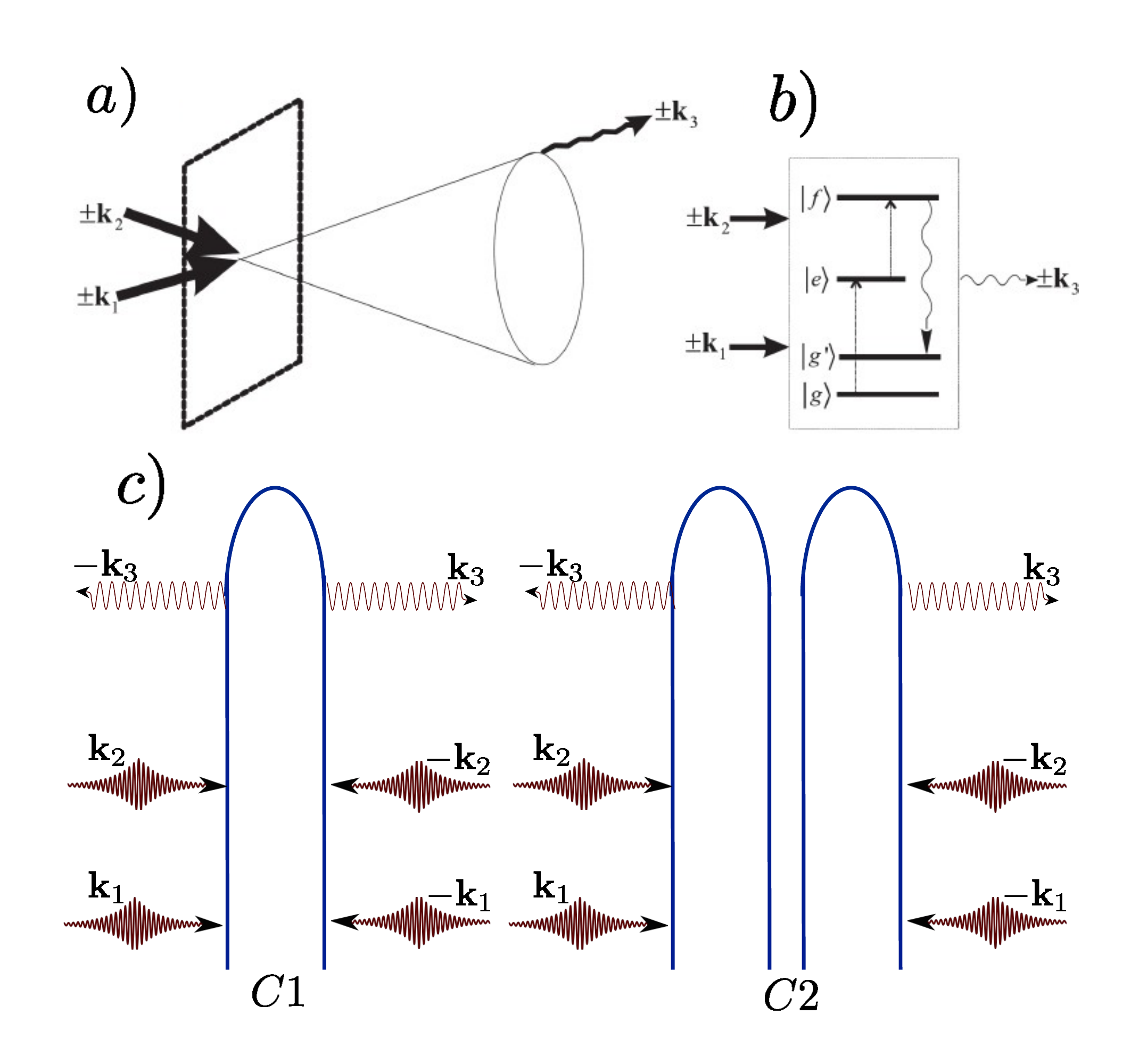}\\
  \caption{Three wave process with two classical and one quantum mode: a) phase-matching condition; b) molecular level scheme; c) loop for the incoherent TPIF (C1) and coherent homodyne SFG (C2)}\label{FIG:3}
\end{figure}

We now turn to techniques involving two classical and one quantum mode where
the initial state of the optical field is:
$\left|{t=-\infty}\right\rangle=
\left|{\beta_1}\right\rangle_1 \left|{\beta_2}\right\rangle_2
\left|{0}\right\rangle_3$. The modes interact with the three level material system (Fig.\ref{FIG:3}(B)). We assume that $\omega_{eg} \ne \omega_{ef}$. This allows to focus on the resonant SNGF's and reduces the number of diagrams.
\par
The two classical modes $\mathbf{k}_1$ and $\mathbf{k}_2$ promote the molecule from its ground state $|g\rangle$ into the intermediate state $|e\rangle$ and to the final state $|f\rangle$. The system then spontaneously decays back into one of the ground state manifold $|g'\rangle$ emitting a photon into the third mode $\mathbf{k}_3$ which is initially in the vacuum state (See Fig.\ref{FIG:3} (B)). The phase-matching condition
$\mathbf{k}_1-\mathbf{k}_1+\mathbf{k}_2-\mathbf{k}_2+\mathbf{k}_3-\mathbf{k}_3=0$
is automatically satisfied for any $\mathbf{k}_3$. Therefore the spontaneous photons are
emitted into a cone (See Fig.\ref{FIG:3} (A)).
\par
We shall calculate the photon flux in the $\mathbf{k}_3$ mode. Since the process involves three different modes and six light/matter interactions.  The signal \eqref{eq:mwmix} must by expanded to fifth order in $H_{int}$. The fifth order signal $S^{(5)}_{CCQ}$ (CCQ implies two classical and one quantum mode) may be written as sum of $2^5$ terms
each given by a product of molecule/field six point SNGF's.
Only the corresponding diagrams shown in Fig.\ref{FIG:3}(C). These satisfy the following conditions:
\begin{enumerate}
  \item {The creation operator of the quantum mode $a^\dag_{3}$ must be accompanied by the corresponding annihilation operator $a_{3}$.}
  \item {The quantum modes de-excite the molecule, which implies that the annihilation operators must act on the bra and the creation operators act on the ket.}
  \item {The coherent optical fields are tuned off resonance with respect to the $\omega_{fg'}$ transition. Hence, the signal is not masked by  stimulated emission.}
\end{enumerate}

\par
To address the collective properties of three wave mixing. 
We consider a collection of $N$ noninteracting molecules positioned at $\mathbf{r}_i$ so that $V=\sum_i V_i \delta(\mathbf{r}-\mathbf{r}_i)$. 
The optical field can interact with different systems at different times. Fig.\ref{FIG:3}(C) 
shows processes involving three different molecules and six light/matter interactions. 
These modes can either interact with the same system (incoherent process, Fig.\ref{FIG:3}(C1)) or 
two different systems (coherent process, Fig.\ref{FIG:3}(C2))\cite{Marx:PhysRevA:08}.

\subsubsection{Two-photon-induced fluorescence (TPIF).}
 This is an incoherent three wave process.
Using the identity $\sum \limits_{i=1}^N \exp{i \left({\mathbf{k}_1-\mathbf{k}_1+\mathbf{k}_2-\mathbf{k}_2+\mathbf{k}_3-\mathbf{k}_3}\right)\left({\mathbf{r}-\mathbf{r}_i}\right)}=N$
the optical field SNGF  yields:
\begin{gather}
\label{E5lrpppp}
\mathbb{E}^{(5)}_{LR++++}\left({t_6,t_5,\ldots,t_1}\right)=\\
\notag
N \mathcal{E}_1(t_1)\mathcal{E}^{*}_1(t_2)\mathcal{E}_2(t_3)\mathcal{E}^{*}_2(t_4)
\notag
\frac{2 \pi \hbar \omega_3}{\Omega}\exp{(i \omega_3(t_6-t_5))}
\end{gather}
\par
The relevant material SNGF is:
\begin{gather}
\label{V5lrpppp}
\mathbb{V}^{(5)}_{LR++++}\left({t_6,t_5,\ldots,t_1}\right)=\\
\notag
\langle \mathcal{T} V^{3}_L(t_6)  V^{3,\dag}_R(t_5) V^{2}_-(t_4) V^{2,\dag}_-(t_3) V^{1}_-(t_2) V^{1,\dag}_-(t_1)\rangle
\end{gather}
Utilizing equations (\ref{E5lrpppp},\ref{V5lrpppp}) the frequency domain signal can be written as:
\begin{gather}
\label{STPIF}
S_{TPIF}(\omega_1,\omega_2)=\frac{N}{\pi \hbar}  \sum \limits_{\mathbf{k}_3} \left| \mathcal{E}_1 \right|^2 \left| \mathcal{E}_2 \right|^2 \frac{2 \pi \hbar \omega_3}{\Omega}\times\\
\notag
\Im \chi^{(5)}_{LR----}\left({-\omega_3;\omega_3,-\omega_2,\omega_2,-\omega_1,\omega_1}\right)
\end{gather}
Note that the above expression is given in the mixed ($L/R, +/-$) representation.  It can be recast into the $+,-$ representation using
$\chi^{(5)}_{LR----}=(\chi^{(5)}_{+-----}+\chi^{(5)}_{++----})/2$. The second term (two "plus" four "minus" indices) arises
since one of the modes is nonclassical.
The  frequency domain material SNGF can be calculated from the diagram of Fig.\ref{FIG:3}(D1):
\begin{eqnarray}
\label{sixpointcorrelation}
\chi^{(5)}_{LR----}\left({-\omega_3;\omega_3,-\omega_2,\omega_2,-\omega_1,\omega_1}\right)=\frac{1}{5! \hbar^5}\\
\nonumber
\langle{g}| V_1 G^\dag(\omega_g+\omega_1) V_2 G^\dag(\omega_g+\omega_1+\omega_2)\\
\nonumber
\times V^{\dag}_3 G^\dag(\omega_g+\omega_1+\omega_2-\omega_3) V_3 \times \\
\nonumber
G(\omega_g+\omega_1+\omega_2) V^\dag_2 G(\omega_g+\omega_1) V^\dag_1 |{g}\rangle
\end{eqnarray}
\par
Expanding equation \eqref{sixpointcorrelation} in molecular states gives:
\begin{gather}
\label{TPIFChi}
\chi_{LR----}^{(5)}(-(\omega_1+\omega_2);\omega_2,\omega_1)=
\sum_{gg^{\prime}} \vert \mu^x_{g'f} \mu^x_{fe} \mu^x_{eg} \vert^2 \\
\notag
\times \dfrac{1}{5!\hbar^5}\frac{1}{[(\omega_1-\omega_{eg})^2+\gamma_{eg}^2][\omega_1+\omega_2-\omega_{fg}+i \gamma_{fg}]}\\
\notag
\times\frac{1}{[\omega_1+\omega_2-\omega_{fg^{\prime}}-i \gamma_{fg^{\prime}}][\omega_1+\omega_2-\omega_3-\omega_{gg'}-i\gamma_{gg^{\prime}}]}
\end{gather}
Provided the energy splitting within the ground state manifold is small comparing to the optical transitions the signal can be recast in the form:
\begin{gather}
\label{TPIF}
S_{TPIF}(\omega_1,\omega_2)= \frac{2N \omega_3}{5! \hbar^5 \Omega} \left| \mathcal{E}_1 \right|^2 \left| \mathcal{E}_2 \right|^2\\
\notag
\left|{T_{g'g}(\omega_1,\omega_2)}\right|^2 \delta(\omega_1+\omega_2-\omega_3-\omega_{gg'})
\end{gather}
where
\begin{equation*}
T_{g'g}(\omega_1,\omega_2)= \frac{\mu_{gf} \mu_{fe} \mu_{eg}}{(\omega_1-\omega_{eg}+i\gamma_{eg})(\omega_1+\omega_2-\omega_{fg}+i\gamma_{fg})}
\end{equation*}
is the transition amplitude. This is similar to the Kramers-Heisenberg form of ordinary (single-photon) fluorescence
\cite{Marx:PhysRevA:08}. As in single photon fluorescence, for a correct description of the TPIF the ground state must not be degenerate.
Otherwise $\gamma_{gg}=0$ (the degenerate ground state the system has infinite life time) and the signal vanishes.
\par
The SNGF in equation
\eqref{sixpointcorrelation} is commonly called as the fluorescence
quantum efficiency (Webb \cite{xu1996mtp}) and two-photon
tensor (Callis \cite{callis_theory_1993}). Our result is
identical to that of Callis, apart from the $\delta(\omega_1+\omega_2-\omega_3-\omega_{gg'})$ factor.
\par
When the two classical coherent modes are degenerate
($\omega_1=\omega_2$) the signal given by equation \eqref{TPIF}
describes non-resonant Hyper-Raman scattering ($\omega_{gg'} \neq
0$) also known as incoherent second harmonic inelastic scattering
\cite{andrews2002ohm,callis_theory_1993}. When $\omega_1=\omega_2$ and
$\omega_{gg'}\rightarrow 0$ (but not equal to) equation \eqref{TPIF}
describes non-resonant Hyper-Rayleigh scattering also known as
incoherent second harmonic elastic scattering. Off-resonant
hyper-scattering is a major complicating factor for TPIF
microscopy \cite{xu_hyper_1997}.

\subsubsection{Homodyne-detected SFG}
Here the
optical field SNGF is given by equation \eqref{E5lrpppp}, but instead of factor $N$ it contains the factor:
factor:
\begin{equation}
\sum \limits_{i=1}^N \sum \limits_{j \ne i}^{N-1} e^{i \Delta \mathbf{k}\left({\mathbf{r}-\mathbf{r}_i}\right) }e^{-i \Delta \mathbf{k}\left({\mathbf{r}-\mathbf{r}_j}\right)} \approx N(N-1)
\end{equation}
where $\Delta \mathbf{k} \approx
\mathbf{k}_1+\mathbf{k}_2-\mathbf{k}_3$. $\approx$ reflects phase uncertainty given by the reciprocal
of the molecular collection length which effectively narrows the
optical cone. For large $N$ the coherent part $\propto N(N-1)$
dominates over the incoherent $\propto N$ response. For a small
sample size, exact calculation of the optical field part of the SNGF
is rather lengthy , but it can be performed in the same fashion as done
by Hong and Mandel\cite{HOM85} for the probability of photon detection.
\par
To calculate the matter SNGF, we must work in the joint space of two molecules $|\rangle_{1,2}=|\rangle_{1}|\rangle_{2}$ interacting with the same field mode. The matter SNGF of the joint system can be factorized into product of each molecule SNGF's:
\begin{gather}
\label{V5Coherent}
\mathbb{V}^{(5)}_{LR++++}\left({t_6,t_5,\ldots,t_1}\right)=\\
\notag
\langle \mathcal{T} V^{3}_+(t) V^{3,\dag}_+(t_5) V^{2}_-(t_4) V^{2,\dag}_-(t_3) V^{1}_-(t_2)V^{1,\dag}_-(t_1)\rangle_{1,2}=\\
\notag
\langle \mathcal{T} V^{3}_+(t) V^{2,\dag}_-(t_3) V^{1,\dag}_-(t_1)\rangle_{1}
\langle \mathcal{T} V^{3,\dag}_+(t_5) V^{2}_-(t_4) V^{1}_-(t_2)\rangle_{2}
\end{gather}
where we have used the fact that the last interaction must be a "plus".
\par
Since the two molecules are identical the following identity holds:
\begin{eqnarray}
\label{identity}
\langle \mathcal{T} V^{3,\dag}_+(t_5) V^{2}_-(t_4) V^{1}_-(t_2)\rangle=\\
\nonumber
\left({\langle \mathcal{T} V^{3}_+(t+\Delta t) V^{2,\dag}_-(t_3+\Delta t) V^{1,\dag}_-(t_1+\Delta t)\rangle}\right)^{*}
\end{eqnarray}
where $v_g \Delta t$ is the optical path length connecting molecules situated at $\mathbf{r}_i$ and $\mathbf{r}_j$. Using this identity and equation \eqref{V5Coherent} The matter SNGF in the frequency domain can be factorized into the square of an ordinary (causal) response function (one ``+'' and several ``-''):
\begin{eqnarray*}
\Im \chi^{(5)}_{++----}\left({-\omega_3;\omega_3,-\omega_2,\omega_2,-\omega_1,\omega_1}\right) = \\
\left|{\chi^{(2)}_{+--} \left({-(\omega_1+\omega_2);\omega_2,\omega_1}\right)}\right|^2
\end{eqnarray*}
The unique factorization of the coherent matter SNGF's can be also obtained in the $L,R$ representation using the diagrammatic technique as shown in Fig.\ref{FIG:3}(C2). The classical modes have to excite the molecules and the quantum mode has to de-excite them. Hence, the interaction with the first molecule ket ($L$) are accompanied by the conjugate interactions with the second molecule bra ($R$). We obtain that the ordinary, causal response function is give by equation \eqref{ORGSFG} as $\chi^{(2)}_{+--} \left({-(\omega_1+\omega_2);\omega_2,\omega_1}\right)=\chi^{(2)}_{LLL} \left({-(\omega_1+\omega_2);\omega_2,\omega_1}\right)$.
\par
The homodyne-detected SFG signal is finally given by:
\begin{align}
\notag
& S_{SFG}=N(N-1)\left| \mathcal{E}_1 \right|^2 \left| \mathcal{E}_2 \right|^2 \frac{2 (\omega_1+\omega_2)}{\Omega} \times \\
& \left| {\chi^{(2)}_{+--}\left({-(\omega_1+\omega_2);\omega_2,\omega_1}\right)} \right|^2
\label{TCPFCoh}
\end{align}
Both homodyne and heterodyne SFG are given by the same causal response function $\chi^{(2)}_{+--}$. The main difference is that the latter satisfies perfect phase matching, while for the former this condition is only approximate. For sufficiently large samples the two techniques are identical.
\par
To conclude, we present the total signal for the three wave process involving two classical and one quantum field ($CCQ)$) which includes both an incoherent and a coherent components:
\begin{gather}
\label{CCQ}
S^{(5)}_{CCQ}(\omega_2,\omega_1)=
\left| \mathcal{E}_1 \right|^2 \left| \mathcal{E}_2 \right|^2 \frac{2 \omega_3}{\Omega}\\
\notag
[N \Im \chi^{(5)}_{++----}\left({-\omega_3;\omega_3,-\omega_2,\omega_2,-\omega_1,\omega_1}\right)+\\
\notag
N(N-1)\left| {\chi^{(2)}_{+--}\left({-(\omega_1+\omega_2);\omega_2,\omega_1}\right)}\right|^2]
\end{gather}

\subsection{Two-photon-emitted fluorescence (TPEF) vs. type-I parametric down conversion (PDC).}

\subsubsection{TPEF}

\begin{figure}
  \includegraphics[width=8.4cm]{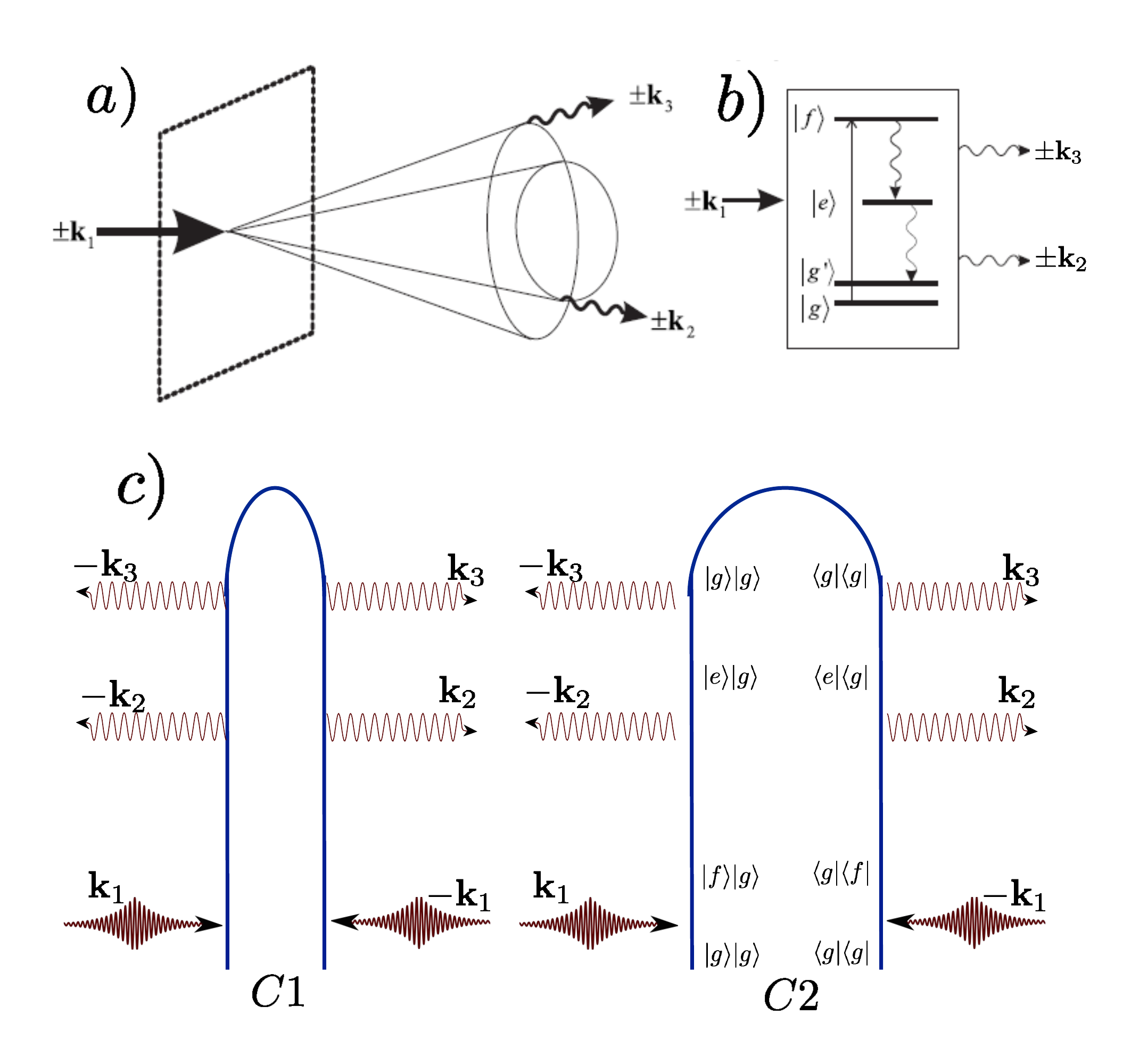}\\
  \caption{Three wave process involuting two quantum and one classical mode: (A) phase-matching condition; (B) molecular level scheme; (C) loop diagrams for the incoherent TPEF (C1) and coherent Type I PDC (C2)}\label{FIG:4}
\end{figure}

We now turn to three wave processes involving one classical and two
quantum modes. We start with the incoherent response of $N$
identical molecules initially in their ground state. The initial
state of the optical field is
$\left|{t=-\infty}\right\rangle=
\left|{\beta_1}\right\rangle_1 \left|{0}\right\rangle_2
\left|{0}\right\rangle_3$. The classical field $\mathbf{k}_1$ pumps
the molecule from its ground state $|g\rangle$ into the excited state $|f\rangle$. The system
then spontaneously emits two photons into modes $\mathbf{k}_2, \;
\mathbf{k}_3$ (See Fig.\ref{FIG:4} (B)) which are initially in the
vacuum state. Such incoherent process which involves one classical
and two quantum modes will be denoted two-photon-emitted
fluorescence (TPEF). To our knowledge there is neither theoretical nor
experimental work concerning this process.
\par
This process is not phase sensitive since the phase matching condition
$\mathbf{k}_1-\mathbf{k}_1+\mathbf{k}_2-\mathbf{k}_2+\mathbf{k}_3-\mathbf{k}_3=0$
is automatically satisfied for any $\mathbf{k}_3$. Therefore
spontaneously generated modes are emitted into two spatial cones
(See Fig.\ref{FIG:4} (A)). For a single molecule the cones are
collinear, as in Type I parametric down conversion (PDC). Due
to this similarity we chose beams polarizations as is usually
done for PDC of this type: the spontaneously generated photons have the
same polarization along the $x$ axis, and orthogonal to
that of the classical mode polarized along the $y$ axis.

\par
The time-averaged photon flux in the $\mathbf{k}_3$ mode  is our TPEF signal.
We again make use of the loop diagrams to identify the relevant SNGF contributing to the signal. Using the initial state of the field the diagrams must satisfy the following conditions:
\begin{enumerate}
  \item {The creation operators of a spontaneously generated modes $a^\dag_{3}, a^\dag_{2}$ acting on the "ket" must be accompanied by the corresponding annihilation operators $a_{3}, a_{2}$ acting on the "bra".}
  \item {The first mode $\omega_1 (\mathbf{k}_1)$  is off-resonance with both $\omega_{eg}$ and $\omega_{fe}$ transitions to avoid stimulated emission contributions.}
\end{enumerate}
Diagram (C1) in Fig.\ref{FIG:4} satisfies the above conditions. The non-resonant diagrams have been omitted.
Using this diagram the optical field SNGF yields:
\begin{gather}
\label{E5llrrpp}
\mathbb{E}^{(5)}_{LLRR++}\left({t_6,t_5,\ldots,t_1}\right)=\\
\notag
N \mathcal{E}_1(t_1)\mathcal{E}^{*}_1(t_2)
\frac{2 \pi \hbar \omega_3}{\Omega} \frac{2 \pi \hbar \omega_2}{\Omega}\times\\
\notag
\exp{(i \omega_3(t_6-t_5))}\exp{(i \omega_3(t_4-t_3))}
\end{gather}
The matter SNGF assumes the form:
\begin{gather}
\label{V5llrrpp}
\mathbb{V}^{(5)}_{LLRR--}\left({t_6,t_5,\ldots,t_1}\right)=\\
\langle \mathcal{T} V^{3}_L(t) V^{3,\dag}_R(t_5) V^{2,\dag}_R(t_4) V^{2}_L(t_3) V^{1}_-(t_2)V^{1,\dag}_-(t_1)\rangle
\end{gather}
Using equations \eqref{E5llrrpp}, \eqref{V5llrrpp} the incoherent part of the frequency domain signal  is given by:
\begin{gather*}
S_{TPEF}(\omega_1)=\frac{N}{\pi \hbar} \left| \mathcal{E}_1 \right|^2 \frac{2 \pi \hbar (\omega_2)}{\Omega} \frac{2 \pi \hbar \omega_3}{\Omega} \times \\
\Im \chi^{(5)}_{LLRR--}\left({-\omega_3;\omega_3,\omega_2,-\omega_2,-\omega_1,\omega_1}\right)
\end{gather*}
The corresponding SNGF can be calculated from the diagram in Fig.\ref{FIG:4}(C1):
\begin{gather*}
\chi^{(5)}_{LLRR--}\left({-\omega_3;\omega_3,\omega_2,-\omega_2,-\omega_1,\omega_1}\right)=\\
\frac{1}{5! \hbar^5}
\langle{g}| V^\dag_1 G^\dag(\omega_g+\omega_1) V^\dag_2 G^\dag(\omega_g+\omega_1-\omega_2) \times\\
V_3 G^\dag(\omega_g+\omega_1-\omega_2-\omega_3) V_3 \times\\
G(\omega_g+\omega_1-\omega_2) V_2 G(\omega_g+\omega_1) V^\dag_1 |{g}\rangle
\end{gather*}
Expansion in the molecular eigenstates brings the response function into the Kramers-Heisenberg form:
\begin{gather}
\label{TPEF}
\chi^{(5)}_{LLRR--}\left({-\omega_3;\omega_3,\omega_2,-\omega_2,-\omega_1,\omega_1}\right)=\\
\notag
\dfrac{1}{5!\hbar^5}\sum_{gg^{\prime}} \vert \mu^x_{g'e} \mu^x_{ef} \mu^y_{fg} \vert^2 \delta (\omega_1-\omega_2-\omega_3)\\
\notag
\dfrac{1}{(\omega_1-\omega_{fg})^2+\gamma_{fg}^2}\vert{\dfrac{1}{\omega_1-\omega_2-\omega_{eg}+i\gamma_{eg}}}\vert^2
\end{gather}
\par
Note that unlike TPIF, the TPEF signal depends on the SNGF other than causal response function $\chi^{(5)}_{+-----}$.

\subsubsection{Type-I PDC.}

We now turn to the coherent response from a collection of identical
molecules which interact with one classical pumping mode and two
spontaneously generated quantum modes (See Fig.\ref{FIG:4} (A,B)).
This is known as Type I parametric down conversion (PDC), which is widely used for producing entangled photon pairs. Hereafter
we assume perfect phase matching $\Delta
\mathbf{k}=\mathbf{k}_1-\mathbf{k}_2-\mathbf{k}_3$ which is the case
for a sufficiently large sample \cite{Ger05}.
\par
The initial condition for PDC are the same as for TPEF and most PDC experiments are well described by the causal response function $\chi^{(2)}_{+--}$.
Therefore one would expect a connection between TPEF and PDC, similar
to that of TPIF and homodyne-detected SFG. However, as we are about to demonstrate, for a complete
description of the PDC process the causal second order response
function is not enough.
\par
We shall establish the corrections to the second order CRF caused by the quantum origin of the spontaneous modes. To do so we again resort to the CTPL diagrams (See Fig.\ref{FIG:4} (C2)). For the coherent response the optical field SNGF is given by equation \eqref{E5llrrpp} with the factor $N$ replaced by $N(N-1)$. The material SNGF \eqref{V5llrrpp} can be factorized as:
\begin{gather}
\label{V5llrrppcoh}
\mathbb{V}^{(5)}_{LLRR--}\left({t_6,t_5,\ldots,t_1}\right)=\\
\langle \mathcal{T} V^{3}_L(t_6) V^{2}_L(t_3) V^{1,\dag}_L(t_1) \rangle \langle\mathcal{T} V^{3,\dag}_R(t_5) V^{2,\dag}_R(t_4) V^{1}_R(t_2)\rangle
\end{gather}
Note that this factorization is unique as $\langle \mathcal{T} V^{3}_L(t) V^{2,\dag}_R(t_4) V^{1,\dag}_L(t_1) \rangle=0$ due to the material initially being in its ground state.
The coherent PDC signal in the frequency domain can then be written as:
\begin{gather}
\label{PDCTypeI}
S_{PDC}(\omega_1)=\\
\notag
\frac{N(N-1)}{4\pi \hbar}\left| \mathcal{E}_1 \right|^2 \frac{2 \pi \hbar \omega_2}{\Omega} \frac{2 \pi \hbar (\omega_1-\omega_2)}{\Omega}\times\\
\notag
|\chi^{(2)}_{LL-}\left({-(\omega_1-\omega_2);\omega_2,\omega_1}\right)|^2
\end{gather}
Here the generalized response function expanded in the eigenstates has form:
\begin{gather*}
\chi^{(2)}_{LL-}\left({-(\omega_1-\omega_2);\omega_2,\omega_1}\right)=\\
\frac{1}{2! \hbar^2}\frac{\mu^y_{gf} \mu^x_{fe} \mu^x_{eg}}{(\omega_1-\omega_{gf}+i\hbar \gamma_{gf})(\omega_1-\omega_2-\omega_{eg}+i\hbar\gamma_{eg})}
\end{gather*}
This mixed representation can be recast in $+,-$ and $L,R$ representations:
\begin{equation}
\label{chi2}
\chi^{(2)}_{LL-}=\chi^{(2)}_{LLL}=\frac{1}{2}(\chi^{(2)}_{+--}+\chi^{(2)}_{++-})
\end{equation}
\par
Comparing the CRF for heterodyne detected DFG \eqref{ORFDFG} and the SNGF's for Type I PDC \eqref{chi2} we see that the latter is described not only by causal response function $\chi^{(2)}_{+--}$ but also by the second moment of material fluctuations $\chi^{(2)}_{++-}$. On the other hand, in the $L,R$ representation it singles out one Liouville pathway $\chi^{(2)}_{LLL}$, while the classical optical fields drive the material system along all possible pathways.

An interesting effects arise in type-I PDC if the detection process is included explicitly. The details of these effects are discussed in Section \ref{sec:PDCmic}.

In summary of this section we give the signal for the three wave process involving one classical and two quantum fields ($CQQ$). This contains both an incoherent and a coherent component:
\begin{gather}
\label{CQQ}
S^{(5)}_{CQQ}(\omega_2,\omega_1)=
\frac{1}{\pi \hbar} \left| \mathcal{E}_1 \right|^2 \frac{2 \pi \hbar \omega_2}{\Omega} \frac{2 \pi \hbar \omega_3}{\Omega}\times\\
\notag
N \Im (\chi^{(5)}_{++----}(-\omega_3;\omega_3,-\omega_2,\omega_2,-\omega_1,\omega_1)+\\
\notag
\chi^{(5)}_{+++---}(-\omega_3;\omega_3,-\omega_2,\omega_2,-\omega_1,\omega_1)+\\
\notag
\chi^{(5)}_{++++--}(-\omega_3;\omega_3,-\omega_2,\omega_2,-\omega_1,\omega_1))+\\
\notag
N(N-1)|\chi^{(2)}_{+--}(-(\omega_1-\omega_2);\omega_2,\omega_1)+\\
\notag
\chi^{(2)}_{++-}(-(\omega_1-\omega_2);\omega_2,\omega_1)|^2
\end{gather}

\subsection{Type II PDC; polarization entanglement.}

\begin{figure}
  \includegraphics[width=0.45\textwidth]{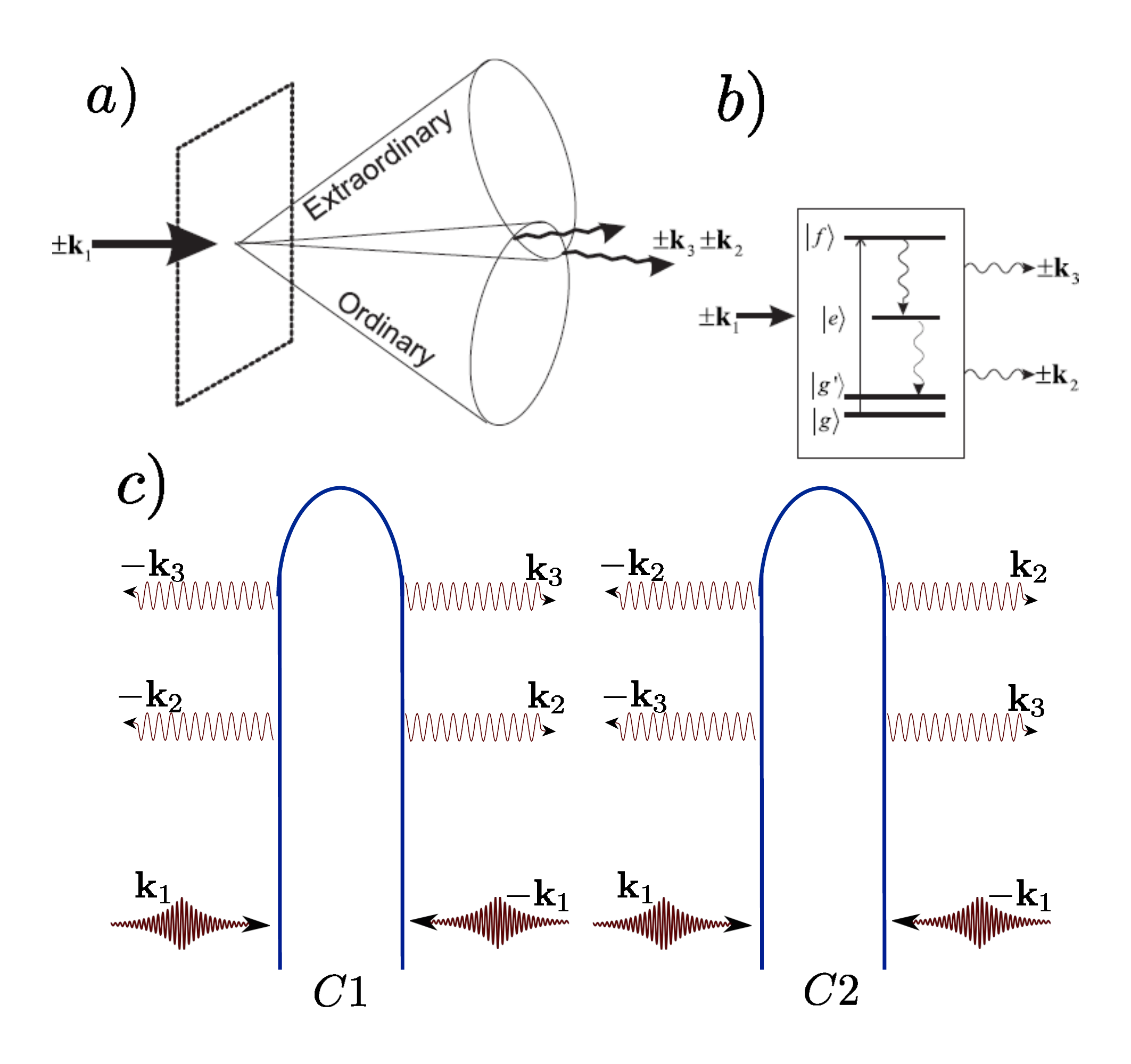}\\
  \caption{Type II PDC: (A) phase matching, (B) molecular level scheme; (C) CTPL diagrams for the signal from $\mathbf{k}_3$ (C1) and $\mathbf{k}_2$ (C2) modes polarized along $x$ direction.}\label{FIG:5}
\end{figure}

In Type II parametric down-conversion, the two spontaneously-generated signals
have orthogonal polarizations. Because of birefringence, the
generated photons are emitted along two non-collinear spatial cones
known as ordinary and extraordinary beams (See
Fig.\ref{FIG:5} (A)). Polarization-entangled light
\cite{Ger05} is generated at the
intersections of the cones. An $x$ polarization filter and a detector are placed at one of the
cones intersections. The detector cannot tell from which beam a photon is obtained. To
describe the process we need five optical modes: one classical
$|1\rangle| y\rangle$ and four quantum modes $\{ |2\rangle
|x\rangle, |2\rangle |y\rangle, |3\rangle |x\rangle,
|3\rangle|y\rangle \}$.
\par
Polarization-entangled signal is described by CTPL diagrams shown in Fig.\ref{FIG:5} (C). The Type II PDC signal consist of two parts $S^{(5)}_{PDCII}=S^{(5)}_{3x}+S^{(5)}_{2x}$. The signal $S^{(5)}_{3x}$ assumes the form of equation \eqref{PDCTypeI}, with the material pathway depicted in Fig.\ref{FIG:5}(C1):
\begin{gather}
\label{PDCTII}
\chi^{(2)}_{LL-}\left({-(\omega_1-\omega_2);-\omega_2,\omega_1}\right)=\\
\notag
\frac{1}{2}(\chi^{(2)}_{+--}\left({-(\omega_1-\omega_2);-\omega_2,\omega_1}\right)+\\
\notag
+\chi^{(2)}_{++-}\left({-(\omega_1-\omega_2);\omega_2,-\omega_1}\right))=\\
\notag
\frac{1}{2! \hbar^2}\frac{C}{(\omega_1-\omega_{gf}+i\hbar \gamma_{gf})(\omega_1-\omega_2-\omega_{eg}+i\hbar\gamma_{eg})}
\end{gather}
where the coefficient $C$ is given by the following equation:
\begin{gather}
\label{C}
C^2=|\mu^y_{gf}|^2(|\mu^x_{fe}|^2|\mu^x_{eg}|^2+2\mu^x_{fe}\mu^x_{eg}\mu^y_{fe}\mu^y_{eg}+\\
\notag
\mu^x_{fe}\mu^x_{eg}\mu^x_{fe}\mu^y_{eg}+\mu^y_{fe}\mu^y_{eg}\mu^y_{fe}\mu^x_{eg}
\end{gather}
The signal $S^{(5)}_{3x}$ is described by the diagram in Fig.\ref{FIG:5} (C2):
\begin{gather}
\label{PDCTIII}
\chi^{(2)}_{LL-}\left({-\omega_2;-(\omega_1-\omega_2),\omega_1}\right)=\\
\notag
\frac{1}{2}(\chi^{(2)}_{+--}\left({-\omega_2;-(\omega_1-\omega_2),\omega_1}\right)+\\
\notag
+\chi^{(2)}_{++-}\left({-\omega_2;-(\omega_1-\omega_2),\omega_1}\right))=\\
\notag
\frac{1}{2! \hbar^2}\frac{C}{(\omega_1-\omega_{gf}+i\hbar \gamma_{gf})(\omega_2-\omega_{eg}+i\hbar\gamma_{eg})}
\end{gather}
The net Type II PDC signal is:
\begin{gather}
\label{PDCTypeII}
S_{PDCII}(\omega_1)=\\
\notag
\frac{N(N-1)}{4\pi \hbar}\left| \mathcal{E}_1 \right|^2 \frac{2 \pi \hbar \omega_2}{\Omega} \frac{2 \pi \hbar (\omega_1-\omega_2)}{\Omega}\times\\
\notag
|\chi^{(2)}_{LL-}\left({-(\omega_1-\omega_2);-\omega_2,\omega_1}\right)|^2+\\
\notag
|\chi^{(2)}_{LL-}\left({-\omega_2;-(\omega_1-\omega_2),\omega_1}\right)|^2.
\end{gather}
\par

\subsection{Time-and-frequency resolved type-I PDC}\label{sec:PDCmic}

Previously we briefly summarized various wave-mixing signals in a very simple toy model without addressing the detection process in detail. In the following we include the photon counting detection described in Section \ref{sec.N-photon-counting} and demonstrate how does the actual generation process in type-I PDC relate to nonlinear response and how different it is compared to semiclassical theory.

 The standard calculation of nonlinear wave mixing assumes that all relevant field frequencies are off resonant with matter. It is then possible to adiabatically eliminate all matter degrees of freedom and describe the process by an effective Hamiltonian for the field that contains a nonlinear cubic coupling of three radiation modes \cite{mandel1995oca}.
For SFG this reads:
\begin{equation}
\label{effectivehamiltonian1}
H_{eff}=-\int d \mathbf{r}
\chi^{(2)} E^{\dag}_3 \left({\omega_1-\omega_2}\right) E^{\dag}_2 \left({\omega_2}\right) E_1 \left({\omega_1}\right)
\end{equation}
and for DFG and PDC:
\begin{equation}
\label{effectivehamiltonian2}
H_{eff}=
-\int d \mathbf{r}
\chi^{(2)} E^{\dag}_3 \left({\omega_1+\omega_2}\right) E_2 \left({\omega_2}\right) E_1 \left({\omega_1}\right)
\end{equation}
All matter information is embedded into a coefficient that is proportional to $\chi^{(2)}$ \cite{Ger05} that is defined by the semiclassical theory of radiation-matter coupling. Langevin quantum noise is added to represent vacuum fluctuations caused by other field modes \cite{Scully97,Glau07,Ave08} and account for photon statistics.

The microscopic theory of type I PDC presented below holds if cascade of two photons in a three level system is generated in both on and off resonance. The resonant case is especially important for potential spectroscopic applications \cite{Mukamel_book}, where unique information about entangled matter \cite{Let10} can be revealed. Other examples are  molecular aggregates and photosynthetic complexes or biological imaging \cite{Saleh98a}. Second, it properly takes into account the quantum nature of the generated modes through a generalized susceptibility that has a very different behavior near resonance than the semiclassical $\chi^{(2)}$. $\chi^{(2)}$ is derived for two classical fields and a single quantum field. While this is true for the reverse process (sum frequency generation) it does not apply for PDC, which couples a single classical and two quantum modes. Third, macroscopic propagation effects are not required for the basic generation of the signal. Fourth,  gated detection  \cite{Dorfman12a} yields the finite temporal and spectral resolution of the coincident photons limited by a Wigner spectrogram. For either time or frequency resolved measurement of the generated field, the signal can be expressed as a modulus square of the transition amplitude that depends on three field modes. This is not the case for photon counting. Shwartz et al. recently reported PDC in diamond, where 18keV pump field generates two X-ray photons \cite{Schwartz12}. Calculation in \cite{Dorfman12b} reproduces experimental data in the entire frequency range without adding Langevin noise. 

  \begin{figure*}[t]
\begin{center}
\includegraphics[trim=0cm 0cm 0cm 0cm,angle=0, width=0.88\textwidth]{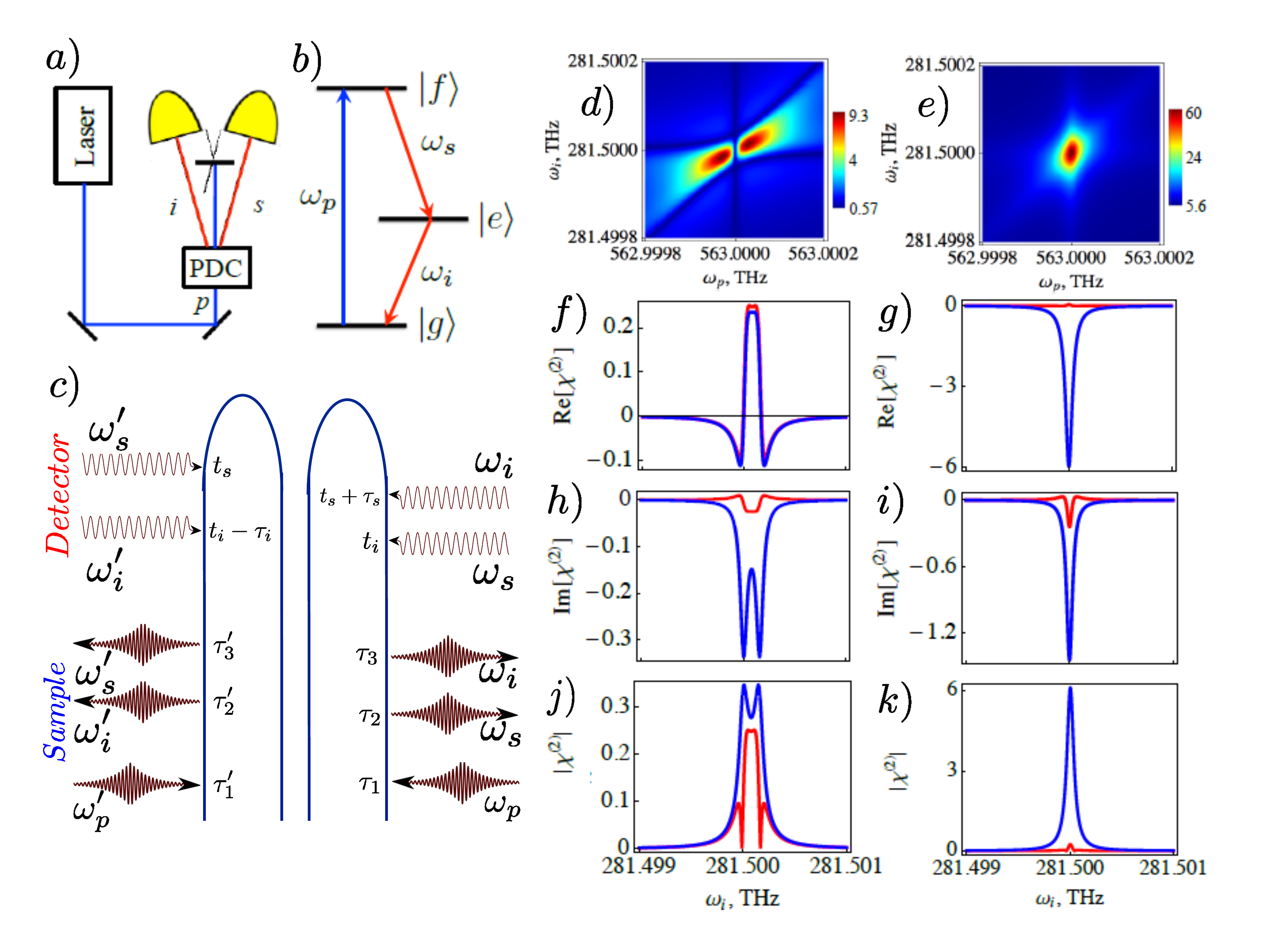}
\end{center}
\caption{(Color online) Left: Schematic of the PDC experiment - (a), the three level model system used in our simulations - (b). One out of four loop diagram (remaining three diagrams are presented in Ref.\cite{Dorfman12b}) for the PCC rate of signal and idler photons generated in type I PDC (Eq. (\ref{eq:Cso})) - (c). The left and right diagrams represent a pair of molecules. Blue (red) arrows represent field-matter interaction with the sample (detectors). There are four possible permutations ($s/i$ and $s'/i'$) which leads to four terms when Eq. (\ref{eq:chiq}) is substituted into Eq. (\ref{eq:Rcb1}). Right: Absolute value of semiclassical susceptibility $|\chi_{+--}^{(2)}(-(\omega_p-\omega_i),-\omega_i,\omega_p)|$ (arb. units) (\ref{eq:chi+--}) - (d), and quantum the susceptibility $|\chi_{LL-}^{(2)}(-(\omega_p-\omega_i),-\omega_i,\omega_p)|$ (\ref{eq:chiq}) - (e) vs pump $\omega_p$ and idler frequency $\omega_i$. We used the standard KTP parameters outlined in the text. Left column: real - (f), imaginary part - (h) and absolute value - (j) of $\chi_{+--}^{(2)}$ - red line and $\chi_{LL-}^{(2)}$ - blue line for off resonant pump $\omega_p-\omega_{gf}=10\gamma_{gf}$. Right column: (g,i,k) - same as (f,h,j) but for resonant pump $\omega_p\simeq\omega_{gf}$.}
\label{fig:PDC1}
\end{figure*}

The nature of entangled light can be revealed by photon correlation measurements that are governed by energy, momentum and/or angular momentum conservation. In PDC, a nonlinear medium  is pumped by electromagnetic field of frequency $\omega_p$ and some of the pump photons are converted into pairs of (signal and idler) photons with frequencies $\omega_s$ and $\omega_i$, respectively (see Fig. \ref{fig:PDC1}a) satisfying $\omega_p=\omega_s+\omega_i$. 

To get  more detailed description of wave-mixing process in PDC we recast an effective semiclassical Hamiltonian in Eq. (\ref{effectivehamiltonian2}) as follows
\begin{align}\label{eq:Hint}
&H_{int}(t)=i\hbar \sum_j\int\frac{d\omega_s}{2\pi}\frac{d\omega_i}{2\pi}\frac{d\omega_p}{2\pi}\chi_{+--}^{(2)}(\omega_s,\omega_i,\omega_p)\notag\\
&\times\hat{a}^{(s)\dagger}\hat{a}^{(i)\dagger}\beta^{(p)}e^{i\Delta\mathbf{k}(\mathbf{r}-\mathbf{r}_j)}e^{-i(\omega_p-\omega_s-\omega_i)t]}+H.c.,
\end{align}
where $\hat{a}^{\dagger(s)}$ and $\hat{a}^{(i)\dagger}$ are creation operators for signal and idler modes, $\beta^{(p)}$ is the expectation value of the classical pump field, $\Delta\mathbf{k}=\mathbf{k}_p-\mathbf{k}_s-\mathbf{k}_i$, $j$ runs over molecules, $\chi^{(2)}_{+--}$, (normally denoted $\chi^{(2)}(-\omega_s;-\omega_i,\omega_p)$) is the second-order nonlinear susceptibility

 \begin{align}\label{eq:chc}
& \chi^{(2)}_{+--}(\omega_s=\omega_p-\omega_i,\omega_i,\omega_p)=\left(\frac{i}{\hbar}\right)^2\int_0^{\infty}\int_{0}^{\infty}dt_2dt_1\times \notag\\
 &e^{i\omega_i(t_2+t_1)+i\omega_pt_1}\langle[[V(t_2+t_1),V(t_1)],V(0)]\rangle+(i\leftrightarrow p).
   \end{align}

The $``+--''$ indices in Eq. (\ref{eq:chc}) signify two commutators followed by an anti commutator. The bottom line of the semiclassical approach is that PDC is represented by 3-point matter-field interaction via the second order susceptibility $\chi^{(2)}_{+--}$ that couples the signal, idler and pump modes. However, it has been realized, that other field modes are needed to yield the correct photon statistics. Electromagnetic field fluctuations are then added as quantum noise (Langevin forces) \cite{Scully97}.

Below we summarize a  microscopic calculation of the photon coincidence counting (PCC) rate in type I PDC \cite{ros09}. We show that PDC is governed by a quantity that resembles but is different from (\ref{eq:chc}). In contrast with the semiclassical approach, PDC emerges as a 6-mode two-molecule rather than 3-mode matter-field interaction process and is represented by convolution of two quantum susceptibilities  $\chi^{(2)}_{LL-}(\omega_s,\omega_i,\omega_p)$ and $\chi^{(2)*}_{LL-}(\omega_s',\omega_i',\omega_p')$ that represent a pair of molecules in the sample interacting with many vacuum modes of the signal  ($s,s'$) and the idler ($i,i'$). Field fluctuations are included self consistently at the microscopic level. Furthermore the relevant nonlinear susceptibility is different from the semiclassical one $\chi^{(2)}_{+--}$ and is given by
 \begin{align}\label{eq:chq}
& \chi^{(2)}_{LL-}(\omega_s=\omega_p-\omega_i,\omega_i,\omega_p)=\left(\frac{i}{\hbar}\right)\int_0^{\infty}\int_{0}^{\infty}dt_2dt_1\times \notag\\
 &e^{i\omega_i(t_2+t_1)+i\omega_pt_1}\langle [V(t_2+t_1)V(t_1),V(0)]\rangle+(i\leftrightarrow s).
   \end{align}
Eq. (\ref{eq:chq}) has a single commutator and is symmetric to a permutation of $\omega_i$ and $\omega_s=\omega_p-\omega_i$, as it should. Eq. (\ref{eq:chc}) has two commutators and lacks this symmetry.

\subsubsection{The bare PCC rate }
The measured PCC rate of signal and idler photons starts with the definition (\ref{eq:S2g}).
The basis for the bare signal in Eq. (\ref{eq:nnVVVV}) is given by the time-ordered product of Green's functions of superoperators in the interaction picture (see \cite{Dorfman12b}).
\begin{align}\label{eq:Cso}
&\langle \mathcal{T}E_R^{\dagger(i)}(t_i')E_R^{\dagger(s)}(t_s'+\tau_s)E_L^{(s')}(t_s')E_L^{(i')}(t_i'-\tau_i)\notag\\
&\times e^{-\frac{i}{\hbar}\int_{-\infty}^{\infty}\sqrt{2}H'_{-}(\tau)d\tau}\rho(-\infty)\rangle.
\end{align}
which represents four spectral modes arriving at the detectors, where modes $s,s', i,i'$ are defined by their frequencies $\omega_s$, $\omega_i'$, $\omega_i=\omega_p-\omega_s$, $\omega_s'=\omega_p'-\omega_i'$.

In type I PDC the sample is composed of $N$ identical molecules initially in their ground state. They interact with one classical pump mode and emit two spontaneously generated quantum modes with the same polarization into two collinear cones. The initial state of the optical field is given by $|0\rangle_s|0\rangle_i|\beta\rangle_p$. A classical pump field promotes the molecule from its ground state $|g\rangle$ to the doubly excited state $|f\rangle$ (see Fig \ref{fig:PDC1}b).

Due to the quantum nature of the signal and the idler fields, the interaction of each of these fields with matter must be at least second order to yield a non vanishing signal.  The leading contribution to Eq. (\ref{eq:Cso}) comes from the four diagrams shown in Fig. \ref{fig:PDC1}c (for rules see Appendix \ref{app:Diag} \cite{ros09}). The coherent part of the signal represented by interaction of two spontaneously generated quantum and one classical modes is proportional to the number of pairs of sites in the sample $\sim N(N-1)$, which dominates the other, incoherent, $\sim N$ response for large $N$. Details of the calculation of the correlation function (\ref{eq:Cso}) are presented in Ref. \cite{Dorfman12b}. We obtain for the ``bare'' frequency domain PCC rate $R_c^{(B)}(\omega_s,\omega_i',\omega_p,\omega_p')\equiv\int dt_s' dt_i' \langle\mathcal{T}\hat{n}_s(t_s',\omega_s')\hat{n}_i(t_i',\omega_i')\rangle_Te^{-i(\omega_s+\omega_i'-\omega_p')t_s'+(\omega_s+\omega_i'-\omega_p)t_i'}$

\begin{align}\label{eq:Rcb1}
&R_c^{(B)}(\omega_s,\omega_i',\omega_p,\omega_p')=N(N-1)\left(\frac{2\pi\hbar}{V}\right)^4\times\notag\\
&\mathcal{E}^{*(p)}(\omega_p)\mathcal{E}^{(p)}(\omega_p')\mathcal{D}(\omega_s)\mathcal{D}(\omega_p-\omega_s)\mathcal{D}(\omega_i')\mathcal{D}(\omega_p'-\omega_i')\times\notag\\
&\chi_{LL-}^{(2)}[-(\omega_p'-\omega_i'),-\omega_i',\omega_p']\chi_{LL-}^{(2)*}[-\omega_s,-(\omega_p-\omega_s),\omega_p],
\end{align}
where $\mathcal{E}^{(p)}(\omega)\equiv E^{(p)}(\omega)\beta^{(p)}$ is a classical field amplitude, $E^{(p)}(\omega)$ is the pump pulse envelope and $\mathcal{D}(\omega)=\omega\tilde{\mathcal{D}}(\omega)$ where $\tilde{\mathcal{D}}(\omega)=V\omega^2/\pi^2c^3$ is the density of radiation modes. For our level scheme (Fig. \ref{fig:PDC1}b) the nonlinear susceptibility $\chi^{(2)}_{LL-}$ (see Eq. (\ref{eq:chq})) is given by

\begin{align}\label{eq:chiq}
&\chi_{LL-}^{(2)}[-(\omega_p'-\omega_i'),-\omega_i',\omega_p']=\frac{1}{2\hbar}\frac{\mu_{gf}^{*}\mu_{fe}\mu_{eg}}{\omega_p'-\omega_{gf}+i\gamma_{gf}}\times\notag\\
&\frac{1}{\omega_p'-\omega_i'-\omega_{eg}+i\gamma_{eg})}+(\omega_i'\leftrightarrow \omega_p'-\omega_i').
\end{align}

Eq. (\ref{eq:Rcb1}) represents a 6-mode  $(\omega_p,\omega_i,\omega_s,\omega_p',\omega_i',\omega_s')$ field-matter correlation function factorized into two generalized susceptibilities each representing the interaction of two quantum and one classical mode with a different molecule. Because of two constraints $\omega_p=\omega_s+\omega_i$, $\omega_p'=\omega_s'+\omega_i'$ that originate from time translation invariance on each of the two molecules that generate the nonlinear response, Eq. (\ref{eq:Rcb1}) only depends on four field modes. Each molecule creates a coherence in the field  between states with zero and one photon. By combining  the  susceptibilities from a pair of molecules we obtain a photon occupation number that can be detected.  Thus, the detection process must be described in the joint space of the two molecules and involves the interference of four quantum pathways (two with bra and two with ket) with different time orderings. Note that this pathway information is not explicit in the Langevin approach.   

For comparison, if all three fields (signal, idler and pump) are classical, the number of material-field interactions is reduced to three - one for each field. Then the leading contribution to the field correlation function  yields the semiclassical nonlinear susceptibility $\chi^{(2)}_{+--}$
\begin{align}\label{eq:chi+--}
&\chi_{+--}^{(2)}[-(\omega_p'-\omega_i'),-\omega_i',\omega_p']=\notag\\
&\frac{1}{2\hbar^2}\langle g|\mathcal{T}V_L\mathcal{G}(\omega_p'-\omega_i')V_L\mathcal{G}(\omega_p')V_L^{\dagger}|g\rangle+\notag\\
&\frac{1}{\hbar^2}\langle g|\mathcal{T}V_L\mathcal{G}^{\dagger}(\omega_i')V_L\mathcal{G}(\omega_p')V_L^{\dagger}|g\rangle=\frac{1}{\hbar^2}\frac{\mu_{gf}^{*}\mu_{fe}\mu_{eg}}{\omega_p-\omega_{gf}+i\gamma_{gf}}\times\notag\\
&\left[\frac{1}{\omega_p-\omega_i'-\omega_{eg}+i\gamma_{eg}}+\frac{1}{\omega_i'-\omega_{eg}-i\gamma_{eg}}\right]+(\omega_i'\leftrightarrow \omega_p-\omega_i'),
\end{align}
where $\mathcal{G}(\omega)=1/(\omega-H_{0-}/\hbar+i\gamma)$ is the retarded Liouville Green's function, and $\gamma$ is lifetime broadening. $\chi^{(2)}_{LL-}$ possesses a permutation symmetry with respect to $s\leftrightarrow i$ (both have $L$ index). In contrast the semiclassical calculation via $\chi^{(2)}_{++-}$ is non symmetric with respect to $s\leftrightarrow i$ ( one $+$ and one $-$ indexes), which results in PCC rate that depends upon whether the signal or idler detector clicks first \cite{Schwartz12}.

\subsubsection{Simulations of typical PDC signals}

\begin{figure}[t]
\centering
\includegraphics[trim=0cm 0cm 0cm 0cm,angle=0, width=0.5\textwidth]{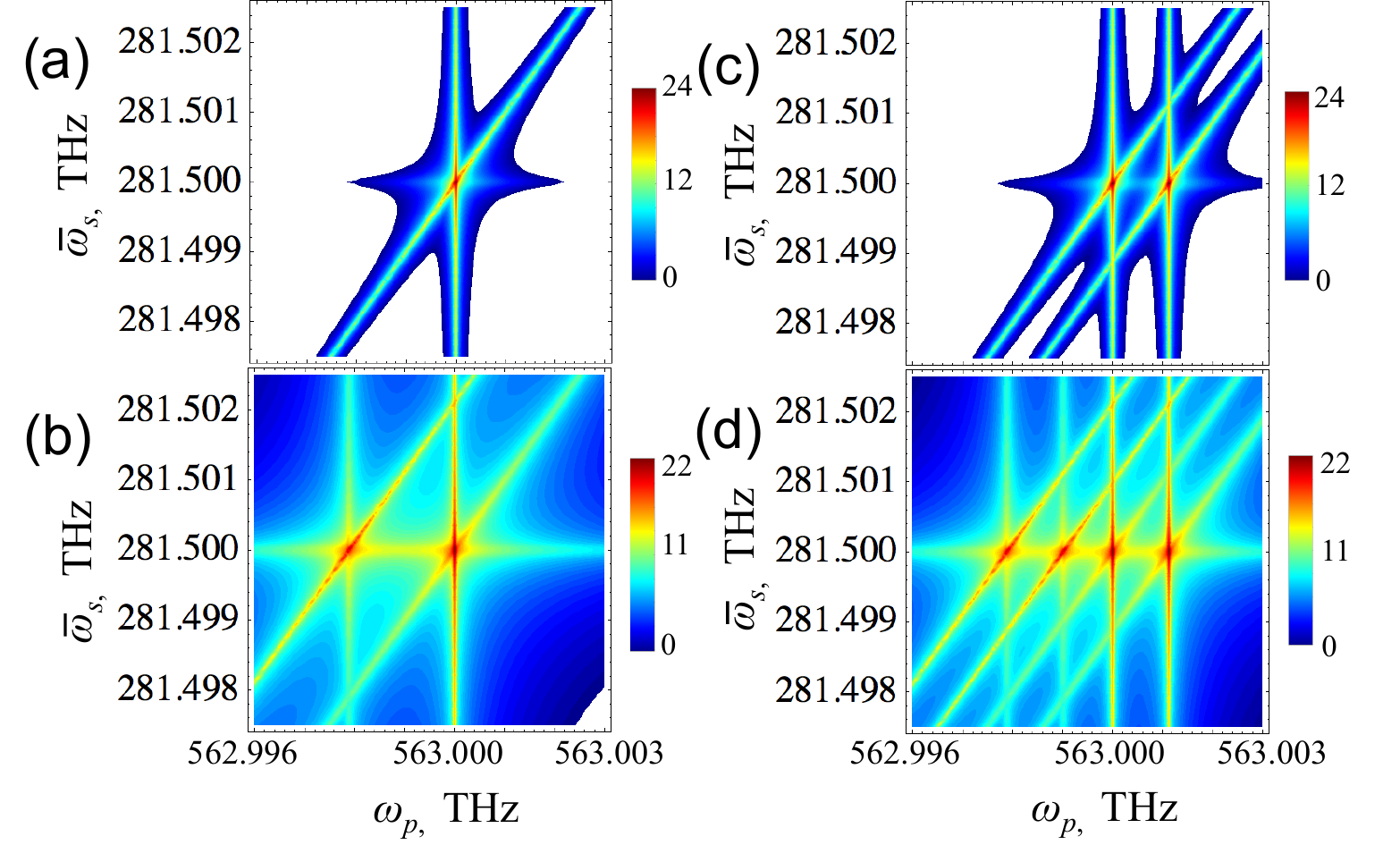}
\caption{(Color online) Left column: two dimensional PCC rate (log scale in arb. units) calculated using quantum theory Eq. (\ref{eq:S2g}) assuming a single monochromatic pump with frequency $\omega_p$. Idler detector resonant with intermediate level $\bar{\omega}_i=\omega_{eg}=282$THz - (a), while $\omega_{eg}-\bar{\omega}_i=2$GHz for (b). Right column: same as left but for a pump made out of two monochromatic beams with frequencies $\omega_p-\omega_p'=2\cdot 10^{-6}\omega_p$.}
\label{fig:PDC2} %
\end{figure}

Fig. \ref{fig:PDC1}d-k compares both susceptibilities calculated for a typical KTP crystal (PPKTP) represented by a degenerate three-level system  with parameters taken from Ref. \cite{Wu86a,Aur08}. Fig. \ref{fig:PDC1}f,h,j show that far from resonances ($\omega_i\neq\omega_{eg}$, $\omega_p\neq\omega_{gf}$) the semiclassical and quantum susceptibilities coincide and depend weakly on the frequencies $\omega_p$ and $\omega_i$. This is the regime covered by the semiclassical theory, where the susceptibility is assumed to be a constant. Similar agreement between classical and quantum susceptibilities can be observed if the pump is resonant with two-photon transition $\omega_p\simeq\omega_{gf}$ but the idler is off resonance $\omega_i\neq\omega_{eg}$ - Fig. \ref{fig:PDC1}g,h,k. However, close to resonance - Fig. \ref{fig:PDC1}d,e the two susceptibilities are very different. The semiclassical susceptibility $\chi_{+--}^{(2)}$ vanishes at resonance, where the quantum susceptibility $\chi_{LL-}^{(2)}$ reaches its maximum.

To put the present ideas into more practical perspective we show in Fig. \ref{fig:PDC2}  the PCC rate for a monochromatic pump $\omega_p$ and mean signal detector frequency $\bar{\omega}_s$. The quantum theory yields one strong resonant peak at $\bar{\omega}_s=\omega_p-\omega_{eg}$ and two weak peaks at $\omega_p=\omega_{gf}$ and $\bar{\omega}_s=\omega_{eg}$ if the idler detector is resonant with the intermediate state $|e\rangle$: $\bar{\omega}_i=\omega_{eg}$ - Fig. \ref{fig:PDC2}a. However, if we tune the idler detector to a different frequency, for instance $\omega_{eg}-\bar{\omega}_i=2$GHz there is an additional peak at $\bar{\omega}_s=\omega_p-\bar{\omega}_i$ - Fig. \ref{fig:PDC2}b. Similarly, when the pump consists of two monochromatic beams $\omega_p\neq\omega_p'$ (panels c,d) the number of peaks are doubled compare to single monochromatic pump. Clearly, one can reproduce the exact same peaks for $\omega_p'$ as for $\omega_p$.

\subsubsection{Spectral diffusion} 

\begin{figure*}[t]
\begin{center}
\includegraphics[trim=0cm 0cm 0cm 0cm,angle=0, width=0.95\textwidth]{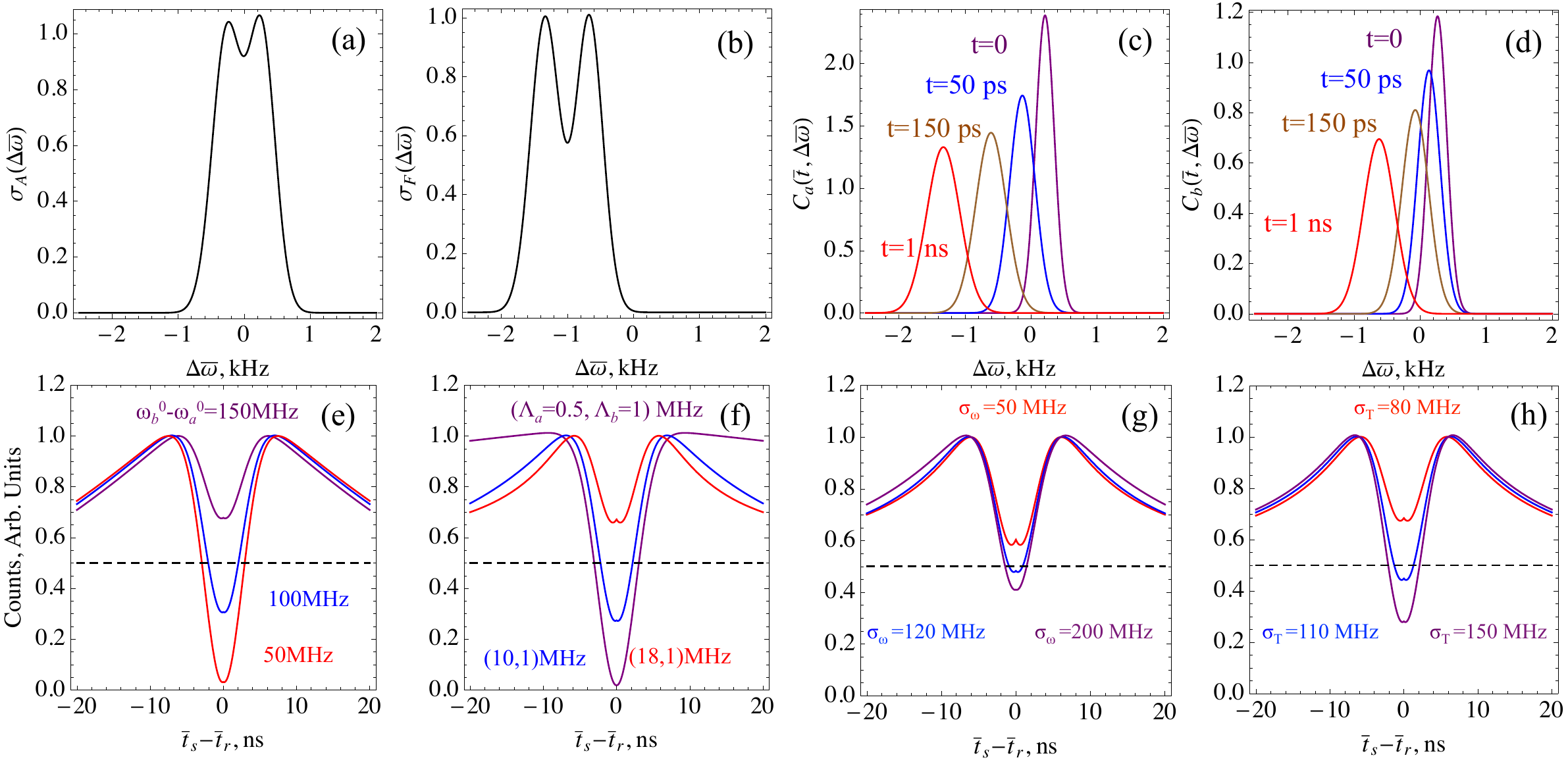}
\end{center}
\caption{(Color online) Top row: Absorption (\ref{eq:sA}) - (a) and fluorescence (\ref{eq:sF}) - (b) line shapes vs displaced frequency $\Delta\bar{\omega}=\omega-\frac{1}{2}(\omega_a+\omega_b)$. Frequency dispersed time-resolved fluorescence (\ref{eq:Ca}) displayed as a snapshot spectra for molecule $a$ - (c) and $b$ - (d). Bottom row: PCC for different transition energies of molecules excited at $\omega_p=\omega_b^0+\lambda_b$ - (e); PCC for different values of the SD time scale $\Lambda_a$ and $\Lambda_b$  and fixed linewidth $\Gamma_a$, $\Gamma_b$ according to Eq. (\ref{eq:Gab}) - (f); PCC for different frequency gate bandwidths - (g) at fixed time gate bandwidth $\sigma_T=100$ MHz and for different time gate bandwidths - (h) at fixed frequency gate bandwidth $\sigma_\omega=100$ MHz. Molecules have distinct SD timescales $\Lambda_a=15$ MHz, $\Lambda_b=1$ MHz and $\omega_b^0-\omega_a^0=1$ MHz.}
\label{fig:PCC}
\end{figure*}

Spectral diffusion (SD) which results from the stochastic modulation of frequencies can manifest itself either as discrete random jumps of the emission frequency \cite{Siy09,San10,Wal12} or as a broadening of a hole burnt in the spectrum by a narrowband pulse \cite{Xie98,Wag12}.
We focus on the latter case assuming that the electronic states of molecule $\alpha=a,b$ are coupled to a harmonic bath described by the Hamiltonian $\hat{H}_B^{\alpha}=\sum_{k}\hbar\omega_{k}(\hat{a}_{k}^{\dagger\alpha}\hat{a}_{k}^{\alpha}+1/2)$ (see Fig. \ref{fig:schemeRc}b). The bath perturbs the energy of state $\nu$. This is represented by the Hamiltonian
\begin{equation}
\hat{H}_{\nu}^{\alpha}=\hbar^{-1}\langle\nu_\alpha|\hat{H}|\nu_\alpha\rangle=\epsilon_{\nu_\alpha}+\hat{q}_{\nu\alpha}+\hat{H}_B^{\alpha},
\end{equation}
where $\hat{q}_{\nu}$ is a collective bath coordinate
\begin{equation}
\hat{q}_{\nu\alpha}=\hbar^{-1}\langle\nu_\alpha|\hat{H}_{SB}|\nu_\alpha\rangle=\sum_{k}d_{\nu_\alpha\nu_\alpha,k}(\hat{a}_{k}^{\dagger}+\hat{a}_{k}),
\end{equation}
$d_{mn,k}$ represents bath-induced fluctuations of the transition energies ($m=n$) and the intermolecular coupling ($m\neq n$). We define the line-shape function
\begin{align}
&g_\alpha(t)\equiv g_{\nu_\alpha\nu'_\alpha}(t)=\int\frac{d\omega}{2\pi}\frac{C''_{\nu_\alpha\nu'_\alpha}(\omega)}{\omega^2}\notag\\
&\times\left[\coth\left(\frac{\beta\hbar\omega}{2}\right)(1-\cos\omega t)+i\sin\omega t-i\omega t\right],
\end{align}
where the bath spectral density is given by
\begin{equation}
C''_{\nu_\alpha\nu'_\alpha}(\omega)=\frac{1}{2}\int_0^{\infty}dte^{i\omega t}\langle[\hat{q}_{\nu\alpha}(t),\hat{q}_{\nu'\alpha}(0)]\rangle,
\end{equation}
$\beta=k_BT$ with $k_B$ being the Boltzmann constant and $T$ is the ambient temperature. We shall use the overdamped Brownian oscillator model for the spectral density, assuming a single nuclear coordinate ($\nu_\alpha=\nu'_\alpha$)
\begin{equation}
C''_{\nu_\alpha\nu_\alpha}(\omega)=2\lambda_{\alpha}\frac{\omega\Lambda_{\alpha}}{\omega^2+\Lambda_{\alpha}^2},
\end{equation}
where $\Lambda_{\alpha}$ is the fluctuation relaxation rate and $\lambda_{\alpha}$ is the system-bath coupling strength. The corresponding lineshape function then depends on two parameters: the reorganization energy $\lambda_\alpha$ denoting the strength of the coupling to the bath and the fluctuation relaxation rate  $\Lambda_{\alpha} = \sqrt{2 \lambda_{\alpha} k_B T / \hbar}$. In the high-temperature limit $k_BT\gg\hbar\Lambda_{\alpha}$ we have
\begin{align}\label{eq:gt}
g_{\alpha}(t)=\left(\frac{\Delta_{\alpha}^2}{\Lambda_{\alpha}^2}-i\frac{\lambda_{\alpha}}{\Lambda_{\alpha}}\right)\left(e^{-\Lambda_{\alpha} t}+\Lambda_{\alpha} t-1\right).
\end{align}
For a given magnitude of fluctuations $\Delta_\alpha$, $\alpha=a,b$ the FWHM of the absorption linewidth  \cite{Mukamel_book}
\begin{align}\label{eq:Gab}
\Gamma_\alpha=\frac{2.355+1.76(\Lambda_\alpha/\Delta_\alpha)}{1+0.85(\Lambda_\alpha/\Delta_\alpha)+0.88(\Lambda_\alpha/\Delta_\alpha)^2}\Delta_\alpha.
\end{align}
The absorption and emission lineshape functions  for a pair of molecules obtained in the slow nuclear dynamics limit: $\Lambda_\alpha\ll\Delta_\alpha$ are given by \cite{Mukamel_book}
\begin{align}\label{eq:sA}
\sigma_A(\omega)=\sum_{\alpha=a,b}(2\pi\Delta_\alpha)^{-1/2}e^{-\frac{(\omega-\omega_\alpha^0-\lambda_\alpha)^2}{2\Delta_\alpha^2}},
\end{align}
\begin{align}\label{eq:sF}
\sigma_F(\omega)=\sum_{\alpha=a,b}(2\pi\Delta_\alpha)^{-1/2}e^{-\frac{(\omega-\omega_\alpha^0+\lambda_\alpha)^2}{2\Delta_\alpha^2}},
\end{align}
where  $2\lambda_{\alpha}$ is the Stokes shift and $\Delta_{\alpha}=\sqrt{2\lambda_{\alpha}k_BT/\hbar}$ is a linewidth parameter. Together with the relaxation rate $\Lambda_\alpha$ (see Eq. (\ref{eq:gt})) these parameters completely describe the SD model and govern the evolution of the emission linewidth between the initial time given by Eq. (\ref{eq:sA}) (see Fig. \ref{fig:PCC}a) and long time given by Eq. (\ref{eq:sF}) (see Fig. \ref{fig:PCC}b). 

The corresponding four-point matter correlation function may be obtained by the second order cumulant expansion \cite{Mukamel_book}
\begin{align}\label{eq:4pt}
F_\alpha(t_1,t_2,t_3,t_4)&=|\mu_{\alpha}|^4e^{-i\omega_{\alpha}(t_1-t_2+t_3-t_4)}e^{\Phi_\alpha(t_1,t_2,t_3,t_4)},
\end{align}
where $\omega_\alpha\equiv\omega_{e_\alpha}-\omega_{g_\alpha}$ is the absorption frequency, $\Phi_\alpha(t_1,t_2,t_3,t_4)$ is the four-point lineshape function $\Phi_\alpha(t_1,t_2,t_3,t_4)=-g_{\alpha}(t_1-t_2)-g_{\alpha}(t_3-t_4)+g_{\alpha}(t_1-t_3)-g_{\alpha}(t_2-t_3)+g_{\alpha}(t_2-t_4)-g_{\alpha}(t_1-t_4)$. 

 We focus on the SD in the ``hole burning'' limit (HBL) which holds under two conditions: First, the dephasing is much faster than the fluctuation timescale, i.e. $t_k'\ll\Lambda_\alpha^{-1}$, $k=1,2,3,4$. Second, if excitation pulse duration $\sigma_p^{-1}$ and the inverse spectral $(\sigma_\omega^{j})^{-1}$, and temporal $(\sigma_T^{j})^{-1}$, $j=r,s$  gate bandwidths are much shorter than the fluctuation time scales, one may neglect the dynamics during the delay between population evolution and its detection. This parameter regime is relevant to crystals which store information in the form of reversible notches that are created in their optical absorption spectra at specific frequencies. Long storage times \cite{Lon05}, high efficiencies \cite{Hed10}, and many photon qubits in each crystal \cite{Sha02} can be achieved in this limit. The HBL limit is natural for long-term quantum memories where entanglement is achieved with telecom photons, proved the possibility of quantum internet \cite{Sag11,Cla11}.

Time-and-frequency resolved fluorescence is the simplest way to observe SD. The molecular transition frequency is coupled linearly to an overdamped Brownian oscillator that represents the bath (see Fig. \ref{fig:schemeRc}b). 
This fluorescence signal given by Eq. (\ref{eq:Ca}) is depicted as a series of the snapshot spectra at different times for molecule $a$  in Fig. \ref{fig:PCC}c. It shows a time dependent frequency redshift $\tilde{\omega}_a(t)$  and time dependent spectral broadening given by $\tilde{\sigma}_{a0}(t)$. Initially $\tilde{\omega}_a(0)=\omega_a^0+\lambda_a$ whereas at long times $\tilde{\omega}_a(\infty)=\omega_a^0-\lambda_a$, where $2\lambda_\alpha$ is the Stokes shift. Same for molecule $b$ is shown in Fig. \ref{fig:PCC}d. Because of the different reorganization energies $\lambda_a$, $\lambda_b$ and relaxation rates $\Lambda_a$, $\Lambda_b$ the Stokes shift dynamics and dispersion are different. Even when the absorption frequencies are the same  $\omega_a=\omega_b$, the fluorescence can show a different  the profile due to SD. This affects the distinguishability of the emitted photons as will be demonstrated below.

\subsection{Generation and  entanglement control of photons produced by two independent molecules by time-and-frequency gated photon coincidence counting (PCC).}

As discussed in section~\ref{sec.quantum light}, entangled photons can be produced by a large variety of $\chi^{(2)}$-processes. Alternatively, photons can be entangled by simultaneous excitation of two remote molecules and mixing the spontaneously emitted photons on the beam splitter. This entangled photon generation scheme is suitable e.g. for spectroscopic studies of single molecules by single photons discussed below. The manipulation of  single photon interference by appropriate time-and-frequency gating discussed below is presented in details in Ref. \cite{Dorfman14c}.
The generated entangled photon pair  can be used in logical operations based on optical measurements that utilize interference between indistinguishable photons. 

\begin{figure*}[t]
\begin{center}
\includegraphics[trim=0cm 0cm 0cm 0cm,angle=0, width=0.75\textwidth]{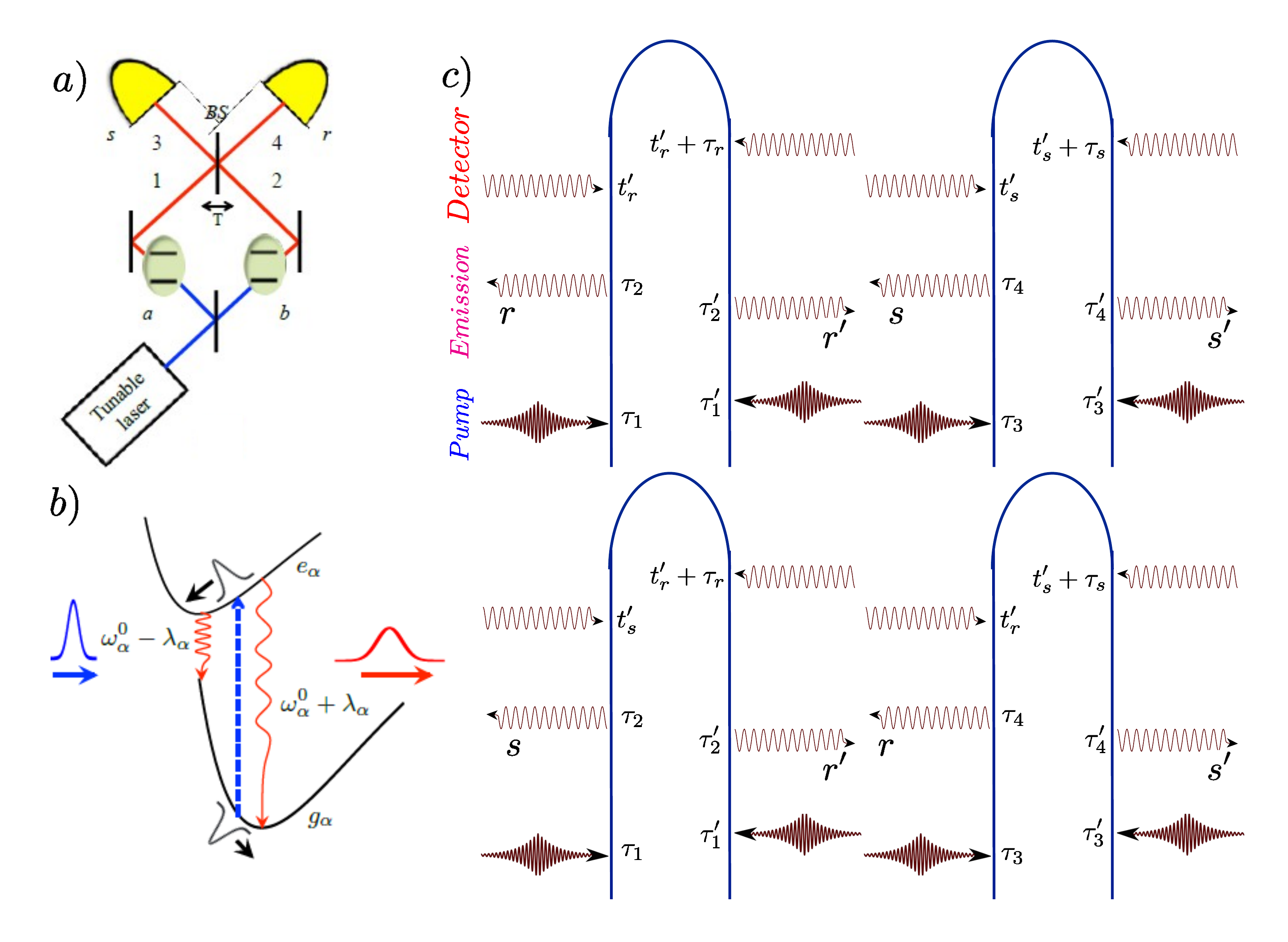}
\end{center}
\caption{(Color online) Time-and-frequency resolved measurement of PCC with spectral diffusion. Schematic of the PCC experiment with two indistinguishable source molecules - (a), the two-level model  of the molecule with SD used in our simulations - (b). (c) - Loop diagrams for the PCC rate of emitted photons from two molecules (for diagram rules see Appendix \ref{app:Diag}). The left and right branches of each diagram represent interactions with ket- and bra- of the density matrix, respectively. Field-matter interactions with the pump pulses $p_1$ and $p_2$ (blue), spontaneously emitted $s,s',r,r'$ photons (red) and detectors (brown).}
\label{fig:schemeRc}
\end{figure*}

We examine photon interference in the setup shown in Fig. \ref{fig:schemeRc}a. A pair of photons is generated by two remote two-level molecules $a$ and $b$  with ground $g_\alpha$ and excited state $e_\alpha$, $\alpha=a,b$. These photons then enter a beam splitter and are subsequently registered by time-and-frequency gated detectors $s$ and $r$. There are three possible outcomes: two photons registered in detector $s$, two photons registered in $r$  or coincidence where one photon is detected in each. The ratios of these outcomes reflects the photon Bose statistics and depends  on their degree of indistinguishability. If the two photons incident on the beam splitter are indistinguishable the photon coincidence counting signal (PCC) vanishes. This causes the Hong-Ou-Mandel (HOM) dip when varying the position of the beam splitter which causes delay $T$ between the two photons. The normalized PCC rate varies between 1 for completely distinguishable photons (large $T$) and 0 when they are totally indistinguishable ($T=0$).  For classical fields and 50:50 beam splitter the PCC rate may not be lower than 1/2. We denote the photons as indistinguishable (distinguishable) if the PCC rate is smaller (larger) than 1/2.

PCC is typically measured by time-resolved detection  \cite{Glau07,Ger05}. Originally performed with entangled photons generated by parametric down conversion (PDC)  \cite{Hong87} the shape of the dip vs delay is usually related to the two-photon state envelope which is governed by an effective PDC Hamiltonian \cite{Dorfman12b}.  Bath effects can become important for remote emitters and have been introduced phenomenologically \cite{Let10} below. We present a microscopic description of PCC with bath fluctuations by formulating the signal in the joint field-matter space measured by simultaneous time-and-frequency resolved detection. This complex measurement can be achieved using high-speed photodiode which converts a fast optical signal into a fast electric current, fast oscilloscopes to observe the waveform, wide bandwidth spectrum analyzers and other elements. Short pulse characterization using time-frequency map such as frequency-resolved optical gating (FROG) \cite{Tre02},  spectral phase interferometry for direct-field reconstruction (SPIDER) \cite{Dor99} are well established tools for ultrafast metrology \cite{Wol07,Wal09}. Extending these techniques to a single photon time and frequency resolved detection is challenging and may be achieved if combined with on-chip tunable detectors \cite{Gus07} or upconversion processes \cite{Gu10,Ma11}.

\subsubsection{PCC  of single photons generated by two remote emitters.}

The time-and-frequency gated PCC signal is described by the two pairs of loop diagrams shown in Fig. \ref{fig:schemeRc}c. Each loop represents one molecule ($a$ or $b$) which undergoes four field-matter interactions and each detector interacts twice with the field. Fig. \ref{fig:schemeRc}c shows that after interacting with the pump (with its ket) at time $t_2$ molecule $a$ evolves in the coherence $\rho_{e_ag_a}$ during time interval $t_2'$. The second interaction of the pump with the bra then brings the molecule into a population state $\rho_{e_ae_a}$ which evolves during interval $t_1$ until the first interaction with spontaneous emission mode occurs with ket. The molecule then evolves into the coherence $\rho_{g_ae_a}$ during $t_1'$ until the second bra- interaction of spontaneous mode. During population and coherence periods, the characteristic timescale of the dynamics is governed by population relaxation and dephasing, respectively.

The relevant single particle information for molecule $\alpha$ is given by the four point dipole correlation function $F_\alpha(t_1,t_2,t_3,t_4)=\langle V_{ge}(t_1)V_{eg}^{\dagger}(t_2)V_{ge}(t_3)V_{eg}^{\dagger}(t_4)\rangle_\alpha$, where $V$ and $V^{\dagger}$ are the lowering and raising dipole transition operators, respectively. Diagrams $i$ in Fig. \ref{fig:schemeRc}c represent non-interfering terms given by a product of two independent fluorescence contributions of the individual molecules. Diagrams $ii$ represent an interference in the joint space of the two molecules and involves the interference of eight quantum pathways (four with the bra and four with the ket) with different time orderings. Each molecule creates a coherence in the field between states with zero and one photon $|0\rangle\langle 1|$ and $|1\rangle\langle 0|$. By combining the contributions from a pair of molecules we obtain a photon population $|1\rangle\langle 1|$ that can be detected \cite{Dorfman12a, Dorfman12b}. For a pair of identical molecules, the beam splitter destroys the pathway information making the molecules indistinguishable and giving rise to quantum interference.

The PCC signal (\ref{eq:S2g}) (see Ref. \cite{Dorfman14c}) has been calculated for the output fields $E_3$ and $E_4$ ($i=3$, $s=4$) of the beam splitter (see Fig. \ref{fig:schemeRc}a). The fields in the output $3,4$ and input $1,2$  ports of the 50:50 beamsplitter are related by
\begin{align}\label{eq:bs}
E_3(t)=\frac{E_1(t)-iE_2(t+T)}{\sqrt{2}},~ E_4(t)=\frac{E_2(t)-iE_1(t-T)}{\sqrt{2}},
\end{align}
where $\pm cT$ is small difference of path length in the two arms. Taking into account that Eq. (\ref{eq:S2g}) should be modified to include the beam splitter position and absorb it into the gating spectrograms, the time-and-frequency resolved PCC in this case is given by
\begin{align}\label{eq:Rc2}
&R_c^{34}(\Gamma_r,\Gamma_s;T)=\frac{1}{(2\pi)^2}\int_{-\infty}^{\infty}d^2\Gamma_r'd^2\Gamma_s'\notag\\
&[W_D^{(r)}(\Gamma_r,\Gamma_r';0)W_D^{(s)}(\Gamma_s,\Gamma_s',0)R_B^{(i)}(\Gamma_r',\Gamma_s')+\notag\\
&W_D^{(r)}(\Gamma_r,\Gamma_r';-T)W_D^{(s)}(\Gamma_s,\Gamma_s',T)R_B^{(ii)}(\Gamma_r',\Gamma_s')]\notag\\
&+(s\leftrightarrow r, T\leftrightarrow -T).
\end{align}
Here $\Gamma_j'=\{t_j',\omega_j'\}$ represents the set of parameters of the matter plus field incident on the detector $j=r,s$. Eq. (\ref{eq:Rc2}) is given by the spectral and temporal overlap of the Wigner spectrograms of detectors $W_D^{(s)}$, $W_D^{(r)}$ and bare signal pathways $R_B^{(i)}$ and $R_B^{(ii)}$ given by
\begin{align}\label{eq:Rbi}
&R_B^{(i)}(t_s',\omega_s';t_r',\omega_r')=\sum_{u,u'}\sum_{v,v'}\int_{-\infty}^{\infty}d\tau_sd\tau_r e^{-i\omega'_s\tau_s-i\omega_r'\tau_r}\notag\\
&\times\langle \mathcal{T} \hat{E}_{u'R}^{\dagger}(t'_s+\tau_s,r_b)\hat{E}_{v'R}^{\dagger}(t'_r+\tau_r,r_a)\hat{E}_{vL}(t_r',r_a)\notag\\
&\times\hat{E}_{uL}(t_s',r_b)e^{-\frac{i}{\hbar}\int_{-\infty}^{\infty}\hat{H}_-'(T)dT}\rangle,
\end{align}
\begin{align}\label{eq:Rbii}
&R_B^{(ii)}(t_s',\omega_s';t_r',\omega_r')=-\sum_{u,u'}\sum_{v,v'}\int_{-\infty}^{\infty}d\tau_sd\tau_r e^{-i\omega'_s\tau_s-i\omega_r'\tau_r}\notag\\
&\times\langle \mathcal{T} \hat{E}_{u'R}^{\dagger}(t'_s+\tau_s,r_b)\hat{E}_{v'R}^{\dagger}(t'_r+\tau_r,r_a)\hat{E}_{uL}(t_s',r_a)\notag\\
&\times\hat{E}_{vL}(t_r',r_b)e^{-\frac{i}{\hbar}\int_{-\infty}^{\infty}\hat{H}_-'(T)dT}\rangle.
\end{align}
Eqs. (\ref{eq:Rbi}) -(\ref{eq:Rbii}) contain all relevant field matter interactions. The modified expressions for the gating spectrograms that include delay $T$ are given in \cite{Dorfman14c}.

\subsubsection{Time-and-frequency gated PCC}   

Under SD and HBL conditions the PCC signal  (\ref{eq:S2g}) is given by 
\begin{align}\label{eq:Rc3}
&R_c^{34}(\Gamma_r,\Gamma_s;T)=R_0C_a^r(\Gamma_r)C_b^s(\Gamma_s)\times\notag\\
&\left[1-\frac{I_a^r(\Gamma_r,\bar{t}_s,-T)I_b^s(\bar{t}_r,\Gamma_s,T)}{C_a^r(\Gamma_r)C_b^s(\Gamma_s)} \cos U(\Gamma_r,\Gamma_s;T)e^{-\tilde{\Gamma}(\bar{t}_s-\bar{t}_r)}\large\right]\notag\\
&+(a\leftrightarrow b,T\leftrightarrow -T),
\end{align}
where expressions in the last line represent permutation of the molecules $a$ and $b$, $\Gamma_j=\{\bar{t}_j,\bar{\omega}_j\}$ represents a set of gating parameters for the detector $j=r,s$. $C_\alpha(\Gamma=\{t,\omega\})$ is the time-and-frequency resolved fluorescence of molecule $\alpha=a,b$ corresponding to diagram $i$ in Fig. \ref{fig:schemeRc}c:
\begin{align}\label{eq:Ca}
C_\alpha^j(t,\omega)=C_{\alpha 0}^j(t)e^{-\frac{(\omega_p-\omega_\alpha^0-\lambda_a)^2}{2\tilde{\sigma}_{p\alpha}^2}-\frac{(\omega-\tilde{\omega}_\alpha(t))^2}{2\tilde{\sigma}_\alpha^{j2}(t)}},
\end{align}
$\omega_\alpha^0=\omega_\alpha-\lambda_\alpha$ is the mean absorption and fluorescence frequency. $I_\alpha^j(\Gamma_1,t_2,\tau)$ and $I_\alpha^j(t_1,\Gamma_2,\tau)$ with $t_1<t_2$ is the interference contribution $\alpha=a,b$, $j=r,s$ corresponding to diagram $ii$ in Fig. \ref{fig:schemeRc}c
\begin{align}\label{eq:Ia}
I_\alpha^j(\Gamma_1,t_2,\tau)&=I_{\alpha 0}^j(t_1,t_2)e^{-\frac{\omega_{ab}^2}{4\sigma_T^{j2}}-\frac{1}{4}\sigma_{\tau\alpha}^j(t_1,t_2)^2\tau^2}\notag\\
\times &e^{-\frac{(\omega_p-\omega_{p\alpha}^j(t_1,t_2))^2}{2\sigma_{p\alpha}^{j2}(t_1,t_2)}}e^{-\frac{(\omega_1-\omega_\alpha^j(t_1,t_2))^2}{2\sigma_\alpha^{j2}(t_1,t_2)}},
\end{align}
$U(\Gamma_r,\Gamma_s;\tau)=\omega_a(\bar{t}_s-\bar{t}_r)+\omega_{\tau a}^r(\bar{t}_r,\bar{t}_s,\bar{\omega}_r)\tau+(\lambda_a/\Lambda_a)(2[F_a(\bar{t}_r)-F_a(\bar{t}_s)]+F_a(\bar{t}_s-\bar{t}_r))-(a\leftrightarrow b,r\leftrightarrow s)$, $\tilde{\Gamma}(t)=\sum_{\alpha=a,b}\frac{\Delta_\alpha^2}{\Lambda_\alpha^2}F_\alpha(t)$ with $F_\alpha(t)=e^{-\Lambda_\alpha t}+\Lambda_\alpha t-1$, $\alpha=a,b$ and all the remaining parameters are listed in  Ref. \cite{Dorfman14c}. The contribution of Eq. (\ref{eq:Ca}) is represented by an amplitude square coming from each molecule in the presence of fluctuations. The interference term (\ref{eq:Ia}) generally cannot be recast as a product of two amplitudes \cite{Mukamel10a}. Eqs. (\ref{eq:Rc3}) - (\ref{eq:Ia})  are simulated below using the typical parameters of the two photon interference experiments \cite{Tre10,Let10,San12,San02,Pat10,Ate12,Coo08,Ber12,Sip12,Wol13}.

\subsubsection{Signatures of gating and spectral diffusion in the HOM dip.} 

Photon indistinguishability depends on the molecular transition frequencies. Fig. \ref{fig:PCC}e shows that for fixed time and frequency gate bandwidths $\sigma_\omega^j$ and $\sigma_T^j$, $j=r,s$ the photons are distinguishabe as long as the transition energy offset $\omega_b^0-\omega_a^0$ is larger than the gate bandwidth and are indistinguishable otherwise. The SD timescale is a key parameter affecting the degree of indistinguishability. Using Eq. (\ref{eq:Gab}) we fixed the absorption linewidth $\Gamma_\alpha$ and varied $\Lambda_\alpha$ and $\Delta_\alpha$. The PCC signal (\ref{eq:Rc3}) depicted in Fig. \ref{fig:PCC}f shows that if the molecules have nearly degenerate transition energy offset for slower fluctuations the photons are indistinguishable. Increasing the SD rate of one of the molecules increases the photon distinguishability when both time and frequency gates are broader than the difference in transition frequencies.

We further illustrate the effect of frequency and time gating on spectral diffusion. Fig. \ref{fig:PCC}g shows that if two molecules have different SD timescales and the frequency gate bandwidth is narrow the photons are rendered distinguishable and HOM dip is 0.6. By increasing the $\sigma_\omega$ the photons become indistinguishable and HOM dip is 0.48 for $\sigma_{\omega}=120$ MHz and 0.35 for $\sigma_\omega=200$ MHz. In all three cases we kept the time gate fixed. Alternatively we change the time gate bandwidth while keeping the frequency gate fixed. Fig. \ref{fig:PCC}h shows that initially indistinguishable photons at $\sigma_T=80$ MHz with HOM dip 0.675 become indistinguishable at $\sigma_T=110$ MHz with HOM dip 0.45 and at $\sigma_T=150$ MHz with HOM dip 0.275. Thus, if the presence of the bath erodes the HOM dip the manipulation of the detection gating allows to preserve the quantum interference

\section{Summary and Outlook}

The term quantum spectroscopy broadly refers to spectroscopy  techniques that make use of the quantum nature of light. Photon counting studies obviously belong to this category. Studies that detect the signal field such as heterodyne detection or fluorescence are obtained by expanding the signals in  powers of the field operators. These depend on multipoint correlation functions of the incoming fields. Classical spectroscopy is then recovered when these can be factorized into products of field amplitudes. Otherwise the technique is quantum. Spectroscopy is classical if all fields are in a coherent state and the observable is given by normally ordered product. 

Another important aspect of quantum spectroscopy, which we had only touched briefly in section~\ref{sec.nl-fluctuation-dissipation}, concerns the nonclassical fluctuations of quantum light and their exploitation as spectroscopic tools \cite{Benatti10, Giovannetti11}. These may also lead to novel features in nonlinear optical signals \cite{Lopez15}.

We now classify and briefly survey the main  features of quantum spectroscopy. First, the unusual time/frequency windows for homodyne, heterodyne and fluorescence detection arising due to the quantum nature of the light generation resulting in the enhanced resolution of the signals not accessible by classical light. Second, photon counting and interferometric detection schemes constitute a class of multidimensional signals that are based on detection and manipulation of a single photons and are parametrized by the emitted photons rather by the incoming fields. Third, the quantum nature of light manifest in collective effects in many-body systems by projecting entanglement of the field to matter. This allows to e.g. prepare  and control of higher excited states in molecular aggregates, access dark multi particle states, etc. Fourth, due to the lack of fluctuation dissipation  relations, quantum light can manifest new combinations of field and corresponding matter correlation functions not governed by semiclassical response functions such as in parametric down conversion, sum- or difference-frequency generation, two-photon-induced fluorescence, etc. Finally, elaborate pulse shaping techniques that have been recently scaled down to single photon level provide an additional tool for multidimensional measurements using delay scanning protocols not available for classical laser experiments.


The reason for the potential advantage of quantum spectroscopy was traced back to the strong time-frequency correlations inherent to quantum light, and the  back-action of the interaction events onto the quantum field's state. The combination of the two effects leads to the excitation of distinct wavepackets, which can be designed to enhance or suppress selected features of the resulting optical signals.

We have described the nonlinear signals in terms of convolutions of multi-time correlation functions of the field and the matter. This approach naturally connects to the established framework of quantum optics, where field correlation functions are analyzed \cite{gla63}, with nonlinear laser spectroscopy, which investigates the information content of matter correlation functions. As such, it provides a flexible platform to explore quantum light interaction with complex systems well beyond spectroscopic applications. This could include coherent control with quantum light \cite{Wu15, Schlawin15c}, or the manipulation of ultracold atoms with light \cite{Mekhov12}.
Entangled quantum states with higher photon numbers \cite{Shalm13a} promise access to the $\chi^{(5)}$-susceptibility and its additional information content. Combination of quantum light with strong coupling  to  intense fields  in optical cavities \cite{Hutchison2012,Schwartz2013,Herrera2014} may possibly result in a new coherent control techniques of chemical reactions.

\section{Acknowledgements}
We gratefully acknowledge the support of the National Science Foundation through Grant No. CHE- 1361516, the Chemical Sciences, Geosciences and Biosciences Division, Office of Basic Energy Sciences, Office of Science, U.S. Department of Energy. KD was supported by the DOE grant. We also thank the support and stimulating environment at FRIAS (Freiburg Institute for Advanced Studies). FS would like to thank the German National Academic Foundation for support.

\appendix

\section{Diagram construction}
\label{app:Diag}

\begin{figure}[t]
\centering
\includegraphics[width=0.25\textwidth]{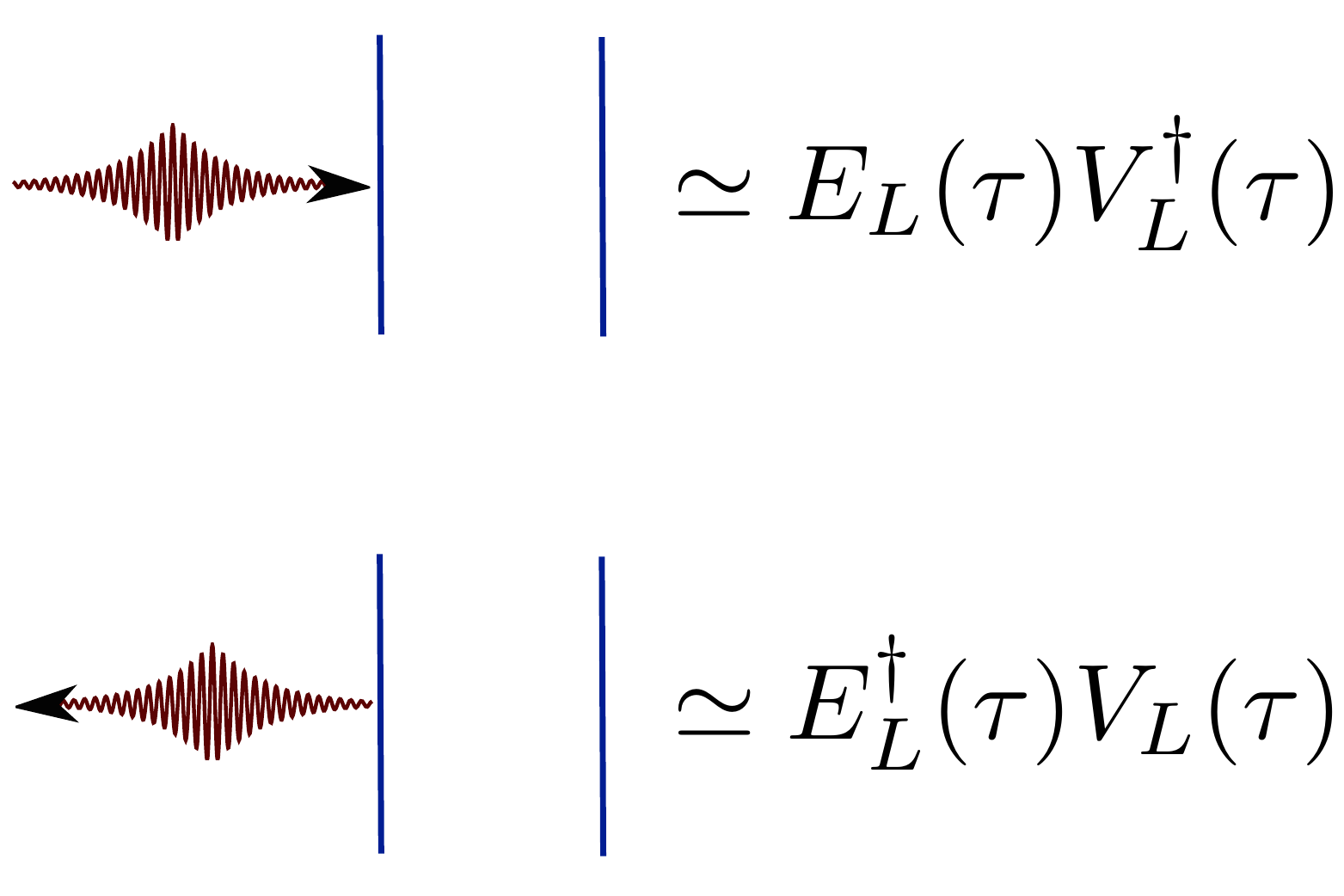}
\caption{Diagram construction in a QED formulation: The vertical blue lines indicate the evolution of the matter density matrix. An arrow pointing towards (away from) it corresponds to a matter excitation $V^{\dagger} (\tau)$ (de-excitation $V (\tau)$), which is accompanied by a photon annihilation $E (\tau)$ [creation $E^{\dagger} (\tau)$]. In the Liouville space formulation, interactions on the left side represent left-superoperators~(\ref{eq.A_L}), and interaction on the right right-superoperators~(\ref{eq.A_R}).}
\label{fig.diagram-construction}
\end{figure}

In leading (fourth-) order perturbation theory, the population of a two-excitation state $f$ is given by
\begin{align}
&p_{\text{f}} (t; \Gamma) = \left( - \frac{i}{\hbar} \right)^4  \int^t_{t_0} \!\! d\tau_4 \int^{\tau_4}_{t_0} \!\!\! d\tau_3 \int^{\tau_3}_{t_0} \!\!\! d\tau_2 \int^{\tau_2}_{t_0} \!\!\! d\tau_1 \notag\\
&\times \big\langle P_{f} (t)  H_{\text{int}, -} (\tau_4)  H_{\text{int}, -} (\tau_3) H_{\text{int}, -} (\tau_2) H_{\text{int}, -} (\tau_1) \varrho (t_0) \big\rangle, \label{eq.f-pop-time}
\end{align} 
where $P_f (t) = \vert f(t) \rangle \langle f(t) \vert$ is the projector onto the final state, the interaction Hamiltonian $H_{\text{int}}$ as given in Eq.~(\ref{eq.H_int}), and $\Gamma$ denotes the set of control parameters in the light field. 

Eq.~(\ref{eq.f-pop-time}) contains $4^4 = 256$ terms, of which only very few contribute to the signal. These contributions can be conveniently found using a diagrammatic representation. Its basic building blocks are shown in Fig.~\ref{fig.diagram-construction}: The evolution of both the bra and the ket side of the density matrix in Eq.~(\ref{eq.signal-DM}) [$\langle \psi \vert$ and $\vert \psi \rangle$ in Eq.~(\ref{eq.signal-WF})] are represented by vertical blue lines, and sample excitations (de-excitations) are represented by arrows pointing towards (away from) the density matrix.

\subsection{Loop diagrams}

\begin{figure*}[t]
\centering
\includegraphics[width=0.7\textwidth]{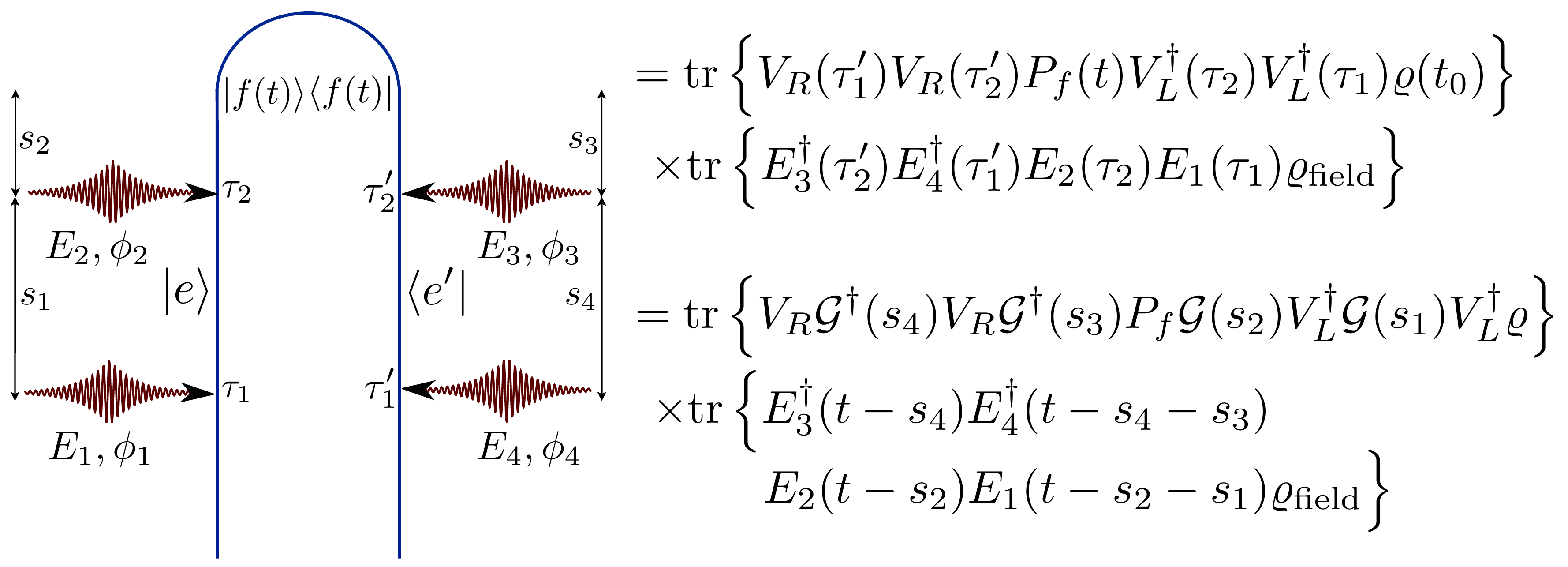}
\caption{Loop diagram construction in a QED formulation: Using the diagram rules given in the text, the diagram translates into the Heisenberg picture expression (top) or the Schr\"{o}dinger picture expression (bottom). The corresponding field correlation functions are reordered as in the loop diagram.}
\label{fig.diagram-construction_loop}
\end{figure*}

The following rules are used to construct the field and matter correlation function from the diagrams \cite{Marx:PhysRevA:08}:
\begin{enumerate}
\item Time runs along the loop clockwise from bottom left to bottom right. \label{rule_time}
\item The left branch of the loop represents the "ket", the right represents the "bra".
\item Each interaction with a field mode is represented by an arrow line on either the right (R-operators) or the left (L-operators).
\item The field is marked by dressing the lines with arrows, where an arrow pointing to the right (left) represents the field annihilation (creation) operator $E_\alpha (t)$ ($E^\dag_\alpha(t)$).
\item Within the RWA, each interaction with the field annihilates the photon $E_\alpha(t)$ and  is accompanied by applying the operator $V^\dag_\alpha(t)$, which leads to excitation of the state represented by ket and deexcitating of the state represented by the bra, respectively. Arrows pointing "inwards" (i.e. pointing to the right on the ket and to the left on the bra) consequently cause absorption of a photon by exciting the system, whereas arrows pointing "outwards" (i.e. pointing to the left on the bra and to the right on the ket) represent deexcitating the system by photon emission.
\item The observation time $t$, is fixed and is always the last. As a convention, it is chosen to occur from the left. This can always be achieved by a reflection of all interactions through the center line between the ket and the bra, which corresponds to taking the complex conjugate of the original correlation function.
\item The loop translates into an alternating product of interactions (arrows) and periods of free evolutions (vertical solid lines) along the loop.
\item Since the loop time goes clockwise along the loop, periods of free evolution on the left branch amount to propagating forward in real time with the propagator give by the retarded Green's function $G$. Whereas evolution on the right branch corresponds to backward propagation (advanced Green's function $G^\dag$).
\item The frequency arguments of the various propagators are cumulative, i.e. they are given by the sum of all "earlier" interactions along the loop. Additionally, the ground state frequency is added to all arguments of the propagators.
\item The Fourier transform of the time-domain propagators adds an additional factor of $i(-i)$ for each retarded (advanced) propagator.
\item The overall sign of the SNGF is given by $(-1)^{N_R}$, where $N_R$ stands for the number of $R$ superoperators.
\end{enumerate}

\subsection{Ladder diagrams}

\begin{figure*}[t]
\centering
\includegraphics[width=0.7\textwidth]{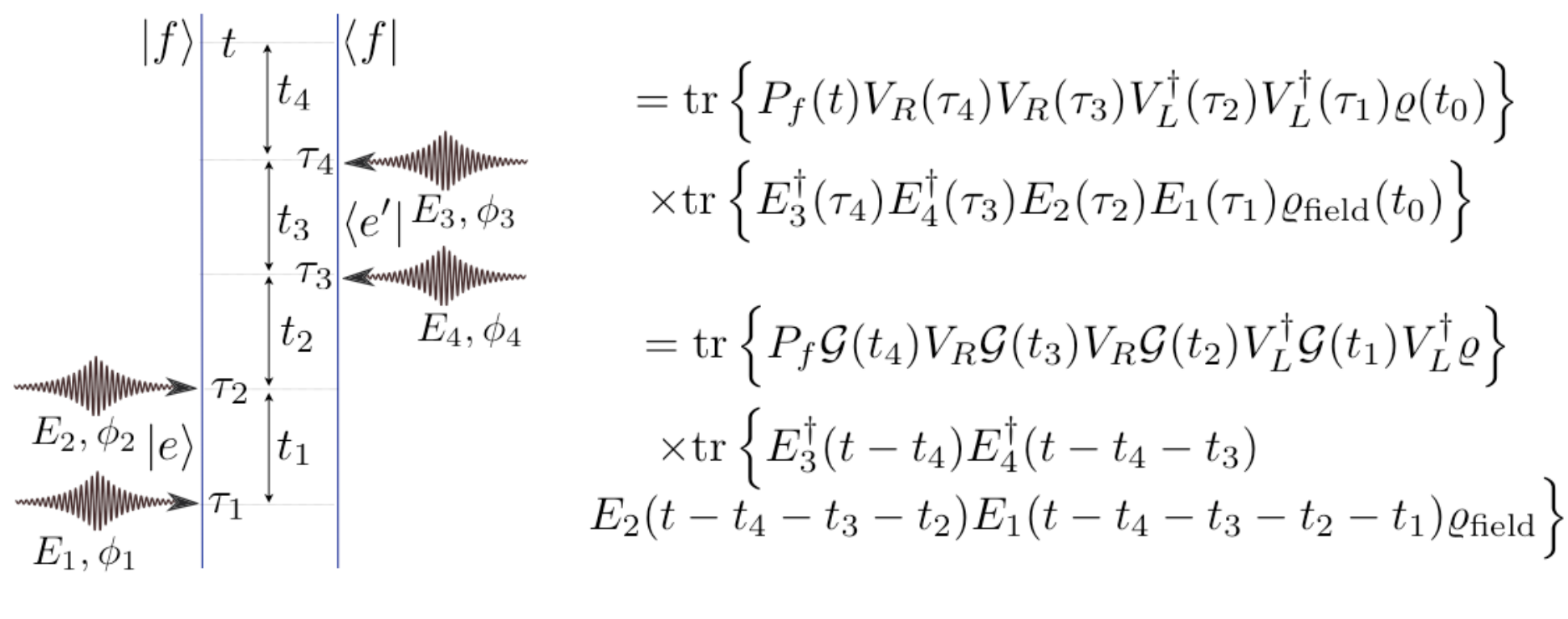}
\caption{Ladder diagram construction in a QED formulation: Using the diagram rules given in the text, the diagram translates into the Heisenberg picture expression (top) or the Schr\"{o}dinger picture expression (bottom). The corresponding field correlation functions are reordered as in the loop diagram.}
\label{fig.diagram-construction_ladder}
\end{figure*}

The same rules may be applied to evaluate ladder diagrams, with the exception of rule~\ref{rule_time}, which reads
\begin{enumerate}
\item Time runs from bottom to top.
\end{enumerate}

\section{Analytical expressions for the Schmidt decomposition}
\label{sec.analytic-decomposition}
Here, we point out how the Schmidt decomposition~(\ref{eq.decomposition}) may be carried out analytically. To this end, we approximate the phase-matching function by a Gaussian,
\begin{align}
\text{sinc} \left( \frac{\Delta k (\omega_a, \omega_b) L}{2} \right) &\approx \exp \left[  - \gamma \left( \Delta k(\omega_a, \omega_b) L \right)^2 \right],
\end{align}
where the factor $\gamma = 0.04822$ is chosen, such that the Gaussian reproduces the central peak of the sinc-function. In combination with the Gaussian pump envelope~(\ref{eq.pump-envelope}), this enables us to use the relation for the decomposition of a bipartite Gaussian \cite{URen01, URen03a}
\begin{align}
- &\frac{i \alpha}{\hbar \sqrt{2 \pi \sigma_p^2}} \exp \left[ - a x^2 - 2 b x y - c y^2 \right] \notag \\
= &\sum_{n = 0}^{\infty} r_n H_n (k_1 x) H^{\ast}_n (k_2 y), 
\end{align}
where we defined the Hermite functions $H_n$. Here, we have $a = 1 / (2 \sigma_p^2) + \gamma T_1^2$, $b = 1 / (2 \sigma_p^2) + \gamma T_1 T_2$, and $c = 1 / (2 \sigma_p^2) + \gamma T_2^2$. The Hermite functions are then given by
\begin{align}
H_n (k_i x) &= \sqrt{\frac{k}{\sqrt{\pi} 2^n n!}} e^{i \frac{3 \pi}{8} - (k_i x)^2}h_n (k_i x), \label{eq.Hermite-fct}
\end{align}
with the Hermite polynomials $h_n$, and we further defined the parameters
\begin{align}
\mu = \frac{- \sqrt{a c} + \sqrt{a c - b^2}}{b}, 
\end{align}
and
\begin{align}
r_n &= \frac{\alpha}{\hbar} \sqrt{\frac{1 + \mu^2}{4 a c \sigma_p^2}} \mu^n, \label{eq.r_n}\\
k_1 &= \sqrt{\frac{2 a (1 - \mu^2)}{1 + \mu^2}}, \\
k_2 &= \sqrt{\frac{2 c (1 - \mu^2)}{1 + \mu^2}}.
\end{align}

\section{Green's functions of  matter}
\label{sec.model-system-appendix}

In the case of the model system described in section~\ref{sec.model-system}, the superoperator correlation functions in Eqs.~(\ref{eq.p_fI})-(\ref{eq.p_fIII}) may be rewritten as sum-over-state expressions,
\begin{widetext}
\begin{align}
p_{f, \text{(I)}} (t; \Gamma) &= \left( - \frac{i}{\hbar} \right)^4 \int_0^{\infty} \!\! dt_4 \int_0^{\infty} \!\! dt_3 \int_0^{\infty} \!\! dt_2 \int_0^{\infty} \!\! dt_1 \notag \\
&\times \sum_{e, e'} \mathcal{G}_{ff} (t_4)  \mu_{e' f} \mathcal{G}_{f e'} (t_3)  \mu_{ge'} \mathcal{G}_{fg} (t_2) \mu_{ef} \mathcal{G}_{eg} (t_1) \mu_{ge} \notag \\
&\times \big\langle E^{\dagger} (t - t_4 - t_3) E^{\dagger} (t - t_4) E (t - t_4 - t_3 - t_2) E (t - t_4 - t_3 - t_2 - t_1) \big\rangle, \label{eq.p_fI_sos} \\
p_{f, \text{(II)}} (t; \Gamma) &= \left( - \frac{i}{\hbar} \right)^4 \int_0^{\infty} \!\! dt_4 \int_0^{\infty} \!\! dt_3 \int_0^{\infty} \!\! dt_2 \int_0^{\infty} \!\! dt_1 \notag \\
&\times \sum_{e, e', e''} \mathcal{G}_{ff} (t_4)  \mu_{e'' f} \mathcal{G}_{f e''} (t_3)  \mu_{e'' f } \mathcal{G}_{e e'; e'' e''} (t_2) \mu_{g e'} \mathcal{G}_{eg} (t_1) \mu_{ge} \notag \\
&\times \big\langle E^{\dagger} (t - t_4 - t_3) E^{\dagger} (t - t_4) E (t - t_4 - t_3 - t_2) E (t - t_4 - t_3 - t_2 - t_1) \big\rangle, \label{eq.p_fII_sos} \\
p_{f, \text{(III)}} (t; \Gamma) &= \left( - \frac{i}{\hbar} \right)^4 \int_0^{\infty} \!\! dt_4 \int_0^{\infty} \!\! dt_3 \int_0^{\infty} \!\! dt_2 \int_0^{\infty} \!\! dt_1 \notag \\
&\times \sum_{e, e', e''} \mathcal{G}_{ff} (t_4)  \mu_{e'' f} \mathcal{G}_{e'' f} (t_3)  \mu_{e'' f } \mathcal{G}_{e e'; e'' e'''} (t_2) \mu_{g e'} \mathcal{G}_{eg} (t_1) \mu_{ge} \notag \\
&\times \big\langle E^{\dagger} (t - t_4 - t_3) E^{\dagger} (t - t_4) E (t - t_4 - t_3 - t_2) E (t - t_4 - t_3 - t_2 - t_1) \big\rangle. \label{eq.p_fIII_sos}
\end{align}
\end{widetext}
Here, we have changed to the time delay variables $t_1, \cdots t_4$. The various propagators are given by 
\begin{align}
\mathcal{G}_{eg} (t) &= \Theta (t) \; e^{- i ( \omega_{eg} - i \gamma_{eg} ) t}, \\
\mathcal{G}_{fe} (t) &= \Theta (t) \; e^{- i ( \omega_{fe} - i \gamma_{fe} ) t}, \\
\mathcal{G}_{fg} (t) &= \Theta (t) \; e^{- i ( \omega_{fg} - i \gamma_{fg} ) t},
\end{align}
and the single-excitation propagator
\begin{align}
\mathcal{G}_{e e'; e'' e'''} (t) &= \Theta (t) \bigg( (1 - \delta_{e e'}) \delta_{e e''} \delta_{e' e'''} e^{- i (\omega_{e e'} - i \gamma_{ee'}) t} \notag \\
&+ \delta_{e e'} \delta_{e'' e'''} \exp \left[ K t \right]_{e e''} \bigg),
\end{align}
with the transport matrix
\begin{align}
K &= \frac{2 \pi}{T_0} \begin{pmatrix}
- 1 / 20 & 1  \\
1 / 20 & -1 \\
\end{pmatrix},
\end{align}
which transfers transfers the single excited state populations to the equilibrium state (with $p_{e_1} = 5 \%$ and $p_{e_2}  = 95 \%$) with the rate $2 \pi / T_0$.

In the frequency domain, the sum-over-state expressions read
\begin{widetext}
\begin{align}
p_{f, \text{(I)}} (t; \Gamma) &= \left( - \frac{i}{\hbar} \right)^4 \int \! \frac{d\omega_a}{2\pi}\int \! \frac{d\omega_b}{2\pi}\int \! \frac{d\omega'_a}{2\pi}\int \! \frac{d\omega'_b}{2\pi}\; \big\langle E^{\dagger} (\omega'_a) E^{\dagger} (\omega'_b) E (\omega_b) E (\omega_a) \big\rangle e^{i (\omega'_a + \omega'_b - \omega_b - \omega_a) t} \notag \\
&\times \sum_{e, e'} \mathcal{G}_{ff} (\omega_a + \omega_b - \omega'_a - \omega'_b)  \mu_{e' f} \mathcal{G}_{f e'} (\omega_a + \omega_b - \omega'_a)  \mu_{ge'} \mathcal{G}_{fg} (\omega_a + \omega_b) \mu_{ef} \mathcal{G}_{eg} (\omega_a) \mu_{ge}, \label{eq.p_fI_freq} \\
p_{f, \text{(II)}} (t; \Gamma) &= \left( - \frac{i}{\hbar} \right)^4 \int \! \frac{d\omega_a}{2\pi}\int \! \frac{d\omega_b}{2\pi}\int \! \frac{d\omega'_a}{2\pi}\int \! \frac{d\omega'_b}{2\pi}\; \big\langle E^{\dagger} (\omega'_a) E^{\dagger} (\omega'_b) E (\omega_b) E (\omega_a) \big\rangle e^{i (\omega'_a + \omega'_b - \omega_b - \omega_a) t} \notag \\
&\times \sum_{e, e', e''} \mathcal{G}_{ff} (\omega_a + \omega_b - \omega'_a - \omega'_b)  \mu_{e'' f} \mathcal{G}_{f e''} (\omega_a + \omega_b - \omega'_a)  \mu_{e'' f } \mathcal{G}_{e e'; e'' e''} (\omega_a - \omega'_a) \mu_{g e'} \mathcal{G}_{eg} (\omega_a) \mu_{ge}, \label{eq.p_fI_freq}  \\
p_{f, \text{(III)}} (t; \Gamma) &= \left( - \frac{i}{\hbar} \right)^4 \int \! \frac{d\omega_a}{2\pi}\int \! \frac{d\omega_b}{2\pi}\int \! \frac{d\omega'_a}{2\pi}\int \! \frac{d\omega'_b}{2\pi}\; \big\langle E^{\dagger} (\omega'_a) E^{\dagger} (\omega'_b) E (\omega_b) E (\omega_a) \big\rangle e^{i (\omega'_a + \omega'_b - \omega_b - \omega_a) t} \notag \\
&\times \sum_{e, e', e''} \mathcal{G}_{ff} (\omega_a + \omega_b - \omega'_a - \omega'_b)  \mu_{e'' f} \mathcal{G}_{e'' f} (\omega_a - \omega'_a - \omega'_b)  \mu_{e'' f } \mathcal{G}_{e e'; e'' e'''} (\omega_a - \omega'_a) \mu_{g e'} \mathcal{G}_{eg} (\omega_a) \mu_{ge}.  \label{eq.p_fIII_freq}
\end{align}
\end{widetext}

\section{Intensity measurements: TPA vs Raman}
\label{app:countingsos}

Expanding the frequency domain signal (\ref{eq.S_1I_dispersed}) - (\ref{eq.S_1IV_dispersed}) \cite{Roslyak09b} in eigenstates we obtain
\begin{widetext}
\begin{align}
S_{\text{1, (I)}} (\omega; \Gamma) &=  \frac{2}{\hbar^4} \Im \int \!\!\frac{d\omega_a}{2 \pi} \int \!\!\frac{d\omega_{\text{sum}}}{2 \pi} \big\langle E^{\dagger} (\omega_{\text{sum}}- \omega) E^{\dagger} (\omega) E (\omega_{\text{sum}} - \omega_a) E (\omega_a) \big\rangle \notag \\
&\times \sum_{e, e', f} \frac{\mu_{ge}}{\omega_a - \omega_e + i \gamma_e} \frac{\mu_{ef}}{\omega_{\text{sum}} - \omega_f + i \gamma_f} \frac{\mu_{fe'}}{\omega_{\text{sum}} - \omega - \omega_{e'} - i \gamma_{e'}} \mu_{e' g}, \label{eq.S_ppI} \\
S_{\text{1, (II)}} (\omega; \Gamma) &=  \frac{2}{\hbar^4} \Im \int \!\!\frac{d\omega_a}{2 \pi} \int \!\!\frac{d\omega_{\text{sum}}}{2 \pi} \big\langle  E^{\dagger} (\omega) E^{\dagger} (\omega_{\text{sum}}- \omega) E (\omega_{\text{sum}} - \omega_a) E (\omega_a) \big\rangle \notag \\
&\times \sum_{e, e', f} \frac{\mu_{ge}}{\omega_a - \omega_e + i \gamma_e} \frac{\mu_{ef}}{\omega_{\text{sum}} - \omega_f + i \gamma_f} \frac{\mu_{fe'}}{\omega - \omega_{e'} + i \gamma_{e'}} \mu_{e' g}. \label{eq.S_ppII}\\
S_{\text{1, (III)}} (\omega; \Gamma) &=  \frac{2}{\hbar^4} \Im \int \!\!\frac{d\omega_a}{2 \pi} \int \!\!\frac{d\omega_b}{2 \pi} \big\langle E^{\dagger} (\omega_a + \omega_b - \omega) E (\omega_b) E^{\dagger} (\omega) E (\omega_a) \big\rangle \notag \\
&\times \sum_{e, e', g'} \frac{\mu_{ge}}{\omega_a - \omega_e + i \gamma_e} \frac{\mu_{eg'}}{\omega_a - \omega - \omega_{g'} - i \gamma_g} \frac{\mu_{g' e'}}{\omega_a + \omega_b - \omega - \omega_{e'} - i \gamma_e} \mu_{e' g}, \label{eq.S_ppIII} \\
S_{\text{1, (IV)}} (\omega; \Gamma) &=  \frac{2}{\hbar^4} \Im \int \!\!\frac{d\omega_a}{2 \pi} \int \!\!\frac{d\omega_b}{2 \pi} \big\langle E^{\dagger} (\omega) E (\omega_b) E^{\dagger} (\omega_a + \omega_b - \omega) E (\omega_a) \big\rangle \notag \\
&\times \sum_{e, e', g'} \frac{\mu_{ge}}{\omega_a - \omega_e + i \gamma_e} \frac{\mu_{eg'}}{\omega - \omega_b - \omega_{g'} + i \gamma_g} \frac{\mu_{g' e'}}{\omega - \omega_{e'} + i \gamma_e} \mu_{e' g}. \label{eq.S_ppIV}
\end{align}
\end{widetext}
Using the definitions of transition operators (\ref{eq.T^1-definition}) - (\ref{eq.T'^3-definition}) we recast the signal as
\begin{align}
\int \!\! \frac{d\omega}{2 \pi} S_{\text{1, (I)}} (\omega; \Gamma) &= \Im \int \!\! \frac{d\omega_{\text{sum}}}{\pi} \sum_f \frac{\big\langle T^{(2) \dagger}_{fg} (\omega_{\text{sum}}) T^{(2) }_{fg} (\omega_{\text{sum}}) \big\rangle}{\omega_{\text{sum}} - \omega_f + i \gamma_f}, \\
\int \!\! \frac{d\omega}{2 \pi} S_{\text{1, (II)}} (\omega; \Gamma) &= \Im \int \!\! \frac{d\omega}{\pi} \sum_{e} \frac{\big\langle T^{(1) \dagger}_{eg} (\omega) T^{(3)}_{eg} (\omega) \big\rangle}{\omega - \omega_e + i \gamma_e},\\
\int \!\! \frac{d\omega}{2 \pi} S_{\text{1, (III)}} (\omega; \Gamma) &= \Im \int \!\! \frac{d\omega_-}{\pi} \sum_{g'} \frac{ \big\langle T^{(2) \dagger}_{g' g} (\omega_-) T^{(2)}_{g' g} (\omega_-) \big\rangle }{\omega_- - \omega_{g'} - i \gamma_g}, \\
\int \!\! \frac{d\omega}{2 \pi} S_{\text{1, (IV)}} (\omega; \Gamma) &=  \Im \int \!\! \frac{d\omega}{\pi} \sum_e \frac{ \big\langle T^{(1) \dagger}_{eg} (\omega) T'^{(3)}_{eg} (\omega) \big\rangle}{\omega - \omega_e + i \gamma_e}.
\end{align}

\section{Time-and-frequency gating}
\label{sec.gating}

\subsection{Simultaneous time-and-frequency gating}

To a good approximation  we can represent an ideal detector by two-level atom that is initially in the ground state $b$ and is promoted to the excited state $a$ by the absorption of a photon (see Fig. \ref{fig:gate}a). The detection of a photon brings the field from its initial state $\psi_i$ to a final state $\psi_f$. The probability amplitude for photon absorption at time $t$ can be calculated in first-order perturbation theory, which yields \cite{Glau07}
\begin{equation}
\langle\psi_f|\mathbf{E}(t)|\psi_i\rangle\cdot\langle a|\mathbf{d}|b\rangle,
\end{equation}
where $\mathbf{d}$ is the dipole moment of the atom and $\mathbf{E}(t)=E^{\dagger}(t)+E(t)$ is the electric field operator (we omit the spatial dependence). Clearly, only the annihilation part of the electric field contributes to the photon absorption process. The transition probability to find the field in state $\psi_f$ at time $t$ is given by the modulus square of the transition amplitude
\begin{align}\label{eq:Glau}
&\sum_{\psi_f}|\langle\psi_f|E(t)|\psi_i\rangle|^2=\langle\psi_i|E^{\dagger}(t)\sum_{\psi_f}|\psi_f\rangle\langle\psi_f|E(t)|\psi_i\rangle\notag\\
&=\langle\psi_i|E^{\dagger}(t)E(t)|\psi_i\rangle.
\end{align}
Since the initial state of the field $\psi_i$ is rarely known with certainty, we must trace over all possible initial states as determined by a density operator $\rho$. Thus, the output of the idealized detector is more generally given by $\text{tr}\left[\rho E^{\dagger}(t)E(t)\right]$. 

Simultaneous time-and-frequency resolved measurement must involve a frequency (spectral) gate combined with time gate - a shutter that opens up for very short interval of time. The combined detector with input located at $r_G$ is represented by a time gate $F_t$ centered at $\bar{t}$ followed by a frequency gate $F_f$ centered at $\bar{\omega}$ \cite{Sto94}. First, the time gate transforms the electric field $E(r_G,t)=\sum_s\hat{E}_s(r_G,t)$ with $\hat{E}_s(r_G,t)=E(r_G,\omega_s)e^{-i\omega_st}$ as follows:
\begin{align}
E^{(t)}(\bar{t}; r_G,t)=F_t(t,\bar{t})E(r_G,t).
\end{align}
Then, the frequency gate is applied $E^{(tf)}(\bar{t},\bar{\omega};r_G,\omega)=F_f(\omega,\bar{\omega})E^{(t)}(\bar{t};r_G,\omega)$ to obtain the time-and-frequency-gated field. We assume that the time gate is applied first. Therefore, the combined detected field at the position $r_D$ can be written as
\begin{align}\label{eq:eft}
&E^{(tf)}(\bar{t},\bar{\omega};r_D,t)=\int_{-\infty}^{\infty}dt'F_f(t-t',\bar{\omega})F_t(t',\bar{t})E(r_G,t'),
\end{align}
where $E(t)$ is the electric field operator~(\ref{eq.E-definition}) in the Heisenberg picture. Similarly, one can apply the frequency gate first  and obtain frequency-and-time-gated field $E^{(ft)}$.
 \begin{align}\label{eq:eft1}
&E^{(ft)}(\bar{t},\bar{\omega};r_D,t)=\int_{-\infty}^{\infty}dt'F_t(t,\bar{t})F_f(t-t',\bar{\omega})E(r_G,t').
\end{align}
The following discussion will be based on Eq. (\ref{eq:eft}). Eq. (\ref{eq:eft1}) can be handled similarly. 

For gaussian gates
\begin{align}
F_t(t',t)=e^{-\frac{1}{2}\sigma_T^2(t'-t)^2},\quad F_f(\omega',\omega)=e^{-\frac{(\omega'-\omega)^2}{4\sigma_\omega^2}},
\end{align}
the detector time-domain and Wigner spectrograms are given by
\begin{align}
D(t,\omega,t',\tau)=\frac{\sigma_\omega}{\sqrt{2\pi}}e^{-\frac{1}{2}\sigma_T^2(t'-t)^2-\frac{1}{2}\tilde{\sigma}_\omega^2\tau^2-[\sigma_T^2(t'-t)+i\omega]\tau}
\end{align}
\begin{align}
W_D(t,\omega;t',\omega')=N_De^{-\frac{1}{2}\tilde{\sigma}_T^2(t'-t)^2-\frac{(\omega'-\omega)^2}{2\tilde{\sigma}_\omega^2}-iA(\omega'-\omega)(t'-t)},
\end{align}
where 
\begin{align}
&\tilde{\sigma}_\omega^2=\sigma_T^2+\sigma_\omega^2,\quad \tilde{\sigma}_T^2=\sigma_T^2+\frac{1}{\sigma_\omega^{-2}+\sigma_T^{-2}}, \notag\\
&N_D=\frac{1}{\sigma^T[\sigma_\omega^2+\sigma_T^2]^{1/2}},\quad A=\frac{\sigma_T^2}{\sigma_T^2+\sigma_\omega^2}.
\end{align}
Note that  $\sigma_T$ and $\sigma_\omega$ can be controlled independently, but the actual time and frequency resolution is controlled by $\tilde{\sigma}_T$ and $\tilde{\sigma}_\omega$, respectively, which always satisfy Fourier uncertainty $\tilde{\sigma}_\omega/\tilde{\sigma}_T>1$.
For lorentzian gates
\begin{align}
F_t(t',t)=\theta(t-t')e^{-\sigma_T(t-t')},\quad F_f(\omega',\omega)=\frac{i}{\omega'-\omega+i\sigma_\omega},
\end{align}
the detector time-domain and Wigner spectrograms are given by
\begin{align}
D(t,\omega,t',\tau)=\frac{i}{2\sigma_\omega}\theta(\tau)\theta(t'-t)e^{-(i\omega+\sigma_\omega+\sigma_T)\tau-2\sigma_T(t'-t)}
\end{align}
\begin{align}
W_D(t,\omega;t',\omega')=-\frac{1}{2\sigma_\omega}\theta(t-t')\frac{e^{-2\sigma_T(t'-t)}}{\omega'-\omega+i(\sigma_T+\sigma_\omega)}.
\end{align}

The gated signal is given by
\begin{equation}\label{eq:S0}
n_{\bar{t},\bar{\omega}}=\int_{-\infty}^{\infty}dt\sum_{s,s'}\langle \hat{E}_{sR}^{(tf)\dagger}(\bar{t},\bar{\omega};r_D,t)\hat{E}_{s'L}^{(tf)}(\bar{t},\bar{\omega};r_D,t)\rangle,
\end{equation}
where the angular brackets denote $\langle ...\rangle\equiv \text{Tr}[\rho(t)...]$. The density operator $\rho(t)$ is defined in the joint field-matter space of the entire system. 
Note, that Eq. (\ref{eq:S0}) represents the observable signal, which is always positive since it can be recast as a modulus square of an amplitude (Eq. (\ref{eq:Glau})).For clarity we hereafter omit the position dependence in the fields assuming that propagation between $r_G$ and $r_D$ is included in the spectral gate function. 

\subsection{The bare signal}

The bare signal assumes infinite spectral and temporal resolution. It is unphysical but carries all necessary information for calculating photon counting measurement. It is given by the closed path time-loop diagram shown in Fig. \ref{fig:gate}\cite{Harbola08a}. We assume an arbitrary field-matter evolution starting from the matter ground state $g$  that promotes the system up to some excited state. The system then emits a photon with frequency $\omega_s$ that leaves the matter in the state $e$. This photon is later absorbed by the detector.


 In the absence of dissipation (unitary evolution)  the matter correlation function can be further factorized into a product of two amplitudes that correspond to unitary evolution of bra- and ket-. This transition amplitude can be recast in Hilbert space without using superoperators and is given by
\begin{align}
&T_{eg}(t)=-\frac{i}{\hbar}\sum_s\frac{2\pi\hbar\omega_s}{\Omega}\int_{-\infty}^tdt_1'e^{-i\omega_s(t-t_1')-i\omega_{eg}t}\notag\\
\times&\langle e(t)|V(t_1')\mathcal{T}\exp\left(-\frac{i}{\hbar}\int_{-\infty}^{t_1'}H'(T)dT\right)|g\rangle
\end{align}
This gives for the bare  Wigner spectrogram
\begin{align}\label{eq:WE11}
n(t',\omega')=&\sum_e\int_0^{\infty}d\tau e^{-i\omega'\tau}\notag\\
\times&T_{eg}(t'-\tau/2)T_{eg}^{*}(t'+\tau/2).
\end{align}

\subsection{Spectrogram-overlap representation for detected signal}
In the standard theory \cite{Sto94}, the detected signal is given by a convolution of the spectrograms of the detector and bare signal. The detector spectrogram is an ordinary function of the gating parameters whereas the bare signal is related to the field prior to detection. We now show that when the process is described in the joint matter plus field space the signal can be brought to the same form, except that now the bare signal is given by a correlation function of matter that further includes a sum over the detected modes. We denote this the spectrogram-overlap (SO) representation of the signal. Alternatively one can
introduce a  spectrogram-superoperator-overlap  (SSO) representation where field modes that interact with detector are included in the detector spectrogram, which becomes a superoperator in the field space. Details of this representation are presented in Ref. \cite{Dorfman12a}. Below we present the signals in the time domain, which can be directly read of the diagram (Fig. \ref{fig:gate}b). We then recast them using Wigner spectrograms, which depict simultaneously temporal and spectral profiles of the signal. We  now define the detector Wigner spectrogram
\begin{align}\label{eq:WD}
&W_D(\bar{t},\bar{\omega};t',\omega')=\int_{-\infty}^{\infty}d\tau\int_{-\infty}^{\infty}\frac{d\omega}{2\pi}e^{i(\omega'-\omega)\tau}\notag\\
&\times|F_f(\omega,\bar{\omega})|^2F_t^{*}(t'+\tau/2,\bar{t})F_t(t'-\tau/2,\bar{t}),
\end{align}
if the spectral gate applied first, using Eq. (\ref{eq:eft1}). The detector spectrogram is alternatively given by
\begin{align}\label{eq:Ftf}
&W_D(\bar{t},\bar{\omega};t',\omega)=\int_{-\infty}^{\infty}d\tau e^{i\omega'\tau}\int_{-\infty}^{\infty}dt\notag\\
&\times|F_t(t,\bar{t})|^2F_f^{*}(t-t'-\tau/2,\bar{\omega})F_f(t-t'+\tau/2,\bar{\omega}).
\end{align}
Combining Eqs. (\ref{eq:eft}) - (\ref{eq:WD}) we obtain that the gated signal is given by the temporal overlap of the bare signal and detector Wigner spectrogram
\begin{equation}\label{eq:S011}
n_{\bar{t},\bar{\omega}}=\int_{-\infty}^{\infty}dt'\frac{d\omega'}{2\pi}W_D(\bar{t},\bar{\omega};t',\omega')n(t',\omega').
\end{equation}

Eq. (\ref{eq:nDn}) can be alternatively recast in terms of Wigner spectrograms
\begin{align}\label{eq:nWDn}
\hat{n}_{t,\omega}=\int dt' \int\frac{d\omega'}{2\pi} W_D(t,\omega;t',\omega')\hat{n}(t',\omega'),
\end{align}
where $W_D(t,\omega,t',\omega')$ is a detector Wigner spectrogram given by
\begin{align}
W_D(t,\omega,t',\omega')=\int d\tau D(t,\omega,t',\tau)e^{i\omega'\tau}
\end{align}
and  Wigner spectrogram for the bare photon number operator is given by
\begin{align}\label{eq:nbarew}
\hat{n}(t',\omega')= \int d\tau e^{-i\omega'\tau}\hat{n}(t',\tau).
\end{align}

This is the conventional form \cite{Sto94} introduced originally for the field space alone. Eq. (\ref{eq:WE11}) contains explicitly the multiple pairs of radiation modes $s$ and $s'$ that can be revealed only in the joint field plus matter space. Eventually this takes into account all the field matter interactions that lead to the emission of the detected field modes. All parameters of $F_f$ and $F_t$ can be freely varied. The spectrogram will always satisfy the Fourier uncertainty $\Delta t\Delta\omega>1$.

Together with the gated spectrogram (\ref{eq:WD}) the bare signal  (\ref{eq:WE11}) represents the time and frequency resolved gated signal. Note, that in the presence of a bath, the signal (\ref{eq:WE11}) is no longer given by a product of two amplitudes. $\hat{T}_{eg}(t)$ is then an operator in the space of the bath degrees of freedom. Therefore, one has to replace product of amplitudes in Eq. (\ref{eq:WE11}) by $ \langle \hat{T}_{eg}(t'-\tau/2)\hat{T}_{eg}^{*}(t'+\tau/2)\rangle$, where $\langle...\rangle$ corresponds to averaging over the bath degrees of freedom. The convolution of two operators  $\hat{T}_{eg}$ reveals the multiple pathways between these initial and final states of matter that allows to observe them through the simultaneous time and frequency resolution.

We now consider two limiting cases. In the absence of a frequency gate, then $F_f(\omega,\bar{\omega})=1$  we get $W_D(\bar{\omega},\bar{t};t,\tau)=\delta(\tau)F_t^{*}(t+\tau/2,\bar{t})F_t(t-\tau/2,\bar{t})$. For the narrow time gate $|F_t(t,\bar{t})|^2=\delta(t-\bar{t})$ we then obtain the time resolved measurement
\begin{equation}\label{eq:St}
n_{\bar{t}}=\sum_e|T_{eg}(\bar{t})|^2.
\end{equation}

In the opposite limit, where there is no time gate, i.e.  $F_t(t,\bar{t})=1$, and the frequency gate is very narrow, such that $F_f(t,\bar{\omega})=\frac{\sqrt{\gamma}}{\pi} e^{-i\bar{\omega}t-\gamma t}\theta(t)$ at $\gamma\to0$, then $W_D(\bar{\omega},\bar{t};t,\tau)=e^{-i\bar{\omega}\tau}$. In this case we obtain the frequency resolved measurement
\begin{equation}\label{eq:Sw}
n_{\bar{\omega}}=\sum_e|T_{eg}(\bar{\omega})|^2,
\end{equation} 
where $T_{eg}(\omega)=\int_{-\infty}^{\infty}dte^{i\omega t}T_{eg}(t)$. Eqs. (\ref{eq:St}) and (\ref{eq:Sw}) indicate that if the measurement is either purely time or purely frequency resolved, the signal can be expressed in terms of the modulus square of a transition amplitude. Interference can then occur only within $T_{eg}$ in Hilbert space but not between the two amplitudes. Simultaneous time and frequency gating also involves interference between the two amplitudes; the pathway is in the joint ket plus bra density matrix space. In the presence of a bath, the signal can be written as a correlation function in the space of bath coordinates $\langle \hat{T}_{eg}^{*}(\bar{t})\hat{T}_{eg}(\bar{t})\rangle$ for Eq. (\ref{eq:St}) and $\langle \hat{T}_{eg}^{*}(\bar{\omega})\hat{T}_{eg}(\bar{\omega})\rangle$ for Eq. (\ref{eq:Sw}).

\subsection{Multiple detections}

The present formalism is modular and may be easily extended to any number of detection events. To that end it is more convenient to use the time domain, rather than Wigner representation. For coincidence counting of two photons measured by first detector with parameters $\bar{\omega}_i,\bar{t}_i$ followed by second detector characterized by $\bar{\omega}_s,\bar{t}_s$ the time-and-frequency resolved measurement in SO representation is given by
\begin{align}\label{}
&S(\bar{t}_s,\bar{\omega}_s;\bar{t}_i,\bar{\omega}_i)=\int_{-\infty}^{\infty}dt_s'd\tau_s\int_{-\infty}^{\infty}dt_i'd\tau_i\notag\\
&\times D^{(s)}(\bar{t}_s,\bar{\omega}_s;t_s',\tau_s)D^{(i)}(\bar{t}_i,\bar{\omega}_i;t_i',\tau_i')B(t_s',\tau_s;t_i',\tau_i')
\end{align}
where the detector spectrogram for mode $\nu=i,s$ reads
\begin{align}\label{eq:Dnu}
&D(\bar{t}_{\nu},\bar{\omega}_{\nu};t_{\nu}',\tau_{\nu})=\int_{-\infty}^{\infty}\frac{d\omega_{\nu}}{2\pi}e^{-i\omega_{\nu} \tau_{\nu}}\notag\\
&\times |F_f(\omega_{\nu},\bar{\omega}_{\nu})|^2F_t^{*}(t_{\nu}'+\tau_{\nu}/2,\bar{t}_{\nu})F_t(t_{\nu}'-\tau_{\nu}/2,\bar{t}_{\nu}).
\end{align}
The bare signal is given by the loop diagram in Fig. \ref{fig:gate}c which reads
\begin{align}\label{eq:Bm1}
&B(t_s',\tau_s;t_i',\tau_i)=\notag\\
&-\sum_eT_{eg}(t_s'-\tau_s/2,t_i'-\tau_i/2)T_{eg}^{*}(t_s'+\tau_s/2,t_i'+\tau_i/2).
\end{align}
The transition amplitude for the ket reads
\begin{align}\label{}
&T_{eg}(t_s,t_i)=\left(-\frac{i}{\hbar}\right)^2\int_{-\infty}^{t_s}dt_1'\int_{-\infty}^{t_i}dt_2'e^{-i\omega_{eg}t_s}\notag\\
\times& \langle\langle e(t_s)g|\hat{V}_L(t_1')\hat{V}_L(t_2')\notag\\
\times&\mathcal{T}\exp\left(-\frac{i}{\hbar}\int_{-\infty}^{\text{max}[t_1',t_2']}\hat{H}_L'(T)dT\right)|gg\rangle\rangle,
\end{align}
and for the bra
\begin{align}\label{}
&T_{eg}^{*}(t_s,t_i)=\left(\frac{i}{\hbar}\right)^2\int_{-\infty}^{t_s}dt_1\int_{-\infty}^{t_i}dt_2e^{i\omega_{eg}t_s}\notag\\
\times&\langle\langle gg|\hat{V}_R^{\dagger}(t_1)\hat{V}_R^{\dagger}(t_2)\notag\\
\times&\mathcal{T}\exp\left(\frac{i}{\hbar}\int_{-\infty}^{\text{max}[t_1,t_2]}\hat{H}_R'(T)dT\right)|e(t_s)g\rangle\rangle.
\end{align}


%

\end{document}